\documentclass[twocolumn,showpacs,prb,citeautoscript,amsmath,amssymb,floatfix,superscriptaddress,eqsecnum]{revtex4-1}
\usepackage{amsmath,amssymb,color}
\usepackage{bm}
\usepackage{graphicx}
\usepackage[export]{adjustbox}
\usepackage{psfrag}
\usepackage{bbold}
\usepackage{bbm}
\usepackage{xfrac}
\usepackage{tikz}
\usepackage{hyperref}
\usepackage{placeins}

\newcommand{\bd}{\bm}
\newcommand{\diff}{\text{d}}
\usepackage{ulem}

\bibpunct{[}{]}{,}{n}{}{}
\newcommand{\bubbleA}{\begin{tikzpicture}[scale=.3, baseline=-3]
	\draw [->,line width=1] (1,0) -- (1.7,0); 
	\draw [line width=1] (1,0) -- (2,0); 
	\draw [line width=1] (3,1) arc [radius=1, start angle=90, end angle=180]; 
	\draw [->,line width=1] (4,0) arc [radius=1, start angle=0, end angle=90]; 
	\draw [->,line width=1] (2,0) arc [radius=1, start angle=180, end angle=270]; 
	\draw [line width=1] (3,-1) arc [radius=1,start angle=270, end angle=360]; 
	\draw [->,line width=1] (4,0) -- (4.7,0); 
	\draw [line width=1] (4,0) -- (5,0); 
	\draw [fill] (2,0) circle [radius=.15];
	\draw [fill] (4,0) circle [radius=.15];
	\end{tikzpicture}
}
\newcommand{\bubbleB}{ \begin{tikzpicture}[scale=.3, baseline=-3]
	\draw [->,line width=1] (1,0) -- (1.7,0); 
	\draw [line width=1] (1,0) -- (2,0); 
	\draw [<-,line width=1] (3,1) arc [radius=1, start angle=90, end angle=180]; 
	\draw [line width=1] (4,0) arc [radius=1, start angle=0, end angle=90]; 
	\draw [->,line width=1] (2,0) arc [radius=1, start angle=180, end angle=270]; 
	\draw [line width=1] (3,-1) arc [radius=1, start angle=270, end angle=360]; 
	\draw [->,line width=1] (4,0) -- (4.7,0); 
	\draw [line width=1] (4,0) -- (5,0); 
	\draw [fill] (2,0) circle [radius=.15];
	\draw [fill] (4,0) circle [radius=.15];
	\end{tikzpicture}
}
\newcommand{\bubbleC}{\begin{tikzpicture}[scale=.3, baseline=-3]
	\draw [->,line width=1] (1,0) -- (1.7,0); 
	\draw [line width=1] (1,0) -- (2,0); 
	\draw [line width=1] (3,1) arc [radius=1, start angle=90, end angle=180]; 
	\draw [->,line width=1] (4,0) arc [radius=1, start angle=0, end angle=90]; 
	\draw [line width=1] (2,0) arc [radius=1, start angle=180, end angle=270]; 
	\draw [<-,line width=1] (3,-1) arc [radius=1, start angle=270, end angle=360]; 
	\draw [->,line width=1] (4,0) -- (4.7,0); 
	\draw [line width=1] (4,0) -- (5,0); 
	\draw [fill] (2,0) circle [radius=.15];
	\draw [fill] (4,0) circle [radius=.15];
	\end{tikzpicture}
}

\begin{document}

\title{Magnon damping in the zigzag phase
of the Kitaev-Heisenberg-$\Gamma$ model on a honeycomb lattice}

\author{R. L. Smit}
\affiliation{Institut f\"{u}r Theoretische Physik, Universit\"{a}t
  Frankfurt,  Max-von-Laue Strasse 1, 60438 Frankfurt, Germany}

\author{S. Keupert}
\affiliation{Institut f\"{u}r Theoretische Physik, Universit\"{a}t
  Frankfurt,  Max-von-Laue Strasse 1, 60438 Frankfurt, Germany}

\author{O. Tsyplyatyev}
\affiliation{Institut f\"{u}r Theoretische Physik, Universit\"{a}t
  Frankfurt,  Max-von-Laue Strasse 1, 60438 Frankfurt, Germany}

\author{P. A. Maksimov}
\affiliation{Department of Physics and Astronomy,
	University of California, Irvine, California 92697, USA}
\affiliation{Bogoliubov Laboratory of Theoretical Physics, Joint Institute for Nuclear Research, 141980 Dubna, Moscow region, Russia}

\author{A. L. Chernyshev}
\affiliation{Department of Physics and Astronomy,
	University of California, Irvine, California 92697, USA}

\author{P. Kopietz}
\affiliation{Institut f\"{u}r Theoretische Physik, Universit\"{a}t
  Frankfurt,  Max-von-Laue Strasse 1, 60438 Frankfurt, Germany}

\date{\today}

 \begin{abstract}

We calculate  magnon dispersions and damping in the Kitaev-Heisenberg model
with an off-diagonal exchange  $\Gamma$ and isotropic third-nearest-neighbor interaction $J_3$ on a honeycomb lattice.
This model is  relevant to a description of the magnetic properties of  iridium oxides
$\alpha$-Li$_2$IrO$_3$ and Na$_2$IrO$_3$, and Ru-based materials such as $\alpha$-RuCl$_3$. 
We use an unconventional parametrization of the 
spin-wave expansion, in which each Holstein-Primakoff boson is represented by two conjugate
hermitian operators. This approach gives us an advantage over the conventional one in 
identifying  parameter regimes where calculations can be performed analytically.
Focusing on the parameter regime with the zigzag spin pattern in the ground state that 
is consistent with experiments, we demonstrate that one such region 
is $\Gamma = K>0$, where $K$ is the Kitaev coupling. Within our approach we are able to obtain explicit  analytical 
expressions for magnon energies and eigenstates and go beyond the standard linear 
spin-wave theory approximation by calculating magnon damping   
and demonstrating its role in the dynamical structure factor. We show that the magnon 
damping effects in both Born and self-consistent approximations are very significant, underscoring the 
importance of non-linear magnon coupling in interpreting  broad features in the neutron-scattering
spectra.
\end{abstract}


\maketitle

\section{Introduction}

Magnetic materials that combine electronic correlations with strong spin-orbit coupling  
attract significant interest as a promising source of topological Mott insulators, 
exotic spin liquids, and unusual magnetically ordered states \cite{Witczak}.  
Due to the crystal field effects, strongly entangled spin and orbital degrees of freedom  
generically result in the  low-energy effective pseudo-spin models with 
bond-dependent anisotropic-exchange interactions 
\cite{Witczak,RauKee_review,Winter_review,Khal_ProgSupp,JK09,JK10}. 
In the last decade, a considerable theoretical and experimental effort has been devoted to 
bringing about a physical realization of the Kitaev spin liquid with fractionalized excitations
\cite{YBKim_review,Takagi_review,Chaloupka,Plumb,Rau14,Kimchi14,Kee,Winter18,kaib19,%
Chaloupka16,Coldea1,Perkins14,NP7,NP19,Vojta2}, 
originally proposed for the tri-coordinated honeycomb lattice with bond-dependent 
Ising-like interactions \cite{Kitaev06}. 
Other studies of the anisotropic-exchange models have revealed a multitude of unconventional 
ordered states \cite{YBKim_review,Rau14a,Rau_tr,NP5,NP6,Ioannis,Trebst_tr,multiQ}, order-by-disorder effects \cite{Baskaran,Avella,NP2}, 
and non-Kitaev spin-liquid states  \cite{Balents_review10,Starykh10,Herfurth13,topography,Tutsch19,PRX,Chen_nonKitaev} in various lattice geometries. 

Strong Kitaev-like bond-dependent couplings between  effective pseudospins-$1/2$  have been 
identified in iridium oxides, such as  $\alpha$-Li$_2$IrO$_3$ and Na$_2$IrO$_3$, $\alpha$-RuCl$_3$, and other materials
\cite{aRu_DFT,Coldea2,Coldea3,Coldea4,Coldea5,Modic,Vojta1,NP8,Mahan90}.
In these systems, magnetic ions form the two-dimensional honeycomb lattices stacked along the $[111]$-direction. 
The magnetic ions in the honeycomb layers are surrounded by an octahedral environment of ligands, which 
provide exchange  pathways facilitating direction-dependent couplings between the pseudospins. 
Importantly, a realistic modeling of these compounds necessitates significant couplings beyond the Kitaev-like ones,
such as the isotropic Heisenberg and off-diagonal exchange interactions
that are allowed by the lattice symmetry \cite{RauKee_review,Winter_review,YBKim_review,aRu_DFT}.
In the theoretical modeling and in real materials,  these couplings appear to be disruptive to the spin-liquid state 
of the pure Kitaev model in favor of the states that are magnetically ordered,  leaving 
a concrete realization of such a spin liquid state elusive as of yet \cite{Winter_review}. 

One school of thought advocates a ``proximate'' spin-liquid scenario for $\alpha$-RuCl$_3$ and
similar systems \cite{Nagler1,Nagler2}. 
In a nutshell, while the ground state of a material may be magnetically
ordered, its excitation spectrum is largely associated with a quantum-disordered spin-liquid state that is nearby 
in the phase diagram. 
This logic seemed to be strongly supported by an observation of the broad features in the 
neutron-scattering dynamical structure factor of $\alpha$-RuCl$_3$. 
At  first glance, these features are hard to reconcile with a  response of a magnetically ordered state, 
which typically yields sharp peaks  associated with magnon excitations.     
One  concern for the proximate spin liquid scenario is that it is necessarily
restricted to a close vicinity of the pure Kitaev phases,  which 
occupy a small fraction of the phase diagram  of the general anisotropic-exchange model,
according to the numerical estimates \cite{numerics1,numerics2,K1K2,Moessner16}. 

A different scenario for the broad features in the spectrum of $\alpha$-RuCl$_3$ 
has been put forward in Ref.~\cite{Winter17}, where it was suggested that the single-magnon excitations 
at  higher energies are short-lived due to strong coupling to, and decay into, the two-magnon continua
of the lower-energy magnons. This scenario was also  argued to be applicable to a vastly wider 
regions of the parameter space of the anisotropic-exchange model,---roughly speaking, to the 
entire phase diagram except  where the off-diagonal exchange terms are artificially suppressed \cite{Winter17}.  

The scenario of Ref.~\cite{Winter17} has advocated the importance of the anharmonic 
couplings in the spin-wave Hamiltonian, which in turn lead to the broad features in the magnon spectrum.
Such broadening effects are well documented,  theoretically and experimentally, in several representatives
of the ordered magnets that include some iconic frustrated magnets, such as triangular- and kagom\'{e}-lattice ones
\cite{tri09,tri06,tri_Sqw,JeGeun1,Moessner_decay,kagome1,kagome2}, 
collinear and non-collinear antiferromagnets  in external field 
\cite{square99,Japanese_5_2,square1,square2,Hong,tri_H,hex_H}, 
spin-phonon coupled systems \cite{JeGeun2}, ferromagnets \cite{FM_DM,FM_YIG}, 
and others \cite{RMP,Zh_SL,Zheludev,Zaliznyak,Kolezhuk,Plumb_ladder,YBKim_decay}.
In many of them, the non-collinearity of the ordered states, whether
due to geometric frustration or field-induced, was crucial  for the anharmonic terms to occur \cite{RMP}.
The persistence of such terms in the {\it collinear} states of the anisotropic-exchange magnets
is due to the omnipresent off-diagonal couplings that make such anharmonic terms virtually
unavoidable, regardless of the region of the phase diagram and the type of magnetic order
assumed by the ground state \cite{Winter17,McClarty18}. 

It turned out that an explicit calculation of the magnon decay rates in the zigzag phase of the 
general Kitaev-Heisenberg-$\Gamma$ model is a challenging problem. Thus,  
in Ref.~\cite{Winter17} the authors have {\it estimated} the effects of magnon broadening  in
$\alpha$-RuCl$_3$ using a simplified form of the anharmonic coupling, which 
will be referred to as the ``constant matrix element approximation'' in this work. 
In spite of this approximation, the results of Ref.~\cite{Winter17} have shown a rather remarkable 
similarity to the experimentally observed features in the neutron-scattering dynamical structure factor
of $\alpha$-RuCl$_3$  and to the numerical exact diagonalization results in small clusters. 

The present work advances the study of Ref.~\cite{Winter17}  in several directions.
We are able to find a parameter space for which  calculations of   magnon damping 
can be performed microscopically, without the simplifying approximations of Ref.~\cite{Winter17}.
For that, we use an unconventional formulation of the spin-wave theory (SWT) 
that is based on the parametrization of each 
Holstein-Primakoff boson in terms of two conjugate hermitian 
operators. The hermitian field parametrization is noteworthy in its own right as it proved to be useful for classifying different types of quantum fluctuations in certain classes of magnetically ordered systems~\cite{Hasselmann06,Kreisel07,Kreisel08,Kreisel11,Kreisel14}.
This approach gives clear criteria that allow us to identify relations between  parameters 
of the anisotropic-exchange model that permit a rigorous analytic solution for the magnon 
eigenenergies, eigenfunctions, and  matrix elements  for the calculation of the damping.
One  such  relation that defines a non-trivial line in the parameter space is
$\Gamma\! =\! K\!>\!0$, where $K$ is the Kitaev coupling and $\Gamma$ is the off-diagonal exchange. 
Although various special symmetry relations have been  previously identified in the
parameter space of the Kitaev-Heisenberg-$\Gamma$ model \cite{Chaloupka},
the  line $\Gamma\! =\! K$ along which  magnon spectrum can be
calculated analytically by solving biquadratic equations has not been noticed before.

We  focus on the regime $\Gamma\! =\! K\!>\!0$ with an additional 
third-nearest-neighbor Heisenberg interaction $J_3$, which is often invoked in the description 
of real materials \cite{aRu_DFT,Winter17}.  The $J_3$ term stabilizes the zigzag-ordered 
ground state for the considered model in a wide parameter space that includes part of the 
$\Gamma\! =\! K$ line.  This ground state  is also consistent with experiments in a broad sense, 
as it is found in several materials of interest \cite{Winter_review,YBKim_review,Chaloupka13}.
While our choice of parameters is not the same as is typically used to 
describe $\alpha$-RuCl$_3$ \cite{aRu_DFT,Winter17},
it allows us to confirm in a quantitative manner the validity of the claims that were put forward in Ref.~\cite{Winter17}. 
Specifically, it gives us an opportunity to demonstrate that 
strong anharmonicities in the magnon description  indeed persist 
throughout the phase diagram of the general anisotropic-exchange model.

We go beyond the standard linear SWT approximation by obtaining explicit
expressions for the anharmonic terms and by using them to calculate magnon damping.
The damping is calculated in the leading-order Born approximation, which inevitably contains 
van Hove singularities of the two-magnon continuum \cite{RMP}. 
To regularize them and to go beyond the Born approximation, 
we use the self-consistent approach based on the solution of the imaginary part of the Dyson's equation,
referred to as the iDE approach, see Refs.~\cite{tri09,tri_H,FM_DM}. 
For the representative values of the model parameters, the magnon damping in both Born and self-consistent iDE
approximations is significant, leading to characteristic broad features
in the dynamical structure factor.
This quantitative result of the present work confirms the assertion of Ref.~\cite{Winter17} that in the 
anisotropic-exchange model, anharmonic  interactions
can lead to  large decay rates such that some of the magnon branches 
cease to be well-defined quasiparticles.
These results underscore the importance of taking into account the nonlinear magnon coupling 
in interpreting  broad features in the neutron-scattering spectra for the general anisotropic-exchange model.  For example, the continuum of excitations far from the low energy region could potentially be described and is a good test-bed for a two-dimensional extension of the recently emerged approaches to this problem in one dimension, such as in Refs.~\cite{Imambekov09a,Imambekov09b,Imambekov12,Jin19} or in Refs.~\cite{Tsyplyatyev15,Tsyplyatyev14,Moreno16}.

In addition, having performed the decay rate calculations using explicit analytical expressions for the matrix elements of the 
magnon couplings, we are also able to verify the validity of the constant matrix element approximation of  
Ref.~\cite{Winter17}, in which the momentum dependence of such magnon vertices was neglected. 
While the momentum dependencies of the Born-approximation damping differ 
rather significantly between these approaches, the agreement becomes more quantitative within the self-consistent iDE approximation,
in agreement with the logic of Ref.~\cite{Winter17}. Still, there are clear differences near certain high-symmetry
points where magnon decays are suppressed by the symmetry requirements, or enhanced due to matrix elements.
These features are lost 
within the constant matrix element approximation of Ref.~\cite{Winter17}.
We also note that the order-of-magnitude estimates of  Ref.~\cite{Winter17} have likely provided a lower 
bound on the damping rates of magnons in $\alpha$-RuCl$_3$, 
and the actual effect of broadening for their model parameters may have been even more significant.

Lastly, while the zigzag phase within the full anisotropic-exchange model 
on the honeycomb lattice generally requires a four-sublattice description,
we have  found that the same logic that yields the reduction of the eigenvalue problem to  
solving biquadratic equations along the $\Gamma\! =\! K$ line
also allows us to reformulate the problem in the two-sublattice language. 
For that alternative formulation, we were able to derive 
a fully analytic form of the Bogoliubov eigenvalues, see Appendix~B.   
For some points along the same $\Gamma\! =\! K$ line, a conventional SWT approach can be used, 
with the details of it to be published elsewhere \cite{future}.

The rest of this paper is organized as follows. 
In Sec.~\ref{sec:model}, we introduce the model and basic notations and present the classical phase
diagram of the model in several projections.
In Sec.~\ref{sec:magzigzag}, we discuss the classical zigzag ground state and derive 
an effective interacting boson model describing fluctuations around this ground state
using the Holstein-Primakoff transformation \cite{Holstein40}. 
In Sec. \ref{sect:spectrum}, we use an unconventional parametrization of the magnon operators in terms
of hermitian operators to show  that on the special line in parameter space $\Gamma =K$ 
the magnon dispersions can be calculated analytically by solving simple biquadratic equations.
In Sec.~V, we compute the magnon damping on the special line $\Gamma =K$ in the 
Born and self-consistent iDE approximations. We also compare our results to the approximate approach of   \cite{Winter17}.
In Sec.~\ref{sec:neutron}, we calculate and plot the corresponding dynamical structure factor and the neutron scattering intensity.
In Sec.~\ref{sec:summary} we summarize our main results and present our conclusions. 
To make this work self-contained, we have added four  appendices.
In Appendix~A we review the conventional algorithm for constructing multi-flavor  
Bogoliubov transformations \cite{Colpa78,Blaizot86}. In  Appendix~B we discuss some details of the two-sublattice approach, and in Appendix~C we give additional technical details about the calculation of  magnon damping 
in the zigzag state. 
Finally, in Appendix~D we provide additional numerical results for the magnon damping and the dynamic structure factor for different parameters of the
Kitaev-Heisenberg-$\Gamma$ model.

\section{Model}
\label{sec:model}

\begin{figure}[t]
\includegraphics[width=80mm]{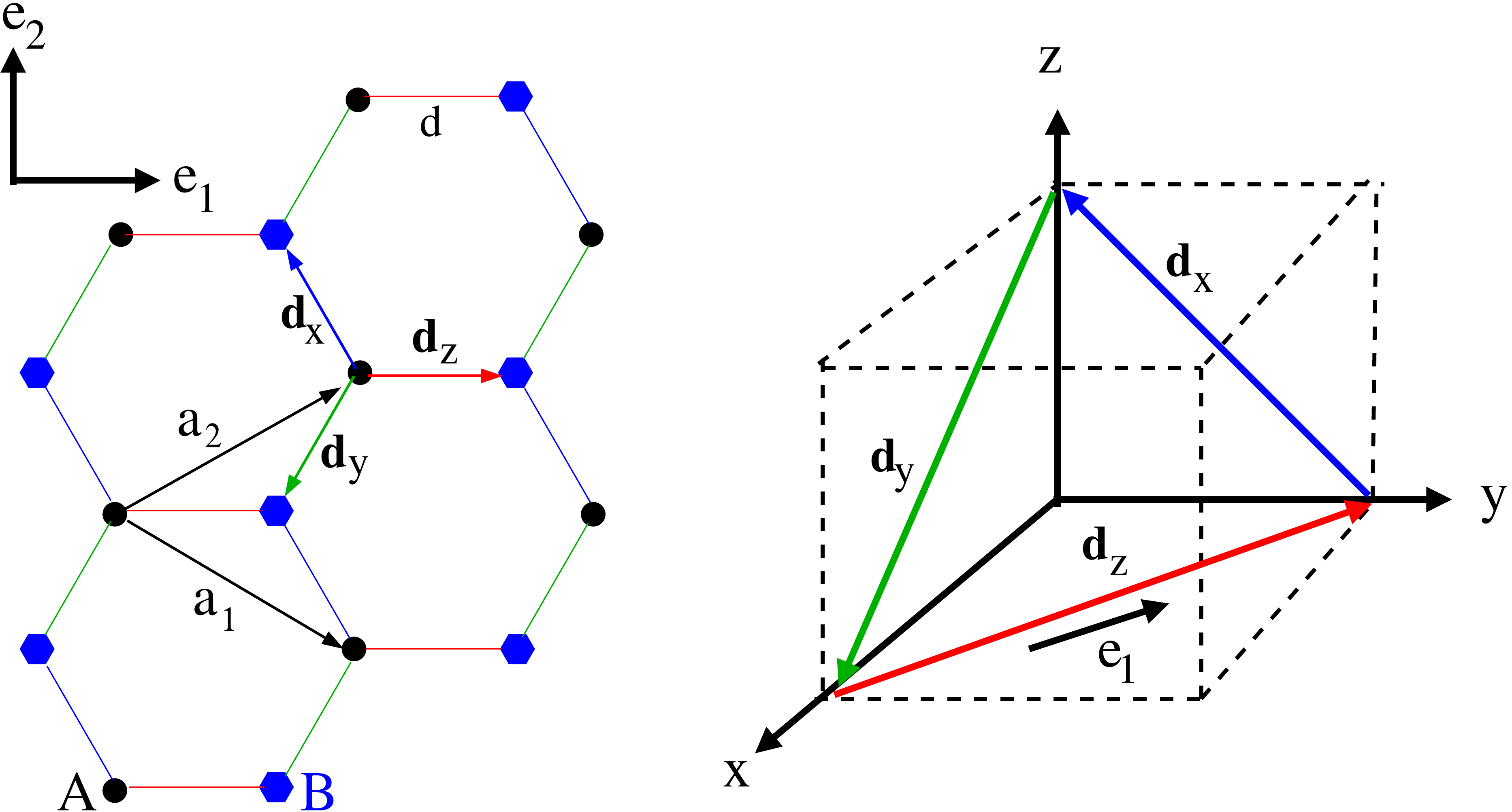}
  \caption{%
Left figure:
The honeycomb lattice can be divided into two triangular sublattices 
labeled A and B
with nearest neighbor distance $d$. 
The underlying Bravais lattice has two sites per unit cell.
We choose a basis $\{ \bd{e}_1, \bd{e}_2, \bd{e}_3 \}$ such that
$\bd{e}_1$ is parallel to the horizontal links and
$\bd{e}_3$ (which is the $[111]$-direction or the underlying cubic lattice 
and is not shown in the figure) is perpendicular to the plane of the lattice.
Nearest neighbors are connected by  
the vectors $ \bd{d}_z = d {\bd{e}}_1$,
$\bd{d}_x =  d [- \frac{1}{2} {\bd{e}}_1 +     \frac{\sqrt{3}}{2}  \hat{\bd{e}}_2    ]$ and
$\bd{d}_y =  d  [   - \frac{1}{2} {\bd{e}}_1  - \frac{\sqrt{3}}{2}  {\bd{e}}_2]$.
We use the same color coding as Ref.~[\onlinecite{Rau14}].
A possible choice for the primitive vectors is $\bd{a}_1 = \bd{d}_z - \bd{d}_x 
 = d  [ \frac{3}{2} {\bd{e}}_1 - \frac{\sqrt{3}}{2} {\bd{e}}_2 ]$ and $\bd{a}_2 =  
 \bd{d}_z - \bd{d}_y  =  d  [ \frac{3}{2} \bd{e}_1 + \frac{\sqrt{3}}{2}  {\bd{e}}_2]$.
Note that $\bd{d}_x + \bd{d}_y + \bd{d}_z =0$, $ \bd{a}_1 + \bd{a}_2 = 3 \bd{d}_z
 =  3 d \bd{e}_1$, and $ \bd{a}_2 - \bd{a}_1 = \bd{d}_x - \bd{d}_y = \sqrt{3} d \bd{e}_2$.
Right figure: In the materials of interest the honeycomb lattice lies 
in the plane perpendicular to the $[111]$-direction. 
The link vectors $\bd{d}_x$, $\bd{d}_y$ and $\bd{d}_z$ connecting 
nearest neighbors of the honeycomb lattice are parallel to
the diagonals of the faces of the cube marked by dashed lines.
We use the same color coding as in the left figure.
The labeling of the link vectors corresponds to the
spin-components in the Kitaev interaction.
}
\label{fig:honeycomb}
\end{figure}

A realistic spin model for the iridium oxides, $\alpha$-RuCl$_3$, and other materials  
containing all relevant nearest-neighbor
couplings allowed by symmetry  is given by the following
effective spin Hamiltonian \cite{Rau14,Winter_review},
\begin{eqnarray}
 {\cal{H}} & = & J \sum_{ \langle ij \rangle} \bd{S}_i \cdot \bd{S}_j
 + K \sum_{ \alpha}  \sum_{ \langle ij \rangle_\alpha } S^{\alpha}_i S^{\alpha}_j
 \nonumber
 \\
 & + & \sum_{ \alpha \beta \gamma} \Gamma^{\alpha}_{ \beta \gamma} 
 \sum_{ \langle ij \rangle_\alpha } 
S^{\beta}_i S^{\gamma}_j  - \sum_{i} \bd{h} \cdot\bd{S}_i,
 \label{eq:hamiltonian}
 \end{eqnarray}
where the $\bd{S}_i$ are (pseudo)spin  $S=1/2$ operators localized at the sites $\bd{R}_i$
of a honeycomb lattice,
$\langle i j \rangle $ enumerates all distinct pairs of the nearest-neighbor sites 
$\bd{R}_i$ and $\bd{R}_j$
of  the lattice, and the labels $\alpha , \beta , \gamma 
 \in \{x,y,z \}$ numerate the three
 link vectors
$\bd{d}_x$, $\bd{d}_y$, and $\bd{d}_z$ which connect a given lattice site to its 
nearest neighbors, as shown in  Fig.~\ref{fig:honeycomb}.
The second term in the right-hand side of Eq.~(\ref{eq:hamiltonian})
is the nearest-neighbor Kitaev interaction. In this term
 $\langle i j \rangle_\alpha$ enumerates all distinct pairs of the nearest neighbors 
whose distance vector  $\bd{R}_i - \bd{R}_j$ is parallel to $\bd{d}_{\alpha}$, and
$S_i^{\alpha}  =  \bd{e}_{\alpha} \cdot \bd{S}_i  $, $\alpha =x,y,z$ are the components 
of the spin operators in the laboratory frame, with  Cartesian basis vectors $
 \{ \bd{e}_x, \bd{e}_y, \bd{e}_z \} \equiv
\{ \hat{\bd{x}}, \hat{\bd{y}}, \hat{\bd{z}} \} $ 
shown in the right part of Fig.~\ref{fig:honeycomb}.
The third term on the right-hand side of Eq.~(\ref{eq:hamiltonian}) is the
symmetric off-diagonal exchange interaction, which arises from spin-orbit coupling 
of the underlying electronic model. The non-zero matrix elements of the 
tensor $\Gamma^{\alpha}_{\beta \gamma}$ are
 \begin{equation}
 \Gamma^x_{yz} = \Gamma^{x}_{ zy} = \Gamma^{y}_{zx} = \Gamma^{y}_{xz} 
= \Gamma^{z}_{xy} = \Gamma^{z}_{yx} = \Gamma.
 \end{equation} 
Finally, the last term in Eq.~(\ref{eq:hamiltonian}) is the 
Zeeman-interaction, where the gyromagnetic tensor is assumed to be diagonal and 
we have absorbed  the values of its 
diagonal elements into the definition of the components of dimensionless magnetic 
field $\bd{h}$.

\begin{figure}[t]
\includegraphics[width=80mm]{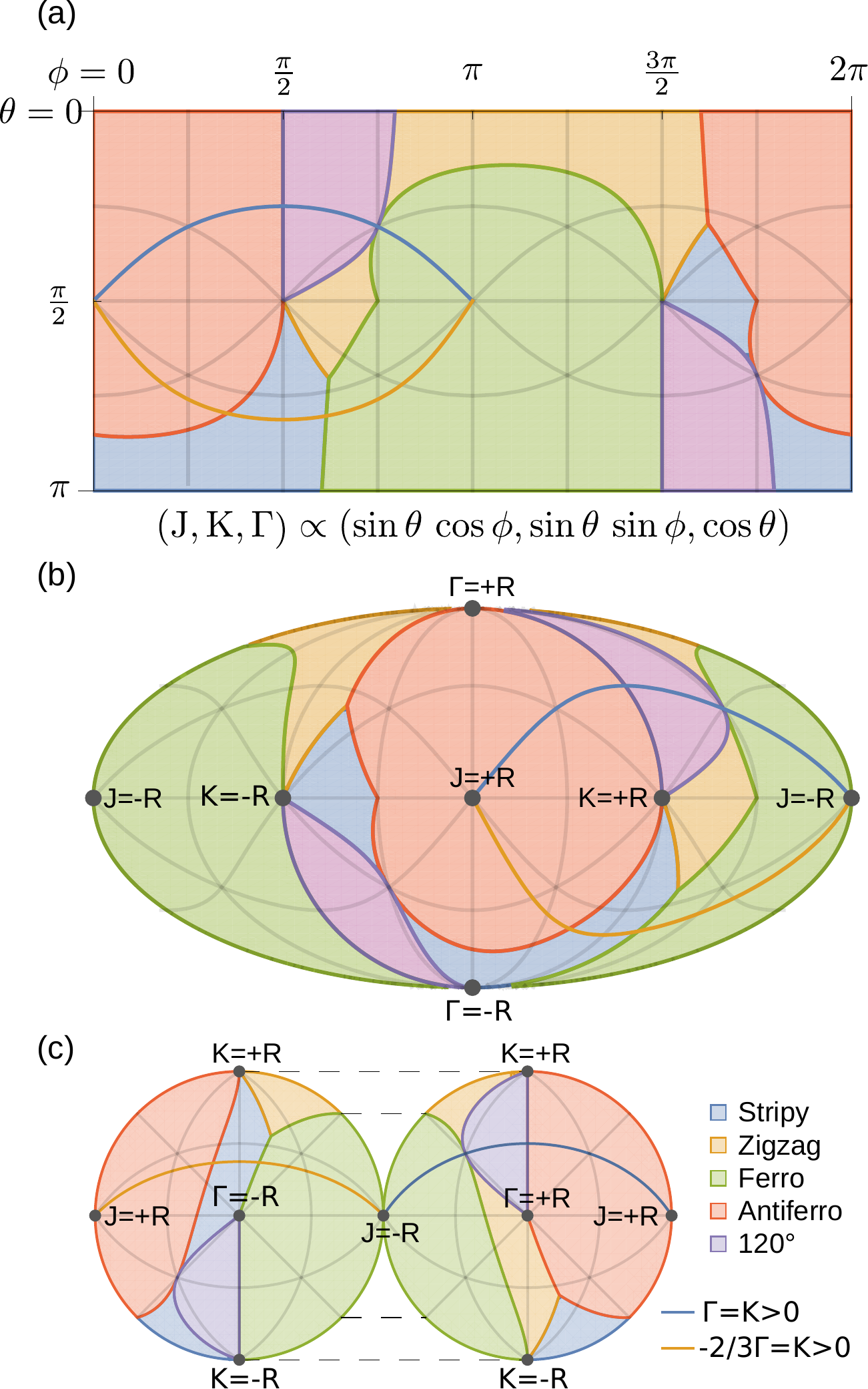}
  \caption{
Three equivalent representations of the
classical phase diagram of the
Kitaev-Heisenberg-$\Gamma$ model for vanishing external magnetic field, omitting incommensurate phases.
The three-dimensional parameter space is spanned by
$J$, $K$, and $\Gamma$. Since the phase diagram depends only on the
relative energy scales, we set
$ J/R = \sin \vartheta \cos \varphi$, $K/R = \sin \vartheta \sin \varphi$, and
$\Gamma /R = \cos \vartheta$, where $R^2 = J^2 + K^2 + \Gamma^2$.
The angles $\vartheta$ and $\varphi$ parametrize the surface of the unit sphere, 
which we project onto the plane in three different ways: (a) equirectangular projection, (b) Mollweide projection,
(c) depiction of the upper ($\Gamma>0$) and the lower ($\Gamma<0$) hemisphere.
The blue line represents the curve
$\Gamma =K$, while the orange line represents $\Gamma = - 3/2 K$ with  $K > 0$.}
\label{fig:phases}
\end{figure}

Since for generic values of the couplings the model (\ref{eq:hamiltonian}) does not have any 
continuous symmetries, it is reasonable to expect that at low temperatures 
the system will exhibit a long-range 
magnetic order, at least for large spin $S$.
In the limit $S \rightarrow \infty$, where the spin operators can be treated as
classical three-component vectors of length $S$, the possible lowest-energy 
spin configurations of the 
model  (\ref{eq:hamiltonian}) have been discussed by several authors \cite{Rau14,RauKee_review}.
Depending on the values of the parameters $J$, $K$ and $\Gamma$,
different spin configurations in the classical ground state are realized, as illustrated
in Fig.~\ref{fig:phases} using three different projections of the three-dimensional
parameter space onto a plane.
In this work we shall focus on the zigzag phase, which is
realized  in the low-temperature regime of the iridium oxides and ruthenates~\cite{Winter_review}.
In this regime, the magnetic ground state further reduces the discrete
translational symmetry of the honeycomb lattice, so that four inequivalent
sublattices are necessary to describe the discrete translational symmetry of the system.
This implies that in the zigzag phase  the spectrum of spin-wave excitations has four different branches,
which have been obtained numerically \cite{Winter17} using the algorithm developed by 
Colpa \cite{Colpa78} (see also Refs.~[\onlinecite{Blaizot86},\onlinecite{Maldonado93},\onlinecite{Serga12}]) that we summarize in Appendix~A.

\section{Magnon Hamiltonian in the zigzag state}
\label{sec:magzigzag}

\subsection{Classical ground states}

To set up the spin-wave expansion, we should first identify the spin configuration in the
classical ground state. In this limit, we treat the  
spin operators $\bd{S}_i$ as classical vectors and minimize the resulting classical Hamiltonian. 
Therefore, it is convenient to work with the coordinate representation of the spins
in the crystallographic basis, where the $\bd{S}_i$ are represented by  the column vectors
 \begin{equation}
 \left( \begin{array}{c} {\bd{e}}_x \cdot \bd{S}_i   \\ {\bd{e}}_y 
\cdot {\bd{S}}_i \\ 
 {\bd{e}}_z \cdot \bd{S}_i \end{array} \right)  = \left( \begin{array}{c} S_i^x   \\ S_i^y \\ S_i^z \end{array} \right)   ,
 \end{equation}
which we call again $\bd{S}_i$ for a notational simplicity.
Then the Kitaev-Heisenberg-$\Gamma$ Hamiltonian 
can  be written as
 \begin{eqnarray}
 {\cal{H}} & =  &   \sum_{\alpha}    \sum_{ < ij>_{\alpha} }
 \bd{S}_i^T {H}_{\alpha} 
 \bd{S}_{ j  } - \sum_i \bd{h} \cdot \bd{S}_i
 \nonumber
 \\
 & = &  \sum_{\alpha} \sum_{ \bd{R} \in A } \bd{S}_{\bd{R}}^T H_{\alpha}
 \bd{S}_{ \bd{R} + \bd{d}_{\alpha} }
 - \sum_{\bd{R}} \bd{h} \cdot \bd{S}_{\bd{R}},
 \end{eqnarray}
where in the second line the symbol  $\sum_{ {\bf{R}} \in A}$ denotes summation over all 
sites of the
$A$-sublattice (see Fig.~\ref{fig:honeycomb}) and
the  $3 \times 3$ -matrices $H_{\alpha}$ are defined by
 \begin{equation}
 H_{\alpha} =  J  \mathbb{1} + K \bd{e}_{\alpha} \bd{e}_{\alpha}^T +
 \sum_{\beta \gamma} \Gamma^{\alpha}_{\beta \gamma} \bd{e}_{\beta} 
 \bd{e}^T_{\gamma},
 \label{eq:Malphadef}
 \end{equation}
or more explicitly
\begin{subequations}
 \begin{eqnarray}
 H_x & = &  \left( \begin{array}{ ccc} J+K & 0 & 0 \\
 0 & J & \Gamma \\
 0 & \Gamma & J \end{array} \right),
 \\
H_y & = &  \left( \begin{array}{ ccc} J & 0 & \Gamma \\
 0 & J + K  & 0  \\
  \Gamma & 0 & J \end{array} \right),
 \\
H_z & = &  \left( \begin{array}{ ccc} J & \Gamma & 0 \\
 \Gamma & J & 0 \\
 0 & 0 & J + K \end{array} \right).
 \end{eqnarray}
\end{subequations}
Introducing the  site-dependent effective magnetic field
 \begin{eqnarray}
 \bd{B}_{\bd{R}}  &  = &  \bd{h} - \sum_{\alpha} H_{\alpha} 
 \bd{S}_{ \bd{R} \pm  \bd{d}_{\alpha} } ,
 \end{eqnarray}
where the  upper sign in $\bd{S}_{ \bd{R} \pm   \bd{d}_{\alpha} } $ should be taken 
for ${\bd{R}} \in A$ and the lower sign for $\bd{R} \in B$,
the conditions for the extremum of the classical energy can be written as \cite{Schuetz03}
 \begin{equation}
 \bd{S}_{\bd{R}} \times \bd{B}_{\bd{R}} =0,
 \label{eq:extremum}
 \end{equation}
which means that for each lattice site $\bd{R}$ 
the effective field $\bd{B}_{\bd{R}}$ must be aligned with $\bd{S}_{\bd{R}}$. 
To obtain an explicit analytical solution of the system (\ref{eq:extremum})  of  non-linear equations we have to make further simplifying assumptions.
Here we restrict ourselves to the spin configurations satisfying
 \begin{equation}
 \bd{S}_{\bd{R} + \bd{d}_{\alpha}} = T_{\alpha} \bd{S}_{\bd{R}  }, 
 \; \; \; \mbox{for $ \bd{R} \in A $},
 \label{eq:condi}
 \end{equation}
where  the $3 \times 3$-matrices $T_{\alpha}$ 
parametrize the relative orientation of the neighboring spins and
depend only on the displacements  $\bd{d}_{\alpha}$ connecting
the spins $\bd{S}_{\bd{R}}$ and  $\bd{S}_{\bd{R} + \bd{d}_{\alpha}}$.
This restriction does not allow for the incommensurate spiral phase, which we ignore in the following analysis 
as  it never crosses the line of our interest $\Gamma=K>0$ and thus does not interfere with our analysis, see  
supplementary notes of Ref.~[\onlinecite{Winter17}].
Renaming  $\bd{R} + \bd{d}_{\alpha} \rightarrow \bd{R}$, the condition
(\ref{eq:condi})  can alternatively be written as
  \begin{equation}
 \bd{S}_{\bd{R} - \bd{d}_{\alpha}} = T^{-1}_{\alpha} \bd{S}_{\bd{R}  }, 
 \; \; \; \mbox{for $ \bd{R} \in B $},
 \end{equation}
which is valid for all sites $\bd{R}$ belonging to the $B$-sublattice shown in
Fig.~\ref{fig:honeycomb}.
For simplicity, we shall from now on consider only the case of vanishing
external magnetic field $\bd{h} =0$. Then, for the $A$-sublattice,
Eq.~(\ref{eq:extremum})  reduces to
 \begin{equation}
 \bd{S}_{\bd{R}} = \pm  S \frac{ \sum_{\alpha} H_{\alpha} \bd{S}_{ \bd{R}  + \bd{d}_{\alpha} } }{
 | \sum_{\alpha} H_{\alpha} \bd{S}_{ \bd{R} +   \bd{d}_{\alpha} } | }
 = \pm  S \frac{ \sum_{\alpha} H_{\alpha} T_{\alpha} \bd{S}_{ \bd{R}   } }{
 | \sum_{\alpha} H_{\alpha} T_{\alpha} \bd{S}_{ \bd{R}  } | },
 \label{eq:minA}
 \end{equation}
and, for the $B$-sublattice, to
\begin{equation}
 \bd{S}_{\bd{R}} = \pm  S \frac{ \sum_{\alpha} H_{\alpha} \bd{S}_{ \bd{R}  - \bd{d}_{\alpha} } }{
 | \sum_{\alpha} H_{\alpha} \bd{S}_{ \bd{R} -   \bd{d}_{\alpha} } | }
 = \pm  S \frac{ \sum_{\alpha} H_{\alpha} T^{-1}_{\alpha} \bd{S}_{ \bd{R}   } }{
 | \sum_{\alpha} H_{\alpha} T_{\alpha}^{-1} \bd{S}_{ \bd{R}  } | }.
 \label{eq:minB}
 \end{equation}
Keeping in mind that the classical energy can be written as
 \begin{equation}
 {\cal{H}}_{0} =  \pm  S  \sum_{ \bd{R} \in A} 
\left| \sum_{\alpha} H_{\alpha} \bd{S}_{ \bd{R}  +  \bd{d}_{\alpha} } \right|
 = \pm  S  \sum_{ \bd{R} \in A} 
\left| \sum_{\alpha} H_{\alpha} T_{\alpha} \bd{S}_{ \bd{R}  } \right|,
 \end{equation}
it is clear that we should choose the minus sign in Eqs.~(\ref{eq:minA}) and
 (\ref{eq:minB}) to minimize the energy.
Note that on the $A$-sublattice the spin $\bd{S}_{\bd{R}}$ must be an eigenvector of the
matrix  $ \sum_{\alpha} H_{\alpha} T_{\alpha}$, while on the $B$-sublattice
$\bd{S}_{\bd{R}}$ must be an eigenvector of
  $ \sum_{\alpha} H_{\alpha} T^{-1}_{\alpha}$.
To construct the minimum of the energy, let 
$\lambda_{\rm max}$ be the eigenvalue of the matrix
$ \sum_{\alpha} H_{\alpha} T_{\alpha}$ with the largest absolute value.
Then the classical ground state energy can be written as
 \begin{equation}
 {\cal{H}}_{0} = - \frac{N}{2} S^2 | \lambda_{\rm max} |.
 \end{equation}
To classify  possible ground states, note that
by successively applying these transformations to the six spins
at the corners of a hexagon we obtain the holonomy condition
 \begin{equation}
 T_x^{-1} T_y T_z^{-1} T_x T_y^{-1} T_z = I,
 \end{equation}
where $I$ is the three-dimensional identity matrix.
If we require that the discrete lattice rotational symmetry should not be broken,
these conditions can be satisfied in five inequivalent ways \cite{Rau14}:
a) ferromagnetic state: $(T_x , T_y, T_z ) = ( I, I , I )$;
b) antiferromagnetic state: $(T_x , T_y, T_z ) = ( - I, - I , - I )$;
c) zigzag states: $(T_x , T_y, T_z ) = ( I, I , - I )$ or
 $( I, - I , I )$ or  $( - I,  I , I )$;
d) stripy states: 
 $(T_x , T_y, T_z ) = ( I, - I , - I )$ or
 $( - I,  I , - I )$ or  $( - I,  -  I , I )$; and
e)  $120^\circ$-state:  $(T_x , T_y, T_z ) = ( I, R_{120} , R_{120}^2 )$,
where $R_{120}$ represents a $120^\circ$-rotation around the $[111]$ direction.

\begin{figure}[t]
\includegraphics[width=80mm]{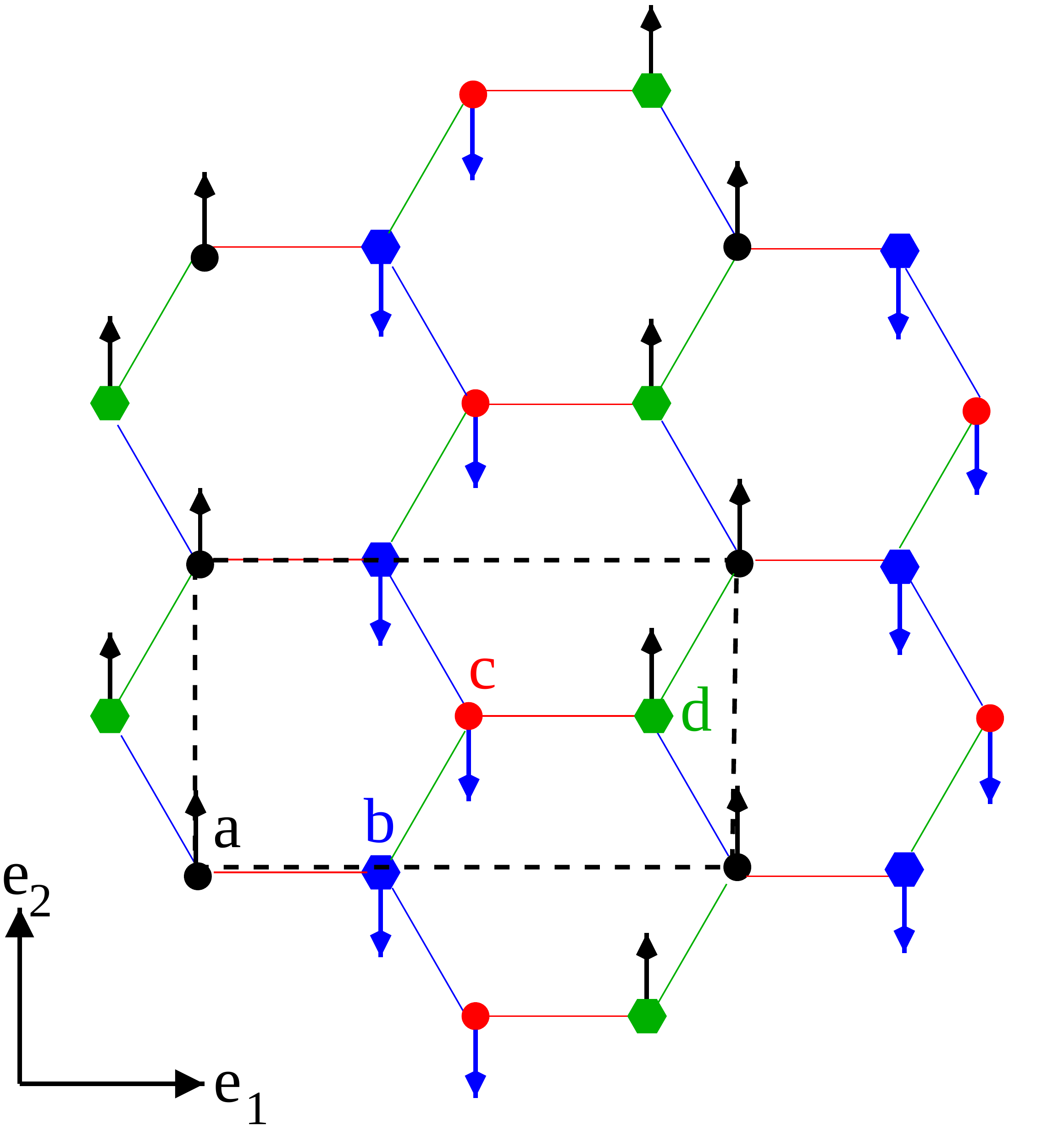}
  \caption{%
Spin configuration in the zigzag state
$\bd{m}_i = \zeta_i \bd{n}_3$ where $\zeta_i = 1$ on sublattices $a$ and $d$ and
$\zeta_i =-1$ on sublattices $b$ and $c$. It turns out that in the special case $\Gamma = K>0$, where the magnon spectrum can be calculated analytically 
as discussed in Sec.~\ref{sec:spectrum}, the local moments are
$\bd{m}_i = \zeta_i \bd{e}_2$, i.e., the magnetization lies in the plane of the
lattice and points in the direction of the stripes.
The dashed rectangle marks the choice of the unit cell of the 
lattice with a four-site basis, with the primitive vectors $\bd{a}_1^{\prime} = 3 d \bd{e}_1$ and
$\bd{a}_2^{\prime} = \sqrt{3} d \bd{e}_2$.
The associated reciprocal lattice basis is $\bd{b}^{\prime}_1 = \frac{2 \pi}{3 d } \bd{e}_1$ and
$\bd{b}^{\prime}_2 = \frac{ 2 \pi}{\sqrt{3} d } \bd{e}_2$.
}
\label{fig:zigzag}
\end{figure}

\subsection{Zigzag state}
\label{subsec:zigzag}

In the rest of this work, we shall focus on the parameter regime where the
magnetization in the classical ground state forms a zigzag pattern
with  $(T_x , T_y, T_z ) = ( \mathbb{1}, \mathbb{1} , -\mathbb{1} )$ as
illustrated in Fig.~\ref{fig:zigzag}.  
Then, the neighboring spins connected by $\bd{d}_z$ are antiparallel,
while the neigboring spins connected by $\bd{d}_x$ and $\bd{d}_y$ are parallel.
This state, which  is  realized in some iridates and  $\alpha$-RuCl$_3$,
breaks the discrete translational symmetry of the honeycomb lattice,
and requires a four-sublattice description. 
We label the sublattices by $a, b, c,$ and $d$, as shown in Fig.~\ref{fig:zigzag}.
The local moments ${\bd{S}}_i $ in the zigzag state are
 \begin{equation}
\bd{S}_i = S \bd{m}_i, \; \; \;  \bd{m}_i =  \zeta_i \bd{n}_3,
 \label{eq:localmoment}
 \end{equation}
where $\zeta_i =1$ for the sites ${\bd{R}}_i $ on the sublattices $a$ and $d$, and $\zeta_i =-1$ 
on the sublattices $b$ and $c$,
and $\bd{n}_3$ is the normalized eigenvector
of the matrix
 \begin{eqnarray}
 &  & \sum_{\alpha}  H_{\alpha} T_{\alpha} =
 H_x + H_y - H_z 
 \nonumber
 \\
 & =  & \left( \begin{array}{ ccc} J+K & - \Gamma & \Gamma \\
 - \Gamma & J +K & \Gamma \\
 \Gamma & \Gamma & J - K  \end{array} \right)
 \label{eq:coupling_matrix}
 \end{eqnarray}
whose eigenvalue $ \lambda_3$ has the largest magnitude.
The eigenvalues of the matrix~(\ref{eq:coupling_matrix})  are
 \begin{subequations}
  \begin{eqnarray}
 \lambda_1 & = & J + K + \Gamma,
 \\
 \lambda_2 & = & J - \frac{\Gamma}{2} +\frac{ R}{2},
 \\
\lambda_{3} & = & J - \frac{\Gamma}{2} - \frac{ R}{2},
 \label{eq:lambda3def}
 \end{eqnarray}
 \end{subequations}
 with
 \begin{equation}
 R = \sqrt{ 4 K^2- 4 K \Gamma + 9 \Gamma^2 } = \sqrt{  ( 2 K - \Gamma )^2 + 8 \Gamma^2 }.
 \end{equation}
The corresponding normalized eigenvectors in the crystallographic basis are
 \begin{subequations}
 \label{eq:eigenvectors}
 \begin{eqnarray}
 \bd{n}_1 & = & \frac{1}{\sqrt{2}} \left( \begin{array}{c} 1 \\ -1 \\ 0 \end{array} \right)   
 = \frac{1}{\sqrt{2}} ( \bd{e}_x  - \bd{e}_y ),
 \label{eq:e1}
 \\
  \bd{n}_2 & = &  \frac{1}{ \sqrt{ 2 +  r^2   } }
 \left( \begin{array}{c} 1 \\  1 \\  r 
 \end{array} \right)  = \frac{1}{ \sqrt{ 2 +  r^2   } }   ( \bd{e}_x  + \bd{e}_y + r \bd{e}_z ),      
 \hspace{7mm}
 \nonumber
 \\
 & & 
 \label{eq:e2}
\\
 \bd{n}_3 & = &  \frac{  {\rm sign} (s)  }{ \sqrt{ 2 + s^2 } }
 \left( \begin{array}{c} 1 \\  1 \\   s
 \end{array} \right)  = \frac{ {\rm sign} (s) }{ \sqrt{ 2 +  s^2   } }  ( \bd{e}_x  + \bd{e}_y + s \bd{e}_z ),     
 \nonumber
 \\
 & & 
 \label{eq:e3}
 \end{eqnarray}
 \label{eq:n123}
 \end{subequations}
where
\begin{equation}
 r =
\frac{ 2 K + 3 \Gamma -   R }{2 K - 3 \Gamma +  R } 
 =- \frac{  3 \Gamma + 2 K -   R }{ 3 \Gamma - 2 K -  R } ,
 \label{eq:smallrdef}
 \end{equation}
and 
 \begin{equation}
 s =  \frac{  2 K + 3 \Gamma +  R  }{ 2 K -  3 \Gamma -  R  }
 =  - \frac{   3 \Gamma +2 K +  R  }{   3 \Gamma -2 K +  R  } = - \frac{2}{r}.
 \label{eq:smallqdef}
 \end{equation}
For later reference, we note that
 \begin{equation}
 r + s = 1 - 2 K/ \Gamma,
\end{equation}
and hence
 \begin{equation}
 K + \Gamma ( r +s ) = \Gamma - K.
 \label{eq:laterref}
 \end{equation}
Recall that the local magnetization in the classical ground state is parallel to 
the eigenvector whose eigenvalue has the largest magnitude.
In the zigzag phase, this is $\bd{n}_3$; the corresponding classical
ground state energy is simply
 \begin{equation}
 {\cal{H}}_{\rm cl} = - \frac{N}{2} S^2 | \lambda_3|.
 \label{eq:Eclassical}
 \end{equation}
For $\Gamma \rightarrow 0$, we may expand
 \begin{equation}
 R = 2 | K | - \Gamma {\rm sign} K + {\cal{O}} ( \Gamma^2 ),
 \end{equation}
so that
 \begin{equation}
 s \sim \frac{ 2 K + 2 | K |  + 3 \Gamma - \Gamma {\rm sign} K + {\cal{O}} ( \Gamma^2 )    }{ 2 K - 2 | K | - 
 3 \Gamma  + \Gamma {\rm sign} K + {\cal{O}} ( \Gamma^2 ) }.
 \end{equation}
For positive $K$, this expression 
diverges as $- 2 K / \Gamma  \rightarrow  \mp \infty $ for $\Gamma \rightarrow \pm 0$. 
In this limit,  $r \rightarrow 0$ 
so that the eigenvector $\bd{n}_2$  reduces to
$( \bd{e}_x + \bd{e}_y ) / \sqrt{2}$, while the eigenvector
$\bd{n}_3$,  which gives the direction of the magnetization, approaches $\bd{e}_z$.
On the other hand, 
for $K < 0$ the parameter $s$ vanishes for $\Gamma \rightarrow 0$ while
$r$ approaches $ \mp \infty$ for $\Gamma \rightarrow \pm 0$; the local
magnetization lies then in the crystallographic $xy$-plane.
Using the relation
\begin{eqnarray}
   r s  & = &     \left( \frac{  3 \Gamma + 2 K -  R }{ 3 \Gamma -2 K -  R } \right) 
\left( \frac{   3 \Gamma + 2 K + R  }{   3 \Gamma - 2 K  + R  }\right) 
 \nonumber
 \\
 & = &  \frac{ (  3 \Gamma + 2 K )^2 - R^2 }{ (  3 \Gamma -2 K )^2 - R^2}
 =  \frac{ 16 K \Gamma}{- 8 K \Gamma} = - 2,
 \end{eqnarray}
one easily verifies that $\bd{n}_1 \times \bd{n}_2 = \bd{n}_3$,
so that $\{ \bd{n}_1 , \bd{n}_2 , \bd{n}_3 \}$ form a right-handed basis
with the third  axis $\bd{n}_3$ matching the direction of the local magnetization
in the zigzag phase.

In Sec.~\ref{sec:spectrum}, we will show that 
for $\Gamma = K > 0$ the magnon spectrum can be calculated analytically,
which will enable us to calculate the magnon damping. 
In this case, $r =1$ and $s = -2$  so that the direction of the
classical magnetization is
 \begin{equation}
 \bd{n}_3 = \frac{1}{\sqrt{6}} \left( \begin{array}{c} -1 \\ -1 \\ 2 \end{array} \right),
 \label{eq:n3gk}
 \end{equation} 
which is the coordinate representation of the vector
$\bd{e}_2$ pointing along the direction of the zigzag pattern
shown in Fig.~\ref{fig:zigzag}.  
Hence, for $\Gamma = K > 0$ the magnetic moments lie in the
plane of the honeycomb lattice and point in the direction of the zigzag pattern.
Note that $\bd{e}_2$  can be combined with 
another 
unit vector $\bd{e}_1$ in the plane of the honeycomb lattice that  
is perpendicular to the zigzag pattern, and with a third unit vector
$\bd{e}_3$ that is  perpendicular to the plane of the honeycomb lattice to form a basis
$\{ \bd{e}_1, \bd{e}_2, \bd{e}_3 \}$, which matches the geometry of the lattice. 
The relation between this honeycomb basis  and the crystallographic basis
 $\{ \bd{e}_x , \bd{e}_y, \bd{e}_z \}$ is
\begin{subequations}
 \begin{eqnarray}
 {\bd{e}}_x & = & - \frac{\bd{e}_1    }{\sqrt{2}}  -
 \frac{ \bd{e}_2  }{\sqrt{6}}  + \frac{\bd{e}_3   }{\sqrt{3}}  ,
 \\
  {\bd{e}}_y & = & \frac{\bd{e}_1    }{\sqrt{2}}  -
 \frac{ \bd{e}_2  }{\sqrt{6}}  + \frac{\bd{e}_3   }{\sqrt{3}},
 \\
 {\bd{e}}_z & = & \sqrt{\frac{2}{3}}  \bd{e}_2  + \frac{ \bd{e}_3}{\sqrt{3}} .
 \end{eqnarray}
 \end{subequations}
The inverse transformations are
 \begin{subequations}
 \label{eq:e123}
 \begin{eqnarray}
 \bd{e}_1 & = & \frac{ - {\bd{e}}_x + {\bd{e}}_y}{\sqrt{2}},
 \\
  \bd{e}_2 & = &    \frac{ - {\bd{e}}_x  -  {\bd{e}}_y   + 2 {\bd{e}}_z }{\sqrt{6} }  ,
 \label{eq:e2vec}
 \\
 \bd{e}_3 & = & \frac{{\bd{e}}_x + {\bd{e}}_y  
 + {\bd{e}}_z }{\sqrt{3}}.
 \end{eqnarray}
 \label{eq:lab_frame_01}
 \end{subequations}
From Eq.~(\ref{eq:e2vec}) it is obvious that $\bm{n}_3$ in Eq.~(\ref{eq:n3gk}) can indeed be identified with $\bm{e}_2$.

\subsection{Projection onto local reference frames}

In this subsection, we consider a general case of the zigzag state   with a finite magnetic field.
To derive the spin-wave spectrum,
we express  spin operators in terms of canonical boson operators using the
Holstein-Primakoff transformation \cite{Holstein40}. Therefore, we project the
operators $\bd{S}_i$ onto the  right-handed basis $\{ \bd{t}_{i1}, \bd{t}_{i2} , \bd{m}_i \}$ with the third direction
 \begin{equation}
 \bd{m}_i = \zeta_i \bd{n}_3
 \label{eq:loc_frame_01}
 \end{equation}
matching the direction defined by the local magnetization in the zigzag state
given in Eq.~(\ref{eq:localmoment}).
The transverse basis vectors 
$\bd{t}_{i1}$ and $\bd{t}_{i2}$ are not unique and
are defined only up to a local
$U(1)$ gauge transformation~\cite{Schuetz03,Spremo05}.
The most general choice of the transverse basis vectors
is
 \begin{subequations}
 \label{eq:loc_frame_02}
 \begin{eqnarray}
 \bd{t}_{i1} & = &     \bd{n}_1    \cos \phi  - \bd{n}_2   \sin \phi,
 \label{eq:n1def}
 \\
{\bd{t}}_{i2} &  = &    \zeta_i 
[ \bd{n}_1   \sin \phi +  \bd{n}_2  \cos \phi ],
  \label{eq:n2def}
\end{eqnarray}
\end{subequations}
where $\bd{n}_1$ and $\bd{n}_2$ are defined in
Eqs.~(\ref{eq:e1}) and (\ref{eq:e2}) and
the angle $\phi$ is arbitrary. The factor $\zeta_i$ is introduced 
such that our local basis  is right-handed.
The corresponding spherical basis vectors are
\begin{equation}
 \bd{t}^p_i = \bd{t}_{i1} + i p \bd{t}_{i2}
=  e^{i p \zeta_i \phi} ( \bd{n}_1 + i p \zeta_i \bd{n}_2 ),
 \; \; \;  p = \pm.
 \label{eq:sphericalbasis}
 \end{equation}
To derive the expansion in powers of $1/S$, we project  spin operators onto our local
basis,
 \begin{equation}
 {\bd{S}}_i =
  S^{\parallel}_i \bd{m}_i + {\bd{S}}^{\bot}_i,
 \label{eq:Sdecomp}
 \end{equation}
with the transverse part given by
  \begin{equation}
  {\bd{S}}^{\bot}_i =\frac{1}{2} \sum_{ p = \pm } S_i^{-p}
  {\bd{t}}^{p}_i.
 \end{equation} 
Then the spin components are  bosonized using the Holstein-Primakoff 
transformation \cite{Holstein40},
\begin{subequations}
\label{eq:Holst_01}
\begin{eqnarray}
S_i^+ &=&  \sqrt{2S} \sqrt{1  - \frac{a^{\dagger}_i a_i}{2S} } \; a_i 
 \approx  \sqrt{2S} \left[  a_i - \frac{a_i^\dagger a_i a_i}{4S} \right],
 \label{eq:HP1}
 \hspace{11mm}
\\
S_i^- &= & 
 \sqrt{2S} 	a_i^{\dagger} \sqrt{1  - \frac{a^{\dagger}_i a_i}{2S} } 
 \approx \sqrt{2S} \left[  a_i^\dagger - \frac{a_i^\dagger a_i^\dagger a_i}{4S}  \right],  
 \label{eq:HP2}
\\
S_i^\parallel &=& S-a_i^\dagger a_i,
\end{eqnarray}
\end{subequations}
where  $a_i$ and $a^{\dagger}_i$ are canonical boson operators satisfying the usual commutation
relations $[ a_i , a^{\dagger}_j ] = \delta_{ij}$. 
To express our Hamiltonian (\ref{eq:hamiltonian})
in terms of the Holstein-Primakoff bosons
it is convenient to write it in the form
 \begin{eqnarray}
 {\cal{H}} & = & \frac{1}{2}  \sum_{ij, \alpha} 
 \Bigl[  J^{\alpha}_{ij}
 S^{\alpha}_i S^{\alpha}_j
 + \sum_{ \beta \gamma} \Gamma^{\alpha}_{ \beta \gamma , ij}
 S^{\beta}_i S^{\gamma}_j \Bigr]
 - \sum_i \bd{h} \cdot  \bd{S}_i,
 \nonumber
 \\
 & & 
 \label{eq:hamiltoniancompact}
 \end{eqnarray}
where $J^{\alpha}_{ij} = J^{\alpha} ( \bd{R}_i - \bd{R}_j )$ and 
$\Gamma^{\alpha}_{ \beta \gamma, ij} 
=  \Gamma^{\alpha}_{ \beta \gamma} ( \bd{R}_i - \bd{R}_j )    $ are only finite if
$\bd{R}_i - \bd{R}_j$  connect nearest neighbor sites on the honeycomb lattice, with
 \begin{subequations}
 \begin{eqnarray}
  J^{\alpha} ( \bd{R}_i - \bd{R}_j = \pm \bd{d}_{\mu} ) & = & J
 +  \delta_{ \alpha \mu} K,
 \\
  \Gamma^{\alpha}_{ \beta \gamma} ( \bd{R}_i - \bd{R}_j 
   =   \pm {\bd{d}}_{\mu}  ) & = &  \delta_{\alpha \mu} \Gamma^{\alpha}_{\beta \gamma}.
 \end{eqnarray}
 \end{subequations}
Substituting the decomposition (\ref{eq:Sdecomp}) into
Eq.~(\ref{eq:hamiltoniancompact}) and setting $S^{\parallel}_i
 = S - a^{\dagger}_i a_i $, our spin Hamiltonian can be written as
 \begin{equation}
 {\cal{H}}  = {\cal{H}}_0 + {\cal{H}}_{2 \parallel}
 + {\cal{H}}_{4 \parallel} + {\cal{H}}_{\bot} 
 + {\cal{H}}_{ \parallel \bot},
 \end{equation}
with
 \begin{eqnarray}
 {\cal{H}}_0 & =  & \frac{S^2}{2} \sum_{ij, \alpha}
 \Bigl[ J_{ij}^{\alpha} m^{\alpha}_i m^{\alpha}_j 
 + \sum_{\beta \gamma} \Gamma^{\alpha}_{ \beta \gamma, ij}
 m^{\beta}_i m^{\gamma}_j \Bigr] 
 \nonumber
 \\
 &   &
 -  S \sum_i \bd{h}
 \cdot {\bd{m}}_i ,
 \label{eq:H0}
 \\
  {\cal{H}}_{2 \parallel} & = & - \frac{S}{2} 
 \sum_{ij, \alpha}
 \Bigl[ J_{ij}^{\alpha}  m_i^{\alpha} m_j^{\alpha} 
 + \sum_{\beta \gamma} \Gamma^{\alpha}_{ \beta \gamma, ij}
 m_i^{\beta} m_j^{\gamma} \Bigr] 
 \nonumber
 \\
  &  & \hspace{6mm} \times ( a^{\dagger}_i a_i + a^{\dagger}_j a_j  )    
 +  \sum_i  \bd{h} \cdot \bd{m}_i a^{\dagger}_i a_i,
 \label{eq:H2lon}
 \\
   {\cal{H}}_{\bot} & = &  \frac{1}{2} 
 \sum_{ij, \alpha}
 \Bigl[ J_{ij}^{\alpha}  (\bd{S}_i^\bot \cdot \bd{e}_{\alpha} ) 
   (\bd{S}_j^\bot \cdot \bd{e}_{\alpha} ) 
 \nonumber
 \\
 & & \hspace{9mm}
 + \sum_{\beta \gamma} \Gamma^{\alpha}_{ \beta \gamma , ij}
 (\bd{S}_i^\bot \cdot \bd{e}_{\beta} ) 
   (\bd{S}_j^\bot \cdot \bd{e}_{\gamma} )  \Bigr]  ,
  \label{eq:Hbot}
 \\
  {\cal{H}}_{ \parallel \bot} & =   &
 - \sum_{ij} \bd{S}^{\bot}_i \cdot \Biggl\{
 \delta_{ij} \bd{h}
 \nonumber
 \\
 &  & \hspace{-10mm} - \sum_{\alpha}
 \Bigl[    J^{\alpha}_{ij} \bd{e}_{\alpha} m_j^{\alpha}
 + \sum_{ \beta \gamma} \Gamma^{\alpha}_{ \beta \gamma , ij} \bd{e}_{\beta} m_j^{\gamma} 
 \Bigr]  (S - a^{\dagger}_j a_j ) \Biggr\}.
 \hspace{7mm}
 \end{eqnarray}
Within these notations, 
the condition (\ref{eq:extremum}) for a spin configuration to be in the classical ground state can be written as
 \begin{equation}
 \bd{m}_i \times \left\{ \bd{h}  - S \sum_{j, \alpha} 
 \Bigl[ J^{\alpha}_{ij} \bd{e}_{\alpha} m_j^{\alpha}
 + \sum_{\beta \gamma} \Gamma^{\alpha}_{ \beta \gamma , ij} \bd{e}_{\beta} m_j^{\gamma} 
 \Bigr] \right\} =0.
 \end{equation}
Using this condition, the part ${\cal{H}}_{\parallel \bot}$ of the Hamiltonian which mixes longitudinal
and transverse fluctuations simplifies to
 \begin{eqnarray}
  {\cal{H}}_{ \parallel \bot} & =   &
 - \sum_{ij, \alpha} \bd{S}^{\bot}_i \cdot
 \Bigl[   J^{\alpha}_{ij} \bd{e}_{\alpha} m_j^{\alpha}
 + \sum_{ \beta \gamma} \Gamma^{\alpha \beta \gamma}_{ij} \bd{e}_{\beta} m_j^{\gamma} 
 \Bigr]  a^{\dagger}_j a_j .
 \nonumber
 \\
 & &
 \label{eq:Hmix}
 \end{eqnarray}
If this term does not vanish by symmetry, it generates
cubic interactions of the Holstein-Primakoff bosons in the  leading order
in the $1/S$ expansion.

\subsection{Quadratic boson Hamiltonian}
\label{sec:quadratic}

From now on we set  $\bd{h}=0$ again. After substituting
the spin projections in the local basis $ \{ \bd{t}_{i}^+, \bd{t}_i^-, \bd{m}_i \}$
of the zigzag state into the
general formulas given in the previous subsection, 
the spin-wave dispersions in the zigzag state can be obtained
from the part ${\cal{H}}_2$ of the Hamiltonian that is quadratic in the boson operators.
For the explicit calculation of ${\cal{H}}_2$, the following
identity for the sum over a function of the nearest-neighbor sites
on the honeycomb lattice is useful,
 \begin{eqnarray}
 \sum_{\langle ij \rangle } f( \bd{R}_i , \bd{R}_j )
 & = & \frac{1}{2} \sum_{\mu = x,y,z}  \Bigl[
 \sum_{ \bd{R}_i \in A }  f( \bd{R}_i , \bd{R}_i + \bd{d}_{\mu} )
 \nonumber
 \\
 & & 
 \hspace{12mm} + \sum_{\bd{R}_i \in B   }  f( \bd{R}_i , \bd{R}_i - \bd{d}_{\mu} )
 \Bigr]
 \nonumber
 \\
 & = & 
\sum_{\mu = x,y,z} 
 \sum_{ \bd{R}_i \in A }  f( \bd{R}_i , \bd{R}_i + \bd{d}_{\mu} ),
 \end{eqnarray}
where $\bd{R}_i \in A  $ means all sites of the sublattice $A = a \cup c$
and  $\bd{R}_i \in B  $ means all sites of the sublattice $B = b \cup d$.
The quadratic contribution ${\cal{H}}_{ 2 \parallel}$ to the longitudinal part of the 
bosonized Hamiltonian defined in Eq.~(\ref{eq:H2lon}) is easily obtained, 
 \begin{equation}
 {\cal{H}}_{2 \parallel} = - F S \sum_{\bd{R}} a^{\dagger}_{\bd{R}} a_{\bd{R}},
 \label{eq:H2lonres}
 \end{equation}
with
 \begin{equation}
 F = J + K ( m_x^2 + m_y^2 - m_z^2) + 2 \Gamma ( m_y m_z + m_z m_x - m_x m_y ).
 \end{equation}
The calculation of the corresponding transverse part ${\cal{H}}_{ 2 \bot}$ is more involved.
For simplicity, we use the gauge $\phi =0$ for  the transverse basis where 
the transverse spherical basis vectors are simply $\bd{t}_i^p = \bd{n}_1 + i p \zeta_i  \bd{n}_2$.
In the zigzag state,
the expansion of the transverse part of the spin operators is 
\begin{subequations}
 \begin{eqnarray}
 \bd{S}^{\bot}_i & = &  \frac{1}{2} \sum_{ p  } S_i^{-p}
  {\bd{n}}^{p}, \;  \; \;  \mbox{for $\bd{R}_i \in a \cup d $},
 \\
 & = &  \frac{1}{2} \sum_{ p } S_i^{p}
  {\bd{n}}^{p}, \;  \; \;   \mbox{for $ \bd{R}_i \in b \cup c$},
 \end{eqnarray}
\end{subequations}
where we  introduced the site-independent spherical basis vectors
 \begin{equation}
 \bd{n}^p = \bd{n}_{1} + i p \bd{n}_{2} .
 \label{eq:npdef}
 \end{equation}
Decomposing the transverse part ${\cal{H}}_{\bot}$
of the spin Hamiltonian defined in Eq.~(\ref{eq:Hbot})
into  contributions from the three 
types of  interactions,
 \begin{equation}
 {\cal{H}}_{  \bot} = {\cal{H}}^J_{  \bot} + {\cal{H}}^K_{  \bot} + {\cal{H}}^\Gamma_{  \bot},
 \end{equation}
we then obtain for the Heisenberg part ($J$-term),
\begin{eqnarray}
{\cal{H}}_{ \bot}^J  
 & = &  \frac{J}{2} \sum_p
\sum_{\bd{R} \in A  }  \left[ 
 S^p_{\bd{R}} S^{-p}_{\bd{R} + \bd{d}_x}
 + S^p_{\bd{R}} S^{-p}_{\bd{R} + \bd{d}_y}
 + S^p_{\bd{R}} S^{p}_{\bd{R} + \bd{d}_z} \right].
 \nonumber
 \\
 & &
 \end{eqnarray}
To explicitly write down the 
 transverse contribution of the Kitaev term in the zigzag state, we 
separate  the  contributions from the four sublattices, 
 \begin{eqnarray}
 & & {\cal{H}}_{ \bot}^{K}  =  \frac{1}{8} \sum_{ p p^{\prime}} 
 \Biggl\{
 \nonumber
 \\
 & & \sum_{ \bd{R} \in a} 
 \left[ K_{xx}^{ \bar{p} \bar{p}^{\prime}}  
 S^p_{\bd{R}} S^{p^{\prime}}_{\bd{R} + \bd{d}_x}
 +  K_{yy}^{ \bar{p} \bar{p}^{\prime}} 
 S^p_{\bd{R}} S^{p^{\prime}}_{\bd{R} + \bd{d}_y}
 + K_{zz}^{ \bar{p} {p}^{\prime}} 
 S^p_{\bd{R}} S^{p^{\prime}}_{\bd{R} + \bd{d}_z} \right]
 \nonumber
 \\
  &+ & \sum_{ \bd{R} \in c } 
 \left[ K_{xx}^{ p p^{\prime}}  
 S^p_{\bd{R}} S^{p^{\prime}}_{\bd{R} + \bd{d}_x}
 +  K_{yy}^{ p p^{\prime}} 
 S^p_{\bd{R}} S^{p^{\prime}}_{\bd{R} + \bd{d}_y}
 + K_{zz}^{ {p} \bar{p}^{\prime}} 
 S^p_{\bd{R}} S^{p^{\prime}}_{\bd{R} + \bd{d}_z} \right]
 \nonumber
 \\
 & + &
\sum_{ \bd{R} \in b } 
 \left[ K_{xx}^{ p p^{\prime}}  
 S^p_{\bd{R}} S^{p^{\prime}}_{\bd{R} - \bd{d}_x}
 +  K_{yy}^{ p p^{\prime}} 
 S^p_{\bd{R}} S^{p^{\prime}}_{\bd{R} - \bd{d}_y}
 + K_{zz}^{ {p} \bar{p}^{\prime}} 
 S^p_{\bd{R}} S^{p^{\prime}}_{\bd{R} - \bd{d}_z} \right]
 \nonumber
 \\
 & + &
\sum_{ \bd{R} \in d } 
 \left[ K_{xx}^{ \bar{p} \bar{p}^{\prime}}  
 S^p_{\bd{R}} S^{p^{\prime}}_{\bd{R} - \bd{d}_x}
 +  K_{yy}^{ \bar{p} \bar{p}^{\prime}} 
 S^p_{\bd{R}} S^{p^{\prime}}_{\bd{R} - \bd{d}_y}
 + K_{zz}^{ \bar{p} {p}^{\prime}} 
 S^p_{\bd{R}} S^{p^{\prime}}_{\bd{R} - \bd{d}_z} \right]
 \Biggr\},
 \nonumber
 \\
 & &
 \label{eq:HKreal}
 \end{eqnarray}
where we have defined
 \begin{equation}
 K_{\alpha \beta}^{p p^{\prime}} = K (   \bd{e}_{\alpha}   \cdot \bd{n}^p )
 ( \bd{e}_{\beta} \cdot    \bd{n}^{p^{\prime}}  ).
 \label{eq:Kppdef}
 \end{equation}
and the superscripts $\bar{p}$ and $\bar{p}^{\prime}$ stand for $-p$ and $- p^{\prime}$.
Similarly, the contribution from the off-diagonal exchange term
to the transverse part of the  spin Hamiltonian can be written as
 \begin{eqnarray}
 & & {\cal{H}}_{ \bot}^{\Gamma}  =  \frac{1}{8} \sum_{ p p^{\prime}} 
 \Biggl\{
 \nonumber
 \\
 & & \sum_{ \bd{R} \in a } 
 \left[ \Gamma_{yz}^{ \bar{p} \bar{p}^{\prime}}  
 S^p_{\bd{R}} S^{p^{\prime}}_{\bd{R} + \bd{d}_x}
 +  \Gamma_{zx}^{ \bar{p} \bar{p}^{\prime}} 
 S^p_{\bd{R}} S^{p^{\prime}}_{\bd{R} + \bd{d}_y}
 + \Gamma_{xy}^{ \bar{p} {p}^{\prime}} 
 S^p_{\bd{R}} S^{p^{\prime}}_{\bd{R} + \bd{d}_z} \right]
 \nonumber
 \\
  &+ & \sum_{ \bd{R} \in c } 
 \left[ \Gamma_{yz}^{ p p^{\prime}}  
 S^p_{\bd{R}} S^{p^{\prime}}_{\bd{R} + \bd{d}_x}
 +  \Gamma_{zx}^{ p p^{\prime}} 
 S^p_{\bd{R}} S^{p^{\prime}}_{\bd{R} + \bd{d}_y}
 + \Gamma_{xy}^{ {p} \bar{p}^{\prime}} 
 S^p_{\bd{R}} S^{p^{\prime}}_{\bd{R} + \bd{d}_z} \right]
 \nonumber
 \\
 & + &
\sum_{ \bd{R} \in b } 
 \left[ \Gamma_{yz}^{ p p^{\prime}}  
 S^p_{\bd{R}} S^{p^{\prime}}_{\bd{R} - \bd{d}_x}
 +  \Gamma_{zx}^{ p p^{\prime}} 
 S^p_{\bd{R}} S^{p^{\prime}}_{\bd{R} - \bd{d}_y}
 + \Gamma_{xy}^{ {p} \bar{p}^{\prime}} 
 S^p_{\bd{R}} S^{p^{\prime}}_{\bd{R} - \bd{d}_z} \right]
 \nonumber
 \\
 & + &
\sum_{ \bd{R} \in d } 
 \left[ \Gamma_{yz}^{ \bar{p} \bar{p}^{\prime}}  
 S^p_{\bd{R}} S^{p^{\prime}}_{\bd{R} - \bd{d}_x}
 +  \Gamma_{zx}^{ \bar{p} \bar{p}^{\prime}} 
 S^p_{\bd{R}} S^{p^{\prime}}_{\bd{R} - \bd{d}_y}
 + \Gamma_{xy}^{ \bar{p} {p}^{\prime}} 
 S^p_{\bd{R}} S^{p^{\prime}}_{\bd{R} - \bd{d}_z} \right]
 \Biggr\},
 \nonumber
 \\
 & &
 \label{eq:HGreal}
 \end{eqnarray}
where we have defined
 \begin{equation}
 \Gamma_{\alpha \beta}^{p p^{\prime}} = \Gamma [
  (  \bd{e}_{\alpha}  \cdot  \bd{n}^p  )
 (  \bd{e}_{\beta}  \cdot   \bd{n}^{p^{\prime}}  )  
 + ( \bd{e}_{\beta} \cdot \bd{n}^p  )
 ( \bd{e}_{\alpha}  \cdot  \bd{n}^{p^{\prime}}     )   ].
 \label{eq:Gammappdef}
 \end{equation}
To obtain the corresponding Hamiltonian 
$ {\cal{H}}_{2 \bot} = {\cal{H}}^J_{2 \bot} +  {\cal{H}}^K_{2 \bot}   + {\cal{H}}^{\Gamma}_{ 2 \bot}$,
which is quadratic in the boson operators, we approximate
the spherical components of the spin operators by the leading terms in the Holstein-Primakoff
transformation, $S^+_i \approx \sqrt{2S} a_i$ and $S^-_i \approx \sqrt{2S} a^{\dagger}_i$, 
see Eqs.~(\ref{eq:HP1}) and (\ref{eq:HP2}). 
Then the resulting quadratic  boson Hamiltonian can  be block-diagonalized
by transforming to the momentum space on each of the
four sublattices separately,
\begin{subequations}
	\label{eq:FT_01}
 \begin{eqnarray}
 a_{\bd{R}} & = & \sqrt{ \frac{4}{N}} {\sum_{\bd{k} }} 
 e^{ i \bd{k} \cdot \bd{R}} a_{\bd{k}} , \; \; \; \bd{R} \in a,
 \\
 & = &\sqrt{ \frac{4}{N}} {\sum_{\bd{k} }}
 e^{ i \bd{k} \cdot \bd{R}} b_{\bd{k}} , \; \; \; \bd{R} \in b,
 \\
 & = & \sqrt{ \frac{4}{N}} {\sum_{\bd{k} }} 
 e^{ i \bd{k} \cdot \bd{R}} c_{\bd{k}} , \; \; \; \bd{R} \in c,
 \\
 & = &\sqrt{ \frac{4}{N}} {\sum_{\bd{k} }} 
 e^{ i \bd{k} \cdot \bd{R}} d_{\bd{k}} , \; \; \; \bd{R} \in d,
 \end{eqnarray}
 \end{subequations}
where the momentum sums are over the reduced (magnetic) Brillouin zone associated with
one of the four sublattices containing $N/4$ lattice sites.
Note that the coordinates of the sites of different sublattices can be transformed into each other
by shifting by a vector that is not a primitive vector of the Bravais lattice, 
 \begin{subequations}
 \begin{eqnarray}
 \bd{R}_b & = & \bd{R}_a + \bd{d}_z,
 \\
  \bd{R}_c & = & \bd{R}_a + \bd{d}_z - \bd{d}_x =  \bd{R}_a + \bd{a}_1,
 \\
 \bd{R}_d & = & \bd{R}_a + \bd{d}_x,
 \end{eqnarray}
 \end{subequations}
where the subscripts indicate the sublattice and the shift vectors $\bd{d}_x$,
$\bd{d}_z$ and $\bd{a}_1$ are defined in the caption
of Fig.~\ref{fig:honeycomb}.
As a consequence, we should distinguish  four types of periodic $\delta$-functions,
 \begin{subequations}
 \label{eq:deltaphase}
 \begin{eqnarray}
  \delta_a (\bd{k} ) & = &  \frac{4}{N} \sum_{\bd{R}_a } e^{ i \bd{k} \cdot \bd{R}_a }  =
 \sum_{ \bd{G}} \delta_{ \bd{k} , \bd{G}},
 \\
 \delta_b (\bd{k} ) & = &  \frac{4}{N} \sum_{\bd{R}_b } e^{ i \bd{k} \cdot \bd{R}_b }  =
 \sum_{ \bd{G}} \delta_{ \bd{k} , \bd{G}} e^{ i \bd{G} \cdot \bd{d}_z },
 \\
 \delta_c (\bd{k} ) & = &  \frac{4}{N} \sum_{\bd{R}_c } e^{ i \bd{k} \cdot \bd{R}_c }  =
 \sum_{ \bd{G}} \delta_{ \bd{k} , \bd{G}} e^{ i \bd{G} \cdot \bd{a}_1 },
 \\
 \delta_d (\bd{k} ) & = &  \frac{4}{N} \sum_{\bd{R}_d } e^{ i \bd{k} \cdot \bd{R}_d }  =
 \sum_{ \bd{G}} \delta_{ \bd{k} , \bd{G}} e^{ i \bd{G} \cdot \bd{d}_x },
 \end{eqnarray}
 \end{subequations}
where $\bd{G}$ are the reciprocal lattice vectors of the honeycomb lattice associated with the $a$-sublattice,
i.e., $e^{i \bd{R}_a \cdot \bd{G}} = 1$. It follows that the Fourier components of the operators 
$a_{\bd{k}}$, $b_{\bd{k}}$, $c_{\bd{k}}$, and $d_{\bd{k}}$
defined via Eq.~(\ref{eq:FT_01})
have the following periodicity properties,
 \begin{subequations}
 \begin{eqnarray}
 a_{ \bd{k} + \bd{G}} & = & a_{\bd{k}},
 \\
  b_{ \bd{k} + \bd{G}} & = & e^{ - i \bd{G} \cdot \bd{d}_z } b_{\bd{k}},
 \\
 c_{ \bd{k} + \bd{G}} & = & e^{ - i \bd{G} \cdot \bd{a}_1 } c_{\bd{k}},
 \\
  d_{ \bd{k} + \bd{G}} & = & e^{ - i \bd{G} \cdot \bd{d}_x } d_{\bd{k}}.
 \end{eqnarray}
 \end{subequations}
These  non-trivial phase factors are crucial for the correct treatment of
Umklapp scattering  in our calculation of magnon damping
presented in Sec.~\ref{sec:damping}.

In the momentum space, the total quadratic part of the boson Hamiltonian 
is of the form
 \begin{eqnarray}
 {\cal{H}}_2 & = &    {\cal{H}}_{2 \parallel} + {\cal{H}}_{2 \bot } = 
 {\sum_{\bd{k}}} \sum_{m n} 
 \Biggl\{  A^{m n}_{\bd{k}}   a^{\dagger}_{\bd{k} m}
 a_{\bd{k} n}
 \nonumber
 \\
 & & 
 + \frac{1}{2} \left[  B_{\bd{k}}^{m n}
 a^{\dagger}_{\bd{k} m} 
 a^{\dagger}_{- \bd{k} n } 
 + (B_{\bd{k}}^{n m})^{\ast}
 a_{-\bd{k} m}   a_{ \bd{k} n } \right]  \Biggr\},
 \hspace{7mm}
 \label{eq:H2res}
 \end{eqnarray}
where the labels $m, n \in  \{ a,b,c,d \}$ refer to the four sublattices
and we have set $a_{\bd{k} a } = a_{\bd{k}}$,
 $a_{\bd{k} b } = b_{\bd{k}}$,
 $a_{\bd{k} c } = c_{\bd{k}}$, and
$a_{\bd{k} d } = d_{\bd{k}}$.
In general,  the hermiticity of the Hamiltonian implies that
 the matrix $\mathbf{A}_{\bd{k}}$ with elements
  $[\mathbf{A}_{\bd{k}}]^{m n} = A_{\bd{k}}^{m n}$ is hermitian, i.e.,
 \begin{eqnarray}
 A_{\bd{k}}^{m n } & = & ( A_{\bd{k}}^{ n m  } )^{\ast} ,
 \; \; \; {\mbox{or}} \; \; \; \mathbf{A}_{\bd{k}} = \mathbf{A}^{\dagger}_{\bd{k}}.
 \end{eqnarray}
Moreover, 
the symmetry under relabeling $ \bd{k} \rightarrow - \bd{k}$ 
in the off-diagonal terms implies
 \begin{eqnarray}
 B_{\bd{k}}^{m n } & = &  B_{- \bd{k}}^{ n m  },
 \; \; \; {\mbox{or}} \; \; \; \mathbf{B}_{\bd{k}} = \mathbf{B}^{T}_{-\bd{k}}.
 \end{eqnarray}
In the zigzag state with the 
local moment given by $\bd{m}_i = \zeta_i {\bd{n}}_3$ 
and general transverse basis vectors given by
Eqs.~(\ref{eq:n1def}) and (\ref{eq:n2def}),
the non-zero elements of the  matrices given above are
 \begin{subequations}
 \begin{eqnarray}
 A_{\bd{k}}^{aa} & = & A_{\bd{k}}^{bb } = A_{\bd{k}}^{cc} 
 = A_{\bd{k}}^{dd } = \lambda,
 \label{eq:A_01}
 \\
  A_{\bd{k}}^{ad} &  = & ( A_{\bd{k}}^{da } )^{\ast} =   A_{\bd{k}}^{cb} 
 =  (A_{\bd{k}}^{bc})^{\ast} = \alpha_{\bd{k}} ,
 \\
 A_{\bd{k}}^{ab} &  = &  ( A_{\bd{k}}^{ba } )^{\ast}  = \beta_{\bd{k}},
 \\
  A_{\bd{k}}^{cd } &  = & ( A_{\bd{k}}^{dc } )^{\ast} =  \beta^{\ast}_{ - \bd{k}},
 \end{eqnarray}
 \end{subequations}
and
 \begin{subequations}
 \begin{eqnarray}
  B_{\bd{k}}^{ab }  & =  &  B_{ - \bd{k} }^{ba}  =
 B_{\bd{k}}^{cd } =  B_{- \bd{k}}^{dc}  = \mu_{\bd{k}}  ,
 \\
   B_{\bd{k}}^{ad }  &  = & B_{ - \bd{k}}^{da} = \nu_{\bd{k}}  ,
 \\
  B_{\bd{k}}^{cb  } &  = &  B_{ - \bd{k}}^{bc} = \nu^{\ast}_{-\bd{k}}  ,
 \end{eqnarray}
 \end{subequations}
where
\begin{eqnarray}
 \lambda & = &  - S \left[ J + K \frac{2 - s^2}{2 + s^2} + \frac{2 \Gamma}{2 + s^2} (  2s -1) \right]
 \nonumber
 \label{eq:lambda_01}
 \\
 & = & S \left[ - J + K \frac{ 2 - r^2}{2 + r^2} +  \Gamma  \frac{ r(4 + r) }{ 2 + r^2} \right],
  \\
 \alpha_{\bd{k}} & = & S \left[ J + \frac{K}{4} 
 \frac{4 + r^2}{2 + r^2} +  \Gamma \frac{r}{ 2 + r^2} \right] 
 \left( e^{ i \bd{k} \cdot \bd{d}_x } + e^{ i \bd{k} \cdot \bd{d}_y } \right),
 \nonumber
 \\
 & &
 \\
 \beta_{\bd{k}} & = & - S e^{ 2 i \phi} 
 \left[ \frac{K}{2}  \frac{ r^2}{ 2 + r^2} + \frac{\Gamma}{2}  \frac{4 + r^2}{2 + r^2} \right] 
e^{ i \bd{k} \cdot \bd{d}_z },
 \label{eq:betakdef}
 \\
 \mu_{\bd{k}} & = &  S   \left[  J +
 \frac{K- \Gamma}{2} \frac{ r^2}{2 + r^2}  \right]  
e^{ i \bd{k} \cdot \bd{d}_z },
 \\
 \nu_{\bd{k}} & = &  S e^{ 2 i \phi} \Biggl\{  \left[
 \frac{K}{4} \frac{r^2}{2 + r^2}   - \Gamma \frac{r}{ 2 + r^2 } \right] 
 \left( e^{ i \bd{k} \cdot \bd{d}_x } + e^{ i \bd{k} \cdot \bd{d}_y } \right)
 \nonumber
 \\
 & & +  i \frac{ K -  \Gamma r}{2} \sqrt{ \frac{2}{2 + r^2 } } \left( e^{ i \bd{k} \cdot \bd{d}_x }
 - e^{ i \bd{k} \cdot \bd{d}_y } \right) \Biggr\}.
 \label{eq:nucomplex}
 \end{eqnarray}
The parameters $r$ and $s = - 2 / r $ are functions of  $K$ and $\Gamma$ 
as given in Eqs.~(\ref{eq:smallrdef}) and (\ref{eq:smallqdef}).
Note that for $K  = \Gamma r$, the last term in Eq.~(\ref{eq:nucomplex}) vanishes so that
$\nu_{\bd{k}} = \nu_{ - \bd{k}}^{\ast}$ for $\phi=0$. It turns out that on this special surface in the 
parameter space, the spin-wave spectrum can be obtained analytically for all $\bd{k}$, as will be 
discussed in Sec.~\ref{sec:spectrum}.
We conclude that in the zigzag state  the
matrices $\mathbf{A}_{\bd{k}}$ and $\mathbf{B}_{\bd{k}}$ 
defined via the quadratic spin-wave Hamiltonian ${\cal{H}}_2$
in Eq.~(\ref{eq:H2res}) have the following structure,
 \begin{eqnarray}
 \mathbf{A}_{\bd{k}} & = &
 \left(
 \begin{array}{cc|cc}
 A^{aa}_{\bd{k}} & A^{ab}_{\bd{k}} & 0 & A^{ad}_{\bd{k}} \\
 (A^{ab}_{\bd{k}})^{\ast} & A^{bb}_{\bd{k}} &  A^{bc}_{\bd{k}} & 0 \\
 \hline
 0 &  (A^{bc}_{\bd{k}})^{\ast} & A^{cc}_{\bd{k}} &  A^{cd}_{\bd{k}} \\
 (A^{ad}_{\bd{k}})^{\ast} & 0 &  (A^{cd}_{\bd{k}})^{\ast} & A^{dd}_{\bd{k}} 
 \end{array}
 \right)
 \nonumber
 \\
 & = &
 \left(
 \begin{array}{cc|cc}
 \lambda & \beta_{\bd{k}}& 0 & \alpha_{\bd{k}} \\
 \beta_{\bd{k}}^{\ast} & \lambda &  \alpha_{\bd{k}}^{\ast} & 0 \\
 \hline
 0 &  \alpha_{\bd{k}} & \lambda &  \beta_{-\bd{k}}^{\ast} \\
 \alpha_{\bd{k}}^{\ast} & 0 &  \beta_{-\bd{k}} & \lambda 
 \end{array}
 \right) ,
 \label{eq:Anew}
 \end{eqnarray}
and
\begin{eqnarray}
 \mathbf{B}_{\bd{k}} & = &
 \left(
 \begin{array}{cc|cc}
 0  & B^{ab}_{\bd{k}} & 0 & B^{ad}_{\bd{k}} \\
 B^{ab}_{ - \bd{k}} & 0  &  B^{bc}_{\bd{k}} & 0 \\
 \hline
 0 &  B_{-\bd{k}}^{bc} & 0  &  B^{cd}_{\bd{k}} \\
 B_{- \bd{k}}^{ad} & 0 &  B^{cd}_{ - \bd{k}} & 0 
 \end{array}
 \right)
 \nonumber
 \\
 & = & \left(
 \begin{array}{cc|cc}
 0 & \mu_{\bd{k}}& 0 & \nu_{\bd{k}} \\
 \mu_{- \bd{k}} & 0 &  \nu_{\bd{k}}^{\ast} & 0 \\
 \hline
 0 &   \nu_{-\bd{k}}^{\ast} & 0 &   \mu_{\bd{k}} \\
 \nu_{ - \bd{k}}  & 0 &  \mu_{ -\bd{k}} & 0 
 \end{array}
 \right) .
 \label{eq:Bnew}
 \end{eqnarray}

 \subsection{Including third-nearest neighbor exchange}

A more realistic model of the spin-orbit coupled iridium oxides and $\alpha$-RuCl$_3$ 
also takes into account an
isotropic third-nearest neighbor Heisenberg exchange interaction $J_3$
connecting spins on the opposite corners in the hexagons of the honeycomb lattice.
Then we should add 
the following term to our Hamiltonian  in Eq.~(\ref{eq:hamiltonian}),
 \begin{equation}
 {\cal{H}}^{J_3} =   \frac{J_3}{2} 
 \sum_{\alpha} \left[ \sum_{ \bd{R} \in A } \bd{S}_{\bd{R} }
 \cdot \bd{S}_{ \bd{R} + \bd{\delta}_{\alpha} } 
 +  \sum_{ \bd{R} \in B } \bd{S}_{\bd{R} }
 \cdot \bd{S}_{ \bd{R} - \bd{\delta}_{\alpha} }  \right], \label{eq:HJ3}
 \end{equation}
where the vectors $\bd{\delta}_{\alpha} = - 2 \bd{d}_{\alpha}$ connect the opposite
sites of the hexagons. 
The classical ground state energy of the zigzag state is then given by
\begin{equation}
 \frac{ {\cal{H}}_{0}}{N S^2} = -  
  \frac{| \lambda_3 |}{2}   - \frac{3}{2}  J_3 
= - \frac{1}{2} \left| J - \frac{\Gamma}{2} - \frac{R}{2} \right| - \frac{3}{2} J_3,
 \label{eq:Eclassical2}
 \end{equation}
where $ \lambda_3$ is given in Eq.~(\ref{eq:lambda3def}). Hence, a
positive $J_3$ stabilizes the zigzag state, similarly to the consideration of Ref.~\cite{Winter17}.
It turns out, that the structure of the matrices $\mathbf{A}_{\bd{k}}$ and $\mathbf{B}_{\bd{k}}$
given in Eqs.~(\ref{eq:Anew}) and (\ref{eq:Bnew}) above does not change with $J_3$; 
we simply have to redefine the diagonal matrix element $\lambda$ of $\mathbf{A}_{\bd{k}}$
as follows,
\begin{equation}
 \lambda = S \left[ 3 J_3 - J + K \frac{2 - r^2}{2 + r^2} 
 + \Gamma \frac{r ( 4 + r ) }{ 2 + r^2} \right],
 \end{equation} 
and replace the off-diagonal element $\mu_{\bd{k}}$ of the matrix
$\mathbf{B}_{\bd{k}}$ by
 \begin{eqnarray}
  B_{\bd{k}}^{ab }  & =  &  B_{ - \bd{k} }^{ba}  =
 B_{\bd{k}}^{cd } =  B_{- \bd{k}}^{dc}  = \mu_{\bd{k}}
 \nonumber
 \\
 & = &   S   \left[  J +
 \frac{K- \Gamma}{2} \frac{ r^2}{2 + r^2}  \right]  
e^{ i \bd{k} \cdot \bd{d}_z }
 \nonumber
 \\
 &  & +  S J_3 \left( e^{ -2 i \bd{k} \cdot \bd{d}_{x}}
 + e^{ -2 i \bd{k} \cdot \bd{d}_{y}}
 + e^{ -2 i \bd{k} \cdot \bd{d}_{z}} \right).
 \hspace{7mm}
 \end{eqnarray}
Note that the additional contribution  to the matrix element $\mu_{\bd{k}}$   involving  $ J_3$
 does not violate the symmetry
$\mu_{- \bd{k}} = \mu_{\bd{k}}^{\ast}$.

\subsection{Cubic boson Hamiltonian}
\label{sec:cubic}

For the calculation of the magnon damping in the zigzag state
presented in Sec.~\ref{sec:damping},  we also need the cubic part ${\cal{H}}_3$ of the 
boson Hamiltonian, which can be obtained
from ${\cal{H}}_{\parallel \bot}$ 
given in Eq.~(\ref{eq:Hmix}) by expanding the transverse components of the
spin operators to linear order in the Holstein-Primakoff bosons. 
In real space we obtain
\begin{eqnarray}
{\cal{H}}_3  & = & - \frac{ \sqrt{2S}}{2}
 \Bigl\{ 
 \sum_{ \bd{R} \in a } a^{\dagger}_{\bd{R}} 
 ( V_x \rho^d_{\bd{R} + \bd{d}_x } 
 + V_y \rho^d_{\bd{R} + \bd{d}_y }
 - V_z \rho^b_{\bd{R} + \bd{d}_z } )
 \nonumber
 \\
 & &  \hspace{10mm}  -   \sum_{ \bd{R} \in c } c^{\dagger}_{\bd{R}} 
 ( V_x^{\ast} \rho^b_{\bd{R} + \bd{d}_x } 
 + V_y^{\ast} \rho^b_{\bd{R} + \bd{d}_y }
 - V_z^{\ast} \rho^d_{\bd{R} + \bd{d}_z } )
 \nonumber 
 \\
 & &  \hspace{10mm}  -   \sum_{ \bd{R} \in b } b^{\dagger}_{\bd{R}} 
 ( V_x^{\ast} \rho^c_{\bd{R} - \bd{d}_x } 
 + V_y^{\ast} \rho^c_{\bd{R} - \bd{d}_y }
 - V_z^{\ast} \rho^a_{\bd{R} - \bd{d}_z } )
 \nonumber
 \\
 & &  \hspace{10mm}  +   \sum_{ \bd{R} \in d } d^{\dagger}_{\bd{R}} 
 ( V_x \rho^a_{\bd{R} - \bd{d}_x } 
 + V_y \rho^a_{\bd{R} - \bd{d}_y }
 - V_z \rho^c_{\bd{R} - \bd{d}_z }  )
 \nonumber
 \\
 & & \hspace{10mm} + {\rm h.c.}
\Bigr\},
\label{eq:H3real}
 \end{eqnarray}
where $\rho^{a}_{\bd{R}} = a^{\dagger}_{\bd{R}} a_{\bd{R}}$,
 $\rho^{b}_{\bd{R}} = b^{\dagger}_{\bd{R}} b_{\bd{R}}$,
$\rho^{c}_{\bd{R}} = c^{\dagger}_{\bd{R}} c_{\bd{R}}$ and
 $\rho^{d}_{\bd{R}} = d^{\dagger}_{\bd{R}} d_{\bd{R}}$ are the number operators 
of the Holstein-Primakoff bosons in the four sublattices $a,b,c$, and $d$,
and
 \begin{subequations}
  \begin{eqnarray}
 V_x & = & \frac{ e^{ i \phi} {\rm sgn} s}{ \sqrt{ 2 + s^2 }}
 \left[ \frac{ K - \Gamma s}{\sqrt{2}} + i \frac{ \Gamma - K }{ \sqrt{ 2 + r^2 }} \right],
 \\
 V_y & = &   \frac{ e^{ i \phi} {\rm sgn} s}{ \sqrt{ 2 + s^2 }}
 \left[ - \frac{ K - \Gamma s}{\sqrt{2}} + i \frac{ \Gamma - K }{ \sqrt{ 2 + r^2 }} \right],
 \\
 V_z & = &  \frac{ 2 e^{ i \phi} {\rm sgn} s}{ \sqrt{ 2 + s^2 } \sqrt{ 2 + r^2 }        }
 ( \Gamma - K ).
 \end{eqnarray}
 \label{eq:Vreal}
 \end{subequations}
Defining the Fourier transform to momentum space as in Eq.~(\ref{eq:FT_01})
and carefully keeping track of the phase factors associated with the Umklapp scattering
using Eq.~(\ref{eq:deltaphase}), we obtain
 \begin{widetext}
 \begin{eqnarray}
 {\cal{H}}_3 & = & 
  \frac{ \sqrt{2S}}{2}    \sqrt{ \frac{4}{N}}
 \sum_{ \bd{k}_1  \bd{k}_2  \bd{k}_3 }  \sum_{\bd{G}} 
\delta_{ \bd{k}_1 + \bd{k}_2 + \bd{k}_3 ,  \bd{G}}
\Biggl\{
 \nonumber
 \\
 & & - \left[  
  \left( V_x e^{i \bd{G} \cdot \bd{d}_x  - i ( \bd{k}_2 + \bd{k}_3 ) \cdot \bd{d}_x } 
   + V_y e^{i \bd{G} \cdot \bd{d}_x  - i ( \bd{k}_2 + \bd{k}_3 ) \cdot \bd{d}_y } 
  \right) d^{\dagger}_{ - \bd{k}_1 } 
+ V_z^{\ast}  e^{i \bd{G} \cdot \bd{d}_z  - i ( \bd{k}_2 + \bd{k}_3 )
 \cdot \bd{d}_z }  b^{\dagger}_{ - \bd{k}_1 }
\right]  a^{\dagger}_{ - \bd{k}_2 } a_{\bd{k}_3 } 
 \nonumber
 \\
 &  & + \left[   
 \left( V_x^{\ast} e^{i \bd{G} \cdot \bd{a}_1  + i ( \bd{k}_2 + \bd{k}_3 ) \cdot \bd{d}_x } 
   + V_y^{\ast} e^{i \bd{G} \cdot \bd{a}_1  + i ( \bd{k}_2 + \bd{k}_3 ) \cdot \bd{d}_y } 
  \right) c^{\dagger}_{ - \bd{k}_1 } 
+ V_z  e^{ i ( \bd{k}_2 + \bd{k}_3 )
 \cdot \bd{d}_z }  a^{\dagger}_{ - \bd{k}_1 }
\right] b^{\dagger}_{ - \bd{k}_2 } b_{\bd{k}_3 }
 \nonumber
 \\
 &  & +  \left[   
\left( V_x^{\ast} e^{i \bd{G} \cdot \bd{d}_z  - i ( \bd{k}_2 + \bd{k}_3 ) \cdot \bd{d}_x } 
   + V_y^{\ast} e^{i \bd{G} \cdot \bd{d}_z  - i ( \bd{k}_2 + \bd{k}_3 ) \cdot \bd{d}_y } 
  \right) b^{\dagger}_{ - \bd{k}_1 } 
+ V_z  e^{ i \bd{G} \cdot \bd{d}_x - i ( \bd{k}_2 + \bd{k}_3 )
 \cdot \bd{d}_z }  d^{\dagger}_{ - \bd{k}_1 }
\right]   c^{\dagger}_{ - \bd{k}_2 } c_{\bd{k}_3 }
 \nonumber
 \\
 &   & - \left[
 \left( V_x e^{ i ( \bd{k}_2 + \bd{k}_3 ) \cdot \bd{d}_x } 
   + V_y e^{i ( \bd{k}_2 + \bd{k}_3 ) \cdot \bd{d}_y } 
  \right) a^{\dagger}_{ - \bd{k}_1 } 
+ V_z^{\ast}  e^{i \bd{G} \cdot \bd{a}_1  + i ( \bd{k}_2 + \bd{k}_3 )
 \cdot \bd{d}_z }  c^{\dagger}_{ - \bd{k}_1 }
   \right]     d^{\dagger}_{ - \bd{k}_2 } d_{\bd{k}_3 }
 + {\rm h.c.}
\Biggr\}.
 \label{eq:H3mom}
\end{eqnarray}
\end{widetext}
In the second line we use 
 $\bd{k}_2 + \bd{k}_3 = \bd{G} - \bd{k}_1 $, 
 $\bd{a}_1 = \bd{d}_z - \bd{d}_x$, $\bd{a}_2 = \bd{d}_z - \bd{d}_y $,
$\bd{d}_x - \bd{d}_y = \bd{a}_2 - \bd{a}_1$, and
$ e^{ i \bd{G} \cdot ( \bd{a}_1 \pm \bd{a}_2 )} = 1$ to simplify the phase factors as follows,
 \begin{subequations}
 \begin{eqnarray}
  e^{i \bd{G} \cdot \bd{d}_x  - i ( \bd{k}_2 + \bd{k}_3 ) \cdot \bd{d}_x } & = & 
   e^{i ( \bd{G} -  \bd{k}_2 - \bd{k}_3 ) \cdot \bd{d}_x }
 = e^{ i \bd{k}_1 \cdot \bd{d}_x },
 \hspace{7mm}
 \\
 e^{i \bd{G} \cdot \bd{d}_x  - i ( \bd{k}_2 + \bd{k}_3 ) \cdot \bd{d}_y } 
 & = & e^{i \bd{G} \cdot ( \bd{d}_x  - \bd{d}_y ) +  i  \bd{k}_1 \cdot \bd{d}_y } 
 \nonumber
 \\
 &= &  e^{i \bd{G} \cdot ( \bd{a}_2  - \bd{a}_1 ) +  i  \bd{k}_1 \cdot \bd{d}_y }
 \nonumber
 \\
 & = & e^{ i \bd{k}_1 \cdot \bd{d}_y },
 \\
 e^{i \bd{G} \cdot \bd{d}_z  - i ( \bd{k}_2 + \bd{k}_3 )
 \cdot \bd{d}_z } & = &  e^{i ( \bd{G} -  \bd{k}_2 - \bd{k}_3 ) \cdot \bd{d}_z }
 = e^{ i \bd{k}_1 \cdot \bd{d}_z }.
 \end{eqnarray}
 \end{subequations}
The phases of the three terms in the third  
line of Eq.~(\ref{eq:H3mom}) can be simplified as follows,
 \begin{subequations}
 \begin{eqnarray}
e^{i \bd{G} \cdot \bd{a}_1  + i ( \bd{k}_2 + \bd{k}_3 ) \cdot \bd{d}_x } & = &
e^{i \bd{G} \cdot ( \bd{a}_1 + \bd{d}_x ) - i \bd{k}_1 \cdot \bd{d}_x } 
 \nonumber
 \\
 & = &  e^{ i \bd{G} \cdot \bd{d}_z }
 e^{ - i \bd{k}_1 \cdot \bd{d}_x },
 \\
e^{i \bd{G} \cdot \bd{a}_1  + i ( \bd{k}_2 + \bd{k}_3 ) \cdot \bd{d}_y } & = &
e^{i \bd{G} \cdot ( \bd{a}_1 + \bd{d}_y ) - i \bd{k}_1 \cdot \bd{d}_y } 
 \nonumber
 \\
 & = & e^{i \bd{G} \cdot ( \bd{a}_2 + \bd{d}_y ) - i \bd{k}_1 \cdot \bd{d}_y } 
\nonumber
 \\
 & = &  e^{ i \bd{G} \cdot \bd{d}_z }
 e^{ - i \bd{k}_1 \cdot \bd{d}_y },
 \\
  e^{ i ( \bd{k}_2 + \bd{k}_3 )
 \cdot \bd{d}_z }  & = &  e^{ i \bd{G} \cdot \bd{d}_z }
 e^{ - i \bd{k}_1 \cdot \bd{d}_z },
\end{eqnarray}
 \end{subequations}
and in the fourth line we can write
 \begin{subequations}
 \begin{eqnarray}
 e^{i \bd{G} \cdot \bd{d}_z  - i ( \bd{k}_2 + \bd{k}_3 ) \cdot \bd{d}_x } & = &
e^{i \bd{G} \cdot ( \bd{d}_z  - \bd{d}_x ) + i  \bd{k}_1 \cdot \bd{d}_x } 
 \nonumber
 \\
 & = &    e^{i \bd{G} \cdot \bd{a}_1 } e^{ i \bd{k}_1 \cdot \bd{d}_x }
 \\
 e^{i \bd{G} \cdot \bd{d}_z  - i ( \bd{k}_2 + \bd{k}_3 ) \cdot \bd{d}_y } & = &
e^{i \bd{G} \cdot ( \bd{d}_z  - \bd{d}_y ) + i  \bd{k}_1 \cdot \bd{d}_y }
 \nonumber 
 \\
 & = &    e^{i \bd{G} \cdot \bd{a}_2 } e^{ i \bd{k}_1 \cdot \bd{d}_y }
 \nonumber
 \\
 & = &  e^{i \bd{G} \cdot \bd{a}_1 } e^{ i \bd{k}_1 \cdot \bd{d}_y },
 \\
 e^{ i \bd{G} \cdot \bd{d}_x - i ( \bd{k}_2 + \bd{k}_3 )  
 \cdot \bd{d}_z } & = &    
 e^{ i \bd{G} \cdot  ( \bd{d}_x - \bd{d}_z )   +  i  \bd{k}_1 \cdot \bd{d}_z }
 \nonumber
 \\
 & = &  e^{ - i \bd{G} \cdot \bd{a}_1 } e^{ i \bd{k}_1 \cdot \bd{d}_z }
 \nonumber
 \\
 & = & e^{  i \bd{G} \cdot \bd{a}_1 } e^{ i \bd{k}_1 \cdot \bd{d}_z } ,
 \end{eqnarray}
 \end{subequations}
where in the last   line we have used the fact that $2 \bd{a}_1$ is a vector of the Bravais lattice so that
 $ e^{- 2 i \bd{G} \cdot \bd{a}_1 } =1$.
Finally, to simplify the phases in the  last line of Eq.~(\ref{eq:H3mom})
we use
$1 = e^{ i \bd{G} \cdot ( \bd{a}_1 - \bd{a}_2 ) } = e^{ - i \bd{G} \cdot \bd{d}_x }
 e^{ i \bd{G} \cdot \bd{d}_y }$ and hence 
$ e^{  i \bd{G} \cdot \bd{d}_x } =
 e^{ i \bd{G} \cdot \bd{d}_y }$ to write
 \begin{subequations} 
\begin{eqnarray}
  e^{ i ( \bd{k}_2 + \bd{k}_3 ) \cdot \bd{d}_x } & = & e^{ i \bd{G} \cdot \bd{d}_x } 
 e^{ - i \bd{k}_1 \cdot {\bd{d}}_x },
 \\
  e^{ i ( \bd{k}_2 + \bd{k}_3 ) \cdot \bd{d}_y } & = & e^{ i \bd{G} \cdot \bd{d}_y } 
 e^{ - i \bd{k}_1 \cdot {\bd{d}}_y } 
 \nonumber
 \\
 & = &  e^{ i \bd{G} \cdot \bd{d}_x } 
 e^{ - i \bd{k}_1 \cdot {\bd{d}}_y } ,
 \\
  e^{i \bd{G} \cdot \bd{a}_1  + i ( \bd{k}_2 + \bd{k}_3 ) \cdot \bd{d}_z } & = & 
  e^{i \bd{G} \cdot \bd{a}_1  + i ( \bd{G} -  \bd{k}_1 ) \cdot \bd{d}_z }
 \nonumber
 \\
 & = & e^{ i \bd{G} \cdot ( \bd{a}_1 + \bd{d}_z ) }  e^{ - i \bd{k}_1 \cdot \bd{d}_z }
 \nonumber
 \\
& = & e^{ i \bd{G} \cdot ( - \bd{a}_1 + \bd{d}_z ) }  e^{ - i \bd{k}_1 \cdot \bd{d}_z }
 \nonumber
 \\
 & = & e^{ i \bd{G} \cdot \bd{d}_x } e^{ - i \bd{k}_1 \cdot \bd{d}_z }.
 \end{eqnarray}
 \end{subequations}
Defining
 \begin{eqnarray}
 V_{\bd{k}} & = &  \frac{ \sqrt{2S}}{2} 
\left( V_x e^{i \bd{k} \cdot \bd{d}_x } 
   + V_y e^{i \bd{k} \cdot \bd{d}_y } 
  \right)
 \nonumber
 \\
 & = & 
\frac{ \sqrt{2S}}{2}  \frac{ e^{ i \phi} {\rm sign} s }{ \sqrt{ 2 + s^2}}
 \biggl[ \frac{ K - \Gamma s }{\sqrt{2}} (  e^{ i \bd{k} \cdot \bd{d}_x } -  
  e^{ i \bd{k} \cdot \bd{d}_y } )
 \nonumber
 \\
 & & \hspace{18mm}
 +
   i \frac{ \Gamma  - K}{\sqrt{ 2 +r^2}} (  e^{ i \bd{k} \cdot \bd{d}_x } + 
  e^{ i \bd{k} \cdot \bd{d}_y } ) \biggr],
 \hspace{7mm}
 \\
 U_{\bd{k}} & = & \frac{ \sqrt{2S}}{2} V_{z} e^{ i \bd{k} \cdot \bd{d}_z }
 \nonumber
 \\
 & = &  \sqrt{2S}  \frac{ e^{ i \phi} {\rm sgn} s}{ \sqrt{ 2 + s^2 } \sqrt{ 2 + r^2 }        }
 ( \Gamma - K )   e^{ i \bd{k} \cdot \bd{d}_z },
 \end{eqnarray}
we finally obtain  the cubic part of the boson Hamiltonian in the zigzag state,
  \begin{eqnarray}
 {\cal{H}}_3 & = &    \sqrt{ \frac{4}{N}}
 \sum_{ \bd{k}_1  \bd{k}_2  \bd{k}_3 }  \sum_{\bd{G}} 
\delta_{ \bd{k}_1 + \bd{k}_2 + \bd{k}_3 ,  \bd{G}}
\Biggl\{ 
 \nonumber
 \\
 &  &  \hspace{9mm} - \left[ V_{ \bd{k}_1 } d^{\dagger}_{ - \bd{k}_1 } + U^{\ast}_{ - \bd{k}_1 } 
b^{\dagger}_{ - \bd{k}_1 } \right]  a^{\dagger}_{ - \bd{k}_2 } a_{\bd{k}_3 } 
 \nonumber
 \\
 & & 
 + e^{ i \bd{G} \cdot \bd{d}_z } \left[   
  V^{\ast}_{ \bd{k}_1}  c^{\dagger}_{ - \bd{k}_1 } 
+ U_{  -  \bd{k}_1 } a^{\dagger}_{ - \bd{k}_1 }
\right] b^{\dagger}_{ - \bd{k}_2 } b_{\bd{k}_3 }
 \nonumber
 \\
 &  & +   e^{i \bd{G} \cdot \bd{a}_1 }    \left[   
V^{\ast}_{ - \bd{k}_1 }  b^{\dagger}_{ - \bd{k}_1 } 
+ U_{\bd{k}_1 }   d^{\dagger}_{ - \bd{k}_1 }
\right]   c^{\dagger}_{ - \bd{k}_2 } c_{\bd{k}_3 }
 \nonumber
 \\
 &   & -   e^{i \bd{G} \cdot \bd{d}_x }  \left[ 
 V_{- \bd{k}_1 } a^{\dagger}_{ - \bd{k}_1 } 
+  U^{\ast}_{\bd{k}_1 }  c^{\dagger}_{ - \bd{k}_1 }
   \right]     d^{\dagger}_{ - \bd{k}_2 } d_{\bd{k}_3 }
 + {\rm h.c.}
 \Biggr\}.
 \nonumber
 \\
 & &
 \label{eq:H3res}
 \end{eqnarray}
Note that the Umklapp processes associated with the 
non-zero vectors $\mathbf{G}$ of the reciprocal lattice
involve non-trivial phase factors.
Below we shall calculate the  magnon damping in the special case
$\Gamma = K > 0$ where $r=1$ and $s = -2$. Then
$U_{\bd{k} } =0$, while 
$V_{\bd{k}}$ reduces to
 \begin{equation}
 V_{\bd{k}} = - \frac{ \sqrt{ 6 S}}{4}  K 
 (  e^{ i \bd{k} \cdot \bd{d}_x } -  
  e^{ i \bd{k} \cdot \bd{d}_y } ),
  \label{eq:V_01}
 \end{equation}
where  we have chosen the gauge $\phi =0$ for simplicity, so that
$V_{- \bd{k}} = V_{\bd{k}}^{\ast}$ .

\section{Magnon spectrum in the zigzag state for $\Gamma = K $}
\label{sect:spectrum}

To obtain the  magnon spectrum, we should diagonalize the
quadratic part ${\cal{H}}_2$ of our boson Hamiltonian
 in Eq.~(\ref{eq:H2res}).
Due to the anomalous terms involving the matrix $\mathbf{B}_{\bd{k}}$, 
this requires a multi-flavor generalization of the Bogoliubov transformation.
A general algorithm for constructing such a transformation has been
described by Colpa \cite{Colpa78} and  by Blaizot and Ripka \cite{Blaizot86}, see also 
more recent discussions in Refs.~[\onlinecite{Maldonado93}, \onlinecite{Serga12}].
We provide a careful review of this algorithm in Appendix~A where we also point out some 
mathematical subtleties \cite{Maldonado93}. 

For a general boson Hamiltonian of the type ~(\ref{eq:H2res}) with $f$ different boson flavors,
Colpa's algorithm transforms the Hamiltonian to a diagonal $2 f \times 2 f$ matrix containing   
magnon energies $\omega_{ \bd{k} n }$ as well as  negative 
magnon energies $- \omega_{ \bd{k} n }$, where $n = 1 , \ldots , f $ labels the magnon bands. 
In the zigzag phase of the Kitaev-Heisenberg-$\Gamma$ model, the number of boson flavors is  $f=4$, 
so  one has to deal with $8 \times 8$ matrices to calculate the magnon spectrum.
Although this can be done numerically, the size of the matrices is too large
for performing analytic calculations beyond the standard linear SWT in a reasonable amount of time.
In this section, we will show that we can avoid this doubling of 
the flavor dimension by using the hermitian-field parametrization of the 
SWT developed in 
Refs.~[\onlinecite{Hasselmann06,Kreisel07,Kreisel08,Kreisel11,Kreisel14}].
Another advantage of this approach is that it allows us to identify special 
regimes in the parameter space of the model where the calculation of the magnon spectrum simplifies.
In fact, we will demonstrate below that for $\Gamma = K>0$ and arbitrary $J$ and $J_3$,
magnon spectrum can be obtained fully analytically, which 
will enable us  to calculate   magnon damping and the
dynamical structure factor for $\Gamma = K$ in Sec.~\ref{sec:damping}.

\subsection{Hermitian field parametrization of spin fluctuations}
 \label{sec:hermitian}

At this point, it is advantageous to work with  the Euclidean action associated with the  
quadratic boson-Hamiltonian (\ref{eq:H2res}),
 \begin{eqnarray}
 S_2 & = &  \beta \sum_{K} 
 \sum_{m n} 
 \Biggl\{  ( A^{m n}_{\bd{k}} - i \omega \delta_{m n} )   \bar{a}_{K m}
 a_{K n}
 \nonumber
 \\
 &  + & 
  \frac{1}{2} \left[  B_{\bd{k}}^{m n}
 \bar{a}_{K m} 
 \bar{a}_{- K n } 
 + (B_{\bd{k}}^{n m})^{\ast}
 a_{-K m}   a_{ K n } \right]  \Biggr\},
 \hspace{7mm}
 \label{eq:S2a}
 \end{eqnarray}
 where $a_{K m}$ are now complex variables labeled by
the momentum-energy index $K = ( \bd{k} , i \omega )$
and the sublattice index $m$. Here  $ i \omega$ is the bosonic Matsubara frequency,
$\beta$ is  inverse temperature, and $\sum_K = \sum_{\bd{k}} \sum_{\omega}$.
For each complex field $a_{K m}$ we now introduce a pair of real fields
$X_{ K m}$ and $P_{K m}$ by setting
\begin{subequations}
 \label{eq:herm_01}
 \begin{eqnarray}
 a_{ K m} & = & \frac{1}{\sqrt{2}} \left[ X_{ K m} + i P_{K m} \right],
 \label{eq:herm_01a}
 \\
 \bar{a}_{ - K m} & = & \frac{1}{\sqrt{2}} \left[ X_{ K m} -  i P_{K m} \right],
 \label{eq:herm_01b}
\end{eqnarray}
\end{subequations}
where $X_{K m}$ and $P_{ K m}$ are the Fourier components of real fields that
satisfy
 \begin{equation}
 X_{ - K m} =  {X}_{ K m}^{\ast} , \; \; \;
 P_{ - K m} =  {P}_{ K m}^{\ast} .
 \end{equation}
In terms of these new variables, the  quadratic part of our spin-wave action can
be written as
 \begin{eqnarray}
 S_2 & = & \frac{\beta}{2} \sum_{K }
 \sum_{m n } \Bigl[ 
 T_{\bd{k}}^{m n} P_{ -K m} P_{K n} +  
 V_{\bd{k}}^{m n} X_{ -K m} X_{K n} 
 \nonumber
 \\
  &  &\hspace{14mm}  + 2 (  \omega \delta^{m n}  + W_{\bd{k}}^{m n} )
 X_{ - K m} P_{K n}
 \Bigr],
 \hspace{7mm}
 \label{eq:SSW}
 \end{eqnarray}
where $T_{\bd{k}}^{m n}$, $V_{\bd{k}}^{m n}$, and $W_{\bd{k}}^{m n}$ are the matrix elements 
of the $f \times f$  matrices  $\mathbf{T}_{\bd{k}} $,
 $\mathbf{V}_{\bd{k}} $, and $\mathbf{W}_{\bd{k}} $ defined by
 \begin{subequations}
 \begin{eqnarray}
 \mathbf{T}_{\bd{k}}  & = & \mathbf{A}_{\bd{k}  }^R - \mathbf{B}_{\bd{k}}^R,
 \\
 \mathbf{V}_{\bd{k}}  & = & 
 \mathbf{A}_{\bd{k}  }^R + \mathbf{B}_{\bd{k}}^R,
 \\
  \mathbf{W}_{\bd{k}}  & = &  
 - \mathbf{A}_{\bd{k}  }^I + \mathbf{B}_{\bd{k}}^I,
 \label{eq:Wmatdef}
  \end{eqnarray}
 \end{subequations}
where we introduced
 \begin{subequations}
 \begin{eqnarray}
 \mathbf{A}_{\bd{k}}^R & = &
 \frac{\mathbf{A}_{\bd{k}  } +\mathbf{A}_{-\bd{k}  }^\ast}{2}
= \frac{ \mathbf{A}_{\bd{k}}  +  \mathbf{A}^{T}_{-\bd{k}} }{2} ,
 \\
\mathbf{A}_{\bd{k}}^I & = &
 \frac{\mathbf{A}_{\bd{k}  } - \mathbf{A}_{-\bd{k}  }^\ast}{2i} 
 = \frac{ \mathbf{A}_{\bd{k}} -  \mathbf{A}^{T}_{-\bd{k}} }{2i} ,
 \\
 \mathbf{B}_{\bd{k}}^R & = &
 \frac{\mathbf{B}_{\bd{k}  } +\mathbf{B}_{-\bd{k}  }^\ast}{2} =
 \frac{ \mathbf{B}_{\bd{k}} + \mathbf{B}^{\dagger}_{\bd{k}} }{2}  ,
 \\
\mathbf{B}_{\bd{k}}^I & = &
 \frac{\mathbf{B}_{\bd{k}  } - \mathbf{B}_{-\bd{k}  }^\ast}{2i}  =   
 \frac{ \mathbf{B}_{\bd{k}} -  \mathbf{B}^{\dagger}_{\bd{k}} }{2i}   .
 \end{eqnarray}
 \end{subequations}
Our notation is motivated by the theory of coupled oscillators in classical mechanics \cite{Goldstein80},
in which the analogue of $\mathbf{T}_{\bm{k}}$ is associated with the kinetic energy of the system and
the analogue of  $\mathbf{V}_{\bm{k}}$ describes potential energy in the harmonic approximation.
Note that the matrix $\mathbf{W}_{\bd{k}}$ can alternatively be written as
 \begin{equation}
\mathbf{W}_{\bd{k}} = \mathbf{W}_{\bd{k},+} +\mathbf{W}_{\bd{k},-},
 \end{equation}
with
 \begin{subequations}
  \begin{eqnarray}
 \mathbf{W}_{\bd{k} , + }  & =  &  \frac{\mathbf{W}_{\bd{k} } + \mathbf{W}_{- \bd{k}}^T }{2}
 = \mathbf{B}_{\bd{k}}^I,
 \\
  \mathbf{W}_{\bd{k} , - }  & =  &  \frac{\mathbf{W}_{\bd{k} } - \mathbf{W}_{- \bd{k}}^T }{2}
  = - \mathbf{A}_{\bd{k}}^I.
 \end{eqnarray}
 \end{subequations}
In these notations,  our action (\ref{eq:SSW})
can be written in a more symmetric form
  \begin{eqnarray}
 S_2 & = & \frac{\beta}{2} \sum_{K }
 \sum_{m n } \Bigl[ 
 T_{\bd{k}}^{m n} P_{ -K m} P_{K n} +  
 V_{\bd{k}}^{m n} X_{ -K m} X_{K n} 
 \nonumber
 \\
  &   &  + (  \omega \delta^{m n}  + W_{\bd{k},-}^{m n} )
 ( X_{ - K  m} P_{K n} - P_{-K  m} X_{K n} )
 \nonumber
 \\
 & &  + W_{\bd{k},+}^{m n}
 ( X_{ - K  m} P_{K n} + P_{-K m} X_{K n} )
 \Bigr].
 \label{eq:SSW2}
 \end{eqnarray}
The symmetry of the fields under the relabeling $K \rightarrow - K$ and $ m \leftrightarrow n$ 
implies 
 \begin{subequations}
 \begin{eqnarray}
 T_{\bd{k}}^{m n} & = &   
 T_{ - \bd{k}}^{n m },
 \; \; \; \mbox{or} \; \; \;   \mathbf{T}_{\bd{k}} = \mathbf{T}^T_{- \bd{k}},
 \\
 V_{\bd{k}}^{m n} & = & 
 V_{ - \bd{k}}^{n m },
 \; \; \; \mbox{or} \; \; \;   \mathbf{V}_{\bd{k}} = \mathbf{V}^T_{- \bd{k}},
  \\
 W_{ \bd{k} , \pm }^{m n}&  = & \pm W_{- \bd{k} , \pm}^{n m},
\; \; \; \mbox{or} \; \; \;   
 \mathbf{W}_{\bd{k}, \pm } = \pm \mathbf{W}^T_{- \bd{k}, \pm}.
 \hspace{7mm}
 \end{eqnarray}
 \end{subequations}
In addition, the hermiticity of the underlying Hamiltonian implies
 \begin{subequations}
  \begin{eqnarray}
 T_{\bd{k}}^{m n} & = &   ( T_{\bd{k}}^{ n m} )^{\ast},
 \; \; \; \mbox{or} \; \; \;   \mathbf{T}_{\bd{k}} = \mathbf{T}^\dagger_{ \bd{k}},
 \\
 V_{\bd{k}}^{m n} & = &   ( V_{\bd{k}}^{ n m} )^{\ast},
\; \; \; \mbox{or} \; \; \;   \mathbf{V}_{\bd{k}} = \mathbf{V}^\dagger_{ \bd{k}},
  \\
 W_{ \bd{k} , \pm }^{m n} &  = & \pm ( W_{ \bd{k} , \pm}^{n m} )^{\ast},
\; \; \; \mbox{or} \; \; \;  
\mathbf{W}_{\bd{k}, \pm } = \pm \mathbf{W}^\dagger_{ \bd{k}, \pm}.
 \hspace{7mm}
 \end{eqnarray}
 \end{subequations}
Hence, the matrices $\mathbf{T}_{\bd{k}}$,  $\mathbf{V}_{\bd{k}}$, and  $\mathbf{W}_{\bd{k},+}$ are hermitian, while  $\mathbf{W}_{\bd{k},-}$ is antihermitian.
Combining the  relations given above, we see that all matrix elements satisfy
 \begin{subequations}
  \begin{eqnarray}
 T_{ \bd{k}}^{m n} & = &   ( T_{ - \bd{k}}^{ m n} )^{\ast},
 \; \; \; \mbox{or} \; \; \;   \mathbf{T}_{\bd{k}} = \mathbf{T}^\ast_{- \bd{k}},
 \\
 V_{ \bd{k}}^{m n} & = &   ( V_{- \bd{k}}^{ m n} )^{\ast},
 \; \; \; \mbox{or} \; \; \;   \mathbf{V}_{\bd{k}} = \mathbf{V}^\ast_{- \bd{k}},
  \\
 W_{  \bd{k} , \pm }^{m n}&  = & ( W_{ - \bd{k} , \pm}^{ m n} )^{\ast},
 \; \; \; \mbox{or} \; \; \;   \mathbf{W}_{\bd{k}, \pm} = \mathbf{W}^\ast_{- \bd{k}, \pm}.
 \hspace{7mm}
 \end{eqnarray}
 \end{subequations}
In the compact matrix notation, our quadratic spin-wave action (\ref{eq:SSW2})
can be written as
 \begin{eqnarray}
 & & S_2 [X , P ] =\frac{\beta}{2} \sum_{K} \Bigl[ {\bd{X}}^{\dagger}_K  \mathbf{V}_{\bd{k}} {\bd{X}}_K 
 + {\bd{P}}^{\dagger}_K  \mathbf{T}_{\bd{k}} {\bd{P}}_K 
 \nonumber
\\
 & & \hspace{20mm}
 + {\bd{X}}^{\dagger}_K  ( \mathbf{W}_{\bd{k}} + \omega )  {\bd{P}}_K 
 + {\bd{P}}^{\dagger}_K  ( \mathbf{W}^{\dagger}_{\bd{k}} - \omega )  {\bd{X}}_K
 \Bigr] 
 \nonumber
 \\
 & = & 
\frac{\beta}{2} \sum_{K}
  ( \bd{X}^T_{-K }, \bd{P}^T_{-K} )
 \left( \begin{array}{cc} \mathbf{V}_{\bd{k}}  & \mathbf{W}_{\bd{k}} +  \omega  \\
  \mathbf{W}_{\bd{k}}^{\dagger} -  \omega & \mathbf{T}_{\bd{k}} 
 \end{array}
 \right) 
 \left( \begin{array}{c} \bd{X}_{K} \\ \bd{P}_K \end{array} \right),
 \nonumber
 \\
 & &
 \label{eq:S2compact}
 \end{eqnarray}
where we have defined the four-component column vectors
 \begin{equation}
 {\bd{X}}_K = \left( \begin{array}{c}
 X_{Ka} \\ X_{Kb} \\ X_{K c} \\ X_{Kd} 
 \end{array}
 \right),
 \; \; \; \;
{\bd{P}}_K = \left( \begin{array}{c}
 P_{Ka} \\ P_{Kb} \\ P_{K c} \\ P_{Kd} 
 \end{array}
 \right).
 \end{equation}
After the analytic continuation to real frequencies ($ \omega = - i i \omega \rightarrow - i \omega$), 
the spin-wave dispersions can be obtained from the roots of the equation
 \begin{equation}
 {\rm det} \left( \begin{array}{cc} \mathbf{V}_{\bd{k}}  & \mathbf{W}_{\bd{k}} - i   \omega  \\
  \mathbf{W}_{\bd{k}}^{\dagger} + i  \omega & \mathbf{T}_{\bd{k}} 
 \end{array}
 \right) =0.
 \label{eq:swdispersion}
 \end{equation}
At  first sight, it seems that one has to calculate the determinant of 
the $2f \times 2f$-matrix in order obtain the $f$~magnon bands,
as in Colpa's algorithm \cite{Colpa78}.
However, we can  
reduce the dimension of the matrices by
performing  Gaussian integration over the  $P$-field.
The resulting effective action for the
$X$-field is
 \begin{eqnarray}
 S_2 [ X ] = \frac{\beta}{2} \sum_K \sum_{m n} X_{-K}^{m} [\mathbf{G}_0^{-1} ( K ) ]^{m n}
 X_{ K}^{n},
 \end{eqnarray}
where the inverse Gaussian propagator of the $X$-field is given by
 \begin{eqnarray}
 & & \mathbf{G}^{-1} _0(  \bd{k} , i \omega  ) = \mathbf{V}_{\bd{k}}  
  -     (  \mathbf{W}_{\bd{k}}  + \omega ) \mathbf{T}_{\bd{k}}^{-1}
 ( \mathbf{W}_{\bd{k}}^{\dagger} - \omega )
 \hspace{7mm}
 \nonumber
 \\
 & = & 
\mathbf{V}_{\bd{k}}   +     ( \omega -  \mathbf{A}_{\bd{k} }^I  + 
 \mathbf{B}_{\bd{k}}^I ) \mathbf{T}_{\bd{k}}^{-1}
  ( \omega - \mathbf{A}_{\bd{k} }^I - 
 \mathbf{B}_{\bd{k}}^I ).
 \label{eq:schur}
 \end{eqnarray}
The  matrix in the right-hand side of Eq.~(\ref{eq:schur})  is the so-called 
Schur complement 
of the block $\mathbf{V}_{\bd{k}}$ in the  matrix 
 \begin{equation}
 \left( \begin{array}{cc} \mathbf{V}_{\bd{k}}  & \mathbf{W}_{\bd{k}} +  \omega  \\
  \mathbf{W}_{\bd{k}}^{\dagger} -  \omega & \mathbf{T}_{\bd{k}} 
 \end{array}
 \right) .
 \end{equation}
 After the analytic continuation to real frequencies ($i \omega \rightarrow \omega$),
the inverse propagator matrix in Eq.~(\ref{eq:schur})
 becomes
\begin{eqnarray}
  & & \mathbf{G}^{-1}_0 (  \bd{k} , \omega  ) 
 \nonumber
 \\
 & = &   \mathbf{V}_{\bd{k}}  
 -   ( \omega + i  \mathbf{W}_{\bd{k}} ) \mathbf{T}_{\bd{k}}^{-1}
  ( \omega  - i  \mathbf{W}_{\bd{k}}^{\dagger} )
\nonumber
 \\
 & = &  \mathbf{V}_{\bd{k}}  
 -   (  \mathbf{W}_{\bd{k}} - i \omega ) \mathbf{T}_{\bd{k}}^{-1}
  (   \mathbf{W}_{\bd{k}}^{\dagger} + i \omega )
\nonumber
  \\
 &   = &
\mathbf{V}_{\bd{k}}   -    ( \omega -  i \mathbf{A}_{\bd{k} }^I  + 
 i \mathbf{B}_{\bd{k}}^I ) \mathbf{T}_{\bd{k}}^{-1}
  ( \omega - i \mathbf{A}_{\bd{k} }^I - i
 \mathbf{B}_{\bd{k}}^I ).
 \hspace{7mm}
 \label{eq:Ginvspinwaves}
 \end{eqnarray}
Note that Eq.~(\ref{eq:Ginvspinwaves})
can also be obtained directly from
Eq.~(\ref{eq:swdispersion}) using  general formula for the determinant
of a block matrix,
 \begin{equation}
 {\rm det} \left( \begin{array}{cc} \mathbf{A} & \mathbf{B} \\
 \mathbf{C} & \mathbf{D} \end{array} \right) = {\rm det} \left( \mathbf{A} - 
 \mathbf{B} \mathbf{D}^{-1} \mathbf{C} \right) {\rm det}  \mathbf{D} .
 \end{equation}
Given that the matrix $\mathbf{A}_{\bd{k} }^I$ is antihermitian, while
$\mathbf{V}_{\bd{k}}$, $\mathbf{T}_{\bd{k}}$ and $\mathbf{B}_{\bd{k} }^I$ are hermitian,
it is obvious that for the real frequencies the inverse propagator matrix
$\mathbf{G}^{-1}_0 (  \bd{k} , \omega  )$ is hermitian, 
so that it can be diagonalized by means of a unitary transformation. Then the spin-wave dispersions
can be obtained from the roots of the equation
 \begin{equation}
 {\rm det}  \mathbf{G}^{-1}_0 (  \bd{k} , \omega  ) =0.
 \label{eq:detGsw}
 \end{equation}
Obviously,  the calculation of the magnon spectrum 
simplifies if the matrix $\mathbf{W}_{\bd{k}}$ vanishes.
In this case the inverse propagator of the $X$-field is simply
\begin{equation}
  \mathbf{G}_0^{-1} ( \bd{k} , i \omega  )  = 
 \mathbf{V}_{\bd{k}} + \omega^2 \mathbf{T}_{\bd{k}}^{-1},
 \label{eq:G0simp}
 \end{equation}
so that Eq.~(\ref{eq:detGsw}) reduces to
 \begin{equation}
 {\rm det}  ( \mathbf{V}_{\bd{k}}   -       \omega^2 \mathbf{T}_{\bd{k}}^{-1}  ) = 0 ,
 \label{eq:detVT}
 \end{equation}
or equivalently
\begin{equation}
 {\rm det}  ( \mathbf{T}_{\bd{k}}   \mathbf{V}_{\bd{k}}   -       \omega^2   ) = 0 .
 \label{eq:detVT2}
 \end{equation}

In summary,  by expressing each Holstein-Primakoff boson in terms of two hermitian operators,
we can reduce the calculation of the
energy bands of a general $f$-flavor  boson Hamiltonian
of the type (\ref{eq:H2res})  to the calculation of a determinant of a
hermitian $f \times f$ matrix.
This is in contrast with the conventional algorithm \cite{Colpa78,Blaizot86,Maldonado93,Serga12} 
reviewed in Appendix~A, within which one has to solve a generalized eigenvalue equation involving a
non-hermitian  $2f \times 2 f$-matrix
$\mathbbm{M}_{\bd{k}}^{\rm dyn}$, see Eq.~(\ref{eq:eigenColpa}).
Another advantage of the hermitian field parametrization is that it allows one to identify special regimes in which
calculations of the magnon spectrum simplifies, significantly easier than in the conventional approach.
In fact, the next subsection shows that within the hermitian field approach, 
we can identify previously unnoticed special surfaces
in the parameter space of the Kitaev-Heisenberg-$\Gamma$ model,
on which the magnon spectrum and   eigenstates of the
Hamiltonian  can be calculated analytically. 
This enables us to go beyond the linear SWT and calculate  magnon 
damping in this regime.

\subsection{Analytically solvable  magnon spectrum in the zigzag state}
 \label{sec:spectrum}

\begin{figure}[t]
 	\includegraphics[width=\linewidth]{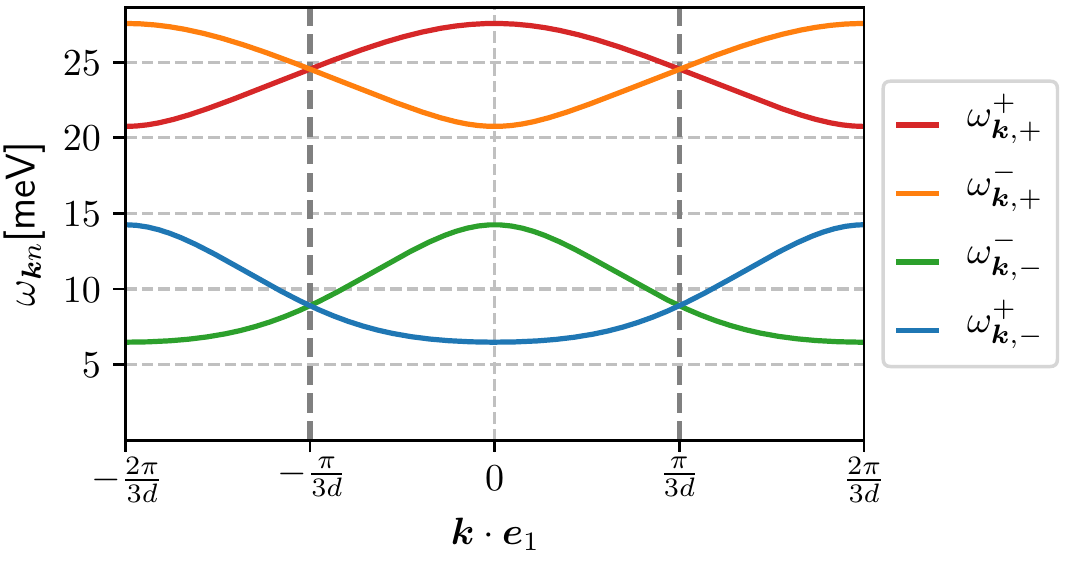}
 	\caption{%
 		  Dispersions of the four magnon branches 
 		in the Kitaev-Heisenberg-$\Gamma$ 
                               model given by Eqs.~(\ref{eq:dispersion3}) and
 		(\ref {eq:dispersion4}) for the model parameters $S=1/2$, $J=-12\text{meV}$, $K=\Gamma =7\text{meV}$,
 		and  $J_3=3\text{meV}$.
 		We show a cut through 
$\bm{k}=0$ along $\bm{e}_1$ perpendicular to the zigzag stripes, see Fig.~\ref{fig:zigzag}. 
The thicker dashed lines mark the boundary of the first magnetic Brillouin zone. Although the individual functions $\omega_{ \bd{k} , \pm }^{ +}$ and
 $\omega_{ \bd{k} , \pm }^{-} $ are not periodic within the first Brillouin zone, the full magnon spectrum is.
}
 \label{fig:magnonHKsec}
 \end{figure}

\begin{figure}[t]
 \includegraphics[width=75mm]{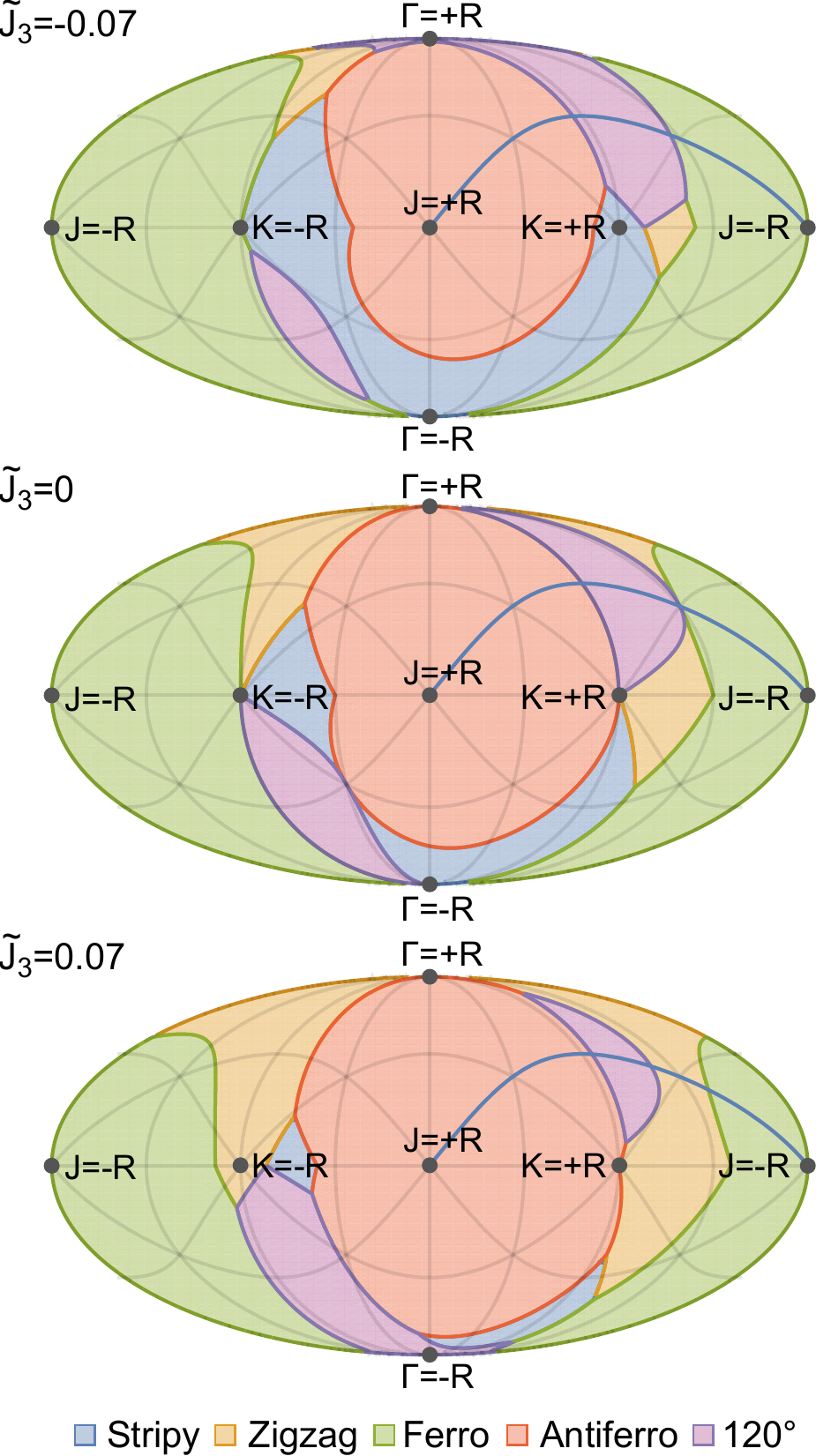}
 \caption{Phase diagram of the Kitaev-Heisenberg-$\Gamma$ model for representative values of $
 \tilde{J}_3 = J_3 / \sqrt{ \Gamma^2 + K^2 + J^2 } $. Note that the zigzag phase is stabilized for $J_3>0$.
The blue line corresponds to  $\Gamma = K > 0$,
which has a finite overlap with the zigzag phase for $J_3>0$.
	}
	\label{fig:phases2}
\end{figure}

At this  point, it is convenient to work with  the gauge  $\phi =0$
in the definition (\ref{eq:loc_frame_02}) of the  local transverse basis. Then the matrix element
 $\beta_{\bd{k}}$ in Eq.~(\ref{eq:betakdef}) has the  symmetry
$\beta_{ \bd{k}} = \beta_{ - \bd{k}}^{\ast}$ so that the matrix $\mathbf{A}_{\bd{k}}$ defined in
Eq.~(\ref{eq:Anew}) can be written as
 \begin{eqnarray}
 \mathbf{A}_{\bd{k}} & = &
 \left(
 \begin{array}{cc|cc}
 \lambda & \beta_{\bd{k}}& 0 & \alpha_{\bd{k}} \\
 \beta_{\bd{k}}^{\ast} & \lambda &  \alpha_{\bd{k}}^{\ast} & 0 \\
 \hline
 0 &  \alpha_{\bd{k}} & \lambda &  \beta_{\bd{k}} \\
 \alpha_{\bd{k}}^{\ast} & 0 &  \beta_{\bd{k}}^{\ast} & \lambda 
 \end{array}
 \right) .
 \label{eq:Anewbeta}
 \end{eqnarray}
Keeping in mind that  $\alpha_{\bd{k}} = \alpha_{ - \bd{k}}^{\ast}$, 
the antisymmetric part
$\mathbf{A}_{\bd{k}}^I $ of the matrix $\mathbf{A}_{\bd{k}}$ vanishes
in the zigzag state, so that
 $\mathbf{W}_{\bd{k}} = \mathbf{B}_{\bf{k}}^I$.
On the other hand, for $\phi =0$ the function $\nu_{\bd{k}}$ 
defined in Eq.~(\ref{eq:nucomplex}) has a part $\nu_{\bd{k} 2}$ violating the symmetry
 $\nu_{\bd{k}} = \nu_{ - \bd{k}}^{\ast}$.
To isolate this part, we write
 \begin{equation}
  \nu_{\bd{k}} = \nu_{\bd{k}1} + i \nu_{\bd{k} 2} ,
 \end{equation}
with
  \begin{eqnarray}
 \nu_{\bd{k}1}  & =  &   \frac{ \nu_{\bd{k}} + \nu^{\ast}_{ - \bd{k}}}{2}
 \nonumber
 \\ 
 & = &  S  \left[
 \frac{K}{4} \frac{r^2}{2 + r^2}   - \Gamma \frac{r}{ 2 + r^2 } \right]
 \left( e^{ i \bd{k} \cdot \bd{d}_x }
 +  e^{ i \bd{k} \cdot \bd{d}_y } \right),
 \hspace{7mm}
 \end{eqnarray}
and
 \begin{eqnarray}
 \nu_{\bd{k}2}  & =  &   \frac{ \nu_{\bd{k}} - \nu^{\ast}_{ - \bd{k}}}{2i}
 \nonumber
 \\ 
 & = & S  \frac{ K -  \Gamma r }{2} \sqrt{ \frac{2}{2 + r^2 } } \left( e^{ i \bd{k} \cdot \bd{d}_x }
 - e^{ i \bd{k} \cdot \bd{d}_y } \right).
 \end{eqnarray}
The two parts of the matrix $\mathbf{B}_{\bd{k}} = \mathbf{B}_{\bd{k}}^R + i  \mathbf{B}_{\bd{k}}^I$ are, therefore,
\begin{eqnarray}
 \mathbf{B}_{\bd{k}}^R 
 & = &  \frac{ \mathbf{B}_{\bd{k}} + \mathbf{B}_{ - \bd{k}}^{\ast}}{2} =
\left(
 \begin{array}{cc|cc}
 0 & \mu_{\bd{k}}& 0 & \nu_{\bd{k}1} \\
 \mu_{ \bd{k}}^{\ast} & 0 &  \nu_{\bd{k}1}^{\ast} & 0 \\
 \hline
 0 &   \nu_{\bd{k}1} & 0 &   \mu_{\bd{k}} \\
 \nu_{  \bd{k}1}^{\ast}  & 0 &  \mu_{ \bd{k}}^{\ast} & 0 
 \end{array}
 \right) ,
 \label{eq:BRnew}
 \end{eqnarray}
and
\begin{eqnarray}
 \mathbf{B}_{\bd{k}}^I 
 & = &  \frac{ \mathbf{B}_{\bd{k}} - \mathbf{B}_{ - \bd{k}}^{\ast}}{2i} = 
\left(
 \begin{array}{cc|cc}
 0 & 0& 0 & \nu_{\bd{k}2} \\
 0 & 0 &  -\nu_{\bd{k}2}^{\ast} & 0 \\
 \hline
 0 &   -\nu_{\bd{k}2} & 0 &  0 \\
 \nu_{  \bd{k}2}^{\ast}  & 0 &  0 & 0 
 \end{array}
 \right)  =  \mathbf{W}_{\bd{k}}  .
 \nonumber
 \\
 &  & 
 \label{eq:BInew}
 \end{eqnarray}
Then the matrices $\mathbf{T}_{\bd{k}}$ and $\mathbf{V}_{\bd{k}}$ are  given by
 \begin{eqnarray}
 & & \mathbf{T}_{\bd{k}}  =   \mathbf{A}_{\bd{k}} - \mathbf{B}_{\bd{k}}^R 
 \nonumber
 \\
 &  & =
 \left(
 \begin{array}{cc|cc}
 \lambda & \beta_{\bd{k}} - \mu_{\bd{k}} & 0 & \alpha_{\bd{k}} - \nu_{\bd{k}1} \\
  \beta_{\bd{k}}^{\ast}  - \mu_{\bd{k}}^{\ast}  & \lambda   & \alpha_{\bd{k}}^{\ast} - \nu_{\bd{k} 1}^{\ast} & 0  \\
 \hline
 0 & \alpha_{\bd{k}} - \nu_{\bd{k}1} &   \lambda &  \beta_{\bd{k}} - \mu_{\bd{k}} \\
 \alpha_{\bd{k}}^{\ast} - \nu_{\bd{k}1}^{\ast} & 0   & \beta_{\bd{k}}^{\ast} - \mu_{\bd{k}}^{\ast}  & \lambda
 \end{array}
 \right) ,
 \label{eq:Tkmatrix}
 \hspace{12mm}
 \end{eqnarray}
and
 \begin{eqnarray}
& &  \mathbf{V}_{\bd{k}}  =  \mathbf{A}_{\bd{k}} + \mathbf{B}_{\bd{k}}^R 
 \nonumber
 \\
 &  & =
 \left(
 \begin{array}{cc|cc}
 \lambda & \beta_{\bd{k}} + \mu_{\bd{k}} & 0 & \alpha_{\bd{k}} + \nu_{\bd{k}1} \\
  \beta_{\bd{k}}^{\ast}  + \mu_{\bd{k}}^{\ast}  & \lambda   & \alpha_{\bd{k}}^{\ast} + \nu_{\bd{k} 1}^{\ast} & 0  \\
 \hline
 0 & \alpha_{\bd{k}} + \nu_{\bd{k}1} &   \lambda &  \beta_{\bd{k}} + \mu_{\bd{k}} \\
 \alpha_{\bd{k}}^{\ast} + \nu_{\bd{k}1}^{\ast} & 0   & \beta_{\bd{k}}^{\ast} + \mu_{\bd{k}}^{\ast}  & \lambda
 \end{array}
 \right) .
 \label{eq:Vkmatrix}
 \hspace{12mm}
 \end{eqnarray}
Now, the crucial point is that  for  $ K = \Gamma r$, the matrix element  $\nu_{\bd{k} 2}$ 
and hence the matrix $\mathbf{W}_{\bd{k}}$ vanishes for all momenta.
In these case, $\nu_{\bd{k}} = \nu^{\ast}_{ - \bd{k}}$ and
the spin-wave dispersions can be obtained from Eq.~(\ref{eq:detVT2}).
The explicit solution of this biquadratic equation gives the squares of the
magnon dispersions,
\vskip 0.5cm
 \begin{widetext}
 \begin{eqnarray}
 & & ( \omega_{\bd{k}, \pm}^{+ })^2  =  \lambda^2 +  | \alpha_{\bd{k}} +  \beta_{\bd{k}} |^2 
- | \mu_{\bd{k}}  + \nu_{\bd{k}} |^2
 \pm  \sqrt{ 2 | \alpha_{\bd{k}} + \beta_{\bd{k}} |^2 ( 2 \lambda^2 - 
 | \mu_{\bd{k}} + \nu_{\bd{k}} |^2 ) + 2 {\rm Re} [ (\alpha_{\bd{k}} + \beta_{\bd{k}} ) ^2 ( \mu_{\bd{k}}^{\ast}
 + \nu_{\bd{k}}^{\ast} )^2 ] },
 \label{eq:dispersion3}
 \\
 & & (\omega_{\bd{k}, \pm}^{- })^2  =  \lambda^2 +  | \alpha_{\bd{k}} -  \beta_{\bd{k}} |^2 
- | \mu_{\bd{k}}  - \nu_{\bd{k}} |^2
 \pm  \sqrt{ 2 | \alpha_{\bd{k}} - \beta_{\bd{k}} |^2 ( 2 \lambda^2 - 
 | \mu_{\bd{k}} - \nu_{\bd{k}} |^2 ) + 2 {\rm Re} [ (\alpha_{\bd{k}} - \beta_{\bd{k}} ) ^2 ( \mu_{\bd{k}}^{\ast}
 - \nu_{\bd{k}}^{\ast} )^2 ] }.
 \label{eq:dispersion4}
 \end{eqnarray}
 \end{widetext}
With $r$ given by Eq.~(\ref{eq:smallrdef}),
the condition $ K = \Gamma r $, under which the spin-wave spectrum can be calculated analytically, can be written as
\begin{equation}
 \frac{ \Gamma}{K} = \frac{1}{r} =
 \frac{ 2 K - 3 \Gamma +  \sqrt{4 K^2 - 4 K \Gamma + 9 \Gamma^2 }       }{2 K + 3 \Gamma -  
  \sqrt{4 K^2 - 4 K \Gamma + 9 \Gamma^2 } }.
 \label{eq:rcondition}
 \end{equation}
For  negative $K$, this equation has only one trivial solution, $\Gamma\!=\! 0$,
but for $K \! >\! 0$ two non-trivial solutions exist,
 \begin{equation}
 \Gamma = K \; \; \; \mbox{and} \; \; \; \Gamma = - \frac{3}{2} K.
 \end{equation}
In Fig.~\ref{fig:magnonHKsec}, we plot the magnon dispersions
(\ref{eq:dispersion3}) and (\ref{eq:dispersion4})
for a representative set of parameters  satisfying $\Gamma = K >0$.
In the projected representations of the
three-dimensional parameter space of the Kitaev-Heisenberg-$\Gamma$ model in Fig.~\ref{fig:phases},
the parameters satisfying $\Gamma = K$ and $ \Gamma = -3 K /2$ with $K > 0$ 
are represented by the blue and orange lines, respectively.
However, one should keep in mind that we have assumed that the zigzag state is the classical ground state.
Therefore, the only meaningful parts of these lines are the ones which overlap with the zigzag phase.
As one can see in Fig.~\ref{fig:phases}, for the $\Gamma = -3 K/2$ line this condition is not met anywhere, while for
the $\Gamma = K > 0$ line there is a single point that touches the zigzag phase, which corresponds to $\Gamma = K=-J$.
Fortunately, adding  the experimentally relevant
third-nearest-neighbor coupling  $J_3 > 0$ to our model, 
the stability region of the zigzag phase is extended, while this extra term does not invalidate  
our analytic calculation of the magnon spectrum. 
In Fig.~\ref{fig:phases2}, we show the phase diagram of the
Kitaev-Heisenberg-$\Gamma$ model for representative values of $J_3 / \sqrt{ \Gamma^2 + K^2 + J^2 }$
in the same projection as in Fig.~\ref{fig:phases}(b) to demonstrate the expansion of the zigzag region for  $J_3 > 0$.

The underlying physical reason for the simplifications in the calculation of the magnon spectrum
 for $\Gamma = K$
is because in this case
the magnetization lies in the plane of the honeycomb lattice and is aligned 
with the direction $\bd{e}_2$
of the zigzag pattern, as  was pointed out in
Sec.~\ref{subsec:zigzag}, see Eq.~(\ref{eq:n3gk}). 
The fact that in this  case the magnon spectrum can be obtained by solving a biquadratic 
equation suggests that for $\Gamma = K$ it must be possible to set up the 
spin-wave expansion such that the magnon spectrum can be obtained from  
 two magnon bands defined in the full Brillouin zone of the honeycomb lattice.
In Appendix B we show that this is indeed the case, because one can simplify the spin Hamiltonian 
to the two-sublattice structure already in real space. Then, one needs only two bosonic flavors 
in order to block-diagonalize the quadratic magnon Hamiltonian 
in momentum space. In the following, we do not follow this path and continue with the original
four-sublattice formulation for the sake of generality.

\section{Magnon damping  in the zigzag state for $\Gamma = K > 0$}
 \label{sec:damping}

In this section, we will present a fully microscopic calculation of the
magnon damping of the Kitaev-Heisenberg-$\Gamma$ model with additional next-nearest-neighbor 
exchange $J_3$ on the  $\Gamma  = K > 0$ line.
Note that in Ref.~[\onlinecite{Winter17}],  matrix elements that determine  magnon damping 
have not been calculated microscopically, but have been estimated on the basis of reasonable
analogies with similar models.  
Here we show that  for $\Gamma = K > 0$, we can perform such calculations explicitly and in a fully microscopic fashion
 because in this case the magnon spectrum and all relevant matrix elements can be obtained analytically.

\subsection{Strategy}

Let us briefly summarize our  strategy.
The first step is to explicitly construct the multi-flavor Bogoliubov transformation 
that diagonalizes the quadratic magnon Hamiltonian. 
In principle, this can be done numerically using the algorithm 
developed by Colpa \cite{Colpa78}, see also
Refs.~[\onlinecite{Blaizot86,Maldonado93,Serga12}].
Fortunately, for $\Gamma = K > 0$ we can construct the Bogoliubov
transformation analytically, which considerably simplifies the numerical effort for the  calculation of 
magnon damping. Here we present a new algorithm to calculate the relevant four-flavor Bogoliubov transformation 
involving only hermitian $4 \times 4$ matrices.
Then, we express the cubic part of the Hamiltonian given in Eq.~(\ref{eq:H3res}) in terms of the Bogoliubov operators,
thus obtaining decay vertices explicitly,   and finally calculate the damping of magnons using perturbation theory.
In the earlier work on the generalized Kitaev-Heisenberg model by Winter {\it{et. al.}} \cite{Winter17}, 
the magnon damping was calculated by approximating momentum-dependent vertices in the cubic part
of the Hamiltonian  by a single momentum-independent constant. 
For $\Gamma = K > 0$, this approximation can be eliminated because explicit analytic expressions 
for the magnon dispersions and all  momentum dependent interaction vertices are available.
We compare the results of the two methods at the end of the section for a representative set of parameters. 
We also calculate the transverse components of the magnetic structure factor and the neutron scattering intensities
to demonstrate the effect of the magnon lifetime on them.

\subsection{Construction of the multi-flavor Bogoliubov transformation}
 \label{sec:construction}

To  diagonalize the quadratic part $S_2$ of the magnon action  defined in Eq.~(\ref{eq:S2a}), we 
first express this  action in terms of the hermitian fields defined in
Eq.~(\ref{eq:herm_01}),  then decouple the momentum modes
by means of a series of canonical transformations, and finally transform back to new complex fields which
completely diagonalize the action. To carry out this program it is convenient to 
use block matrix notations and write the quadratic magnon action $S_2$
defined in Eq.~(\ref{eq:S2a}) as
\begin{eqnarray}
S_2 & = & \frac{\beta}{2} \sum_{K}
\begin{pmatrix} \bm{a}_K \\ \bar{\bm{a}}_{-K} \end{pmatrix}^{\dagger}
\begin{pmatrix} \mathbf{A}_{\bm{k}} - i\omega  & \mathbf{B}_{\bm{k}} \\ \mathbf{B}_{\bm{k}}^\dagger & \mathbf{A}_{-\bm{k}}^T + i\omega \end{pmatrix}
\begin{pmatrix} \bm{a}_K \\ \bar{\bm{a}}_{-K} \end{pmatrix},
\nonumber \\
 \label{eq:S2four}
\end{eqnarray}
where the four-component vector
 \begin{equation}
\bm{a}_K = \left( \begin{array}{c} a_K \\ b_K \\ c_K \\ d_K 
 \end{array} \right) 
 \end{equation}
contains the four flavors of the Holstein-Primakoff magnons introduced in
Eq.~(\ref{eq:FT_01}). 

\subsubsection{Parametrization in terms of hermitian fields}

To begin, we express  each complex field in 
terms of  two real fields as in  Eq.~(\ref{eq:herm_01}). For our four-flavor theory 
the transformation can be written in a matrix form as
\begin{equation}
\begin{pmatrix} \bm{a}_{K} \\ \bar{\bm{a}}_{-K} \end{pmatrix}
=
\mathbb{N}
\begin{pmatrix} \bm{X}_{K} \\ \bm{P}_{K} \end{pmatrix}.
\label{eq:N_01}
\end{equation}
Here we have defined the $8 \times 8$ matrix
\begin{equation}
\mathbb{N}
=
\frac{1}{\sqrt{2}}
\begin{pmatrix} \bm{1} & i \bm{1} \\ \bm{1} & -i \bm{1} \end{pmatrix}   ,
\end{equation}
where $\mathbf{1}$ is the $4 \times 4$ identity matrix.
Then the  action $S_2$ in Eq.~(\ref{eq:S2four}) can  be written as
\begin{eqnarray}
&S_2& =
\frac{\beta}{2} \sum_{K}
\begin{pmatrix} \bm{X}_{-K}^T, & \bm{P}_{-K}^T \end{pmatrix}
\begin{pmatrix} \mathbf{V}_{\bm{k}} & \omega \\ - \omega & \mathbf{T}_{\bm{k}} \end{pmatrix}
\begin{pmatrix} \bm{X}_K \\ \bm{P}_{K} \end{pmatrix},
\label{eq:S2_02}
 \hspace{7mm}
\end{eqnarray}
with
\begin{subequations}
	\begin{eqnarray}
	\mathbf{V}_{\bm{k}} &= \mathbf{A}_{\bm{k}} + \mathbf{B}_{\bm{k}},
	\\
	\mathbf{T}_{\bm{k}} &= \mathbf{A}_{\bm{k}} - \mathbf{B}_{\bm{k}}.
	\end{eqnarray}
\end{subequations}
Here we have used that for  $\Gamma = K > 0$, 
the matrix $\mathbf{W}_{\bm{k}}$ that encodes the imaginary parts of the matrices 
$\mathbf{A}_{\bd{k}}$ and $\mathbf{B}_{\bd{k}}$ and is defined in Eq.~(\ref{eq:Wmatdef})
vanishes identically. This is the key for the following diagonalization as it simplifies the calculation significantly.
Note also that the $4 \times 4$ matrices  $\mathbf{V}_{\bm{k}}$ and  $\mathbf{T}_{\bm{k}}$ are
hermitian.

\subsubsection{Transformation to normal modes}

We now follow  the theories of coupled oscillators \cite{Goldstein80} and 
phonons \cite{Mahan90} and perform a series of canonical transformations
to decouple degrees of freedom with different momenta.
As a first step, we define new fields such that the ``kinetic energy matrix'' $\mathbf{T}_{\bd{k}}$
is transformed to the identity matrix.
Since $\mathbf{T}_{\bm{k}}$ is hermitian, we can construct a hermitian matrix 
$\mathbf{T}_{\bm{k}}^{\sfrac{1}{2}}$ with the property $( \mathbf{T}_{\bm{k}}^{\sfrac{1}{2}} )^2 = \mathbf{T}_{\bm{k}}$.
Therefore, we diagonalize  $\mathbf{T}_{\bm{k}}$ via  a unitary transformation,
	\begin{equation}
 \mathbf{U}_{\bm{k}}^\dagger \mathbf{T}_{\bm{k}} \mathbf{U}_{\bm{k}} =  \mathbf{D}_{\bm{k}}    
  \text{\ \ \ diagonal},
	\end{equation}
and define the square root $\mathbf{D}_{\bm{k}}^{\sfrac{1}{2}}$ of  $\mathbf{D}_{\bm{k}}$ in terms of the
square roots of the diagonal elements of $\mathbf{D}_{\bm{k}}$ such that 
 $(\mathbf{D}_{\bm{k}}^{\sfrac{1}{2}} )^2 = \mathbf{D}_{\bm{k}}$.
 The matrix $\mathbf{T}_{\bm{k}}^{\sfrac{1}{2}}$ is then defined by
	\begin{equation}
	\mathbf{T}_{\bm{k}}^{\sfrac{1}{2}} = \mathbf{U}_{\bm{k}} \mathbf{D}_{\bm{k}}^{\sfrac{1}{2}} \mathbf{U}_{\bm{k}}^\dagger.
	\label{eq:Thalf_definition}
	\end{equation}
Note that the inverse of $ \mathbf{T}_{\bm{k}}^{\sfrac{1}{2}}$ is given by
\begin{equation}
\mathbf{T}_{\bm{k}}^{\sfrac{-1}{2}}
\equiv
\left( \mathbf{T}_{\bm{k}}^{\sfrac{1}{2}} \right)^{-1}
=
\mathbf{U}_{\bm{k}} \mathbf{D}_{\bm{k}}^{\sfrac{-1}{2}} \mathbf{U}_{\bm{k}}^\dagger,
\end{equation}
An explicit expression for $\mathbf{T}_{\bm{k}}^{\sfrac{1}{2}}$ in our 4-flavor case is given in 
Eq.~(\ref{eq:Thalf_01}) of Appendix~C. 
With the  canonical transformation
\begin{equation}
\begin{pmatrix} \bm{X}_K \\ \bm{P}_K \end{pmatrix}
=
\begin{pmatrix}
\mathbf{T}_{\bm{k}}^{\sfrac{1}{2}} & 0 \\
0 & \mathbf{T}_{\bm{k}}^{-\sfrac{1}{2}}
\end{pmatrix}
\begin{pmatrix}
\bm{\tilde{X}}_K \\ \bm{\tilde{P}}_K
\end{pmatrix},
\end{equation}
the ``kinetic energy matrix''   in the action~(\ref{eq:S2_02}) is transformed to the identity matrix,
\begin{eqnarray}
S_2
=
\frac{\beta}{2} \sum_{K}
\begin{pmatrix} \bm{\tilde{X}}_{-K}^T, & \bm{\tilde{P}}_{-K}^T \end{pmatrix}
\begin{pmatrix} \mathbf{\tilde{V}}_{\bm{k}} & \omega  \\ - \omega 
& \bm{1} \end{pmatrix}
\begin{pmatrix} \bm{\tilde{X}}_K \\ \bm{\tilde{P}}_{K} \end{pmatrix},
 \hspace{7mm}
\label{eq:s2_03}
\end{eqnarray}
where the transformed ``potential energy matrix'' is
\begin{equation}
\mathbf{\tilde{V}}_{\bm{k}} = \mathbf{T}_{\bm{k}}^{\sfrac{1}{2}} \mathbf{V}_{\bm{k}} \mathbf{T}_{\bm{k}}^{\sfrac{1}{2}}.
\end{equation}
By construction,  $\mathbf{\tilde{V}}_{\bm{k}}$ is hermitian, so that it can be diagonalized
by means of another  unitary matrix $\mathbf{S}_{\bm{k}}$ 
\begin{equation}
\mathbf{S}_{\bm{k}} \mathbf{\tilde{V}}_{\bm{k}} \mathbf{S}_{\bm{k}}^{\dagger}
=
\mathbf{\Omega}_{\bm{k}}^2\ \ \text{diagonal},
\label{eq:Sunitary_definition}
\end{equation}
where the elements of the diagonal matrix $\mathbf{\Omega}_{\bm{k}}$ are the magnon 
energies, see Eq.~(\ref{eq:Omega}) of Appendix C.
An explicit expression for $\mathbf{S}_{\bm{k}}$ for the discussed case is given in Eq.~(\ref{eq:Sunit_01}).
With the canonical transformation
\begin{equation}
\begin{pmatrix} \bm{\tilde{X}}_K \\ \bm{\tilde{P}}_K \end{pmatrix}
=
\begin{pmatrix}
\mathbf{S}_{\bm{k}}   \mathbf{\Omega}_{\bd{k}}^{\sfrac{1}{2}}  & 0 \\
0 & \mathbf{S}_{\bm{k}}  \mathbf{\Omega}_{\bd{k}}^{-  \sfrac{1}{2}}
\end{pmatrix}
\begin{pmatrix}
\bm{X}_K^{\prime} \\ \bm{P}_K^{\prime}
\end{pmatrix},
\end{equation}
our quadratic magnon action~(\ref{eq:s2_03}) assumes the form
\begin{eqnarray}
S_2
=
\frac{\beta}{2} \sum_{K}
\begin{pmatrix} \bm{X}^{\prime T}_{-K}, & \bm{P}^{\prime T}_{-K} \end{pmatrix}
\begin{pmatrix} \mathbf{\Omega}_{\bm{k}}  & \omega \\ - \omega  &    \mathbf{\Omega}_{\bm{k}}  \end{pmatrix}
\begin{pmatrix} \bm{X}^{\prime}_K \\ \bm{P}^{\prime}_{K} \end{pmatrix}.
 \hspace{7mm}
\end{eqnarray}
As a result, the fluctuations with different momenta are now decoupled.

\subsubsection{Complete diagonalization via  complex fields}
Finally, we use the 
inverse transformation of Eq.~(\ref{eq:N_01}) to introduce  a new four-component
complex field $\bm{b}_K$ via 
\begin{eqnarray}
\begin{pmatrix} \bm{{X}}^{\prime}_{K} \\ \bm{{P}}^{\prime}_{K} \end{pmatrix}
=
\mathbb{N}^{-1}
\begin{pmatrix} \bm{b}_{K} \\ \bar{\bm{b}}_{-K} \end{pmatrix}
=
\frac{1}{\sqrt{2}}
\begin{pmatrix} \mathbf{1} &  \bm{1} \\ -i \bm{1} & i \bm{1} \end{pmatrix}
\begin{pmatrix} \bm{b}_{K} \\ \bar{\bm{b}}_{-K} \end{pmatrix}.
\nonumber\\
\end{eqnarray}
With this transformation, our quadratic magnon action is completely diagonalized
\begin{eqnarray}
&S_2&
= 
\frac{\beta}{2} \sum_{K}
\begin{pmatrix} \bm{{b}}_{K} \\ \bm{\bar{b}}_{-K} \end{pmatrix}^\dagger
\begin{pmatrix} \mathbf{\Omega}_{\bm{k}} - i \omega & 0 \\ 0 & \mathbf{\Omega}_{\bm{k}} 
 + i \omega  \end{pmatrix}
\begin{pmatrix} \bm{b}_K \\ \bm{\bar{b}}_{-K} \end{pmatrix}.
\nonumber \\
\end{eqnarray}
The described  chain of the canonical transformations defines the multi-flavor
Bogoliubov transformation and
can be expressed in terms of a single block matrix $\mathbb{T}_{\bd{k}}$ as,
\begin{equation} 
\begin{pmatrix} \bm{a}_{K} \\ \bar{\bm{a}}_{-K} \end{pmatrix}
=
\mathbb{T}_{\bd{k}}
\begin{pmatrix} \bm{b}_{K} \\ \bar{\bm{b}}_{-K} \end{pmatrix},
\label{eq:M_01}
\end{equation}
where
\begin{eqnarray} 
\mathbb{T}_{\bd{k}}
& = &
\mathbb{N}
\begin{pmatrix}
  \mathbf{T}_{\bm{k}}^{\sfrac{1}{2}}\ \mathbf{S}_{\bm{k}}       \ \mathbf{\Omega}_{\bm{k}}^{-\sfrac{1}{2}} & 0 \\
0 &   \mathbf{T}_{\bm{k}}^{-\sfrac{1}{2}}\ \mathbf{S}_{\bm{k}}   \ \mathbf{\Omega}_{\bm{k}}^{\sfrac{1}{2}}
\end{pmatrix}
\mathbb{N}^{-1}
 \nonumber
 \\
 & = &  \left( \begin{array}{cc}  
\mathbf{Q}_{\bd{k}} &  \mathbf{R}_{\bd{k}} \\
\mathbf{R}_{\bd{k}} &  \mathbf{Q}_{\bd{k}} 
 \end{array} \right),
\label{eq:M_02}
\end{eqnarray}
with the 
$4 \times 4$ matrices $\mathbf{Q}_{\bd{k}}$ and $\mathbf{R}_{\bd{k}}$
are given by
 \begin{eqnarray}
 \mathbf{Q}_{\bd{k}} & = & 
 \frac{1}{2} \left[
\mathbf{T}_{\bm{k}}^{\sfrac{1}{2}}\ \mathbf{S}_{\bm{k}}       \ \mathbf{\Omega}_{\bm{k}}^{-\sfrac{1}{2}} +
 \mathbf{T}_{\bm{k}}^{-\sfrac{1}{2}}\ \mathbf{S}_{\bm{k}}   \ \mathbf{\Omega}_{\bm{k}}^{\sfrac{1}{2}} \right],
 \label{eq:ublock}
 \\
\mathbf{R}_{\bd{k}} & = &  \frac{1}{2}
 \left[  \mathbf{T}_{\bm{k}}^{\sfrac{1}{2}}\ \mathbf{S}_{\bm{k}}       \ \mathbf{\Omega}_{\bm{k}}^{-\sfrac{1}{2}} -
 \mathbf{T}_{\bm{k}}^{-\sfrac{1}{2}}\ \mathbf{S}_{\bm{k}}   \ \mathbf{\Omega}_{\bm{k}}^{\sfrac{1}{2}} \right].
 \label{eq:vblock}
 \end{eqnarray}
Note that in the case considered here the
matrices $\mathbf{Q}_{\bd{k}}$ and $\mathbf{R}_{\bd{k}}$
satisfy $\mathbf{Q}_{\bd{k}} =   \mathbf{Q}^{\ast}_{- \bd{k}}$ and
 $\mathbf{R}_{\bd{k}} =   \mathbf{R}^{\ast}_{- \bd{k}}$,
so that the parametrization (\ref{eq:M_02}) agrees with
the general structure of the transformation matrix 
$\mathbb{T}_{\bd{k}}$ of a multi-flavor Bogoliubov tranformation 
given Eq.~(\ref{eq:TQRgeneral}) of Appendix~A. 
The explicit expressions for the $4 \times 4$ matrices $\mathbf{L}_{\bd{k}} = 
\mathbf{T}_{\bm{k}}^{\sfrac{1}{2}}\ \mathbf{S}_{\bm{k}}$
and $\mathbf{Y}_{\bd{k}}   =   \mathbf{T}_{\bm{k}}^{-\sfrac{1}{2}}\ \mathbf{S}_{\bm{k}}$ 
are given in Eqs.~(\ref{eq:L_01})~and~(\ref{eq:Y_01}) of Appendix C.
Moreover, in Appendix B we use the procedure described above to calculate 
magnon spectrum of the Kitaev-Heisenberg-$\Gamma$  model for
$\Gamma = K > 0$ in an alternative two-sublattice approach where
$\mathbf{T}_{\bd{k}}^{\sfrac{1}{2}}$ and $\mathbf{S}_{\bd{k}}$ are 
only $2 \times 2 $ matrices.

\subsection{Transformation of the cubic interaction}

For the  calculation of   magnon damping,
we have to express the Euclidean action 
associated with the cubic Hamiltonian ${\cal{H}}_3$
given in Eq.~(\ref{eq:H3res})
in terms of the components of the Bogoliubov fields $\bd{b}_{K}$ and
$\bar{\bd{b}}_K$, which diagonalize the quadratic part of the action.
For $\Gamma = K > 0$, the Euclidean action associated with 
${\cal{H}}_3$  in terms of the original Holstein-Primakoff fields is given by
   \begin{eqnarray}
 {\cal{S}}_3 & = &    \beta \sqrt{ \frac{4}{N}}
 \sum_{ K_1  K_2  K_3 }  \sum_{\bd{G}} 
\delta_{ \bd{k}_1 + \bd{k}_2 + \bd{k}_3 ,  \bd{G}} \delta_{ \omega_1 + \omega_2 + \omega_3 , 0 }
\Biggl\{ 
 \nonumber
 \\
 &  &  \hspace{9mm} -  V_{ \bd{k}_1 }    \left[  
 \bar{d}_{ - K_1 }   \bar{a}_{ - K_2 } a_{K_3 } 
 + d_{K_1} a_{K_2} \bar{a}_{ - K_3 } \right]
 \nonumber
 \\
& &  + e^{ i \bd{G} \cdot \bd{d}_z } 
  V^{\ast}_{ \bd{k}_1} \left[     \bar{c}_{ - K_1} \bar{b}_{ - K_2} b_{K_3} 
 + c_{K_1} b_{ K_2} \bar{b}_{ - K_3} \right]
 \nonumber
 \\
 &  & +   e^{i \bd{G} \cdot \bd{a}_1 }    V_{  \bd{k}_1 }    \left[
 \bar{b}_{ -K_1} \bar{c}_{ -K_2} c_{ K_3}  
 + b_{ K_1} c_{K_2} \bar{c}_{ - K_3}
 \right]
 \nonumber
 \\
 &   & -   e^{i \bd{G} \cdot \bd{d}_x }   V_{\bd{k}_1 }^{\ast}   \left[ 
  \bar{a}_{ - K_1 } \bar{d}_{ - K_2} d_{K_3} 
 + a_{K_1} d_{ K_2} \bar{d}_{ - K_3 } \right]
 \Biggr\}.
 \nonumber
 \\
 & &
 \label{eq:S3res}
 \end{eqnarray}
Defining the $8$-component field
 \begin{equation}
 ( \phi_K^{\mu} ) =
 \left( \begin{array}{c}
 \phi_K^1 \\ \phi^2_K \\ \phi^3_K  \\ \phi^4_K \\
 \phi^5_{  K } \\ \phi^6_{  K } \\ \phi^7_{  K } \\ \phi^8_{  K } 
 \end{array}
 \right)
=
 \left( \begin{array}{c}
 a_K \\ b_K \\ c_K  \\ d_K \\
 \bar{a}_{ - K } \\ \bar{b}_{ - K } \\ \bar{c}_{ - K } \\ \bar{d}_{ - K } 
 \end{array}
 \right),
 \end{equation} 
where the index $\mu \in \{ 1, 2,3,4,5,6,7,8 \}$ labels the 
eight different field types,
we can write the cubic part of the action in tensor notation
    \begin{eqnarray}
 {\cal{S}}_3 & = &    \beta \sqrt{ \frac{4}{N}}
 \sum_{ K_1  K_2  K_3 }  \sum_{\bd{G}} 
\delta_{ \bd{k}_1 + \bd{k}_2 + \bd{k}_3 ,  \bd{G}} \delta_{ \omega_1 + \omega_2 + \omega_3 , 0 }
 \nonumber
 \\
 & \times & \frac{1}{3!}  \sum_{ \mu \nu \lambda} \Gamma^{\mu \nu \lambda} ( \bd{k}_1 ,
 \bd{k}_2 , \bd{k}_3 )  \phi^{\mu}_{ K_1 } \phi^{\nu}_{ K_2 } \phi^{\lambda}_{ K_3 }.
 \label{eq:S3tensor}
 \end{eqnarray} 
The vertex 
$\Gamma^{\mu \nu \lambda} ( \bd{k}_1 ,
 \bd{k}_2 , \bd{k}_3 ) $ is fully symmetric with respect to the exchange of any of its  three index pairs.
In Eqs.~(\ref{eq:threevertices1})--(\ref{eq:threevertices4}) of Appendix C we list the  $48$ non-zero index combinations.
Next, we express $S_3$ in terms of the Bogoliubov   fields $\bd{b}_K$ and
$\bar{\bd{b}}_{-K}$ defined via the transformation (\ref{eq:M_01}), which we write as
 \begin{equation}
 \phi^{\mu}_{K} = \sum_{\mu^{\prime}} 
  \mathbb{T}_{\bd{k}}^{\mu \mu^{\prime}} \psi^{\mu^{\prime}}_{K},
 \end{equation}
where the $8 \times 8$ transformation matrix $\mathbb{T}_{\bd{k}}$ is
defined in Eq.~(\ref{eq:M_02}) and
 \begin{equation}
  (\psi^{\mu^{\prime}}_{K}) =\begin{pmatrix} \bm{b}_{K} \\ \bar{\bm{b}}_{-K} \end{pmatrix}
 \end{equation}
is an $8$-component field that contains  Bogoliubov bosons
$\bd{b}_K$ and their conjugates
$\bar{\bd{b}}_{-K}$.
Then, the cubic part of the action assumes the form
    \begin{eqnarray}
 {\cal{S}}_3 & = &    \beta \sqrt{ \frac{4}{N}}
 \sum_{ K_1  K_2  K_3 }  \sum_{\bd{G}} 
\delta_{ \bd{k}_1 + \bd{k}_2 + \bd{k}_3 ,   \bd{G}} 
\delta_{ \omega_1 + \omega_2 + \omega_3 , 0 }
 \nonumber
 \\
 & \times & \frac{1}{3!}  \sum_{ \mu \nu \lambda} \tilde{\Gamma}^{\mu \nu \lambda} ( \bd{k}_1 ,
 \bd{k}_2 , \bd{k}_3 )  \psi^{\mu}_{ K_1 } \psi^{\nu}_{ K_2 } \psi^{\lambda}_{ K_3 },
  \label{eq:S3tensor1}
\end{eqnarray} 
with
\begin{eqnarray}
\tilde{\Gamma}^{\mu \nu \lambda} ( {\bm{k}}_1,  \bm{k}_2 , \bm{k}_3 ) 
& = &
\sum_{\mu'\nu'\lambda'}
{\Gamma}^{\mu^{\prime} \nu^{\prime} \lambda^{\prime}} ( {\bm{k}}_1,  \bm{k}_2 , \bm{k}_3 ) 
 \nonumber
 \\
 & & \times
  \mathbb{T}_{\bm{k}_1}^{ \mu'\mu}
 \mathbb{T}_{\bm{k}_2}^{ \nu'\nu}
  \mathbb{T}_{\bm{k}_3}^{\lambda'\lambda}.
 \hspace{7mm}
 \label{eq:tildeGammadef}
\end{eqnarray}
The explicit  analytic expressions for the elements of the 
transformed tensor
$\tilde{\Gamma}^{\mu \nu \lambda} ( {\bm{k}}_1,  \bm{k}_2 , \bm{k}_3 ) $
are rather lengthy, but can be easily implemented in the symbolic manipulation software  MATHEMATICA.

\subsection{Magnon damping in Born approximation}
 \label{sec:magnondamp}

To the leading order in the  expansion in powers of $1/S$, 
the magnon damping is determined by the magnon self-energies that involve squares of the cubic vertices,
 \begin{equation}
 \Sigma_n ( K ) = \Sigma_n^{(a)} ( K ) + 
\Sigma_n^{(b)} ( K ) + \Sigma_n^{(c)} ( K ),
  \label{eq:abc_Sigma}
\end{equation}
where the index $n=1,2,3,4$ and the indices $n^{\prime}$ and $m$ below
 label  the four magnon bands. The three contributions $(a)$, $(b)$, and $(c)$ in (\ref{eq:abc_Sigma})
 can be represented by the following Feynman diagrams,
 \begin{subequations}
\begin{eqnarray}
\Sigma_{ n}^{(a)} ( K ) &=& -\frac{2}{2\beta} (3!)^2 \; \bubbleA, \\
\Sigma_{ n}^{(b)} (K ) &=& -\frac{1}{2\beta} (3!)^2 \;  \bubbleB, \\
\Sigma_{ n}^{(c)} (K )  &=& -\frac{1}{2\beta} (3!)^2 \; \bubbleC,
\end{eqnarray}
 \end{subequations}
where arrows denote magnon propagators and dots represent cubic vertices.
The combinatorial factor of $(3!)^2$ is due to the fact that we have 
unified all  cubic vertices as symmetric tensors of dimension 8 instead of splitting them into different 
decay and absorption channels.

Let us now discuss the explicit evaluation of the first contribution
 $\Sigma_{ n}^{(a)} ( K ) $ to the self-energy,
\begin{eqnarray}
& & \Sigma_{ n}^{(a)} ( K )
=
-\frac{2 }{2\beta} (3!)^2 \;  \bubbleA 
\nonumber\\
& & = 
-\frac{4 \beta}{N(3!)^2} \sum_{K'} \sum_{ n^{\prime} m } (3!)^2 
\frac{1}{\beta} G^0_{ n^{\prime}} ( K^{\prime})  \frac{1}{\beta} 
G^0_{m} (G_1-K-K') 
\nonumber
  \\
 & &  \hspace{22mm} \times
\left| 
 \tilde{\Gamma}^{n n^{\prime}  m+4} ( \bm{k} , \bm{k'} ,\bm{G}_{1} -\bm{k}- \bm{k'} ) \right|^2 
\nonumber\\
& & =
-\frac{4}{N}  \sum_{\bm{k'}}  \sum_{n^{\prime} m }   \left| 
\tilde{\Gamma}^{n n^{\prime}  m+4} ( \bm{k} , \bm{k'} ,\bm{G}_{1} -\bm{k}- \bm{k'} )
\right|^2 
\nonumber\\
& & \hspace{10mm} \times  \frac{1}{\beta} \sum_{\omega'} \frac{1}{\omega_{\bm{k'} n^{\prime}}-i\omega'} \frac{1}{\omega_{\bm{G}_1 - \bm{k} - \bm{k'},  m }-i \omega - i \omega'},
\nonumber
\\
& &
 \label{eq:dampa}
\end{eqnarray}
where $G^0_n ( K ) = [ i \omega - \omega_{\bd{k} n} ]^{-1}$ is a non-interacting 
propagator of magnons with the band index $n$ and the reciprocal lattice vector
$\bd{G}_{1} = \bd{G}_{1} ( \bd{k} + \bd{k}^{\prime} )$
has to be chosen such that $\bd{k} + \bd{k}^{\prime} - \bd{G}_{1}$ lies in the
first Brillouin zone. We also defined $G_1 = (0,\bd{G}_{1})$.
The  superscript $m+4$ can have  values from  $5$ to $8$ corresponding to the
last four possible values of the index $\lambda$
in $\tilde{\Gamma}^{\mu \nu \lambda} ( {\bm{k}}_1,  \bm{k}_2 , \bm{k}_3 ) $.
The sum over the Matsubara frequencies in the last line of 
Eq.~(\ref{eq:dampa}) 
can be performed analytically using the identity
\begin{eqnarray}
F^{(a)}(E_1, E_2, i\omega)
&\equiv &  \frac{1}{\beta}
\sum_{\omega'} \frac{1}{E_1-i\omega'} \frac{1}{E_2-i \omega- i \omega'} 
\nonumber\\
&=&
\frac{   n(E_1)-n(E_2)   }{E_1-E_2+i\omega} ,
\end{eqnarray}
where
\begin{equation}
n(E) = \frac{1}{e^{\beta E}-1}
\end{equation}
is the Bose function.
After analytic continuation to real frequencies
$i\omega\to\omega+i0^+$  we can extract the imaginary part of $F^{(a)} (E_1,E_2, \omega + i 0^+)$,
\begin{eqnarray}
& & \text{Im} \ F^{(a)}(E_1,E_2,\omega + i0^+) \nonumber\\
& = &-\pi  \left[ n(E_1)-n(E_2) \right] \delta \left( \omega + E_1 - E_2 \right).
\end{eqnarray}
From this, we obtain the imaginary part of the self-energy for the external frequencies
infinitesimally above the real frequency axis,
\begin{eqnarray}
 \text{Im} \ \Sigma_{n}^{(a)}(& &\bm{k},\omega + i 0^+) =
 \nonumber
 \\
& & 
\pi \frac{4}{N}  \sum_{\bm{k'}} \sum_{n^{\prime} m }
\left| 
 \tilde{\Gamma}^{n n^{\prime}  m+4} ( \bm{k} , \bm{k'} ,\bm{G}_{1} -\bm{k}- \bm{k'} ) \right|^2 
\nonumber\\
& &\times \left[ n(\omega_{\bm{k'}n'})-n(\omega_{\bm{G}_1 - \bm{k} - \bm{k'} m } ) \right] 
\nonumber
\\
& &\times
\delta \left( \omega+\omega_{\bm{k'}n^{\prime}}-
 \omega_{ \bm{G}_1 - \bm{k} - \bm{k'} m} \right).
\label{eq:imSigA_01}
 \nonumber
 \\
 & &
\end{eqnarray}
The second contribution to the self-energy can be evaluated 
analogously with the result
\begin{eqnarray}
\text{Im} \ \Sigma_{n}^{(b)}(& &\bm{k},\omega + i 0^+) = 
 \nonumber
\\
& & - \frac{\pi}{2} \frac{4}{N}  \sum_{\bm{k'}} \sum_{n^{\prime} m}  
\left| \tilde{\Gamma}^{n+4 \; n^{\prime} m} ( -\bm{k},\bm{k'} , \bd{G}_{2} + \bm{k}-\bm{k'} ) \right|^2 
 \nonumber 
\\ & & \times 
\left[ 1+ n(\omega_{\bm{k'} n^\prime })+n(\omega_{\bm{G}_2 + \bm{k}-\bm{k'} m } ) \right]  
\nonumber
\\
& &\times
\delta \left( \omega-\omega_{ \bm{k'} n^{\prime} }   
 - \omega_{\bm{G}_2 + \bm{k}-\bm{k'}  m} \right),
 \nonumber
 \\
 & &
\label{eq:imSigB_01}
\end{eqnarray}
where the reciprocal lattice vector $\bd{G}_{2} = \bd{G}_2 ( \bd{k}^{\prime} - \bd{k} )$ should be chosen such that  $\bd{k}^{\prime} - \bd{k} - \bd{G}_2$ is in the first Brillouin zone. Finally, the third contribution is
\begin{eqnarray}
\text{Im} \ \Sigma_{n}^{(c)}(& &\bm{k},\omega + i 0^+)= 
\nonumber
\\
 & & \frac{\pi}{2}
\frac{4}{N} \sum_{\bm{k'}}  \sum_{n^{\prime} m}    \left| 
\tilde{\Gamma}^{n n^{\prime}  m} ( \bm{k} ,\bm{k'},   \bd{G}_1 - \bm{k}- \bm{k'} ) \right|^2  \nonumber 
\\
& &
\times \left[ 1+n(\omega_{\bm{k'}n^{\prime}})+n(\omega_{ \bd{G}_1 - \bm{k}- \bm{k'} m} ) \right]
\nonumber
\\
& &\times
 \delta \left( \omega+\omega_{\bm{k'}n'}+\omega_{\bd{G}_1 - \bm{k}- \bm{k'} m} \right).
 \nonumber
 \\
 & &
\label{eq:imSigC_01}
\end{eqnarray}
Then, the magnon damping in band $n$ is  given by
\begin{equation}
\gamma_{\bm{k} n}
=
- \text{Im} \ \Sigma_{n}(\bm{k}, \omega_{\bm{k} n} + i 0^+).
\end{equation}
The Bose functions appearing in Eqs.~(\ref{eq:imSigA_01})--(\ref{eq:imSigC_01}) vanish at $T=0$. Moreover, 
the argument of the  delta-function in  (\ref{eq:imSigC_01}) for 
$\text{Im} \ \Sigma_{n}^{(c)}(\bm{k},\omega + i 0^+)$  is always finite so 
this term does not contribute to the magnon damping. 
Therefore, at $T=0$ we  obtain for the magnon damping in the lowest-order (Born) approximation
\begin{eqnarray}
\gamma_{\bm{k}n}
&=& -
\text{Im} \ \Sigma_{n}^{(b)}(\bm{k}, \omega_{\bm{k}n} + i 0^+)
\nonumber\\
&=&
\frac{\pi}{2}  \frac{4}{N}  \sum_{\bm{k'}} \sum_{n' m}   
\left| \tilde{\Gamma}^{n+4 \; n^{\prime} m} ( -\bm{k},\bm{k'} , \bd{G}_{2} + \bm{k}-\bm{k'} ) \right|^2 
\nonumber\\
& &\times \delta \left( \omega_{\bm{k}n}-\omega_{\bm{k'}n'}-
 \omega_{\bd{G}_{2} + \bm{k}-\bm{k'} m} \right).
\label{eq:damp_01}
\end{eqnarray}

\subsection{Numerical evaluation of the damping}
\label{sec:numerical_Born}

\begin{figure}
	\includegraphics[width=0.9\linewidth]{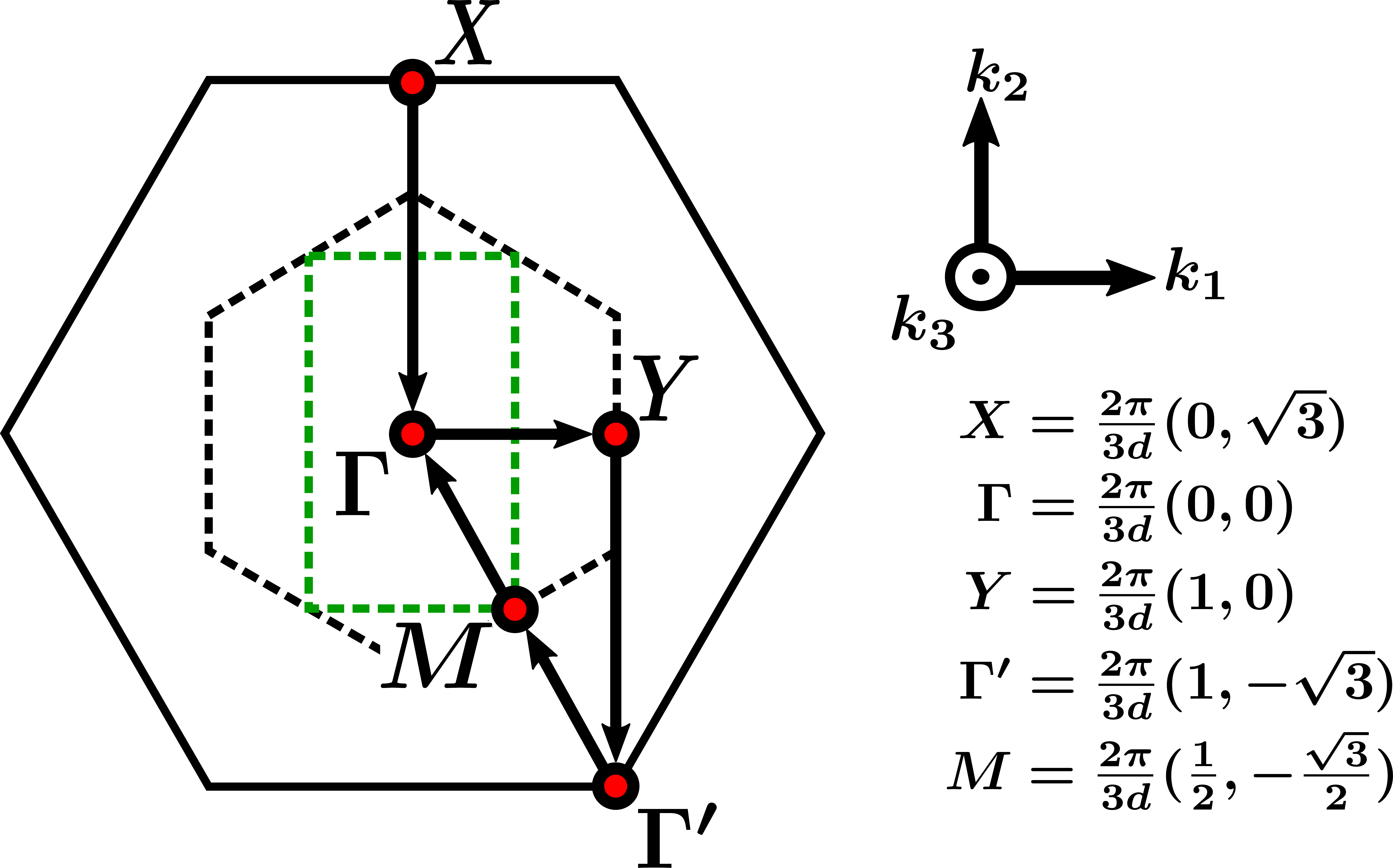}
	\caption{
		The momentum-space path used in our numerical evaluation of the magnon damping.
		The dashed black hexagon indicates the Brillouin zone of the honeycomb lattice.
		The dashed green rectangle indicates the first magnetic Brillouin zone of the zigzag state, see Fig.~\ref{fig:zigzag}. 
		The component $k_1$ is perpendicular to the zigzag stripes. 
	}
	\label{fig:contour}
\end{figure}

In the thermodynamic limit  ($N \rightarrow \infty$), the momentum sums can be converted to the integrals over the first Brillouin zone.
Furthermore, one can omit the reciprocal lattice vector $\bd{G}_{2}$ in Eq.~(\ref{eq:damp_01}) because $S_2$ in Eq.~(\ref{eq:S2four}) and $S_3$ in Eq.~(\ref{eq:S3tensor}) are invariant if one shifts one of the summation momenta by a reciprocal lattice vector.
Eq.~(\ref{eq:damp_01}) can then be written as
\begin{eqnarray}
\gamma_{\bm{k}n}
&=&
 \frac{\pi}{2}
\int_{\text{BZ}} \frac{\diff^2 k' }{V_{\text{BZ}}}
\sum_{n' m}
\left| \tilde{\Gamma}^{n+4 \; n^{\prime} m} ( -\bm{k},\bm{k'} , \bm{k}-\bm{k'} ) \right|^2 
\nonumber\\
& &\times \delta \left( \omega_{\bm{k}n}-\omega_{\bm{k'}n'}-
\omega_{\bm{k}-\bm{k'} m} \right),
\label{eq:damping_integral}
\end{eqnarray}
where $V_{\text{BZ}}$ is the area of the first magnetic Brillouin zone marked  
by the dashed green line in Fig.~\ref{fig:contour}. We evaluate this expression for 
a representative momentum $\bm{k}$ cut along the path shown in Fig.~\ref{fig:contour}. 
Note that for the calculations of the neutron-scattering structure factor in the next section,
we also choose  a finite out-of-plane momentum component $k_3 = \sqrt{3} \pi /d$.

\begin{figure}
\includegraphics[width=0.80\linewidth]{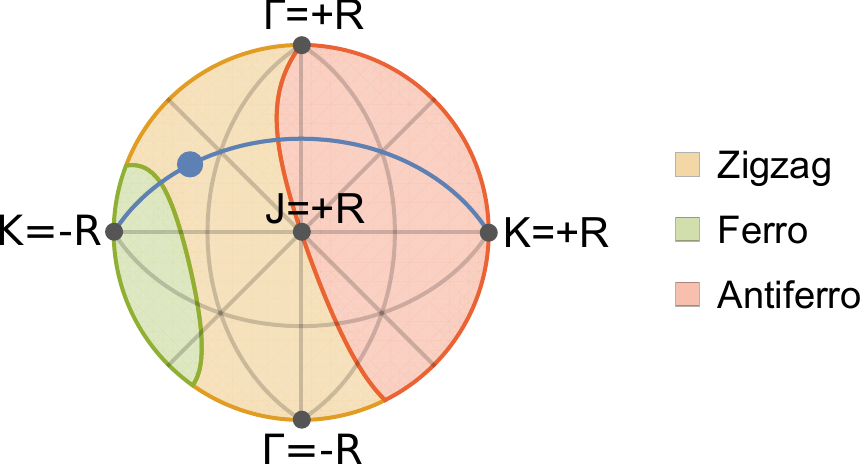}
\caption{Phase diagram of the Kitaev-Heisenberg-$\Gamma$ model for 
$J^2+K^2+\Gamma^2 = 242\text{\text{meV}}^2$ with additional third nearest neighbor Heisenberg exchange  
$J_3 = 3\text{\text{meV}}$. We use the same parametrization and projection as in Fig.~\ref{fig:phases}(c). 
We highlight the line $\Gamma = K > 0$ and the point~(\ref{eq:point}) in the parameter space, for which the 
magnon damping is calculated.}
\label{fig:point}
\end{figure}

For the numerical calculations, we use a representative set of the model parameters
\begin{subequations}
\label{eq:point}
\begin{eqnarray}
S &=& \frac{1}{2}, \\
J &=& -12\text{\text{meV}}, \\
K=\Gamma &=& \hphantom{-}7\text{\text{meV}}, \\
J_3 &=& \hphantom{-}3\text{\text{meV}}.
\end{eqnarray}
\end{subequations}
This set of parameters is shown in Fig.~\ref{fig:point} as a blue dot 
along the $K=\Gamma$ line.
The integration procedure for the two-dimensional integral over the Brillouin zone was 
implemented using the standard routines in MATHEMATICA with the $\delta$-function 
in Eq.~(\ref{eq:damping_integral}) represented as a  Lorentzian of width 
$w= \frac{3}{4} |J| \times 10^{-3}$, which is 
much smaller than all the characteristic features produced by the calculation.

\begin{figure}
	\includegraphics[width=0.95\linewidth]{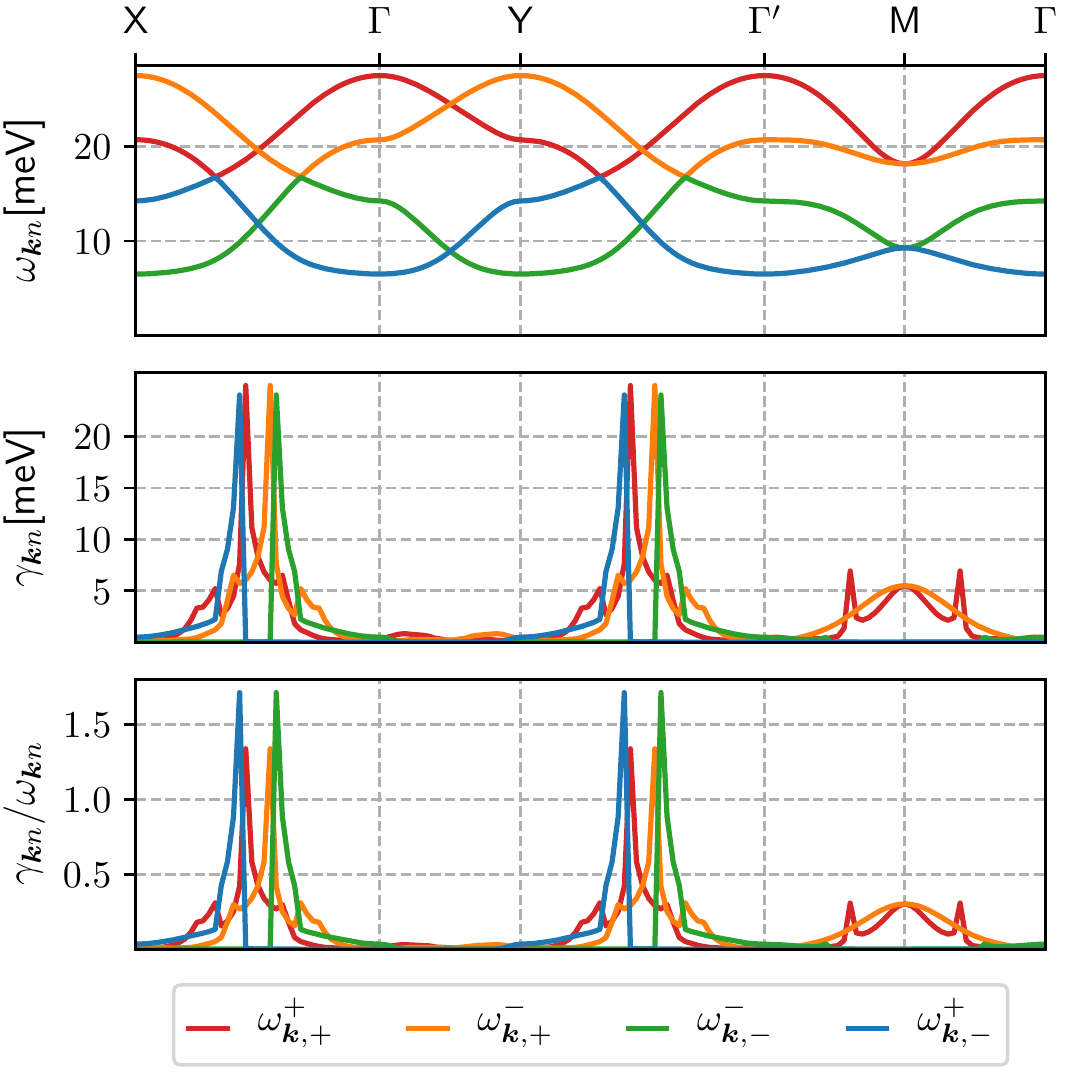}
	\caption{
Magnon damping in the Born approximation (\ref{eq:damping_integral}). 
The panels show the magnon energies $\omega_{\bm{k}n}$ (top),
the magnon damping $\gamma_{\bm{k}n}$ (middle),
and the relative magnon damping $\gamma_{\bm{k}n}/\omega_{\bm{k}n}$ (bottom) for all four magnon branches.
The model parameters are given in Eqs.~(\ref{eq:point}) and the momentum path is 
shown in Fig.~\ref{fig:contour}. The color-coding of the damping is the same as for the magnon energies.}
\label{fig:damping_born}
\end{figure}

The resulting magnon damping in Born approximation is plotted in Fig.~\ref{fig:damping_born}.
The overall magnon decay rates are rather significant.
The most striking features are the peaks between the $X$ and $\Gamma$ points and between the $Y$ and $\Gamma'$ points
that occur in a proximity of the magnon band crossings. 
These features are due to the van Hove singularities in the density of two-magnon states
that are also enhanced by the decay matrix elements, which facilitate transitions between the nearby branches.
It is also interesting to note that the lower magnon bands experience as much of a damping as the upper ones, 
despite the naive expectation for them having less kinematic phase space for decays.  
Such van Hove singularities are expected \cite{RMP} and need to be regularized, as we do in the next subsection.

\subsection{Beyond Born approximation: self-consistent imaginary Dyson equation}
\label{sec:iDE}

\begin{figure}[b]
	\includegraphics[width=0.95\linewidth]{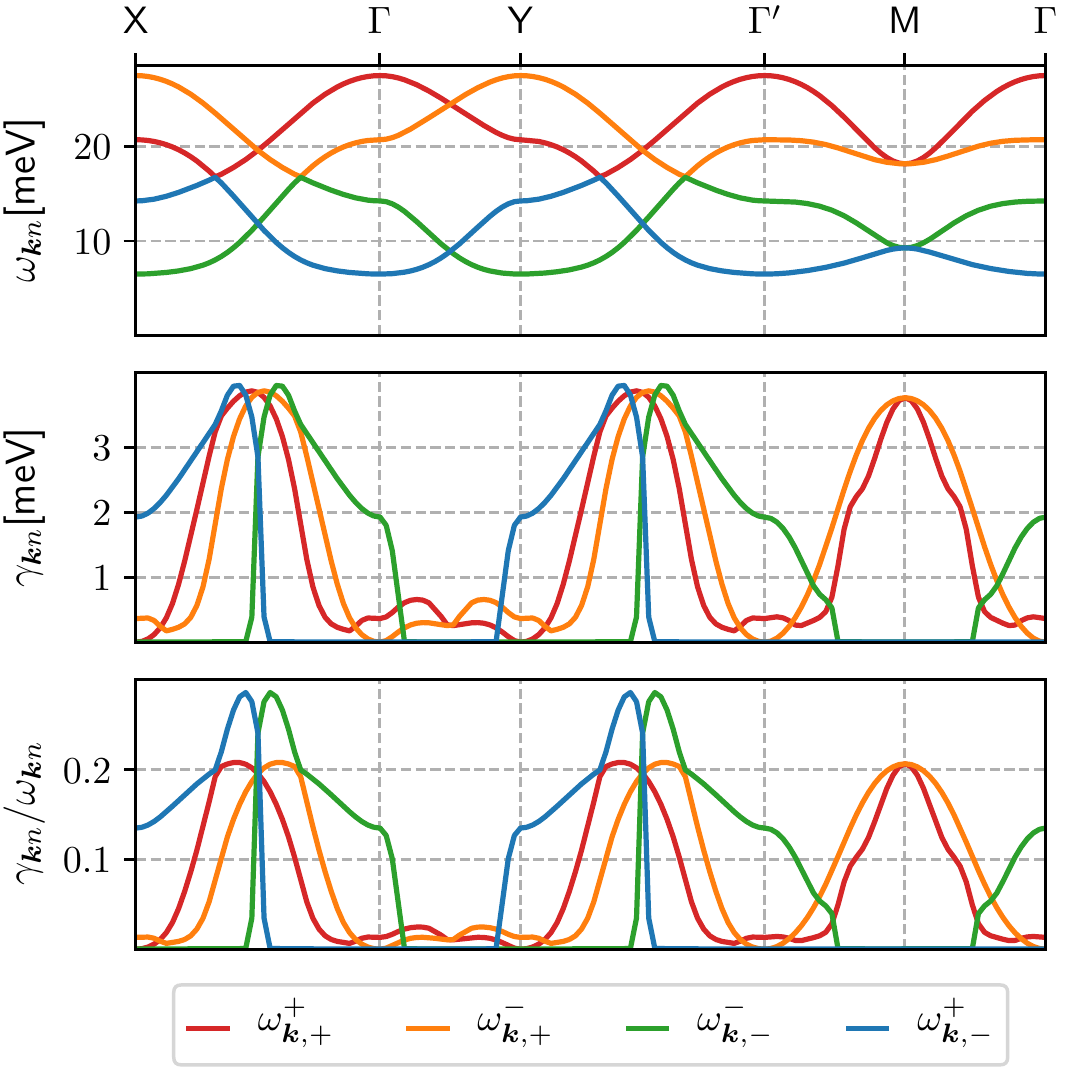}
	\caption{
Same as in Fig.~\ref{fig:damping_born} in the self-consistent iDE approximation (\ref{eq:dampingiDE0}). 
}
	\label{fig:damping}
\end{figure}

Given the well-pronounced van Hove singularities and that the 
Born-approximation damping shown in Fig.~\ref{fig:damping_born} 
is comparable to the magnon energies in large regions of the Brillouin zone, the validity of the
Born approximation can be questioned. A simple way to go beyond  Born approximation and regularize singularities 
is to self-consistently take into account the imaginary part of the self-energy
of the initial-state magnon, the damping of which  we calculate.
This  procedure amounts to solving the Dyson's equation for the self-energy and
retaining only its imaginary part (hence the abbreviation iDE) in a self-consistency 
loop \cite{tri09,tri_H,FM_DM}, 
\begin{eqnarray}
\gamma_{\bm{k}n}=-\text{Im} \ \Sigma_{n}(\bm{k}, \omega_{\bm{k}n} + i \gamma_{\bm{k}n}).
\label{eq:dampingiDE0}
\end{eqnarray} 
In practice, this can be achieved by iterating the recursion relation
\begin{eqnarray}
\gamma_{\bm{k}n}^{(i)}
&=&
 \frac{\pi}{2}
\int_{\text{BZ}} \frac{\diff^2 k' }{V_{\text{BZ}}}
\sum_{n' m}
\left| \tilde{\Gamma}^{n+4 \; n^{\prime} m} ( -\bm{k},\bm{k'} , \bm{k}-\bm{k'} ) \right|^2 
\nonumber\\
& &\times \delta \left( \omega_{\bm{k}n}  + i \gamma_{\bm{k}n}^{(i-1)}     -\omega_{\bm{k'}n'}-
\omega_{\bm{k}-\bm{k'}m} \right),
\label{eq:dampingiDE}
\end{eqnarray}
until $\gamma_{\bm{k}n}^{(i)}$ converges. Here, the $\delta$-function with the complex argument is a shorthand 
for a Lorentzian.
For the given model parameters, it took about 30 iterations for $\gamma_{\bm{k}n}^{(i)}$ to converge at 
all points along the momentum path shown in Fig.~\ref{fig:contour}.
The resulting self-consistent iDE results for damping are presented in Fig.~\ref{fig:damping}.

The van Hove singularities are regularized by the iDE procedure. 
We note that for the range of momenta between the $\Gamma$ and $Y$ points, i.e., 
for the momenta perpendicular to the spin magnetization, the damping rate is comparatively small. 
In the other parts of the momentum path, one finds that some of the magnons are still significantly damped 
with the typical damping rate $\sim 0.2$ of the magnon energies. 
This implies that the neutron scattering experiments 
will show well-defined magnon branches in some regions of the momentum space  as well as 
broadened excitation continua in the others. These are the characteristic features observed in $\alpha$-RuCl$_3$.
We will further elaborate on this discussion in Sec.~\ref{sec:neutron} where we present our results for the 
dynamical structure factor and the neutron scattering intensity.

\subsection{Comparison with the constant matrix element approximation}

While the preceding  discussion outlines a fully analytical approach and demonstrates its power for the  
problem of magnon damping, it is applicable only along the special $\Gamma\!=\!K$ line in the parameter space.
A general set of parameters of the same model would require numerical diagonalization and 
manipulations with the transformation matrix from Eq.~\eqref{eq:M_01} to obtain damping rate at the 
potentially prohibitive  computational cost. Therefore, it would be useful to have a justifiable approximate method that is 
less technically demanding, but  is able to produce magnon damping that is qualitatively correct 
or at least give an overall reasonable estimate of the effect. 

Such a method has been proposed in Ref.~\cite{Winter17}, which is referred  to as the ``constant matrix element'' 
approximation. In this approximation, the momentum dependence of the magnon interaction is accounted for in an effective way by a coupling strength $V_{\text{eff}}^{(3)}$ and a phenomenological average momentum dependence $f$ as defined later in the text.
Having an explicit analytic solution presented in this work offers us an opportunity to 
verify the overall validity and expose possible shortcomings of the constant matrix element approximation 
of Ref.~\cite{Winter17} for the same Kitaev-Heisenberg-$\Gamma$ model and for the same set of parameters. 
Let us briefly describe the nature of this approximation.

The first step is to find the three-magnon coupling strength. 
The  Holstein-Primakoff bosonization yields the three-boson Hamiltonian  $\mathcal{H}_3$ in Eq.~\eqref{eq:H3real}. 
For the $\Gamma\!=\!K$ line and in the zigzag phase
the real-space three-magnon coupling for bonds ${x,y,z}$ are given by
\begin{align}
\big|V^{(3)}_{x}\big|=\big|V^{(3)}_{y}\big|=\frac{\sqrt{6S}K}{4},  \quad   \big|V^{(3)}_{z}\big|=0,
\label{eq:kh_vertex}
\end{align}
see also Eq.~\eqref{eq:V_01}. 
Introducing the sum of these real-space vertices  over nearest bonds yields the overall scale 
\begin{equation}
V^{(3)}_{\rm eff}=\big|V^{(3)}_{x}\big|+\big|V^{(3)}_{y}\big|+\big|V^{(3)}_{z}\big|=\frac{\sqrt{6S}K}{2},
\label{eq:3_scale}
\end{equation}
that can be used as a definition of the three-magnon coupling strength. 
This definition is consistent with the one previously used in Ref.~\cite{Winter17}. 

 Then, one can redefine the symmetrized three-magnon vertex function 
 $\tilde{\Gamma}^{\mu \nu \lambda} ( {\bm{k}}_1,  \bm{k}_2 , \bm{k}_3 )$
 introduced in Sec.~\ref{sec:damping}~C  
\begin{eqnarray}
\tilde{\Gamma}^{\mu \nu \lambda} ( {\bm{k}}_1,  \bm{k}_2 , \bm{k}_3 )  \equiv V^{(3)}_{\rm eff} \,
\widetilde{\Phi}^{\mu \nu \lambda} ( {\bm{k}}_1,  \bm{k}_2 , \bm{k}_3 ),
\label{eq:gamma_to_phi}
\end{eqnarray}
where the  dimensionless vertices $\widetilde{\Phi}^{\mu \nu \lambda} ( {\bm{k}}_1,  \bm{k}_2 , \bm{k}_3 )$ 
include all the necessary transformations and 
symmetrizations of Eq.~\eqref{eq:tildeGammadef} and $V^{(3)}_{\rm eff}$ is the three-magnon coupling 
strength introduced in Eq.~\eqref{eq:3_scale}. Note that such a redefinition is independent of whether the vertex is
derivable analytically or requires a numerical diagonalization of  $\mathcal{H}_2$ 
via a generalized Bogoliubov transformation \eqref{eq:M_01}, 
which is needed to transform the Holstein-Primakoff three-magnon 
Hamiltonian \eqref{eq:H3real} 
to  the cubic Hamiltonian for the magnon quasiparticles in the 
form of Eq.~\eqref{eq:S3tensor1}.  Substituting the parametrization 
\eqref{eq:gamma_to_phi} for the interaction vertices into
the lowest Born approximation decay rate 
given in Eq.~\eqref{eq:damp_01} we obtain
\begin{eqnarray}
\gamma_{\bm{k}n} &=&
  \frac{\big|V^{(3)}_{\rm eff}\big|^2}{2}   \frac{4 \pi}{N}
\sum_{\bm{k'}} \sum_{n' m} 
 \delta \left( \omega_{\bm{k}n}-\omega_{\bm{k'}n'}-
\omega_{\bm{k}-\bm{k'} m} \right),
\nonumber\\
& &\times \left|
\widetilde{\Phi}^{n+4 \; n^{\prime} m} ( -\bm{k},\bm{k'} , \bm{k}-\bm{k'} )\right|^2 ,
\label{eq5_gammak}
\end{eqnarray}
with the three-magnon coupling explicitly factored out.
Then, it is tempting to relate the decay rate to the on-shell two-magnon density of states (DoS)
 \begin{align}
D_{\bm{k}n}=D_{\bm{k}}(\omega_{\bm{k}n})
=\frac{4\pi}{N}  \sum_{\bm{k'}} \sum_{n' m}\delta \left( \omega_{\bm{k}n}-\omega_{\bm{k'}n'}-
\omega_{\bm{k}-\bm{k'} m} \right),
\label{eq5_dos}
\end{align}
which quantifies the overlap of the single-magnon excitations  of the branch $n$ 
with the two-magnon continuum along the energies $\omega=\omega_{\bm{k}n}$
and characterizes the kinematic phase space for decays of the $n$th mode.

The main idea of the constant matrix element approach is exactly that: to approximate 
the decay rate (\ref{eq5_gammak}) as  proportional 
to the on-shell two-magnon DoS (\ref{eq5_dos}),
\begin{align}
\gamma_{\bm{k}n} \approx \frac{f}{2} \big|V^{(3)}_{\rm eff}\big|^2 D_{\bm{k}n},
\label{eq5_gamma_app}
\end{align}
where the constant $f$ is used as a phenomenological parameter. This parameter can be thought of 
as a result of the averaging of the dimensionless vertex, 
 \begin{equation}
f=\langle\big| 
\widetilde{\Phi}^{n+4 \; n^{\prime} m} ( -\bm{k},\bm{k'} , \bm{k}-\bm{k'} )\big|^2 \rangle,
 \end{equation}
where  brackets represent averaging over all momenta.
This approximation leads to a drastic simplification, because all one now needs for the 
decay rate calculation are the magnon energies  $\omega_{\bm{k}n}$ from the  harmonic
theory and the three-magnon coupling scale $V^{(3)}_{\rm eff}$, 
skipping the need for costly calculation and manipulation of the eigenvectors and vertices altogether. 

There are two justifications for the use of this approximation. 
First, the singularities in the Born decay rates are {\it always} due to the corresponding van Hove singularities 
in the two-magnon DoS \cite{RMP}, although their strength can be reduced or magnified by the matrix element effect
in the ``full-vertex'' 
calculation of Eq.~\eqref{eq5_gammak}. This relation should already make the results of Eq.~\eqref{eq5_gamma_app} 
similar to that of Eq.~\eqref{eq5_gammak}. Second, the self-consistent iDE approach of Eq.~\eqref{eq:dampingiDE}
involves an effective averaging over the decay vertex, thus suggesting that the 
constant matrix element approximation in combination with the iDE should give 
a better agreement with the iDE results  of Eq.~\eqref{eq:dampingiDE} obtained with the full vertex. 

The iDE scheme, described in Sec.~\ref{sec:iDE}, as applied to the constant matrix element approximation, 
is given by the self-consistent equation
\begin{align}
\gamma_{\bm{k}n} = \frac{f}{2} \big|V^{(3)}_{\rm eff}\big|^2 D_{\bm{k}}(\omega_{\bm{k}n}+i\gamma_{\bm{k}n}),
\label{eq:cv_ide}
\end{align}
where the $\delta$-function with the complex argument is a shorthand for a Lorentzian as before.

\begin{figure}[t]
	\includegraphics[width=0.95\linewidth]{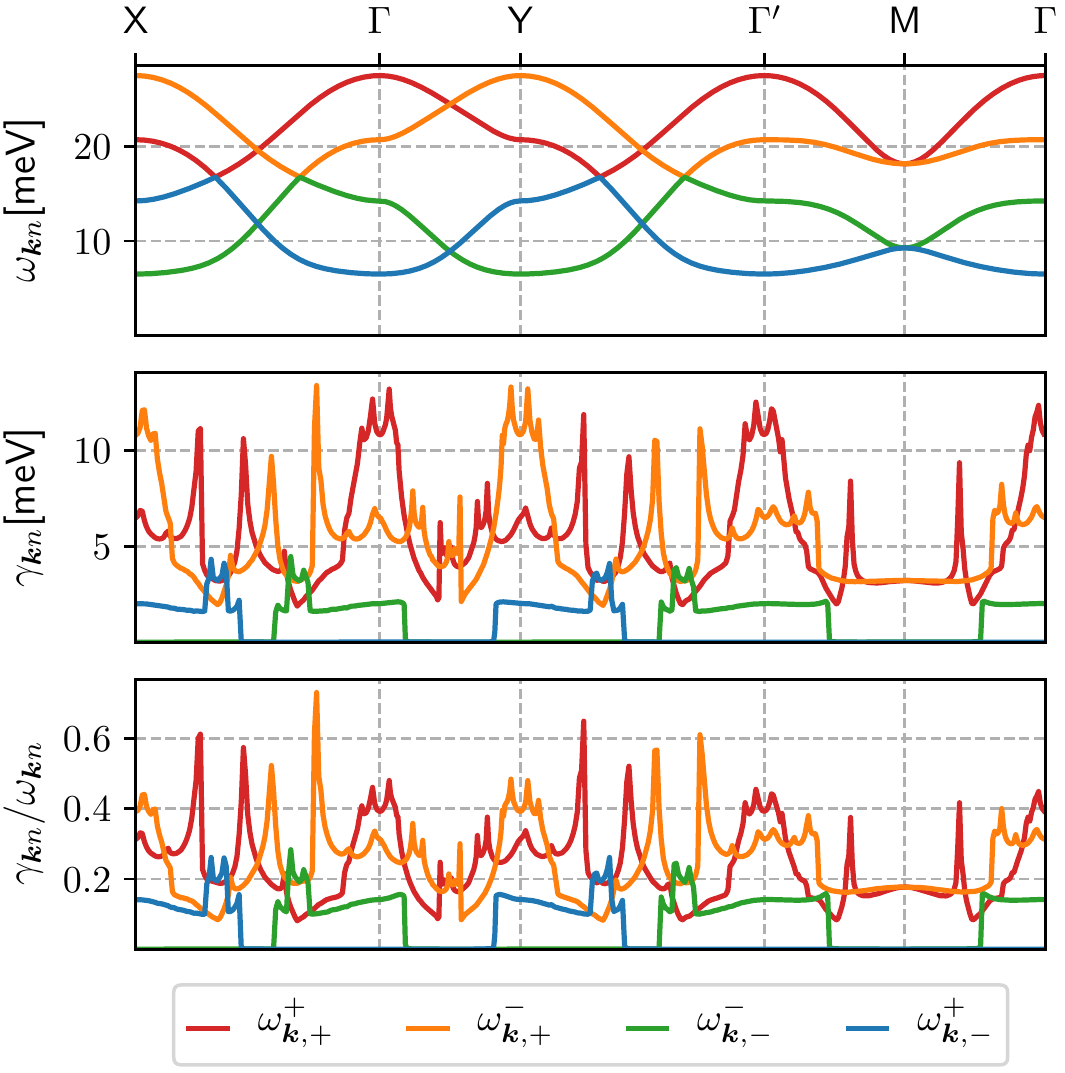}
	\caption{
Same as in Fig.~\ref{fig:damping_born} in the constant matrix element Born approximation (\ref{eq:dampingiDE0}).
The three-magnon coupling strength is
$V^{(3)}_{\rm eff}=6.06$ meV  and the parameter $f=0.2$. 
To be compared with Fig.~\ref{fig:damping_born}. 
}
	\label{fig:cv_damping}
\end{figure}

\begin{figure}[t]
	\includegraphics[width=0.95\linewidth]{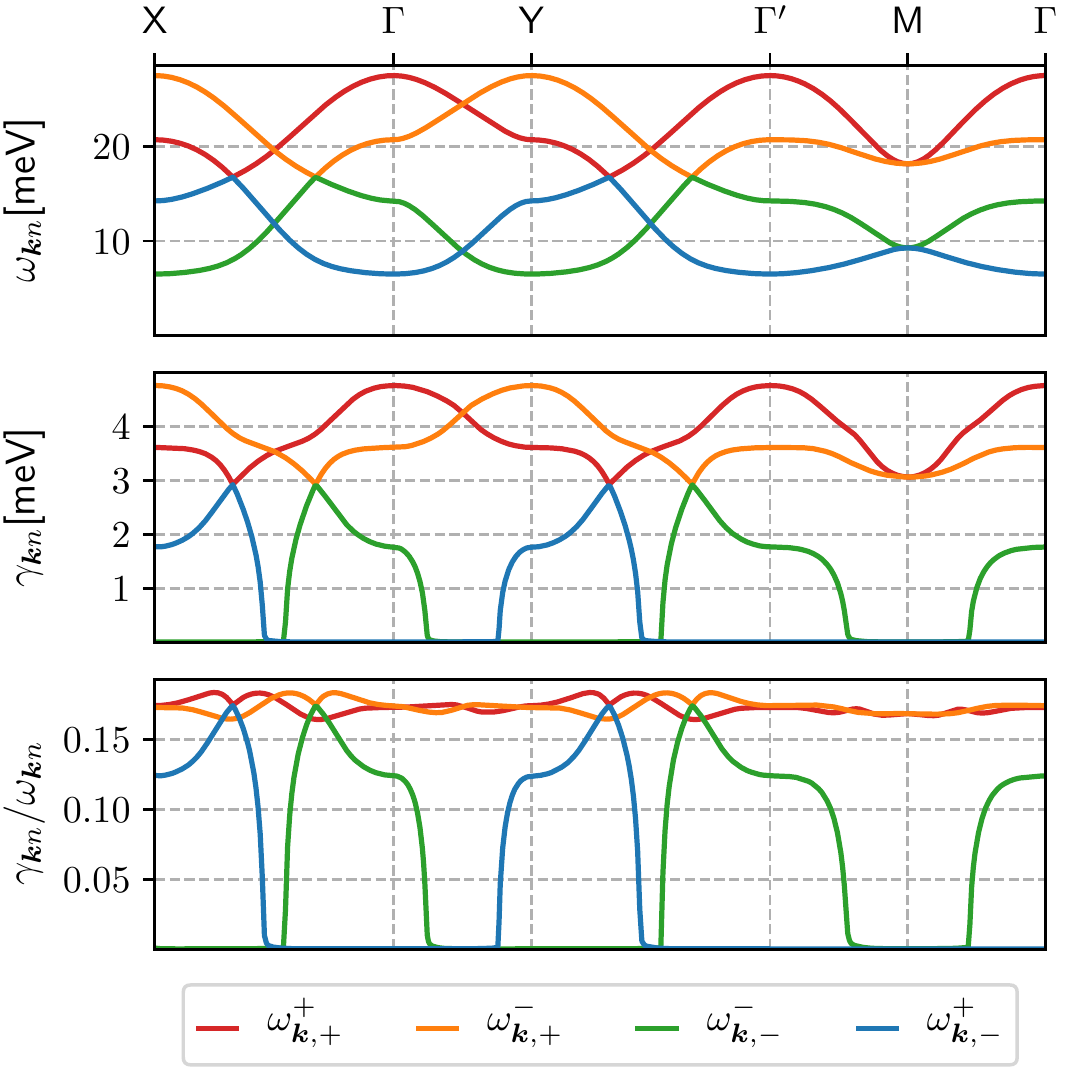}
	\caption{
Same as in Fig.~\ref{fig:cv_damping} in the constant matrix element iDE approximation of Eq.~\eqref{eq:cv_ide}. 
To be compared with Fig.~\ref{fig:damping}.  
}
	\label{fig:cv_damping_iDE}
\end{figure}

The only remaining problem is the educated choice of the phenomenological parameter $f$. 
Ref.~\cite{Winter17} has considered Kitaev-Heisenberg-$\Gamma$-$J_3$ model
for the choice of parameters associated with the  description of $\alpha$-RuCl$_3$ \cite{Winter_review}.
In  that work, the constant $f$ has been estimated to be $f\approx 1/9$ 
on the basis of a comparison with the constant matrix element calculations for the Born decay rates 
\eqref{eq5_gamma_app} in the honeycomb-lattice $XXZ$ model in external field, 
for which magnon decay rates have been calculated fully microscopically in Ref.~\cite{hex_H}. 
The present work allows us to determine the $f$-parameter based on the 
damping calculation directly for the Kitaev-Heisenberg-$\Gamma$ model, albeit in a different part of the phase diagram. 

The results of the constant matrix element approach  
for the Born and  self-consistent iDE approximations are shown in Figs.~\ref{fig:cv_damping} and \ref{fig:cv_damping_iDE},
 respectively. 
We use the same set of parameters \eqref{eq:point} as for the results in Figs.~\ref{fig:damping_born} and \ref{fig:damping}.
For $S\!=\!1/2$ and $K\!=\!7\text{ meV}$, the three-magnon coupling strength in Eq.~(\ref{eq:3_scale}) is 
$V^{(3)}_{\rm eff}\!=\!6.06$~meV. 
Comparisons with the overall values of the decay rates in 
Figs.~\ref{fig:damping_born} and \ref{fig:damping} suggest  an estimate for the $f$-parameter near $f\approx 0.2$,
somewhat higher than estimated in Ref.~\cite{Winter17}.

The results of the Born approximation constant matrix element approach in Fig.~\ref{fig:cv_damping} 
correctly reproduce  some of the qualitative features of the ``full vertex'' calculations in Fig.~\ref{fig:damping_born}. 
As expected, they include positions of the van Hove singularities as well as the regions where magnon modes are stable
because  decays are kinematically forbidden for them, e.~g., the region for the lower modes between $\Gamma$ and Y points.
However, some other qualitative and quantitative features are not properly reproduced.
For instance, in the full-vertex results of Fig.~\ref{fig:damping_born} there is a clear enhancement 
of the singularities due to the matrix element effect 
in the proximity of the magnon band crossings along the X-Y and $\Gamma$-Y directions.
Another inconsistency is in the lack of a suppression of  decays near  $\Gamma$ and Y points for 
the upper modes that is missing in Fig.~\ref{fig:cv_damping} but is obvious in Fig.~\ref{fig:damping_born}.
It clearly stems from the symmetries of interaction vertex that are missing in the constant matrix element approximation. 
Lastly, the overall decay rate of the upper modes is higher in the constant matrix element  
approximation than it is in a full-vertex calculations. 

Some of these differences are mitigated within the self-consistent iDE approximation, 
with the overall agreement  of Fig.~\ref{fig:cv_damping_iDE} and Fig.~\ref{fig:damping} becoming  more quantitative,
in accord with the expectations of Ref.~\cite{Winter17}. The overall scale of the damping is  similar to the 
full-vertex result, although the constant matrix element approach continues to overestimate the damping of 
the upper modes and underestimates the damping of the lower modes. 
Similarly to the Born approximation, there is also a lack of decay suppression near the high-symmetry $\Gamma$ and Y points. 

Last but not the least, we also note that  the phenomenological $f$-parameter in the present analysis is larger than 
in Ref.~\cite{Winter17}, $f\!\approx\! 1/5$ versus $f\!\approx\! 1/9$. 
Therefore, the calculations of   Ref.~\cite{Winter17} 
have likely provided a lower bound on the damping rates of magnons in $\alpha$-RuCl$_3$,
while the actual effect of broadening for the model parameters of that work may have been even more significant.

\section{Dynamical structure factor and neutron scattering intensity}
 \label{sec:neutron}

Having obtained the magnon energies and the dampings, we 
can calculate the dynamical structure factor  $S^{\alpha\beta}(\bm{k},\omega)$,
which determines the experimentally measured neutron scattering intensity,
\begin{equation} 
\mathcal{I}( \bm{k}, \omega )
=
F^2(\bm{k}) \sum_{\alpha\beta} \left( \delta_{\alpha\beta} - k_\alpha k_\beta / k^2 \right) S^{\alpha\beta}(\bm{k},\omega),
\label{eq:I_01}
\end{equation}
where $F(\bm{k})$ is the material-dependent formfactor and
the dynamical structure factor is defined as the Fourier transform of the
two-spin correlation function,
\begin{eqnarray}
S^{\alpha\beta}(\bm{k},\omega)
& = &
\int_{- \infty}^{\infty} \frac{dt}{2\pi}
 \frac{1}{N}
\sum_{i j} \langle S_i^\alpha(t) S_j^\beta(0) \rangle e^{-i\bm{k} \cdot (\bm{R}_i-\bm{R}_j) + i\omega t}
\nonumber
 \\ &=& \int \frac{dt}{2\pi} \langle S_{ \bm{-k} }^\alpha(t) S_{ \bm{k} }^\beta(0) \rangle e^{i\omega t}	.
\end{eqnarray}
Here we have introduced the Fourier components of the spin operators via
\begin{equation}
\bm{S}_{\bm{k}}(t) = \frac{1}{\sqrt{N}} \sum_i \bm{S}_i(t) e^{-i\bm{k}\cdot\bm{R}_i}.
\end{equation}
\begin{figure*}
	\includegraphics[width=1.\textwidth]{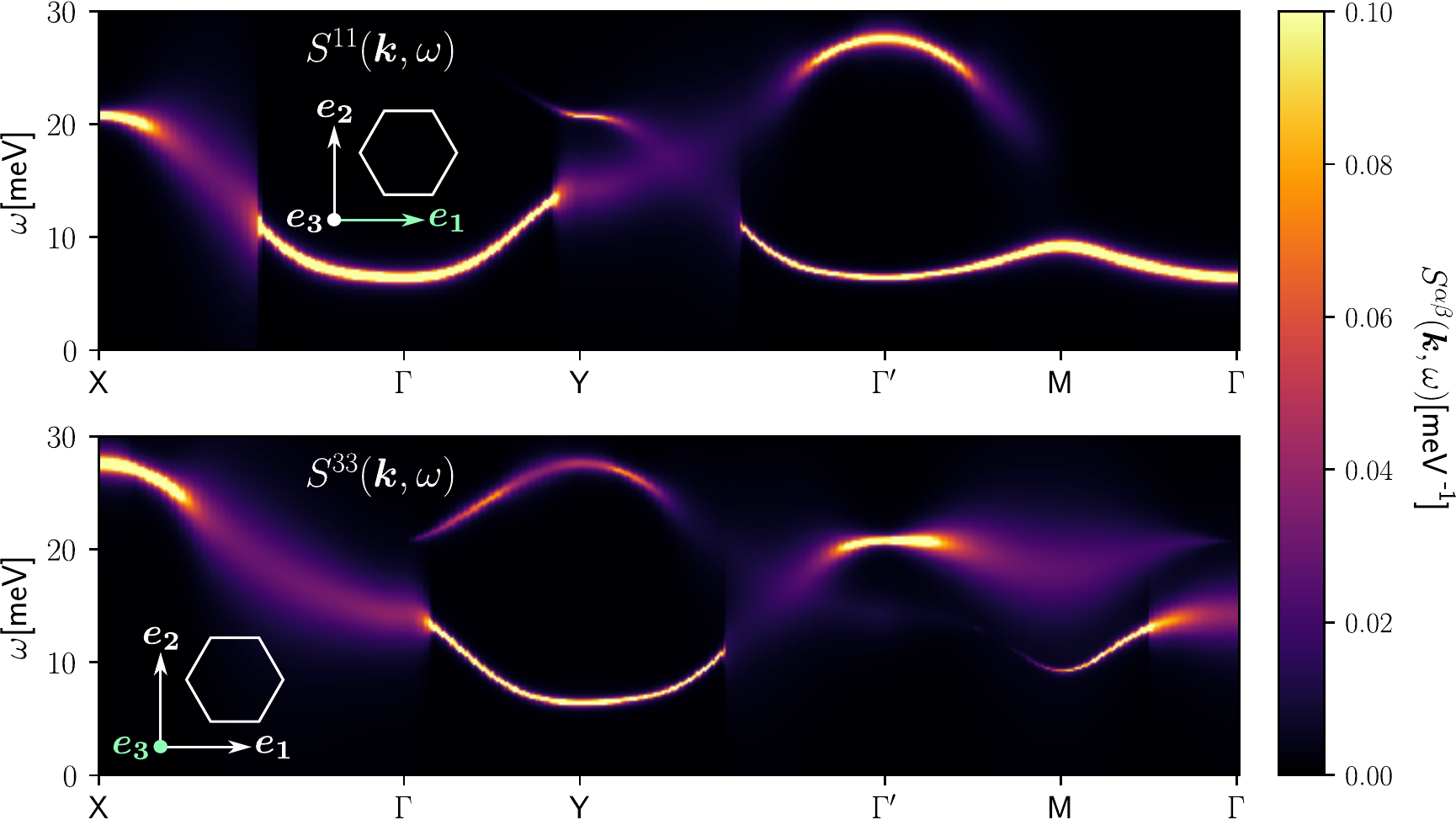}
	\caption{
		The diagonal components of the transverse part of the dynamical structure factor $S^{\alpha\beta} (\bm{k}, \omega)$
		that include magnon lifetime effects in  the iDE approximation (\ref{eq:dampingiDE})
		as given by  Eq.~(\ref{eq:structure_factor}) in the laboratory frame $\{\bm{e}_1,\bm{e}_2,\bm{e}_3\}$. 
		The  momentum $\bm{k}$  follows the same representative path as in Fig.~\ref{fig:contour}.
		The upper (lower) panel: $S^{11(33)} (\bm{k}, \omega)$. All other components are higher order in the 
		$1/S$-expansion. An artificial broadening of $0.1$meV is included in the imaginary part of the Green function. The plot range is cut at 0.1meV$^{-1}$ in order to emphasize details of the structure factor.
	}
	\label{fig:structure_factor}
\end{figure*}
The superscripts $ \alpha$ and $\beta$ label the three Cartesian 
components of the spins in the honeycomb 
basis $ \{ \bd{e}_1 , \bd{e}_2, \bd{e}_3 \}$,
which is aligned with the geometry of the honeycomb lattice,
see Figs.~\ref{fig:honeycomb}~and~\ref{fig:zigzag}.
Staying within the leading order in $1/S$, we consider
only the components of the structure factor
transverse to the magnetization; the longitudinal components can be neglected because they are of the higher order in $1/S$. 
To calculate the transverse components for $\Gamma = K > 0$ in the zigzag state, we note that in this case
the magnetization of the ordered moments is aligned with the direction $\bd{e}_2$ of the zigzag pattern, so that
the basis  $\{ \bd{n}_1 , \bd{n}_2, \bd{n}_3 \}$ defined in Eq.~(\ref{eq:n123}),
onto which the spin operators are projected, is related to the
honeycomb basis $\{\bm{e}_1,\bm{e}_2,\bm{e}_3\}$ defined in Eq.~(\ref{eq:e123}) via
\begin{eqnarray}
\bm{n}_1 &=& -\bm{e}_1, \; \; \; \; \; 
\bm{n}_2 = \bm{e}_3, \; \; \; \; \;
\bm{n}_3 = \bm{e}_2.
\end{eqnarray}
To calculate the dynamical structure factor within our spin-wave expansion, 
we   express the transverse components of $\bm{S}_{\bm{k}}$ 
in terms of the local spin frame defined via Eqs.~~(\ref{eq:loc_frame_01}) and~(\ref{eq:loc_frame_02}). 
Choosing the gauge $\phi=0$  for the transverse basis,
we obtain for the two components transverse to the magnetization,
 \begin{subequations}
\begin{eqnarray}
& & S_{\bm{k}}^1 = 
\bm{e}_1 \cdot \bm{S}_{\bm{k}}
=
-\bm{n}_1 \cdot \bm{S}_{\bm{k}}
= -\frac{1}{\sqrt{N}} 
 \sum_i e^{-i \bm{k} \cdot \bm{R}_i } \bm{n}_1 \cdot \bm{S}_i 
 \nonumber
 \\
 & & =  -\frac{1}{\sqrt{N}} 
 \left[
  \sum_{i \in a,d} e^{-i\bm{k} \cdot \bm{R}_i}\ \bm{t}_{i1} \cdot \bm{S}_i
 +  \sum_{i \in b,c} e^{-i\bm{k} \cdot \bm{R}_i}\ \bm{t}_{i1} \cdot \bm{S}_i
 \right],
 \nonumber
 \\
 & &
\end{eqnarray}
\begin{eqnarray}
& & S_{\bm{k}}^3
=
\bm{e}_3 \cdot \bm{S}_{\bm{k}}
=
\bm{n}_2 \cdot \bm{S}_{\bm{k}}
= 
\frac{1}{\sqrt{N}} \sum_i e^{-i \bm{k} \cdot \bm{R}_i } \bm{n}_2 \cdot \bm{S}_i 
\nonumber\\
& & =
\frac{1}{\sqrt{N}} \left[ 
 \sum_{i \in a,d} e^{-i\bm{k} \cdot \bm{R}_i}\ \bm{t}_{i2} \cdot \bm{S}_i
  - \sum_{i \in b,c} e^{-i\bm{k} \cdot \bm{R}_i}\ \bm{t}_{i2} \cdot \bm{S}_i \right].
 \nonumber
 \\
 & &
\end{eqnarray}
 \end{subequations}
Next,  we approximate the spin components in the local reference frames by the Holstein-Primakoff 
transformation~(\ref{eq:Holst_01}) to the leading order,
 \begin{subequations}
\begin{eqnarray}
\bm{t}_{i1} \cdot \bm{S}_i &  \approx & \frac{\sqrt{2S}}{2} \left( a_i + a_i^\dagger \right),
\\
\bm{t}_{i2} \cdot \bm{S}_i & \approx & \frac{\sqrt{2S}}{2i} \left( a_i - a_i^\dagger \right),
\end{eqnarray}
 \end{subequations}
and obtain
 \begin{subequations}
\begin{eqnarray}
S_{\bm{k}}^1
&=&
-\frac{1}{\sqrt{N}} \sum_{i \in a,d} e^{-i\bm{k} \cdot \bm{R}_i} \frac{\sqrt{2S}}{2} \left( a_i + a_i^\dagger \right)
\nonumber\\ & &
-\frac{1}{\sqrt{N}} \sum_{i \in b,c} e^{-i\bm{k} \cdot \bm{R}_i} \frac{\sqrt{2S}}{2} \left( a_i + a_i^\dagger \right),
\end{eqnarray}
\begin{eqnarray} 
S_{\bm{k}}^3
&=&
\frac{1}{\sqrt{N}} \sum_{i \in a,d} e^{-i\bm{k} \cdot \bm{R}_i} \frac{\sqrt{2S}}{2i} \left( a_i - a_i^\dagger \right)
\nonumber\\& &
-\frac{1}{\sqrt{N}} \sum_{i \in b,c} e^{-i\bm{k} \cdot \bm{R}_i} \frac{\sqrt{2S}}{2i} \left( a_i - a_i^\dagger \right).
\end{eqnarray}
 \end{subequations}
Using the sublattice Fourier transform~(\ref{eq:FT_01}) of the 
Holstein-Primakoff bosons, we obtain
 \begin{subequations}
\begin{eqnarray}
S_{\bm{k}}^1
&=&
-\frac{\sqrt{2S}}{4} \bigg( a_{\bm{k}} + a_{-\bm{k}}^\dagger + b_{\bm{k}} + b_{-\bm{k}}^\dagger 
\nonumber\\ & & \hspace{11mm}
+ c_{\bm{k}} + c_{-\bm{k}}^\dagger + d_{\bm{k}} + d_{-\bm{k}}^\dagger \bigg)
\nonumber\\ 
 & \equiv &  -\frac{\sqrt{2S}}{4} \sum_{ \mu } \phi_{ \bm{k}}^{ \mu },
 \hspace{7mm}
\end{eqnarray}
\begin{eqnarray}
S_{\bm{k}}^3
&=&
\frac{\sqrt{2S}}{4i} \bigg( a_{\bm{k}} - a_{-\bm{k}}^\dagger - b_{\bm{k}} + b_{-\bm{k}}^\dagger 
\nonumber\\ & & \hspace{11mm}
- c_{\bm{k}} + c_{-\bm{k}}^\dagger + d_{\bm{k}} - d_{-\bm{k}}^\dagger \bigg)
\nonumber\\ & \equiv &
\frac{\sqrt{2S}}{4i} \sum_{ \mu } \sigma_{\mu}\ \phi_{ \bm{k}}^{ \mu },
\end{eqnarray}
 \end{subequations}
where the symbols $\sigma_{\mu}= \pm 1$ 
determine the signs of the field components according to 
the following rule
 \begin{subequations}
\begin{eqnarray}
 & & \sigma_1 = \sigma_4 = \sigma_6 = \sigma_7 = 1,  
  \\
 & &
 \sigma_2 = \sigma_3 = \sigma_5 = \sigma_8 = -1.
\end{eqnarray}
 \end{subequations}
Therefore, the off-diagonal part of the transverse
structure factor is 
\begin{eqnarray}
S^{13}(\bm{k},\omega)
&=&
- \int \frac{dt}{2\pi} e^{i \omega t} \langle S_{-\bm{k}}^1 (t) S_{\bm{k}}^3(0) \rangle
\nonumber\\ &=&
-\int \frac{dt}{2\pi} e^{i \omega t} \frac{2S}{16i} \sum_{ \mu \nu } \sigma_{ \nu }
\langle \phi_{ -\bm{k}  }^{\mu} (t) \phi_{ \bm{k}  }^{\nu} (0) \rangle
\nonumber\\ &=&
-\frac{2S}{16i} \sum_{ \mu \nu } \sigma_{ \nu } 
\sum_{ \mu' \nu' } {\mathbb{T}}_{ -\bm{k} }^{ \mu \mu' } \mathbb{T}_{ \bm{k}}^{ \nu \nu' }
\nonumber\\&  &
\times
\int \frac{dt}{2\pi} e^{i \omega t} \langle \psi_{ -\bm{k}}^{ \mu' } (t) \psi_{ \bm{k}}^{ \nu' }(0) \rangle.
\end{eqnarray}
Here, $\mathbb{T}_{ \bm{k}}^{ \mu \mu' }$ 
are the components of the $8\times 8$ transformation matrix $\mathbb{T}_{\bm{k}}$ given in Eq.~(\ref{eq:M_01}) and the components of the operators $\psi_{\bd{k}}^{\mu}$ contain 
Bogoliubov bosons associated with the four magnon bands,
 \begin{equation}
 ( \psi_{\bd{k}}^{\mu} ) =
 \left( \begin{array}{c}
 \psi_{\bd{k}}^1 \\ \psi^2_{\bd{k}} \\ \psi^3_{\bd{k}}  \\ \psi^4_{\bd{k}} \\
 \psi^5_{  \bd{k} } \\ \psi^6_{  \bd{k} } \\ \psi^7_{  \bd{k} } \\ \psi^8_{  \bd{k} } 
 \end{array}
 \right)
=
 \left( \begin{array}{c}
 b_{\bd{k} 1} \\ b_{\bd{k} 2} \\  b_{\bd{k} 3}  \\ b_{\bd{k} 4} \\
{b}^{\dagger}_{ - \bd{k} 1 } \\ {b}^{\dagger}_{ - \bd{k} 2 } \\ {b}^{\dagger}_{ -  \bd{k} 3 } 
 \\ {b}^{\dagger}_{ - \bd{k} 4 } 
 \end{array}
 \right).
 \label{eq:bog2a}
 \end{equation} 
Recall that    the range of the field-type labels $\mu$ and $\nu$ is 
$\{ 1,2, \ldots, 8 \}$, while the band label $n$ assumes values in the range $ \{ 1,2,3,4 \}$.
\begin{figure*}
	\includegraphics[width=1.\linewidth]{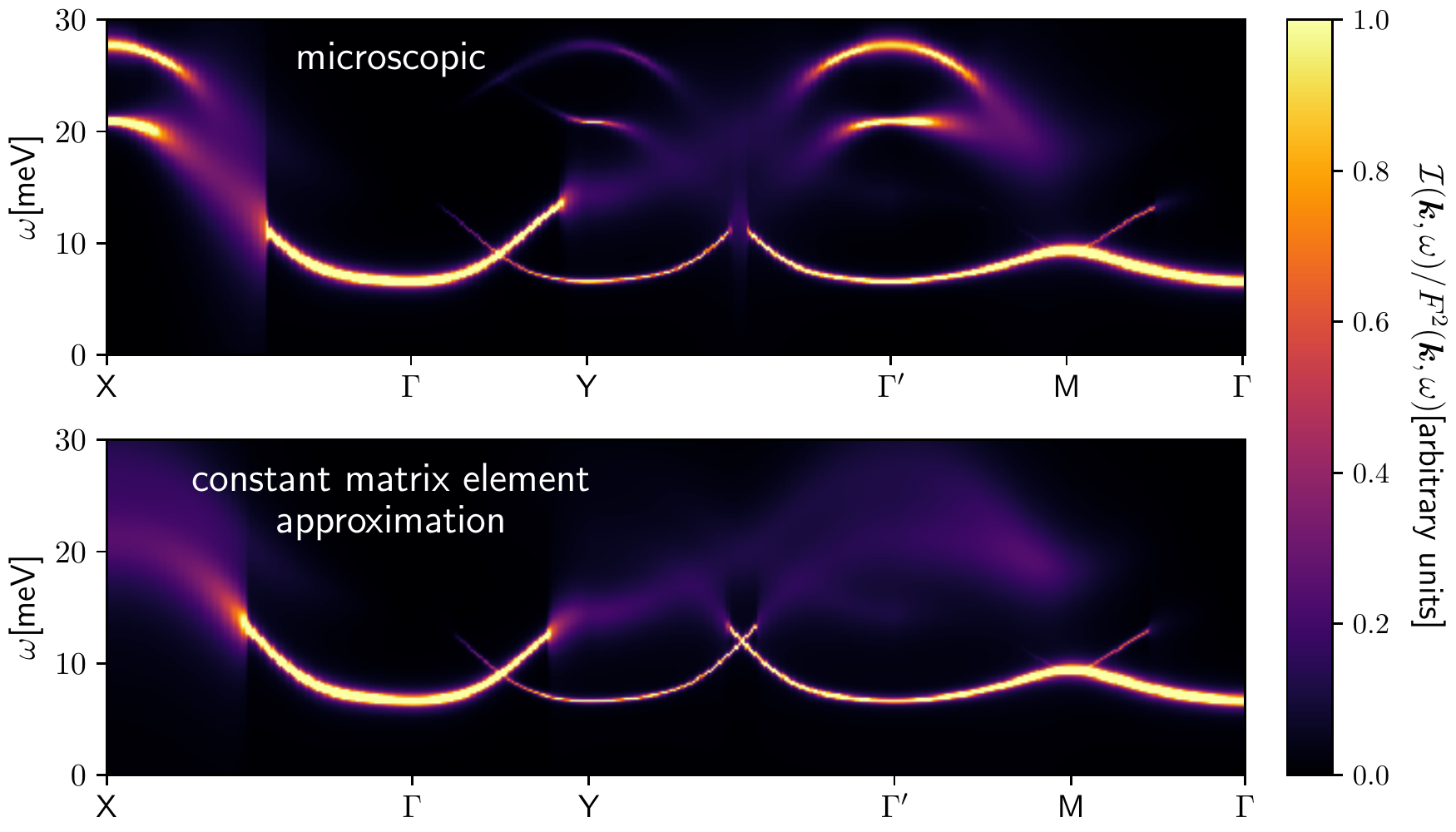}
	\caption{
		Upper panel: neutron-scattering intensity $\mathcal{I}( \bm{k},\omega )$ 
		normalized by  the square of the atomic formfactor as given in Eq.~(\ref{eq:I_div_f_01}) with magnon lifetime effects in the iDE approximation given by Eq.~(\ref{eq:dampingiDE}).
		The presented $\bm{k}$-path also involves a finite out-of-plane component $k_3 = \sqrt{3} \pi /d$, see the text. 
		Lower panel: same with the decay rates from the constant matrix element approximation.
		An artificial broadening of $0.1$meV is included in the imaginary part of the Green function.
	}
	\label{fig:intensity}
\end{figure*}
The transformation to the Bogoliubov bosons (\ref{eq:bog2a}) block-diagonalizes the expectation values and we obtain
\begin{eqnarray}
S^{13} (\bm{k},\omega)
&=&
-\frac{2S}{16i} \sum_{ \mu \nu } \sum_{n}
\sigma_{ \nu } \int \frac{dt}{2\pi} e^{i \omega t}  
\bigg[
\nonumber
\\ &  &
 \mathbb{T}_{ -\bm{k}}^{ \mu \, n+4 } \mathbb{T}_{ \bm{k}}^{ \nu n }
\langle b_{ \bm{k} n }^\dagger(t) b_{ \bm{k} n }(0) \rangle
\nonumber
\\ & + &   
 \mathbb{T}_{ -\bm{k} }^{\mu n } \mathbb{T}_{ \bm{k}}^{ \nu \, n+4 }
\langle b_{ -\bm{k} n }(t) b_{ -\bm{k} n }^\dagger(0) \rangle
\bigg].
\end{eqnarray}
The time-integrals can be expressed in terms of the retarded magnon Green functions,
 \begin{subequations}
\begin{eqnarray}
 & & \int \frac{dt}{2 \pi} e^{i \omega t} \langle b_{ \bm{k} n }^\dagger (t) b_{ \bm{k} n } (0) \rangle
 \nonumber
 \\
& & 
= \frac{1}{ e^{-\beta \omega} - 1 } \frac{1}{\pi} \text{Im} \ G_n ( \bm{k}, -\omega + i 0^+),
\label{eq:int_neg}
\end{eqnarray}

\begin{eqnarray}
 & & \int \frac{dt}{2 \pi} e^{i \omega t} \langle b_{ -\bm{k} n } (t) b_{ -\bm{k} n }^\dagger (0) \rangle
 \nonumber
 \\
& & = \frac{1}{ e^{-\beta \omega} - 1 } \frac{1}{\pi} \text{Im} \ G_n ( \bm{k}, \omega + i 0^+).
\end{eqnarray}
 \end{subequations}
At $T=0$, the expectation value in Eq.~(\ref{eq:int_neg}) vanishes, leaving only
\begin{equation}
S^{13}( \bm{k}, \omega )
=
\frac{1}{\pi} \sum_n W_{ \bm{k} n }^{13} \ 
 \text{Im} \ G_n ( \bm{k}, \omega + i 0^+),
\end{equation}
with
\begin{equation}
W_{ \bm{k} n }^{13} =
\frac{S}{8i} \sum_{ \mu \nu } \mathbb{T}_{ -\bm{k}}^{ \mu n }\ \sigma_{ \nu }\ 
 \mathbb{T}_{ \bm{k}}^{  \nu  \, n+4 }.
\end{equation}
Analogous calculations for the remaining transverse components of the structure factor lead to
\begin{equation}
S^{\alpha\beta} (\bm{k}, \omega ) = \frac{1}{\pi} \sum_n W_{ \bm{k} n }^{\alpha\beta} \text{Im}\ G_n ( \bm{k}, \omega + i 0^+),
\label{eq:structure_factor}
\end{equation}
\begin{subequations}
with the ``envelope'' functions
\begin{eqnarray} 
W_{ \bm{k} n }^{11} &=&
\frac{S}{8} \sum_{ \mu \nu }  \mathbb{T}_{ -\bm{k}}^{\mu n }  
 \mathbb{T}_{ \bm{k}}^{\nu \, n+4 },
\\
W_{ \bm{k} n }^{13} = (W_{ \bm{k} n }^{31} )^{\ast} &=&
\frac{S}{8i} \sum_{ \mu \nu } \mathbb{T}_{ -\bm{k} }^{ \mu n } \ 
 \sigma_{ \nu }\ \mathbb{T}_{ \bm{k} }^{ \nu \, n+4 },
\\
W_{ \bm{k} n }^{33} &=&
-\frac{S}{8} \sum_{ \mu \nu } \sigma_{ \mu } \mathbb{T}_{ -\bm{k}}^{ \mu n }\ 
 \sigma_{ \nu }\ 
 \mathbb{T}_{ \bm{k}}^{ \nu \, n+4 },\quad \quad \quad
\end{eqnarray}
\end{subequations}
Within our approximations,  the imaginary part of the magnon propagator is
\begin{equation}
\text{Im} \ G_n (\bm{k},\omega + i 0^+ )
\approx
\frac{ \gamma_{ \bm{k}n } }{ ( \omega - \omega_{ \bm{k} n }  )^2 + \gamma_{\bm{k}n}^2}.
\end{equation}
In Fig.~\ref{fig:structure_factor}, we plot the
diagonal components of the transverse structure factor
for the same representative set of the model parameters given 
in Eq.~(\ref{eq:point}). For the magnon damping $\gamma_{\bm{k}n}$, we used our 
results obtained within the iDE approach in Sec.~\ref{sec:iDE}.
Note that within our approximation, 
the  off-diagonal components of the structure factor vanish identically.

To analyze the effect of magnon interactions 
in the neutron scattering intensity, we normalize the intensity
in Eq.~(\ref{eq:I_01}) by the square of the material-dependent formfactor,
\begin{eqnarray}
\frac{ \mathcal{I}( \bm{k},\omega ) }{ F^2( \bm{k} ) }
& = &
\sum_{\alpha\beta} \left( \delta_{\alpha\beta} - k_\alpha k_\beta / k^2 \right) S^{\alpha\beta}(\bm{k},\omega)
\nonumber\\
&=&
\frac{1}{\pi} \sum_n
I_{\bm{k}n}
\frac{ \gamma_{ \bm{k}n } }{ ( \omega - \omega_{ \bm{k} n }  )^2 
 + \gamma_{\bm{k}n}^2},
\nonumber\\
\label{eq:I_div_f_01}
\end{eqnarray}
where we defined the  $\bm{k}$-dependent weights $ I_{\bm{k}n}$
associated with a given magnon band $n$ as
\begin{equation}
I_{\bm{k}n} = \sum_{\alpha\beta} \left( \delta_{\alpha\beta} - k_\alpha k_\beta / k^2 \right) W_{ \bm{k} n }^{\alpha\beta}.
\label{eq:weight_01}
\end{equation}
The intensity defined in Eq.~(\ref{eq:I_div_f_01}) is plotted in Fig.~\ref{fig:intensity}.
Note that while the in-plane component of the momentum $\bm{k}$ follows the same
representative path shown in Fig.~\ref{fig:contour}, for the neutron-scattering intensity in Fig.~\ref{fig:intensity}, 
the contour also has a finite out-of-plane component $k_3 = \sqrt{3} \pi /d$
to avoid artificial singularities.
One can clearly distinguish sharp excitation branches in wide regions of the ${\bm{k}}$-space, indicating well-defined 
magnon quasiparticles. However, for a significant range of the ${\bm k}$--$\omega$ space,  
the quasiparticles cease to exist and are replaced instead by  a broad continuum of excitations. 
This result justifies the claim put forward in Ref.~\cite{Winter17} that the anharmonic magnon couplings can destroy 
the quasiparticle character of the magnetic excitation spectrum in the zigzag phase of the 
Kitaev-Heisenberg-$\Gamma$ model in a large part of the Brillouin zone.
In the same Fig.~\ref{fig:intensity}, the lower panel offers a comparison of the 
effects of the ``full-vertex'' calculations of the magnon damping in the iDE approximation
with that of the constant matrix element approximation. The damping rates in the latter approach are from 
Fig.~\ref{fig:cv_damping_iDE}. It is again visible that the constant matrix element approximation overestimates the damping of the higher energy magnon branches around the $\Gamma$ and $Y$ points. The overall form
of decays is however very similar, showing a coexistence of the ${\bm k}$-$\omega$-regions with 
well-defined quasiparticles with the regions where they are absent.

\section{Summary and conclusions}
 \label{sec:summary}

The present work advances the studies of the  Kitaev-Heisenberg-$\Gamma$ 
model  in several directions.
First of all, we have found a special line in the parameter space of the Kitaev-Heisenberg-$\Gamma$ model along which the magnon spectrum and all matrix elements needed for the calculation of the magnon damping can be obtained analytically for the physically relevant zigzag phase. This line is defined by $\Gamma = K > 0$,  arbitrary nearest-neigbor exchange $J$, and third-nearest-neighbor exchange $J_3$.
This enormously reduces the complexity of the evaluation
of the perturbative expressions for the
magnon damping and has enabled us to calculate  the magnon damping
in this regime without additional simplifying assumptions.
Although special points in the parameter space of the 
Kitaev-Heisenberg-$\Gamma$ model characterized by additional symmetries 
have been identified in the past \cite{Chaloupka}, 
the fact that on the line $\Gamma =K > 0$ the magnon spectrum and all interaction vertices
in the zigzag state can be obtained analytically has not been noticed before.
Physically, the origin for the simplifications for   $\Gamma = K >0$ is that
on this line the magnetic moments in the zigzag state lie in the plane of the honeycomb lattice
and point in the direction of the zigzag pattern.

Next, we would like to emphasize that
our explicit calculation of the magnon damping for $\Gamma = K > 0$
within the leading order Born approximation and the self-consistent 
iDE approach based on the solution of the
imaginary part of the Dyson's  equation  
is at the cutting edge of what can be done analytically within spin-wave theory.
To carry out this calculation, it was crucial to work with an
unconventional parameterization of the spin-wave theory where each Holstein-Primakoff boson
is expressed in terms of two conjugate  hermitian operators~\cite{Hasselmann06,Kreisel07,Kreisel08,Kreisel11,Kreisel14}. The advantages of this approach as compared with the conventional procedure outlined in Appendix~A are (a) that it simplifies the identification of special 
points in parameter space where the calculations simplify,  (b)
that the explicit diagonalization of the quadratic spin-wave Hamiltonian obtained after Holstein-Primakoff transformation can be mapped on the well-known diagonalization procedure
for coupled harmonic oscillators \cite{Goldstein80,Mahan90},
and (c) that for the implementation of this procedure for a system with $f$ boson flavors
one has to manipulate only hermitian $f \times f$ matrices.
In Appendix B we give another example for the  ``hermitian field formulation'' 
of spin-wave theory by calculating the magnon spectrum and the relevant Bogoliubov transformation
of the 
Kitaev-Heisenberg-$\Gamma$ model  for $\Gamma = K > 0$ in a two-sublattice approach.
Somewhat surprisingly, we could not find such an explicit analytic construction in the literature, although in this case one only has $f = 2$ boson flavors.

We have demonstrated for the representative values of the model parameters, 
that the magnon damping in approximations 
based on the Born and the self-consistent iDE approaches
is significant, leading to characteristic broad features
in the dynamical structure factor. 
These results underscore the importance of taking into account the nonlinear magnon coupling 
in interpreting  broad features in the neutron-scattering spectra for the general Kitaev-Heisenberg-$\Gamma$ model.
The present work thus confirms the assertion of Ref.~\cite{Winter17} that 
anharmonic  interactions can lead to  large decay rates such that some of the magnon branches 
cease to be well-defined quasiparticles, as is possibly observed in $\alpha$-RuCl$_3$.
By focusing our attention on the regime $\Gamma\! =\! K\!>\!0$ with an additional 
third-nearest-neighbor Heisenberg interaction $J_3$ to stabilize the zigzag-ordered 
state, we have been able to confirm in a 
quantitative manner the validity of the claims regarding the importance of the anharmonic magnon coupling 
terms that were put forward in Ref.~\cite{Winter17}. 
In particular, we have shown that the phenomenological 
constant matrix element approximation used in  Ref.~\cite{Winter17}
can indeed be used to estimate semi-quantitatively the magnitude of the decay rates 
in a large part of the Brillouin zone.  On the other hand, in some parts  of the Brillouin 
zone the momentum-dependence of the interaction vertex is important, 
so that the constant matrix element approximation cannot reliably predict the order of magnitude of
magnon damping and the spectral line-shape of the dynamic structure factor.
This is especially true for momenta in the proximity of magnon band crossings along the
X-Y and $\Gamma$-Y directions. Moreover, as shown in Appendix D,
the damping becomes even stronger for all modes in certain areas on the momentum plane 
when the third-nearest-neighbor exchange interaction is smaller than all other interactions.

Finally, let us emphasize that this work contains technical advances 
in spin-wave theory that can also be useful for other spin models.
First of all, the hermitian field parametrization
of spin-wave theory developed in Sec.~\ref{sec:construction} (see also Appendix~B)
is an efficient alternative to Colpa's 
algorithm \cite{Colpa78,Blaizot86,Maldonado93,Serga12} in the magnetically ordered phase
of any spin-model with a complicated magnon spectrum consisting of several bands.
Moreover, for the calculation of the magnon damping in multi-band 
magnon systems it is crucial to carefully keep track of all phase factors in the 
interaction vertices generated by Umklapp scattering processes. 
In Sec.~\ref{sec:cubic} we have carefully derived the proper phase factors
for the cubic interaction vertices in the zigzag state of the 
Kitaev-Heisenberg-$\Gamma$ model. Similar considerations should be used to derive
Umklapp phase factors
in other models with multiple magnon bands.

\begin{acknowledgments}

This work was financially supported by the German Science Foundation (DFG) through the program SFB/TRR 49 (P. K. and O.~T.) and by the U. S. Department of Energy,
Office of Science, Basic Energy Sciences under Award No. DE-FG02-04ER46174 (P. A. M. and A. L. C.).
P.~K. and R.~S. acknowledge the hospitality of the 
Department of Physics and Astronomy of the University of California, Irvine,
where this work was initiated during a sabbatical stay. 
A.~L.~C. would like to thank Aspen Center for Physics and the Kavli Institute for Theoretical Physics 
where different stages of this work were advanced. 
The work at Aspen was supported in part by NSF Grant No. PHY-1607611 and the research at 
KITP was supported in part by NSF Grant No. NSF PHY-1748958.
\end{acknowledgments}

\begin{appendix}

\section*{APPENDIX A: Construction of multi-flavor 
Bogoliubov transformations}
\setcounter{equation}{0}
\renewcommand{\theequation}{A\arabic{equation}}
 \label{sec:Bogoliubov}

In this appendix, we review the method
for reducing the problem of diagonalizing a  general $f$-flavor quadratic 
boson Hamiltonian of the form [see Eq.~(\ref{eq:H2res})]
\begin{eqnarray}
{\cal{H}}_2 & = &  {\sum_{\bd{k}}} \sum_{n m = 1}^{f} 
\Biggl\{  A^{n m}_{\bd{k}}   a^{\dagger}_{\bd{k} n}
a_{\bd{k} m}
\nonumber
\\
& & 
+ \frac{1}{2} \left[  B_{\bd{k}}^{n m}
a^{\dagger}_{\bd{k} n} 
a^{\dagger}_{- \bd{k} m } 
+ (B_{\bd{k}}^{m n})^{\ast}
a_{-\bd{k} n}   a_{ \bd{k} m } \right]  \Biggr\}
 \hspace{7mm}
\label{eq:H2res2}
\end{eqnarray}
 to a $2f$-dimensional generalized eigenvalue problem.
Note that the hermiticity of the Hamiltonian implies that
\begin{equation}
A_{\bd{k}}^{n m } = ( A_{\bd{k}}^{m n } )^{\ast},
\label{eq:symA}
\end{equation}
and the symmetry under relabeling $\bd{k} \rightarrow - \bd{k} $
in the off-diagonal terms implies that the coefficients $B_{\bd{k}}^{n m }$ can be chosen such that
\begin{equation}
B_{\bd{k}}^{n m } =  B_{- \bd{k}}^{m n }.
\label{eq:symB}
\end{equation}
For $f=1$ the Hamiltonian (\ref{eq:H2res2})
can be diagonalized by the 
the usual Bogoliubov transformation. For arbitrary  $f$, a general algorithm
for diagonalizing this type of Hamiltonian has been constructed by
Colpa \cite{Colpa78}. A discussion of this algorithm can also be found in the textbook
by Blaizot and Ripka \cite{Blaizot86} and in 
Refs.~[\onlinecite{Maldonado93}, \onlinecite{Serga12}]. 
Here we review some mathematical subtleties of this treatment, 
as presented by Maldonado \cite{Maldonado93}, which are often ignored in the literature.

It is convenient to define the $f$-component column vectors
\begin{equation}
\bd{a}_{\bd{k}} = \left(  \begin{array}{c} a_{\bd{k}1}  \\ \vdots \\ a_{\bd{k} f} \end{array} \right),
\; \; \; 
\bd{a}^{\ast}_{\bd{k}} = ( \bd{a}_{\bd{k}}^{T} )^\dagger =
\left(  \begin{array}{c} a^{\dagger}_{\bd{k}1}  \\ \vdots \\ a^{\dagger}_{\bd{k} f} \end{array} \right),
\end{equation}
and the adjoint  row vectors
\begin{equation}
\bd{a}^{\dagger}_{\bd{k}} =
( a^{\dagger}_{\bd{k}1}  \ldots  a^{\dagger}_{\bd{k} f} ), \; \; \;  \bd{a}^T_{\bd{k}} = ( a_{\bd{k} 1} \ldots a_{\bd{k} f }).
\end{equation}
These vectors can be combined to vectors with $2 f$ components containing both
annihilation and creation operators,
\begin{equation}
{\bd{\phi}}_{\bd{k}} =  \left( \begin{array}{c} \bd{a}_{\bd{k}} \\ \bd{a}^{\ast}_{-\bd{k}} \end{array} \right)
= 
\left( \begin{array}{c} a_{\bd{k}1}  \\ \vdots \\ a_{\bd{k} f} \\ a^{\dagger}_{-\bd{k}1}  \\ \vdots \\ a^{\dagger}_{-\bd{k} f} \end{array} \right),
\label{eq:phikdef}
\end{equation}
\begin{equation}
{\bd{\phi}}^{\dagger}_{\bd{k}} =  \left(  \bd{a}^{\dagger}_{\bd{k}} ,  \bd{a}^{T}_{-\bd{k}}  \right)
= 
\left( a^{\dagger}_{\bd{k}1}   \hdots   a^{\dagger}_{\bd{k} f} , a_{-\bd{k}1}    \hdots   a_
{-\bd{k} f}  \right).
\end{equation}
Then, our quadratic boson Hamiltonian (\ref{eq:H2res2}) can  be written in a matrix form as follows
\begin{equation}
{\cal{H}}_2 =    \frac{1}{2}  \sum_{\bd{k}} \left[ \bd{\phi}_{\bd{k}}^{\dagger}  \mathbbm{M}_{\bd{k}} 
\bd{\phi}_{\bd{k}} -  {\rm Tr} 
\mathbf{A}_{\bd{k}} \right],
\label{eq:H2mat}
\end{equation}
where the $2f \times 2f$-matrix $\mathbbm{M}_{\bd{k}}$ is of the form
\begin{equation}
\mathbbm{M}_{\bd{k}} = \left( \begin{array}{cc} \mathbf{A}_{\bd{k}} & \mathbf{B}_{\bd{k}} \\
\mathbf{B}^{\dagger}_{\bd{k}} &  \mathbf{A}_{- \bd{k}}^ {T}
\end{array}
\right) = \left( \begin{array}{cc} \mathbf{A}_{\bd{k}} & \mathbf{B}_{\bd{k}} \\
\mathbf{B}^{\ast}_{- \bd{k}} &  \mathbf{A}_{- \bd{k}}^ {\ast}
\end{array}
\right),
\label{eq:MABdef}
\end{equation}
with the $f \times f$ blocks $\mathbf{A}_{\bd{k}}$ and $\mathbf{B}_{\bd{k}}$ defined by
$ [ \mathbf{A}_{\bd{k}} ]^{n m } = A_{\bd{k}}^{n m }$ and
$ [ \mathbf{B}_{\bd{k}} ]^{n m } = B_{\bd{k}}^{n m }$.
In the second equality in Eq.~(\ref{eq:MABdef}), we have used the
symmetries (\ref{eq:symA}) and (\ref{eq:symB}) which imply that
\begin{eqnarray}
\mathbf{A}_{\bd{k}}  & = & \mathbf{A}_{\bd{k}}^{\dagger},
\\
\mathbf{B}_{\bd{k}} & = & \mathbf{B}_{- \bd{k}}^T.
\end{eqnarray}
We would like to construct a new set of boson operators $b_{\bd{k} 1}, \ldots , b_{\bd{k} f }$,
which diagonalize the Hamiltonian.
We combine these operators and their adjoints $b^{\dagger}_{\bd{k} n}$ to form
a $2f$-component column vector with the same structure as $\bd{\phi}_{\bd{k}}$ in 
Eq.~(\ref{eq:phikdef}),
\begin{equation}
{\bd{\psi}}_{\bd{k}} =  \left( \begin{array}{c} \bd{b}_{\bd{k}} \\ \bd{b}^{\ast}_{-\bd{k}} \end{array} \right)
= 
\left( \begin{array}{c} b_{\bd{k}1}  \\ \vdots \\ b_{\bd{k} f} \\ b^{\dagger}_{-\bd{k}1}  \\ \vdots \\ b^{\dagger}_{-\bd{k} f} \end{array}
\right).
\label{eq:psikdef}
\end{equation}
Let us make the following ansatz for the desired transformation
\begin{equation}
\bd{\phi}_{\bd{k}} = \mathbbm{T}_{\bd{k}} \bd{\psi}_{\bd{k}},
\end{equation}
where  $\mathbbm{T}_{\bd{k}}$ is an invertible $2f \times 2f$ matrix.
Substituting this ansatz into the Hamiltonian (\ref{eq:H2mat}) we obtain
\begin{equation}
{\cal{H}}_2 =    \frac{1}{2}  \sum_{\bd{k}} \left[ \bd{\psi}_{\bd{k}}^{\dagger} 
\mathbbm{T}_{\bd{k}}^{\dagger}
\mathbbm{M}_{\bd{k}}  \mathbbm{T}_{\bd{k}}
\bd{\psi}_{\bd{k}} -  {\rm Tr} 
\mathbf{A}_{\bd{k}} \right],
\label{eq:H2mat2}
\end{equation}
The transformation matrix $\mathbbm{T}_{\bd{k}}$ should be constructed such that the
matrix
\begin{equation}
\mathbbm{D}_{\bd{k}}= 
\mathbbm{T}_{\bd{k}}^{\dagger}
\mathbbm{M}_{\bd{k}}  \mathbbm{T}_{\bd{k}}
\label{eq:diag}
\end{equation}
is diagonal.
In addition, the matrix $\mathbbm{T}_{\bd{k}}$
has to satisfy the following two conditions:

\begin{enumerate}
	
	\item {\textit{Boson condition:}} the new operators $b_{\bd{k} n}$
	should satisfy canonical bosonic commutation relations. 
	This implies that  only those transformations $\mathbbm{T}_{\bd{k}}$ are allowed, 
	which are pseudo-orthogonal in the sense that
	\begin{equation}
	\mathbbm{T}_{\bd{k}}^{\dagger} \mathbbm{G}  \mathbbm{T}_{\bd{k}}
	=  \mathbbm{G} =
	\mathbbm{T}_{\bd{k}} \mathbbm{G}  \mathbbm{T}_{\bd{k}}^{\dagger},
	\label{eq:bosoncond}
	\end{equation}
	where the metric matrix $\mathbbm{G}$ has the block structure
	\begin{equation}
	\mathbbm{G} = \left( \begin{array}{cc} \mathbf{1} & 0 \\
	0 & - \mathbf{1} \end{array} \right).
	\end{equation}
	Here $\mathbf{1}$ is the $f$-dimensional identity matrix.
	
	\item {\textit{Permutation condition:}} this condition follows from the fact that
	the second $f$ components of the vectors 
	$\bm{\phi}_{\bd{k}}$ and $\bm{\psi}_{\bd{k}}$ cannot be chosen independently of the
	first $f$-components, because they are related by a permutation as follows,
	\begin{equation}
	\left( \begin{array}{c} a^{\dagger}_{- \bd{k}1}  \\ \vdots \\ a^{\dagger}_{- \bd{k} f} 
	\\ a_{\bd{k}1}  \\ \vdots \\ a_{\bd{k} f} \end{array}
	\right) = \left( \begin{array}{cc} 0 & \mathbf{1} \\
	\mathbf{1} & 0 \end{array} \right) 
	\left( \begin{array}{c} a_{\bd{k}1}  \\ \vdots \\ a_{\bd{k} f} \\ a^{\dagger}_{-\bd{k}1}  \\ \vdots \\ a^{\dagger}_{-\bd{k} f} \end{array} \right). 
	\label{eq:phikcond2}
	\end{equation}
	Introducing the permutation matrix
	\begin{equation}
	\mathbbm{P} = \left( \begin{array}{cc} 0 & \mathbf{1} \\
	\mathbf{1} & 0 \end{array} \right) ,
	\end{equation}
	the condition (\ref{eq:phikcond2}) and the anologous condition for the new boson
	operators $\bd{\psi}_{\bd{k}}$ imply that
	\begin{eqnarray}
	\bd{\phi}^{\ast}_{ - \bd{k}} & = & \mathbbm{P} \bd{\phi}_{\bd{k}},
	\\
	\bd{\psi}^{\ast}_{ - \bd{k}} & = & \mathbbm{P} \bd{\psi}_{\bd{k}}.
	\end{eqnarray}
	Hence,
	\begin{eqnarray}
	\mathbbm{P} \mathbbm{T}_{\bd{k}}  \bd{\psi}_{\bd{k}} & = &
	\mathbbm{P}  \bd{\phi}_{\bd{k}} =   \bd{\phi}^{\ast}_{ - \bd{k}} 
	\nonumber
	\\
	& = &   \mathbbm{T}^{\ast}_{- \bd{k}}  \bd{\psi}^{\ast}_{- \bd{k}} =    
	\mathbbm{T}^{\ast}_{- \bd{k}} \mathbbm{P}  \bd{\psi}_{\bd{k}},
	\end{eqnarray}
	which implies
	\begin{equation}
	\mathbbm{P} \mathbbm{T}_{\bd{k}} = \mathbbm{T}^{\ast}_{- \bd{k}} \mathbbm{P}.
	\label{eq:permu}
	\end{equation}
	Using $ \mathbbm{P}^2 = \mathbbm{1}$, this relation can also be written as
	\begin{equation}
	\mathbbm{P} \mathbbm{T}_{\bd{k}} \mathbbm{P}   = \mathbbm{T}^{\ast}_{- \bd{k}}.
	\end{equation} 
	It follows that the matrix $ \mathbbm{T}_{\bd{k}}$ must have the
	following block structure,
	\begin{equation}
	\mathbbm{T}_{\bd{k}} = \left( \begin{array}{cc} 
	\mathbf{Q}_{\bd{k}} & \mathbf{R}_{\bd{k}} \\
	\mathbf{R}^{\ast}_{ - \bd{k}} & \mathbf{Q}^{\ast}_{ - \bd{k}}
	\end{array} \right),
                \label{eq:TQRgeneral}
	\end{equation}
	with two independent $f \times f$ matrices $\mathbf{Q}_{\bd{k}}$ and $\mathbf{R}_{\bd{k}}$.

\end{enumerate}
The boson condition (\ref{eq:bosoncond}) as well as the permutation condition 
(\ref{eq:permu}) define two different  groups. The intersection of these groups, i.e., 
the set of matrices $\mathbbm{T}_{\bd{k}}$ satisfying both conditions 
(\ref{eq:bosoncond}) and (\ref{eq:permu}),  defines the group of $f$-flavor Bogoliubov transformations. 
We are looking for a matrix of this type, 
which diagonalizes the matrix $\mathbbm{M}_{\bd{k}}$ according to Eq.~(\ref{eq:diag}).
Note that the matrix $\mathbbm{M}_{\bd{k}}$ in Eq.~(\ref{eq:MABdef})
also satisfies the permutation condition (\ref{eq:permu}).
In fact, given the Hamiltonian (\ref{eq:H2res}), there is some redundancy in the definition
of the matrix $\mathbbm{M}_{\bd{k}}$ because we can use the commutation relations
$    a_{\bd{k} m} a^{\dagger}_{\bd{k} n}  = 
a^{\dagger}_{ \bd{k} n}   a_{\bd{k} m }     +    \delta_{ n m } $ to rewrite
Eq.~(\ref{eq:H2mat}) in the form
\begin{equation}
{\cal{H}}_2 =    \frac{1}{2}  \sum_{\bd{k}} \left[ \bd{\phi}_{\bd{k}}^{\dagger}  
\mathbbm{M}^{\prime}_{\bd{k}}  
\bd{\phi}_{\bd{k}} -  {\rm Tr} 
\mathbf{A}_{\bd{k}} -     {\rm Tr}  \mathbf{A}^{\prime}_{\bd{k}} \right],
\label{eq:H2mat3}
\end{equation}
where now
\begin{equation}
\mathbbm{M}^{\prime}_{\bd{k}} = 
\mathbbm{M}_{\bd{k}}  +
\left( \begin{array}{cc} \mathbf{A}^{\prime}_{\bd{k}} &  \mathbf{B}^{\prime}_{\bd{k} 1} \\
\mathbf{B}^{\prime}_{\bd{k}2} & - ( \mathbf{A}^{\prime}_{ - \bd{k}} )^T \end{array}
\right).
\end{equation}
Here the matrix $\mathbbm{M}_{\bd{k}}$ is  the same as in
Eq.~(\ref{eq:MABdef}),  the $f \times f$ matrix $\mathbf{A}^{\prime}_{\bd{k}}$  
is arbitrary and the
$f \times f $ matrices $\mathbf{B}^{\prime}_{\bd{k}1}$ and
$\mathbf{B}^{\prime  }_{\bd{k}2}$ are antisymmetric in the sense that
$  \mathbf{B}^{\prime}_{\bd{k} i } = -   ( \mathbf{B}^{\prime}_{ - \bd{k} i} )^T   $ for
$i =1,2$,
which guarantees that the corresponding contributions in Eq.~(\ref{eq:H2mat3}) cancel 
after summation.
Our  choice above, $\mathbf{A}^{\prime}_{\bd{k}}  = 0= \mathbf{B}_{\bd{k} i }^{\prime}$, 
is unique because the matrix $\mathbbm{M}_{\bd{k}}$
satisfies the permutation condition, which also guarantees that 
$\mathbbm{M}_{\bd{k}}^{\prime} =  \mathbbm{M}_{\bd{k}}$ is hermitian.   
Then the Heisenberg equation of motion for the original boson operators can be written as
\begin{equation}
i \partial_t \bd{\phi}_{\bd{k}}  = [ \bd{\phi}_{\bd{k}} , {\cal{H}}_2 ] =
\mathbbm{G} \mathbbm{M}_{\bd{k}} \bd{\phi}_{\bd{k}} \equiv
\mathbbm{M}_{\bd{k}}^{\rm dyn} \bd{\phi}_{\bd{k}},
\end{equation}
where we have introduced the dynamical matrix
\begin{equation}
\mathbbm{M}_{\bd{k}}^{\rm dyn} =  \mathbbm{G} \mathbbm{M}_{\bd{k}}.
\end{equation}
Note that this definition differs from what Colpa calls the dynamical matrix in 
Ref.~[\onlinecite{Colpa78}]. The new bosons then satisfy
\begin{eqnarray}
i \partial_t \bd{\psi}_{\bd{k}}  & = &  [ \bd{\psi}_{\bd{k}} , {\cal{H}}_2 ] =  
[ \mathbbm{T}_{\bd{k}}^{-1} \bd{\phi}_{\bd{k}} , {\cal{H}}_2 ]
=   \mathbbm{T}_{\bd{k}}^{-1} [ \bd{\phi}_{\bd{k}} , {\cal{H}}_2 ]
\nonumber
\\
& = &  \mathbbm{T}_{\bd{k}}^{-1}  \mathbbm{M}_{\bd{k}}^{\rm dyn} 
\mathbbm{T}_{\bd{k}}  \bd{\psi}_{\bd{k}}.
\end{eqnarray}
The linear transformation $ \bd{\phi}_{\bd{k}} \rightarrow
\bd{\psi}_{\bd{k}} = \mathbbm{T}_{\bd{k}} \bd{\phi}_{\bd{k}}$ maps 
the matrix $\mathbbm{M}_{\bd{k}}$ onto $\mathbbm{T}^{\dagger}_{\bd{k}} 
\mathbbm{M}_{\bd{k}} \mathbbm{T}_{\bd{k}} $, while the dynamical matrix
$\mathbbm{M}^{\rm dyn}_{\bd{k}}$ transforms differently, $  
\mathbbm{M}^{\rm dyn}_{\bd{k}} \rightarrow
\mathbbm{T}^{-1}_{\bd{k}} 
\mathbbm{M}_{\bd{k}} \mathbbm{T}_{\bd{k}} $.
Mathematically, the different transformation behavior of
$\mathbbm{M}_{\bd{k}}$ and
$\mathbbm{M}_{\bd{k}}^{\rm dyn}$ is due to the fact that
$\mathbbm{M}_{\bd{k}}$ is the matrix representation  of a bilinear form (i.e., a rank $(0,2)$-tensor), while $\mathbbm{M}^{\rm dyn}_{\bd{k}}$ represents the linear mapping
$ [ \ldots , {\cal{H}}_2 ]$, which is a rank $(1,1)$-tensor.
Multiplication by the pseudo-metric $\mathbbm{G}$ establishes the transformation
between these two objects, similar to the dualization of Lorentz-vectors in relativity.

To explicitly calculate the spectrum 
of ${\cal{H}}_2$,  
we write the transformation matrix $\mathbbm{T}_{\bd{k}}$
in the form
\begin{equation}
\mathbbm{T}_{\bd{k}} = ( \bd{v}_{\bd{k} 1 }, \bd{v}_{\bd{k} 2}, \ldots ,
\bd{v}_{\bd{k} 2f } ),
\end{equation}
i.e., the  columns of the matrix  $\mathbbm{T}_{\bd{k}}$ are identified
with the column vectors $\bd{v}_{\bd{k} i }$, $i =1 , \ldots , 2f $.
Then the  diagonalization condition (\ref{eq:diag}) can  be written as
\begin{equation}
\left[   \mathbbm{T}_{\bd{k}}^{\dagger}
\mathbbm{M}_{\bd{k}}  \mathbbm{T}_{\bd{k}} \right]_{ij} =
\bd{v}_{ \bd{k} i }^{\dagger} \mathbbm{M}_{\bd{k}}   \bd{v}_{\bd{k} j } =
\delta_{ij} d_{\bd{k} i },
\label{eq:condiag}
\end{equation}
where $ d_{\bd{k} i }$ are the diagonal elements of the diagonal matrix
$\mathbbm{D}_{\bd{k}}$ in  Eq.~(\ref{eq:diag}).
The boson  condition  (\ref{eq:bosoncond}) implies that 
the column vectors $\bd{v}_{\bd{k} i }$ satisfy the pseudo-orthogonality condition
	\begin{equation}
	\bd{v}_{ \bd{k} i }^{\dagger} \mathbbm{G}   \bd{v}_{\bd{k} j } =
	g_{ij}, 
	\label{eq:pseudodiag}
	\end{equation}
	where $g_{ij} = \delta_{ij}$ for $i =1, \ldots , f$, and
	$g_{ij} = -\delta_{ij}$ for $i = f+1, \ldots, 2f$.

To explicitly construct the vectors $\bd{v}_{\bd{k} i }$ with the above properties, 
it is useful to consider the solutions $\bd{v}_{\bd{k}}$
of the generalized eigenvalue equation \cite{Colpa78},
\begin{equation}
\mathbbm{M}_{\bd{k}} \bd{v}_{\bd{k}} = \omega  \mathbbm{G} \bd{v}_{\bd{k}}.
\label{eq:genee}
\end{equation}
Multiplying both sides by  $\mathbbm{G}$ and using $\mathbbm{G}^2 = 1$
 makes Eq.~(\ref{eq:genee}) equivalent to the conventional eigenvalue equation for
$\mathbbm{M}_{\bd{k}}^{\rm dyn} = \mathbbm{G} \mathbbm{M}_{\bd{k}}$,
\begin{equation}
\mathbbm{M}^{\rm dyn}_{\bd{k}} \bd{v}_{\bd{k}} = \omega  \bd{v}_{\bd{k}}.
\label{eq:eigendyn}
\end{equation}
Although, in general,  $\mathbbm{M}^{\rm dyn}_{\bd{k}}$ is  not hermitian,
let us assume  that the eigenvalue equation 
(\ref{eq:eigendyn})
 indeed has $2 f$ linearly independent eigenvectors
$\bd{v}_{ \bd{k} 1 }, \ldots , \bd{v}_{ \bd{k} 2f }$ with eigenvalues 
$\omega_{\bd{k} n }$.
If we can normalize the eigenvectors such that
the pseudo-orthogonalization condition (\ref{eq:pseudodiag}) is satisfied,
we have by construction
\begin{equation}
\bd{v}_{ \bd{k} i }^{\dagger} \mathbbm{M}_{\bd{k}}   \bd{v}_{\bd{k} j } 
= \omega_{\bd{k} j}  
\bd{v}_{ \bd{k} i }^{\dagger} \mathbbm{G}   \bd{v}_{\bd{k} j }
= \delta_{ij} \omega_{\bd{k} i } g_{ii} ,
\end{equation} 
so that we may identify $d_{\bd{k} i } = \omega_{\bd{k} i } g_{ii}$.
Assuming that the Hamiltonian ${\cal{H}}_2$ describes a stable magnon system,
the hermitian matrix  $\mathbbm{M}_{\bd{k}} $ must be positive definite, which means that
\begin{equation}
\omega_{\bd{k} i}  
\bd{v}_{ \bd{k} i }^{\dagger} \mathbbm{G}   \bd{v}_{\bd{k} i } > 0.
\label{eq:positive}
\end{equation}
We refer to 
%
$\bd{v}_{ \bd{k} i }^{\dagger} \mathbbm{G}   \bd{v}_{\bd{k} i }$ as  pseudo-norm of $\bd{v}_{\bd{k} i }$.
It remains to be shown that the solutions of the
eigenvalue equation (\ref{eq:eigendyn}) can indeed by constructed such that
they satisfy the pseudo-orthogonality condition (\ref{eq:pseudodiag}).
The hermiticity of $\mathbbm{M}_{\bd{k}}$ implies that for any two
eigenvectors $\bd{v}_{\bd{k}  i}$ und 
$\bd{v}_{\bd{k}  j}$,
\begin{eqnarray}
0 & = & ( \mathbbm{M}_{\bd{k}} \bd{v}_{\bd{k} i } )^{\dagger} 
\bd{v}_{\bd{k} j } - \bd{v}_{\bd{k} i }^{\dagger}  ( 
\mathbbm{M}_{\bd{k}} \bd{v}_{\bd{k} j } )
= ( \omega^{\ast}_{\bd{k} i } -   \omega_{ \bd{k} j } )  
\bd{v}_{\bd{k} i }^{\dagger}  
\mathbbm{G} \bd{v}_{\bd{k} j } ,
\nonumber
\\
& &
\end{eqnarray}
so that either $\omega_{\bd{k} i }^{\ast} = \omega_{\bd{k} j }$ or
$\bd{v}_{\bd{k} i }^{\dagger}  
\mathbbm{G} \bd{v}_{\bd{k} j } =0$.
In particular, the eigenvalues of all eigenvectors with 
$\bd{v}_{\bd{k} i }^{\dagger}  
\mathbbm{G} \bd{v}_{\bd{k} i } \neq 0$ are real. 
If all eigenvalues are real and pairwise distinct, the matrix 
$\mathbbm{T}_{\bd{k}}^{\dagger} \mathbbm{M}_{\bd{k}} 
\mathbbm{T}_{\bd{k}}$ is diagonal and by properly normalizing the
eigenvectors we can satisfy the pseudo-orthogonality condition 
(\ref{eq:pseudodiag}). 
Given the fact that the metric $\mathbbm{G}$ has $f$ positive and 
$f$ negative eigenvalues, this must also be true for
the unitarily equivalent matrix $\mathbbm{T}_{\bd{k}}^{\dagger}
\mathbbm{G} \mathbbm{T}_{\bd{k}}$, so that exactly $f$ eigenvectors
can be normalized such that
$  \bd{v}_{\bd{k} i }^{\dagger} \mathbbm{G} \bd{v}_{\bd{k} i }  = 1$ and the remaining
$f$ eigenvectors can be normalized such that
$ \bd{v}_{\bd{k} i }^{\dagger} \mathbbm{G}  \bd{v}_{\bd{k} i } = -1$.
However, according to Eq.~(\ref{eq:positive}), we have $\omega_{\bd{k} i } 
\bd{v}_{\bd{k} i }^{\dagger} \mathbbm{G} \bd{v}_{\bd{k} i }  > 0$,
so that the eigenvectors $\bd{v}_{\bd{k} i }$  with positive pseudo-norm  have positive 
eigenvalues $\omega_{\bd{k} i } > 0$, while the 
eigenvectors with negative pseudo-norm have negative eigenvalues $\omega_{\bd{k} i } < 0$.

In case of degeneracy of eigenvalues,  
the corresponding linearly independent eigenvectors do not necessarily satisfy the pseudo-orthogonality condition
(\ref{eq:pseudodiag}). 
However, by means of a generalized Gram-Schmidt orthogonalization 
procedure, we can construct linear combinations of the eigenvectors in the
degenerate subspaces with the desired pseudo-orthogonalization.
For example, if the eigenvectors $\bd{v}_{\bd{k} 1 }, \ldots , \bd{v}_{\bd{k} m }$ all have the same eigenvalue, we should replace the first $m-1$ eigenvectors by
\begin{equation}
\bd{v}_{\bd{k} i } \rightarrow \bd{v}_{\bd{k} i } - \sum_{j = i +1}^m 
\bd{v}_{\bd{k} j }  ( \bd{v}_{\bd{k} i }^{\dagger} \mathbbm{G} \bd{v}_{\bd{k} j } ),
\; \;  \; i = 1, \ldots , m-1.
\end{equation}

The eigenvalues of $\mathbbm{M}_{\bd{k}}^{\rm dyn}$
 always appear in pairs: if $\omega_{\bd{k}}$ is an eigenvalue
wth eigenvector $\bd{v}_{\bd{k}}$, then
$- \omega^{\ast}_{-\bd{k}}$ is an eigenvalue with
eigenvector $\mathbbm{P} \bd{v}_{-\bd{k}}^{\ast}$, which follows from the following chain of identities,
\begin{eqnarray}
\mathbbm{M}_{\bd{k}} ( \mathbbm{P} \bd{v}^{\ast}_{-\bd{k}} ) & = &
\mathbbm{P} \mathbbm{M}_{\bd{-k}}^{\ast}  
\bd{v}^{\ast}_{-\bd{k}}
=  \mathbbm{P}  ( \omega_{\bd{-k}} \mathbbm{G}  
\bd{v}_{-\bd{k}} )^{\ast}
\nonumber
\\
& = & \omega_{\bd{-k}}^{\ast}   \mathbbm{P} \mathbbm{G}  
\bd{v}^{\ast}_{-\bd{k}} =
-  \omega_{\bd{-k}}^{\ast}   \mathbbm{G} ( \mathbbm{P}  
\bd{v}^{\ast}_{-\bd{k}} ).
\hspace{7mm}
\end{eqnarray}
After suitable relabeling, the eigenvalues can always be arranged such that
$\omega_{\bd{k} n+i } = - \omega_{ - \bd{k} i }$ and $\bd{v}_{\bd{k} n+i} = 
\mathbbm{P} \bd{v}^{\ast}_{ - \bd{k} i }$ for $i =1, \ldots , f$.
Then the matrix $\mathbbm{T}_{\bd{k} }$ can   be written in terms of column vectors as follows,
\begin{equation}
\mathbbm{T}_{\bd{k}} = ( \bd{v}_{ \bd{k} 1 } , \ldots ,  \bd{v}_{ \bd{k} f }, 
\mathbbm{P}   \bd{v}^{\ast}_{ -\bd{k} 1 } , \ldots , \mathbbm{P}   \bd{v}^{\ast}_{ -\bd{k} f }
),
\end{equation}
which satisfies the
permutation condition (\ref{eq:permu}).
The diagonalized  Hamiltonian can be written as
\begin{equation}
{\cal{H}}_2 = \frac{1}{2} \sum_{\bd{k}} 
\left[ \sum_{ i =1}^f \omega_{\bd{k} i } ( b^{\dagger}_{\bd{k} i } b_{\bd{k} i } +
b_{\bd{k} i } b^{\dagger}_{\bd{k} i } ) - {\rm{Tr}} \mathbf{A}_{\bd{k}} \right].
\end{equation}
The magnon spectrum can be obtained directly from the positive roots of
\begin{equation}
{\rm det} ( \mathbbm{M}^{\rm dyn}_{\bd{k}} - \omega \mathbbm{1} ) =0.
\label{eq:eigenColpa}
\end{equation}
The new boson annihilation operators $b_{\bd{k} i }$ can be obtained with the help of
$\mathbbm{T}_{\bd{k}}^{-1} = \mathbbm{G} \mathbbm{T}_{\bd{k}}^{\dagger}
\mathbbm{G}$ from the components of the first $f$ columns of the matrix 
$\mathbbm{T}_{\bd{k}}$,
\begin{equation}
b_{\bd{k} i } = ( \mathbbm{T}_{\bd{k}}^{-1} \bd{\phi}_{\bd{k}} )_i =
\sum_{ j =1}^{f} \left[ ( \bd{v}_{\bd{k} i }^{\dagger} )_j a_{\bd{k} j }
- ( \bd{v}_{\bd{k} i }^{\dagger} )_{f+j} a^{\dagger}_{- \bd{k} j } \right],
\end{equation}
where $i =1, \ldots , f$.

In  case when $\mathbbm{M}_{\bm{k}}$ is only positive {\it semi}definite, 
the Hamiltonian may still be representable as a sum of number operators, 
but this requires a careful handling of zero modes that will not be discussed here.	
The combined eigenvalue/Gram-Schmidt procedure gives an explicit construction for a 
pseudo-unitary $\mathbb P$-consistent diagonalization of an arbitrary hermitian positive definite matrix 
$\mathbbm{M}_{\bm{k}}$ as long $\mathbb G\mathbbm{M}_{\bm{k}}$ has only real eigenvalues. However, the generalized 
Gram-Schmidt procedure presented above is not numerically 
stable, similarly to the ordinary Gram-Schmidt procedure. 
One could adapt the known algorithms for unitary diagonalization to use the indefinite form 
$\bd{v}^\dagger \mathbb G \bd{w}$ as a scalar product, which would require handling edge cases and peculiarities 
of the particular algorithm used.

In Ref.~\cite{Colpa78}, Colpa takes a different approach.
By reducing the problem to a Cholesky decomposition and an ordinary unitary diagonalization, one can make 
use of the known efficient and/or stable algorithms for these well-researched problems, which are already 
available in software libraries or computer algebra systems. Colpa's algorithm is executed as follows:
\begin{enumerate}
	\item Find an upper triangular Matrix $\mathbb{H}_{\bm{k}}$ with $\mathbb{M}_{\bm{k}}=\mathbb{H}_{\bm{k}}^\dagger \mathbb{H}_{\bm{k}}$ (Cholesky decomposition).
	\item Find a unitary matrix $\mathbb{U}_{\bm{k}}$ such that $\mathbb{L}_{\bm{k}}=\mathbb{U}_{\bm{k}}^\dagger \mathbb{H}_{\bm{k}}\mathbb{G}\mathbb{H}_{\bm{k}}^\dagger \mathbb{U}_{\bm{k}}$ is diagonal (unitary diagonalization of $\mathbb{H}\mathbb{G}\mathbb{H}^\dagger$).
	\item Order the columns of $\mathbb{U}_{\bm{k}}$ such that $\mathbb{L}_{\bm{k}}$ has the signature $(+,\dots,+,-,\dots,-)$.
	\item Solve the equation $\mathbb{H}_{\bm{k}}\mathbb{T}_{\bm{k}}=\mathbb{U}_{\bm{k}}\mathbb{\Omega}^{1/2}_{\bm{k}}$ with $\mathbb{\Omega}_{\bm{k}}:=\mathbb{G}\mathbb{L}_{\bm{k}}$ for the components of $\mathbb{T}_{\bm{k}}$.
\end{enumerate}
Note that the fourth step in Colpa's procedure requires only a trivial $2f$-step recursion because $\mathbbm{H}_{\bm{k}}$ is triagonal.

The algorithm constructed by Colpa is the method of choice 
when the  magnon spectrum and the Bogoliubov transformation 
are  calculated numerically. 
On the other hand, for the analytic calculation of the  magnon spectrum presented in this work, 
the hermitian field approach developed  
in  Sec.~\ref{sec:construction} of the main text  is more convenient. 

In Appendix~B we shall give another application of the
hermitian field approach by  diagonalizing the quadratic magnon Hamiltonian 
for the Kitaev-Heisenberg-$\Gamma$ model for $\Gamma = K$ using only
two sublattices. 
\newline

\section*{APPENDIX B: Two-sublattice approach for $\Gamma = K$}
\setcounter{equation}{0}
\renewcommand{\theequation}{B\arabic{equation}}
 \label{sec:twosub}
As is mentioned in the last paragraph  of Sec.~\ref{sect:spectrum},
for $\Gamma = K >0$ it is possible to diagonalize the quadratic magnon Hamiltonian by using only  
two sublattices $A$ and $B$ of the honeycomb lattice, thus avoiding an additional 
complexity of the four-sublattice formulation.
In this appendix we show why and how this two-sublattice approach works and
construct the corresponding Bogoliubov transformation using 
the hermitian field method developed in Sec.~\ref{sec:construction}.

Let us  go back to the derivation of the quadratic magnon Hamiltonian
in the zigzag state presented in Sec.~\ref{sec:quadratic}, where we 
projected spin operators on each site onto  local axes that match the direction
of the local magnetization of the zigzag state.
In this basis, the contributions  from the Kitaev part and the off-diagonal exchange part to the transverse
part ${\cal{H}}_{\bot}$ 
 of the spin Hamiltonian, ${\cal{H}}^K_{\bot}$ and ${\cal{H}}^\Gamma_{\bot}$,  are given in 
Eqs.~(\ref{eq:HKreal}) and (\ref{eq:HGreal}). Adding these two contributions, we obtain
\begin{widetext}
 \begin{eqnarray}
   {\cal{H}}_{ \bot}^{K} +   {\cal{H}}_{ \bot}^{\Gamma}  & = &
 \frac{1}{8} \sum_{ p p^{\prime}} 
 \Biggl\{  
\sum_{ \bd{R} \in a } 
 \left[ 
( K_{xx}^{ \bar{p} \bar{p}^{\prime}}  + \Gamma_{yz}^{ \bar{p} \bar{p}^{\prime}} )  
 S^p_{\bd{R}} S^{p^{\prime}}_{\bd{R} + \bd{d}_x}
 +   ( K_{yy}^{ \bar{p} \bar{p}^{\prime}} +   \Gamma_{zx}^{ \bar{p} \bar{p}^{\prime}} )
 S^p_{\bd{R}} S^{p^{\prime}}_{\bd{R} + \bd{d}_y}
 +   ( K_{zz}^{ \bar{p} {p}^{\prime}} +   \Gamma_{xy}^{ \bar{p} {p}^{\prime}}  ) 
 S^p_{\bd{R}} S^{p^{\prime}}_{\bd{R} + \bd{d}_z} \right]
 \nonumber
 \\
  & & \hspace{8mm} + \sum_{ \bd{R} \in c } 
 \left[   ( K_{xx}^{ p p^{\prime}}    +   \Gamma_{yz}^{ p p^{\prime}}  )
 S^p_{\bd{R}} S^{p^{\prime}}_{\bd{R} + \bd{d}_x}
 +       ( K_{yy}^{ p p^{\prime}} +   \Gamma_{zx}^{ p p^{\prime}} )
 S^p_{\bd{R}} S^{p^{\prime}}_{\bd{R} + \bd{d}_y}
 +   ( K_{zz}^{ {p} \bar{p}^{\prime}}  +    \Gamma_{xy}^{ {p} \bar{p}^{\prime}} )
 S^p_{\bd{R}} S^{p^{\prime}}_{\bd{R} + \bd{d}_z} \right]
 \nonumber
 \\
 &  & \hspace{8mm} + 
\sum_{ \bd{R} \in b } 
 \left[ (    K_{xx}^{ p p^{\prime}}  + \Gamma_{yz}^{ p p^{\prime}}  )
 S^p_{\bd{R}} S^{p^{\prime}}_{\bd{R} - \bd{d}_x}
 +  (  K_{yy}^{ p p^{\prime}}    + \Gamma_{zx}^{ p p^{\prime}} )
 S^p_{\bd{R}} S^{p^{\prime}}_{\bd{R} - \bd{d}_y}
 + (  K_{zz}^{ {p} \bar{p}^{\prime}}   + \Gamma_{xy}^{ {p} \bar{p}^{\prime}} )
 S^p_{\bd{R}} S^{p^{\prime}}_{\bd{R} - \bd{d}_z} \right]
 \nonumber
 \\
 &  & \hspace{8mm} +
\sum_{ \bd{R} \in d } 
 \left[ (  K_{xx}^{ \bar{p} \bar{p}^{\prime}}   + \Gamma_{yz}^{ \bar{p} \bar{p}^{\prime}}   )
 S^p_{\bd{R}} S^{p^{\prime}}_{\bd{R} - \bd{d}_x}
 +  (   K_{yy}^{ \bar{p} \bar{p}^{\prime}}  + \Gamma_{zx}^{ \bar{p} \bar{p}^{\prime}}  )
 S^p_{\bd{R}} S^{p^{\prime}}_{\bd{R} - \bd{d}_y}
 + (    K_{zz}^{ \bar{p} {p}^{\prime}} + \Gamma_{xy}^{ \bar{p} {p}^{\prime}}  )
 S^p_{\bd{R}} S^{p^{\prime}}_{\bd{R} - \bd{d}_z} \right]
 \Biggr\}.
 \label{eq:HKGreal}
 \end{eqnarray}
Here the coefficients
$K^{ p p^{\prime}}_{\alpha \beta}$ and
$\Gamma^{ p p^{\prime}}_{\alpha \beta}$
are defined in Eqs.~(\ref{eq:Kppdef}) and (\ref{eq:Gammappdef})
of the main text.
In general, all these  coefficients are complex and the coefficients
in the $c$-sublattice sum (second line) 
are the complex conjugates of the
coefficients in the $a$-sublattice sum (first line); similarly, the
coefficients in the $d$-sublattice sum (last line) are the
complex conjugates of the coefficients in the $b$-sublattice sum (third line).
It turns out, however, that for $\Gamma = K > 0$, the imaginary parts
of all coefficients in the sums
$K^{ p p^{\prime}}_{\alpha \beta} + 
\Gamma^{ p p^{\prime}}_{\alpha \beta}$ cancel, so that 
the coefficients in the  $a$-sum  are identical to the
coefficients in  the  $c$-sum,  while the coefficients in the $b$-sum match
those of the $d$-sum. As a consequence,
 it is sufficient to work only with two sublattices 
$A= a \cup c$ and $B = b \cup d$ in this case. By explicitly evaluating the coefficients for
$ \Gamma = K > 0$ using Eqs.~(\ref{eq:eigenvectors}) and (\ref{eq:npdef})
we obtain
 \begin{eqnarray}
   {\cal{H}}_{ \bot}^{K} +   {\cal{H}}_{ \bot}^{\Gamma}  & = &
 \frac{K}{8} 
\Biggl\{  
\sum_{ \bd{R} \in A } 
 \left[ \frac{3}{2}  ( S^+_{\bd{R}} S^{-}_{\bd{R} + \bd{d}_x}
 + S^+_{\bd{R}} S^{-}_{\bd{R} + \bd{d}_y} )
 - 2 S^+_{\bd{R}} S^{-}_{\bd{R} + \bd{d}_z}
 -  \frac{1}{2} ( S^+_{\bd{R}} S^{+}_{\bd{R} + \bd{d}_x}
 + S^+_{\bd{R}} S^{+}_{\bd{R} + \bd{d}_y} )   
 + {\rm h.c.}
 \right]
 \nonumber
 \\
 & & \hspace{4mm} + 
\sum_{ \bd{R} \in B } 
 \left[ \frac{3}{2}  ( S^+_{\bd{R}} S^{-}_{\bd{R} - \bd{d}_x}
 + S^+_{\bd{R}} S^{-}_{\bd{R} - \bd{d}_y} )
 - 2 S^+_{\bd{R}} S^{-}_{\bd{R} - \bd{d}_z}
 -  \frac{1}{2} ( S^+_{\bd{R}} S^{+}_{\bd{R} - \bd{d}_x}
 + S^+_{\bd{R}} S^{+}_{\bd{R} - \bd{d}_y} ) + {\rm h.c.}
 \right] 
 \Biggl\}.
 \label{eq:H2lines}
 \end{eqnarray}
Actually, keeping in mind that the nearest neighbor vectors $\bd{d}_{\alpha}$ connect different
sublattices and shifting   $\bd{R}_b - \bd{d}_{\alpha} = \bd{R}_a$ 
(where the subscript indicates the sublattice) in the second line, we see that
the contribution from the two  lines in Eq.~(\ref{eq:H2lines}) are identical,
so that in the special care $\Gamma = K > 0$ we may write
 \begin{eqnarray}
   {\cal{H}}_{ \bot}^{K} +   {\cal{H}}_{ \bot}^{\Gamma}  & = &
 \frac{K}{4} 
\sum_{ \bd{R} \in A } 
 \left[ \frac{3}{2}  ( S^+_{\bd{R}} S^{-}_{\bd{R} + \bd{d}_x}
 + S^+_{\bd{R}} S^{-}_{\bd{R} + \bd{d}_y} )
 - 2 S^+_{\bd{R}} S^{-}_{\bd{R} + \bd{d}_z}
 -  \frac{1}{2} ( S^+_{\bd{R}} S^{+}_{\bd{R} + \bd{d}_x}
 + S^+_{\bd{R}} S^{+}_{\bd{R} + \bd{d}_y} )   
 + {\rm h.c.}
 \right].
 \label{eq:H1lines}
 \end{eqnarray}
\end{widetext}
In the special case  $\Gamma = K$, it is, therefore, possible 
to diagonalize the quadratic magnon Hamiltonian by introducing only
two sublattices,
which simplifies the calculation of magnon spectrum and
the construction of the Bogoliubov transformation.
After expressing spin operators in terms of the Holstein-Primakoff bosons
and retaining all terms quadratic in bosons, 
we find that the Hamiltonian  
(\ref{eq:hamiltonian}) with additional next-next nearest neighbor Heisenberg 
exchange $J_3$
leads to the following quadratic boson Hamiltonian  for $\Gamma = K > 0$,
 \begin{eqnarray}
 {\cal{H}}_2 & = & (3 J_3 - J + 2 K ) S \sum_{\bd{R}} a^{\dagger}_{\bd{R}} a_{\bd{R}}  
 \nonumber
 \\
 & + & JS \sum_{\bd{R} \in A} \left[ a^{\dagger}_{ \bd{R}  }  a_{ \bd{R} + \bd{d}_x } 
    + a^{\dagger}_{ \bd{R}  }  a_{ \bd{R} + \bd{d}_y } +  a^{\dagger}_{ \bd{R}  }  a^{\dagger}_{ \bd{R} + \bd{d}_z }
 + {\rm h.c.} \right]
 \nonumber
 \\
 & + & J_3 S \sum_{\bd{R} \in A} \sum_{\alpha =x,y,z}
\left[ a^{\dagger}_{ \bd{R}  }  a^{\dagger}_{ \bd{R} -  \bd{d}_\alpha } 
 + {\rm h.c.} \right]
 \nonumber
 \\
 & + & K S \sum_{\bd{R} \in A} \biggl[ \frac{3}{4} \left( a^{\dagger}_{\bd{R}} a_{\bd{R} + \bd{d}_x } 
 + a^{\dagger}_{\bd{R}} a_{\bd{R} + \bd{d}_y }  \right) -    a^{\dagger}_{\bd{R}} a_{\bd{R} + \bd{d}_z }
 \nonumber
 \\
 & & \hspace{11mm}  -  \frac{1}{4} \left( a^{\dagger}_{\bd{R}} a^{\dagger}_{\bd{R} + \bd{d}_x } 
 + a^{\dagger}_{\bd{R}} a^{\dagger}_{\bd{R} + \bd{d}_y }  \right) 
   + {\rm h.c.} \biggr].
 \end{eqnarray}
Defining
 \begin{subequations}
 \begin{eqnarray}
  a_{\bd{R}} & = & \sqrt{\frac{2}{N}} \sum_{\bd{k}} e^{ i \bd{k} \cdot \bd{R}} a_{\bd{k}}, \; \; \; \; {\bd{R}} \in A,
 \\
& = & \sqrt{\frac{2}{N}} \sum_{\bd{k}} e^{ i \bd{k} \cdot \bd{R}} b_{\bd{k}}, \; \; \; \; {\bd{R}} \in B,
\end{eqnarray}
\end{subequations}
where the sums are over the first Brillouin zone of the honeycomb lattice, and $N$ is the total number of lattice sites, we obtain
 \begin{eqnarray}
 {\cal{H}}_2 & = & - JS \sum_{\bd{k}} \biggl\{  A ( a^{\dagger}_{\bd{k}} a_{\bd{k}} + b^{\dagger}_{\bd{k}} b_{\bd{k}} )
 \nonumber
 \\
 & & \hspace{12mm} + \left[ B_{\bd{k}} a^{\dagger}_{\bd{k}} b_{\bd{k}} - C_{\bd{k}} a^{\dagger}_{\bd{k}} b^{\dagger}_{ - \bd{k}} 
 + {\rm h.c.} \right] \biggr\},
 \label{eq:H2spec}
 \end{eqnarray}
where
 \begin{subequations}
 \begin{eqnarray}
 A & = & 1 - \frac{2 K}{J} - \frac{ 3 J_3}{J},
 \\
 B_{\bd{k}} & = & - \left( 1 + \frac{3 K}{4J} \right) \gamma^{xy}_{\bd{k}} + \frac{K}{J} \gamma^z_{\bd{k}},
 \\
 C_{\bd{k}} & = &  - \frac{K}{4J} \gamma^{xy}_{\bd{k}} + \gamma^z_{\bd{k}} +     \frac{ 3 J_3}{J} \gamma^{(3)}_{\bd{k}},
 \end{eqnarray}
 \end{subequations}
with
 \begin{subequations}
 \begin{eqnarray}
 \gamma^{xy}_{\bd{k}} & = & e^{ i \bd{k} \cdot \bd{d}_x } +    e^{ i \bd{k} \cdot \bd{d}_y },
 \\
 \gamma^{z}_{\bd{k}} & = & e^{ i \bd{k} \cdot \bd{d}_z } ,
 \\
 \gamma^{(3)}_{\bd{k}} & = & \frac{1}{3} \sum_{\alpha =x,y,z} e^{ - 2 i \bd{k} \cdot \bd{d}_{\alpha} }.
 \end{eqnarray}
 \end{subequations}
Obviously, the Hamiltonian (\ref{eq:H2spec}) is of the form
(\ref{eq:H2res2}) with $f=2$ boson flavors, so that
we could  use  Colpa's algorithm to calculate  
magnon spectrum. The matrices 
$\mathbf{A}_{\bd{k}}$ and $\mathbf{B}_{\bd{k}}$ in this case are  given by
 \begin{eqnarray}
 \mathbf{A}_{\bd{k}} & = &  \left( \begin{array}{cc} A^{aa}_{\bd{k}} & A^{ab}_{\bd{k}} \\
 A^{ba}_{\bd{k}} & A^{bb}_{\bd{k}} \end{array} \right)
 = (- JS ) \left( \begin{array}{cc} A & B_{\bd{k}} \\
 B^{\ast}_{\bd{k}} & A \end{array} \right),
 \\
 \mathbf{B}_{\bd{k}} & = &  \left( \begin{array}{cc} B^{aa}_{\bd{k}} & B^{ab}_{\bd{k}} \\
 B^{ba}_{\bd{k}} & B^{bb}_{\bd{k}} \end{array} \right)
 = JS  \left( \begin{array}{cc} 0 & C_{\bd{k}} \\
 C^{\ast}_{\bd{k}} & 0 \end{array} \right).
 \end{eqnarray}
We have not been able to find  in the existing literature an explicit analytic construction
of the Bogoliubov transformation that  would diagonalize the quadratic boson Hamiltonian (\ref{eq:H2spec}) 
in a general case of non-commuting matrices $\mathbf{A}_{\bd{k}}$ and $\mathbf{B}_{\bd{k}}$.  
We, therefore, provide an explicit construct of such a transformation using the hermitian-field 
method developed in  Sec.~\ref{sec:hermitian} and in Sec.~\ref{sec:construction} instead of the Colpa's approach.

First of all, we note that in the case of our interest,
the matrices satisfy 
$\mathbf{A}_{\bd{k}} = \mathbf{A}_{ - \bd{k}}^{\ast}$ and
$\mathbf{B}_{\bd{k}} = \mathbf{B}_{ - \bd{k}}^{\ast}$, so that
the matrix $\mathbf{W}_{\bd{k}}$ 
defined in Eq.~(\ref{eq:Wmatdef}) vanishes identically.
Then, the  magnon spectrum can be obtained from the roots of
[see Eq.~(\ref{eq:detVT2}) of the  main text]  
\begin{equation}
 {\rm det} (  \mathbf{T}_{\bd{k}} \mathbf{V}_{\bd{k}} - \omega^2 \mathbf{1} ) =0,
 \label{eq:detcon}
 \end{equation}
where
 \begin{eqnarray}
  \mathbf{T}_{\bd{k}} & = & \mathbf{A}_{\bd{k}} - \mathbf{B}_{\bd{k}} 
 = (- JS ) \left( \begin{array}{cc} A & B_{\bd{k}} + C_{\bd{k}} \\
 B^{\ast}_{\bd{k}} + C_{\bd{k}}^{\ast} & A \end{array} \right),
 \nonumber
 \\
 &&
 \label{eq:Tkdef}
 \\
 \mathbf{V}_{\bd{k}} & = & \mathbf{A}_{\bd{k}} + \mathbf{B}_{\bd{k}} 
 = (- JS ) \left( \begin{array}{cc} A & B_{\bd{k}} - C_{\bd{k}} \\
 B^{\ast}_{\bd{k}} - C_{\bd{k}}^{\ast} & A \end{array} \right).
 \nonumber
 \\
 & &
 \label{eq:Vkdef}
  \end{eqnarray}
Eq.~(\ref{eq:detcon}) can be reduced to the biquadratic equation
 \begin{eqnarray}
  &  & 0= \left( \frac{ \omega^2 }{ (JS)^2 } \right)^2 - 2  
 [ A^2 + | B_{\bd{k}} |^2 -  | C_{\bd{k}} |^2 ]  \frac{ \omega^2 }{ (JS)^2 }
 \nonumber
 \\
 & & 
 + [ A^2 + | B_{\bd{k}} |^2 -  | C_{\bd{k}} |^2 ]^2
 - ( B_{\bd{k}} C_{\bd{k}}^{\ast} - B_{\bd{k}}^{\ast} C_{\bd{k}} )^2 - 4 A^2
    | {B}_{\bd{k}} |^2,
 \nonumber
 \\
 & &
 \end{eqnarray}
which has  positive roots
 \begin{equation}
 \omega_{\bd{k} \pm} = | J | S \sqrt{ A^2  + | B_{\bd{k}} |^2 -  | C_{\bd{k}} |^2
 \pm R_{\bd{k}} },
 \label{eq:dispersions}
 \end{equation}
with
 \begin{eqnarray}
 R_{\bd{k}} & = & \sqrt{ 4 A^2 | B_{\bd{k}} |^2 +  ( B_{\bd{k}} C_{\bd{k}}^{\ast} - B_{\bd{k}}^{\ast} C_{\bd{k}} )^2 }
 \nonumber
 \\
 & = & 2 \sqrt{  A^2 | B_{\bd{k}} |^2 -  [ {\rm Im} (   B_{\bd{k}} C_{\bd{k}}^{\ast} ) ]^2  }.
 \label{eq:Rkdef}
 \end{eqnarray}
Keeping in mind that in the two-sublattice approach
the momentum $\bd{k}$ belongs to  the first Brillouin zone of the honeycomb lattice 
(black dashed hexagon in Fig.~\ref{fig:contour}), while in the
four-sublattice approach the corresponding first Brillouin zone is only half as large 
(green dashed rectangle in  Fig.~\ref{fig:contour}), 
we see that the magnon spectrum $\{ \omega_{\bd{k} +} , \omega_{\bd{k} -} \}$
obtained in the   two-sublattice approach is indeed identical to the
magnon spectrum $\{ \omega_{\bd{k} + }^{+} , \omega_{\bd{k} +}^- ,
 \omega_{\bd{k} - }^{+} , \omega_{\bd{k} -}^- \}$ 
obtained in the four-sublattice approach, see Eqs.~(\ref{eq:dispersion3}) and
(\ref{eq:dispersion4}).

To construct the explicit Bogoliubov transformation that 
diagonalizes the Hamiltonian (\ref{eq:H2spec}),
we use the hermitian field algorithm described in  Sec.~\ref{sec:construction}, which consists of the following steps:
\begin{enumerate}

\item Calculate the square root  $\mathbf{T}_{\bd{k}}^{1/2}$    of the ``kinetic energy matrix'' $\mathbf{T}_{\bd{k}}$ and its inverse
 $\mathbf{T}_{\bd{k}}^{-1/2}$.

\item Calculate the transformed ``potential energy matrix''
$\tilde{\mathbf{V}}_{\bd{k}} = \mathbf{T}_{\bd{k}}^{1/2} \mathbf{V}_{\bd{k}}
  \mathbf{T}_{\bd{k}}^{1/2}$.

\item Calculate the  unitary matrix $\mathbf{S}_{\bd{k}}$ that diagonalizes 
$\tilde{\mathbf{V}}_{\bd{k}}$:
\begin{equation}
\mathbf{S}_{\bm{k}} \mathbf{\tilde{V}}_{\bm{k}} \mathbf{S}_{\bm{k}}^{\dagger}
=
\mathbf{\Omega}_{\bm{k}}^2\ \ \text{diagonal}.
\end{equation}

\item Then, the two-flavor Bogoliubov transformation to the new 
operators $b_{\bd{k} 1}$ and $b_{\bd{k} 2}$
that diagonalize the Hamiltonian
can be  expressed in terms of a single $4 \times 4 $ block 
matrix $\mathbb{T}_{\bd{k}}$ as follows,
\begin{equation} 
\left( \begin{array}{c} a_{\bd{k}}  \\  b_{\bd{k}} \\ a^{\dagger}_{- \bd{k} }     
\\  
 b^{\dagger}_{ - \bd{k}} \end{array} \right)
=
 \mathbb{T}_{\bd{k}} 
\left( \begin{array}{c} b_{\bd{k}1}   \\ b_{\bd{k} 2} \\ 
 b^{\dagger}_{- \bd{k} 1}   \\ 
 b^{\dagger}_{ - \bd{k} 2} \end{array}
\right) 
=  \left( \begin{array}{cc}  
\mathbf{Q}_{\bd{k}} &  \mathbf{R}_{\bd{k}} \\
\mathbf{R}_{\bd{k}} &  \mathbf{Q}_{\bd{k}} 
 \end{array} \right)
\left( \begin{array}{c} b_{\bd{k}1}   \\ b_{\bd{k} 2} \\ 
 b^{\dagger}_{- \bd{k} 1}   \\ 
 b^{\dagger}_{ - \bd{k} 2} \end{array}
\right) ,
\label{eq:bog2}
\end{equation}
where the $2 \times 2$ blocks $\mathbf{Q}_{\bd{k}}$ and 
$\mathbf{R}_{\bd{k}}$ are given by
 \begin{eqnarray}
 \mathbf{Q}_{\bd{k}} & = & \frac{1}{2} \left[
\mathbf{T}_{\bm{k}}^{\sfrac{1}{2}}\ \mathbf{S}_{\bm{k}}       \ \mathbf{\Omega}_{\bm{k}}^{-\sfrac{1}{2}} +
 \mathbf{T}_{\bm{k}}^{-\sfrac{1}{2}}\ \mathbf{S}_{\bm{k}}   \ \mathbf{\Omega}_{\bm{k}}^{\sfrac{1}{2}} \right],
 \label{eq:ublock}
 \\
\mathbf{R}_{\bd{k}} & = & 
 \frac{1}{2} \left[
\mathbf{T}_{\bm{k}}^{\sfrac{1}{2}}\ \mathbf{S}_{\bm{k}}       \ \mathbf{\Omega}_{\bm{k}}^{-\sfrac{1}{2}} -
 \mathbf{T}_{\bm{k}}^{-\sfrac{1}{2}}\ \mathbf{S}_{\bm{k}}   \ \mathbf{\Omega}_{\bm{k}}^{\sfrac{1}{2}} \right].
 \label{eq:vblock}
 \end{eqnarray}
\end{enumerate}
Let us now explicitly construct the matrices above for the specific two-flavor 
Hamiltonian ${\cal{H}}_2$ given in Eq.~(\ref{eq:H2spec}).
Writing the ``kinetic energy matrix'' $\mathbf{T}_{\bd{k}}$ introduced 
in Eq.~(\ref{eq:Tkdef}) as
 \begin{eqnarray}
 \mathbf{T}_{\bd{k}} & = & \left( \begin{array}{cc} a & t_{\bd{k}} \\
t_{\bd{k}}^{\ast} & a \end{array} \right),
 \end{eqnarray}
where $a = - JSA $ and $t_{\bd{k}} = - JS ( B_{\bd{k}} + C_{\bd{k}} )$,
the eigenvalues and normalized eigenvectors of $\mathbf{T}_{\bd{k}}$ are
 \begin{eqnarray}
 \bd{t}_{\bd{k} + } & = & \frac{1}{\sqrt{2}} \left(
 \begin{array}{c} \lambda_{\bd{k}} \\ 1 \end{array} \right) , \; \; \; \mbox{eigenvalue $a + | t_{\bd{k}} | $},
 \\
 \bd{t}_{\bd{k} - } & = & \frac{1}{\sqrt{2}} \left(
 \begin{array}{c} -1  \\ \lambda^{\ast}_{\bd{k}}  \end{array} \right) , \; \; \; \mbox{eigenvalue 
$a  - | t_{\bd{k}} | $},
 \end{eqnarray}
where we have introduced the phase factor 
 \begin{equation}
 \lambda_{\bd{k}} = t_{\bd{k}} / | t_{\bd{k}} |.
 \end{equation}
The matrix $\mathbf{T}_{\bd{k}}$ is, therefore, diagonalized by the following unitary matrix
 \begin{equation}
 \mathbf{U}_{\bd{k}}  =  ( \bd{t}_{\bd{k} +} , \bd{t}_{\bd{k} - } ) =
 \frac{1}{\sqrt{2}} \left( \begin{array}{cc} \lambda_{\bd{k}} & -1  \\ 1 & \lambda_{\bd{k}}^{\ast} \end{array} \right).
 \end{equation}
Explicitly,
  \begin{equation}
 \mathbf{U}_{\bd{k}}^{\dagger} \mathbf{T}_{\bd{k}}   \mathbf{U}_{\bd{k}}=
 \left( \begin{array}{cc} a + | t_{\bd{k}} | & 0 \\
 0 & a - | t_{\bd{k}} | \end{array} \right).
 \end{equation}
We conclude that the  square root of $\mathbf{T}_{\bd{k}}$ and its inverse 
can be written as
 \begin{eqnarray}
 \mathbf{T}_{\bd{k}}^{1/2} & = &   \mathbf{U}_{\bd{k}}
 \left( \begin{array}{cc} \sqrt{ a + | t_{\bd{k}} | } & 0 \\
 0 &  \sqrt{ a - | t_{\bd{k}} | } \end{array} \right) \mathbf{U}^{\dagger}_{\bd{k}}
\nonumber
 \\
 & = &   \left( \begin{array}{cc}  x_{\bd{k}} &  z_{\bd{k}} \\
   z_{\bd{k}}^{\dagger} & x_{\bd{k}} \end{array} \right),
 \\
\mathbf{T}_{\bd{k}}^{- 1/2} & = &   \frac{1}{ \sqrt{ a^2 - | t_{\bd{k}} |^2 }}
 \left( \begin{array}{cc}  x_{\bd{k}} &  - z_{\bd{k}} \\
  - z_{\bd{k}}^{\ast} & x_{\bd{k}} \end{array} \right),
 \end{eqnarray}
where
 \begin{eqnarray}
 x_{\bd{k}} & = & \frac{1}{2} \left[ \sqrt{a + | t_{\bd{k} } | }
 +   \sqrt{ a - | t_{\bd{k} } | } \right],
 \\
  z_{\bd{k}} & = & \frac{\lambda_{\bd{k}}}{2} \left[ \sqrt{ a + | t_{\bd{k} } | }
 -    \sqrt{ a - | t_{\bd{k} } | } \right],
 \end{eqnarray}
and we have used $ x_{\bd{k}}^2 - | z_{\bd{k}} |^2 = \sqrt{a^2 - | t_{\bd{k}} |^2}$.
Writing
 \begin{eqnarray}
 \mathbf{V}_{\bd{k}} 
& = & \left( \begin{array}{cc} a & v_{\bd{k}} \\
v_{\bd{k}}^{\ast} & a \end{array} \right),
 \end{eqnarray}
where $v_{\bd{k}} = - JS ( B_{\bd{k}} - C_{\bd{k}} )$,
the transformed ``potential energy matrix'' can be written as
 \begin{equation}
 \tilde{\mathbf{V}}_{\bd{k}} = \mathbf{T}_{\bd{k}}^{1/2} \mathbf{V}_{\bd{k}}
  \mathbf{T}_{\bd{k}}^{1/2} = \left( \begin{array}{cc} \tilde{a}_{\bd{k}} &
 \tilde{v}_{\bd{k}} \\
 \tilde{v}_{\bd{k}}^{\ast} & \tilde{a}_{\bd{k}} 
 \end{array} \right),
 \end{equation}
with 
 \begin{eqnarray}
 \tilde{a}_{\bd{k}} & = & a ( x_{\bd{k}}^2 + | z_{\bd{k}} |^2 ) +
 ( v_{\bd{k}} z_{\bd{k}}^{\ast} + v_{\bd{k}}^{\ast} z_{\bd{k}} ) x_{\bd{k}}
 \nonumber
 \\
 & = & ( JS )^2 [ A^2 + | B_{\bd{k}} |^2 - | C_{\bd{k}} |^2 ],
 \\
 \tilde{v}_{\bd{k}} & = &  v_{\bd{k}} x_{\bd{k}}^2 + v_{\bd{k}}^{\ast} z_{\bd{k}}^2 
+ 2 a x_{\bd{k}} z_{\bd{k}}
 \nonumber
 \\
 & = & 
\lambda_{\bd{k}} \left[ a + {\rm Re} ( v_{\bd{k}} 
\lambda_{\bd{k}}^{\ast} )  + i \sqrt{ a^2 - | {t}_{\bd{k}} |^2}
 {\rm Im} ( v_{\bd{k}}  \lambda_{\bd{k}}^{\ast} )
 \right].
 \nonumber
 \\
 & & \label{eq:tildevk}
 \end{eqnarray}
In terms of the dimensionless coefficients $B_{\bd{k}}$ and $C_{\bd{k}}$ defined above we
can write
 \begin{eqnarray}
 {\rm Re} ( v_{\bd{k}}  \lambda_{\bd{k}}^{\ast} ) & = & \frac{1}{2} \left[ v_{\bd{k}}  \lambda_{\bd{k}}^{\ast}
 + v_{\bd{k}}^{\ast}  \lambda_{\bd{k}} \right]
 \nonumber
 \\
 & = & \frac{ ( JS)^2}{ | t_{\bd{k}} | } (  | B_{\bd{k} } |^2 - | C_{\bd{k}} |^2 )
 \nonumber
 \\
 & = & | J |  S \frac{   | B_{\bd{k} } |^2 - | C_{\bd{k}} |^2 }{ | B_{\bd{k}}  + C_{\bd{k} } | }.
 \end{eqnarray}
and
 \begin{eqnarray}
 {\rm Im } ( v_{\bd{k}}  \lambda_{\bd{k}}^{\ast} ) & = & \frac{1}{2i} \left[ v_{\bd{k}}  \lambda_{\bd{k}}^{\ast}
 -  v_{\bd{k}}^{\ast}  \lambda_{\bd{k}} \right]
 \nonumber
 \\
 & = & \frac{ ( JS)^2}{ i | t_{\bd{k}} | } (   B_{\bd{k} } C_{\bd{k}}^{\ast}  - B_{\bd{k}}^{\ast} C_{\bd{k}} )
 \nonumber
 \\
 & = & 2 | J |  S \frac{  {\rm Im} ( B_{\bd{k}} C_{\bd{k}}^{\ast} ) }{ | B_{\bd{k}}  + C_{\bd{k} } | }.
 \end{eqnarray}
By construction, the eigenvalues of the hermitian matrix $\tilde{\mathbf{V}}_{\bd{k}}$ are the
squares $\omega_{\bd{k} , \pm}^2$ of the spin-wave dispersions
given in Eq.~(\ref{eq:dispersions}),
 \begin{equation}
 \omega_{\bd{k} \pm}^2 = \tilde{a}_{\bd{k}} \pm | \tilde{v}_{\bd{k}} | =
 ( JS )^2 [ A^2 + | B_{\bd{k}} |^2 - | C_{\bd{k}} |^2 ] \pm R_{\bd{k}},
  \end{equation}
where $R_{\bd{k}}$ is given in Eq.~(\ref{eq:Rkdef}).
To see explicitly that indeed $ | \tilde{v}_{\bd{k}} | = R_{\bd{k}}$,
we take the squared absolute value of Eq.~(\ref{eq:tildevk}) and obtain
 \begin{eqnarray}
 | \tilde{v}_{\bd{k}} |^2 & = & a^2 [ | t_{\bd{k}} | +    {\rm Re} ( v_{\bd{k}}  \lambda_{\bd{k}}^{\ast} ) ]^2
 + ( a^2 - | t_{\bd{k}} |^2 ) [  {\rm Im } ( v_{\bd{k}}  \lambda_{\bd{k}}^{\ast} ) ]^2
 \nonumber
 \\
 & = & a^2 [ | t_{\bd{k}} |^2 + | v_{\bd{k}} |^2 + 2 {\rm Re} ( v_{\bd{k}} t^{\ast}_{\bd{k}} ) ]
 -   [  {\rm Im } ( v_{\bd{k}}  t_{\bd{k}}^{\ast} ) ]^2
 \nonumber
 \\
 & = & 
 4  A^2 | B_{\bd{k}} |^2 -  4 [ {\rm Im} (   B_{\bd{k}} C_{\bd{k}}^{\ast} ) ]^2  = R_{\bd{k}}^2.
 \end{eqnarray}
The normalized eigenvectors and eigenvalues of the matrix $\tilde{\mathbf{V}}_{\bd{k}}$ are
 \begin{eqnarray}
 \bd{v}_{\bd{k} + } & = & \frac{1}{\sqrt{2}} \left(
 \begin{array}{c} \tilde{\lambda}_{\bd{k}} \\ 1 \end{array} \right) , \; \; \; 
 \mbox{eigenvalue $\tilde{a}_{\bd{k}} + | \tilde{v}_{\bd{k}} | = \omega^2_{\bd{k} +} $},
 \hspace{9mm}
 \\
 \bd{v}_{\bd{k} - } & = & \frac{1}{\sqrt{2}} \left(
 \begin{array}{c} -1  \\ \tilde{\lambda}^{\ast}_{\bd{k}}  \end{array} \right) , \; \; \; \mbox{eigenvalue 
$\tilde{a}_{\bd{k}}  - | \tilde{v}_{\bd{k}} | = \omega^2_{\bd{k} -} $},
 \end{eqnarray}
where we have introduced the phase factor 
 \begin{equation}
 \tilde{\lambda}_{\bd{k}} = \tilde{v}_{\bd{k}} / | \tilde{v}_{\bd{k}} |.
 \end{equation}
With
 \begin{equation}
 \mathbf{S}^{\dagger}_{\bd{k}}  =  ( \bd{v}_{\bd{k} +} , \bd{v}_{\bd{k} - } ) =
 \frac{1}{\sqrt{2}} \left( \begin{array}{cc} \tilde{\lambda}_{\bd{k}} & -1  \\ 1 & \tilde{\lambda}_{\bd{k}}^{\ast} \end{array} \right)
 \end{equation}
we obtain
  \begin{equation}
 \mathbf{S}_{\bd{k}}\tilde{\mathbf{V}}_{\bd{k}}   \mathbf{S}^{\dagger}_{\bd{k}}=
 \mathbf{\Omega}_{\bd{k}}^2 = 
 \left( \begin{array}{cc} \omega^2_{\bd{k} +} & 0 \\
 0 & \omega^2_{\bd{k} -} \end{array} \right).
 \end{equation}
With that, all matrices that are necessary to  calculate the 
transformation matrix $\mathbb{T}_{\bd{k}}$ in 
Eq.~(\ref{eq:bog2}) are now explicitly constructed.

\section*{APPENDIX C: Technical details of the 
calculation of the
magnon damping for $\Gamma = K$}
\setcounter{equation}{0}
\renewcommand{\theequation}{C\arabic{equation}}
 \label{sec:technical}

In this appendix we give additional technical details of
our calculation of the magnon damping for $\Gamma = K$ using the
four-sublattice formulation presented in Sec.~\ref{sec:damping}.

\subsection{Propagator matrices at $\Gamma = K > 0$}
For $\Gamma = K > 0$, the parameters $r$ and $s$ defined in Eq.~(\ref{eq:smallrdef}) and Eq.~(\ref{eq:smallqdef}) have the values $r=1$ and $s=-2$. Choosing the gauge angle $\phi$ introduced in Eq.~(\ref{eq:loc_frame_02}) as $\phi=0$, the matrix elements of the quadratic Hamiltonian defined in Eqs.~(\ref{eq:lambda_01})-(\ref{eq:nucomplex}) reduce to
 \begin{subequations}
 \label{eq:matspecial}
\begin{eqnarray}
\lambda &=& S \left( -J + 2K + 3J_3 \right),
\\
\alpha_{\bm{k}} &=& S \left( J + \frac{3}{4} K \right) \left(e^{i\bm{k}\cdot\bm{d}_x}+e^{i\bm{k}\cdot\bm{d}_y}\right),
\\
\beta_{\bm{k}} &=& -S K e^{i\bm{k}\cdot\bm{d}_z},
\\
\mu_{\bm{k}} &=& S J e^{i\bm{k}\cdot\bm{d}_z} + S J_3 \sum_{\alpha = x,y,z} e^{-2i \bm{k} \cdot \bm{d_\alpha}},
\\
\nu_{\bm{k}} &=& -\frac{1}{4} S K \left(e^{i\bm{k}\cdot\bm{d}_x}+e^{i\bm{k}\cdot\bm{d}_y}\right).
\end{eqnarray}
\end{subequations}
\subsection{Transformation matrices}
For the construction of the multi-flavor Bogoliubov transformation
by means of the hermitian-field approach of Sec.~\ref{sec:construction}, we have to calculate
the square root of the hermitian matrix 
$\mathbf{T}_{\bd{k}}$ in Eq.~(\ref{eq:Tkmatrix}) for the matrix elements given in Eqs.~(\ref{eq:matspecial}).
From the definition~(\ref{eq:Thalf_definition}), we find that the square root of the
``kinetic energy matrix'' has the structure
\begin{equation}
\mathbf{T}_{\bm{k}}^{\sfrac{1}{2}}
=
 \left(
\begin{array}{cc|cc}
t_{1,\bm{k}} & t_{3,\bm{k}} & t_{2,\bm{k}} & t_{4,\bm{k}} \\
t_{3,\bm{k}}^* & t_{1,\bm{k}} & t_{4,\bm{k}}^* & t_{2,\bm{k}} \\
 \hline
t_{2,\bm{k}}^* & t_{4,\bm{k}} & t_{1,\bm{k}} & t_{3,\bm{k}} \\
t_{4,\bm{k}}^* & t_{2,\bm{k}}^* & t_{3,\bm{k}}^* & t_{1,\bm{k}}
 \end{array}
\right).
\label{eq:Thalf_01}
\end{equation}
Defining
 \begin{subequations}
\begin{eqnarray}
\eta_{1,\bm{k}} &=& \alpha_{\bm{k}} + \beta_{\bm{k}} - \mu_{\bm{k}} - \nu_{\bm{k}}, \\
\eta_{2,\bm{k}} &=& \alpha_{\bm{k}} - \beta_{\bm{k}} + \mu_{\bm{k}} - \nu_{\bm{k}},
\end{eqnarray}
 \end{subequations}
the matrix elements of 
$\mathbf{T}_{\bm{k}}^{\sfrac{1}{2}}$
can be written as
 \begin{subequations}
\begin{eqnarray}
t_{1,\bm{k}} = \frac{1}{4} \bigg\{& & \sqrt{\lambda-|\eta_{1,\bm{k}}|} + \sqrt{\lambda+|\eta_{1,\bm{k}}|}
\nonumber
\\&+& \sqrt{\lambda-|\eta_{2,\bm{k}}|} + \sqrt{\lambda+|\eta_{2,\bm{k}}|} \bigg\},
\end{eqnarray}
\begin{eqnarray}
t_{2,\bm{k}} = \frac{1}{4} \bigg\{& \sqrt{\lambda-|\eta_{1,\bm{k}}|} + \sqrt{\lambda+|\eta_{1,\bm{k}}|}
\nonumber
\\ &- \sqrt{\lambda-|\eta_{2,\bm{k}}|} - \sqrt{\lambda+|\eta_{2,\bm{k}}|} \bigg\},
\end{eqnarray}
\begin{eqnarray}
t_{3,\bm{k}} = \frac{1}{4} \bigg\{ &-& \text{sgn} (\eta_{1,\bm{k}}) \left[ \sqrt{\lambda-|\eta_{1,\bm{k}}|} - \sqrt{\lambda+|\eta_{1,\bm{k}}|} \right]
\nonumber
\\ &+& \text{sgn} (\eta_{2,\bm{k}}) \left[ \sqrt{\lambda-|\eta_{2,\bm{k}}|} - \sqrt{\lambda+|\eta_{2,\bm{k}}|} \right] \bigg\},
\nonumber
\\
\end{eqnarray}
\begin{eqnarray}
t_{4,\bm{k}} = \frac{1}{4} \bigg\{ &-& \text{sgn} (\eta_{1,\bm{k}}) \left[ \sqrt{\lambda-|\eta_{1,\bm{k}}|} - \sqrt{\lambda+|\eta_{1,\bm{k}}|} \right]
\nonumber
\\ &-& \text{sgn} (\eta_{2,\bm{k}}) \left[ \sqrt{\lambda-|\eta_{2,\bm{k}}|} - \sqrt{\lambda+|\eta_{2,\bm{k}}|} \right] \bigg\}.
\nonumber
\\
\end{eqnarray}
 \end{subequations}
For the notational simplicity, we define
 \begin{equation}
{\rm sgn}(z)=z /|z|
 \end{equation}
for any complex number $z \neq 0$. 
For the unitary matrix $\mathbf{S}_{\bm{k}}$ 
defined in Eq.~(\ref{eq:Sunitary_definition}) that diagonalizes the 
modified ``potential energy matrix'' ${\mathbf{\tilde{V}}_{\bd{k}}}$
we find
\begin{eqnarray}
 \mathbf{S}_{\bm{k}}
 &=&
 \frac{1}{2}
 \left(
 \begin{array}{cc|cc}
 s_{1,\bm{k}} & -s_{1,\bm{k}} & -s_{2,\bm{k}} & s_{2,\bm{k}} \\
 -1 & -1 & 1 & 1\\
 \hline
 -s_{1,\bm{k}} & s_{1,\bm{k}} & -s_{2,\bm{k}} & s_{2,\bm{k}} \\
 1 & 1 & 1 & 1
 \end{array} \right),
 \label{eq:Sunit_01}
\end{eqnarray}
with
 \begin{subequations}
\begin{eqnarray}
 s_{1,\bm{k}} &=& \text{sgn}(\eta_{2,\bm{k}}) \; \text{sgn} \bigg( 2 \lambda\ \text{Re} \left\{ (-\alpha_{\bm{k}}+\beta_{\bm{k}})\text{sgn}(\eta_{2,\bm{k}}^*) \right\}
 \nonumber \\
  & &+ 2i \sqrt{\lambda^2-|\eta_{2,\bm{k}}|^2}\ \text{Im} \left\{ (-\alpha_{\bm{k}}+\beta_{\bm{k}})\text{sgn}(\eta_{2,\bm{k}}^*) \right\} \bigg),
   \nonumber \\
\end{eqnarray}
\begin{eqnarray}
 s_{2,\bm{k}} &=& \text{sgn}(\eta_{1,\bm{k}}) \;  \text{sgn} \bigg( 2 \lambda\ \text{Re} \left\{ (\alpha_{\bm{k}}+\beta_{\bm{k}})\text{sgn}(\eta_{1,\bm{k}}^*) \right\} 
 \nonumber \\
 & &+ 2i \sqrt{\lambda^2-|\eta_{1,\bm{k}}|^2}\ \text{Im} \left\{ (\alpha_{\bm{k}}+\beta_{\bm{k}})\text{sgn}(\eta_{1,\bm{k}}^*) \right\} \bigg).
 \nonumber \\
\end{eqnarray}
 \end{subequations}
The diagonal matrix $\mathbf{\Omega}_{\bd{k}}$ introduced in Eq.~(\ref{eq:Sunitary_definition}) contains the magnon energies given in Eqs.~(\ref{eq:dispersion3}) and~(\ref{eq:dispersion4}) with the following ordering,
\begin{equation}
\mathbf{\Omega}_{\bm{k}} =
 \begin{pmatrix}
\omega_{{\bm{k}},-}^{-} & 0 & 0 & 0 \\
0 & \omega_{{\bm{k}},+}^{-} & 0 & 0\\
0 & 0 & \omega_{{\bm{k}},-}^{+} & 0 \\
0 & 0 & 0 & \omega_{{\bm{k}},+}^{+}
\label{eq:Omega}
\end{pmatrix}.
\end{equation}
Given the matrices $\mathbf{T}_{\bd{k}}^{\sfrac{1}{2}}$ and 
and $\mathbf{S}_{\bd{k}}$, we can explcitly construct the matrices
$\mathbf{L}_{\bm{k}}  =  \mathbf{T}_{\bm{k}}^{\sfrac{1}{2}} \mathbf{S}_{\bm{k}}$ and $\mathbf{Y}_{\bm{k}} = \mathbf{T}_{\bm{k}}^{-\sfrac{1}{2}} \mathbf{S}_{\bm{k}}$, which are the building blocks of the $4 \times 4$ matrices
$\mathbf{Q}_{\bd{k}}$ and $\mathbf{R}_{\bd{k}}$ 
that appear in the  $8 \times 8$ Bogoliubov transformation matrix  $\mathbb{T}_{\bd{k}}$ in Eq.~(\ref{eq:M_02}). 
We, thus, obtain

\begin{widetext}
\begin{equation}
\mathbf{L}_{\bm{k}} = \mathbf{T}_{\bm{k}}^{\sfrac{1}{2}} \mathbf{S}_{\bm{k}}
=
 \left(
\begin{array}{cc|cc}
s_{1,\bm{k}} \vartheta_{\bm{k}}^{+}-\vartheta_{\bm{k}}^{-} & -\vartheta_{\bm{k}}^{-}-s_{1,\bm{k}} \vartheta_{\bm{k}}^{+} & -\varphi_{\bm{k}}^{-}-s_{2,\bm{k}} \varphi_{\bm{k}}^{+} & s_{2,\bm{k}} \varphi_{\bm{k}}^{+}-\varphi_{\bm{k}}^{-} \\
s_{1,\bm{k}} \vartheta_{\bm{k}}^{-*}-\vartheta_{\bm{k}}^{+} & -\vartheta_{\bm{k}}^{+}-s_{1,\bm{k}} \vartheta_{\bm{k}}^{-*} & \varphi_{\bm{k}}^{+}+s_{2,\bm{k}} \varphi_{\bm{k}}^{-*} & \varphi_{\bm{k}}^{+}-s_{2,\bm{k}} \varphi_{\bm{k}}^{-*} \\
 \hline
\vartheta_{\bm{k}}^{-}-s_{1,\bm{k}} \vartheta_{\bm{k}}^{+} & \vartheta_{\bm{k}}^{-}+s_{1,\bm{k}} \vartheta_{\bm{k}}^{+} & -\varphi_{\bm{k}}^{-}-s_{2,\bm{k}} \varphi_{\bm{k}}^{+} & s_{2,\bm{k}} \varphi_{\bm{k}}^{+}-\varphi_{\bm{k}}^{-} \\
\vartheta_{\bm{k}}^{+}-s_{1,\bm{k}} \vartheta_{\bm{k}}^{-*} & \vartheta_{\bm{k}}^{+}+s_{1,\bm{k}} \vartheta_{\bm{k}}^{-*} & \varphi_{\bm{k}}^{+}+s_{2,\bm{k}} \varphi_{\bm{k}}^{-*} & \varphi_{\bm{k}}^{+}-s_{2,\bm{k}} \varphi_{\bm{k}}^{-*} \\
\end{array}
 \right),
\label{eq:L_01}
\end{equation}
and 
\begin{equation}
\mathbf{Y}_{\bm{k}} = \mathbf{T}_{\bm{k}}^{-\sfrac{1}{2}} \mathbf{S}_{\bm{k}}
= \left(
 \begin{array}{cc|cc}
q_{2,\bm{k}} \left(\vartheta_{\bm{k}}^{-}+s_{1,\bm{k}} \vartheta_{\bm{k}}^{+} \right) &
q_{2,\bm{k}}\left(\vartheta_{\bm{k}}^{-}-s_{1,\bm{k}} \vartheta_{\bm{k}}^{+} \right) &
q_{1,\bm{k}}\left(\varphi_{\bm{k}}^{-}-s_{2,\bm{k}}
\varphi_{\bm{k}}^{+} \right) &
q_{1,\bm{k}}\left(\varphi_{\bm{k}}^{-}+s_{2,\bm{k}} \varphi_{\bm{k}}^{+}\right) \\
-q_{2,\bm{k}}\left(\vartheta_{\bm{k}}^{+}+s_{1,\bm{k}} \vartheta_{\bm{k}}^{-*}\right) & -q_{2,\bm{k}}\left(\vartheta_{\bm{k}}^{+}-s_{1,\bm{k}} \vartheta_{\bm{k}}^{-*}\right) &
q_{1,\bm{k}}\left(\varphi_{\bm{k}}^{+}-s_{2,\bm{k}} \varphi_{\bm{k}}^{-*}\right) &
q_{1,\bm{k}}\left(\varphi_{\bm{k}}^{+}+s_{2,\bm{k}} \varphi_{\bm{k}}^{-*}\right) \\
 \hline
-q_{2,\bm{k}}\left(\vartheta_{\bm{k}}^{-}+s_{1,\bm{k}} \vartheta_{\bm{k}}^{+}\right) & -q_{2,\bm{k}}\left(\vartheta_{\bm{k}}^{-}-s_{1,\bm{k}} \vartheta_{\bm{k}}^{+}\right) &
q_{1,\bm{k}}\left(\varphi_{\bm{k}}^{-}-s_{2,\bm{k}} \varphi_{\bm{k}}^{+}\right) & q_{1,\bm{k}}\left(\varphi_{\bm{k}}^{-}+s_{2,\bm{k}} \varphi_{\bm{k}}^{+}\right) \\
q_{2,\bm{k}}\left(\vartheta_{\bm{k}}^{+}+s_{1,\bm{k}} \vartheta_{\bm{k}}^{-*}\right) & q_{2,\bm{k}}\left(\vartheta_{\bm{k}}^{+}-s_{1,\bm{k}} \vartheta_{\bm{k}}^{-*}\right) &
q_{1,\bm{k}}\left(\varphi_{\bm{k}}^{+}-s_{2,\bm{k}} \varphi_{\bm{k}}^{-*}\right) &
q_{1,\bm{k}}\left(\varphi_{\bm{k}}^{+}+s_{2,\bm{k}} \varphi_{\bm{k}}^{-*}\right) \\
\end{array} \right),
\label{eq:Y_01}
\end{equation}
where 
\begin{equation}
q_{i,\bm{k}} = \frac{1}{\sqrt{\lambda^2-|\eta_{i,\bm{k}}|^2}},\quad i=1,2,
\end{equation}
and
\begin{subequations}
\begin{eqnarray}
\varphi_{\bm{k}}^{+} &=& \frac{1}{4} \left( \sqrt{\lambda-|\eta_{1,\bm{k}}|} + \sqrt{\lambda+|\eta_{1,\bm{k}}|} \right),
\ \ \ \
\varphi_{\bm{k}}^{-} = \frac{1}{4} \left( \sqrt{\lambda-|\eta_{1,\bm{k}}|} - \sqrt{\lambda+|\eta_{1,\bm{k}}|} \right) \text{sgn} (\eta_{1,\bm{k}}),
\\
\vartheta_{\bm{k}}^{+} &=& \frac{1}{4} \left( \sqrt{\lambda-|\eta_{2,\bm{k}}|} + \sqrt{\lambda+|\eta_{2,\bm{k}}|} \right),
\ \ \ \
\vartheta_{\bm{k}}^{-} = \frac{1}{4} \left( \sqrt{\lambda-|\eta_{2,\bm{k}}|} - \sqrt{\lambda+|\eta_{2,\bm{k}}|} \right) \text{sgn} (\eta_{2,\bm{k}}).\quad \quad
 \hspace{7mm}
\end{eqnarray}
 \end{subequations}

\subsection{Cubic vertices at $\Gamma = K > 0$}
For the calculation of magnon damping in 
Sec.~\ref{sec:magnondamp}, we need the cubic interaction vertices in the
Bogoliubov basis, in which the quadratic part of the bosonized Hamiltonian is diagonal.
Therefore, we first have to derive the cubic part of the Hamiltonian
in the Holstein-Primakoff basis. The corresponding Euclidean action
is given in Eq.~(\ref{eq:S3res}). It is convenient to symmetrize the vertices and
write the action in the symmetrized form (\ref{eq:S3tensor}). In this notation 
the  following $48$ vertices are non-zero,
 \begin{subequations}
 \label{eq:threevertices1}
 \begin{eqnarray}
 \Gamma^{ \bar{d} \bar{a} a } ( \bd{1} , \bd{2} , \bd{3} ) & = & 
 \Gamma^{ \bar{d} a \bar{a}  } ( \bd{1} , \bd{2} , \bd{3} ) = 
  \Gamma^{ {d} a \bar{a}  } ( \bd{1} , \bd{2} , \bd{3} )  =  
 \Gamma^{  {d} \bar{a} a  } ( \bd{1} , \bd{2} , \bd{3} )
 = - V_{ \bd{1}},
 \\
 \Gamma^{ \bar{a} \bar{d}  a } ( \bd{1} , \bd{2} , \bd{3} ) & = & 
 \Gamma^{ a \bar{d}  \bar{a}  } ( \bd{1} , \bd{2} , \bd{3} )  = 
\Gamma^{  {a}  {d}  \bar{a} } ( \bd{1} , \bd{2} , \bd{3} )  =  
 \Gamma^{ \bar{a}  {d}   {a}  } ( \bd{1} , \bd{2} , \bd{3} ) =
 - V_{ \bd{2}},
 \\
 \Gamma^{ \bar{a} a \bar{d}   } ( \bd{1} , \bd{2} , \bd{3} ) & = & 
 \Gamma^{ a   \bar{a} \bar{d}  } ( \bd{1} , \bd{2} , \bd{3} )  =
 \Gamma^{ a \bar{a}   {d}   } ( \bd{1} , \bd{2} , \bd{3} )  = 
 \Gamma^{  \bar{a} a  {d}  } ( \bd{1} , \bd{2} , \bd{3} ) 
=  - V_{ \bd{3}},
 \end{eqnarray}
 \end{subequations}
 \begin{subequations}
 \begin{eqnarray}
 \Gamma^{ \bar{c} \bar{b} b } ( \bd{1} , \bd{2} , \bd{3} ) & = & 
 \Gamma^{ \bar{c} b \bar{b}  } ( \bd{1} , \bd{2} , \bd{3} ) = 
  \Gamma^{ {c} b \bar{b}  } ( \bd{1} , \bd{2} , \bd{3} )  =  
 \Gamma^{  {c} \bar{b} b  } ( \bd{1} , \bd{2} , \bd{3} )
 =  e^{ i ( \bd{k}_1 + \bd{k}_2 + \bd{k}_3 )  \cdot \bd{d}_z } V^{\ast}_{ \bd{1}},
 \\
 \Gamma^{ \bar{b} \bar{c}  b } ( \bd{1} , \bd{2} , \bd{3} ) & = & 
 \Gamma^{ b \bar{c}  \bar{b}  } ( \bd{1} , \bd{2} , \bd{3} )  = 
\Gamma^{  {b}  {c}  \bar{b} } ( \bd{1} , \bd{2} , \bd{3} )  =  
 \Gamma^{ \bar{b}  {c}   {b}  } ( \bd{1} , \bd{2} , \bd{3} ) =
 e^{ i ( \bd{k}_1 + \bd{k}_2 + \bd{k}_3 )  \cdot \bd{d}_z } V^{\ast}_{ \bd{2}},
 \\
 \Gamma^{ \bar{b} b \bar{c}   } ( \bd{1} , \bd{2} , \bd{3} ) & = & 
 \Gamma^{ b   \bar{b} \bar{c}  } ( \bd{1} , \bd{2} , \bd{3} )  =
 \Gamma^{ b \bar{b}   {c}   } ( \bd{1} , \bd{2} , \bd{3} )  = 
 \Gamma^{  \bar{b} b  {c}  } ( \bd{1} , \bd{2} , \bd{3} ) 
 =
 e^{ i ( \bd{k}_1 + \bd{k}_2 + \bd{k}_3 )  \cdot \bd{d}_z } V^{\ast}_{ \bd{3}},
 \end{eqnarray}
 \end{subequations}
 \begin{subequations}
 \begin{eqnarray}
 \Gamma^{ \bar{b} \bar{c} c } ( \bd{1} , \bd{2} , \bd{3} ) & = & 
 \Gamma^{ \bar{b} c \bar{c}  } ( \bd{1} , \bd{2} , \bd{3} ) = 
  \Gamma^{ {b} c \bar{c}  } ( \bd{1} , \bd{2} , \bd{3} )  =  
 \Gamma^{  {b} \bar{c} c } ( \bd{1} , \bd{2} , \bd{3} )
 =  e^{ i ( \bd{k}_1 + \bd{k}_2 + \bd{k}_3 )  \cdot \bd{a}_1 } V_{ \bd{1}},
 \\
 \Gamma^{ \bar{c} \bar{b}  c } ( \bd{1} , \bd{2} , \bd{3} ) & = & 
 \Gamma^{ c \bar{b}  \bar{c}  } ( \bd{1} , \bd{2} , \bd{3} )  = 
\Gamma^{  {c}  {b}  \bar{c} } ( \bd{1} , \bd{2} , \bd{3} )  =  
 \Gamma^{ \bar{c}  {b}   {c}  } ( \bd{1} , \bd{2} , \bd{3} ) =
 e^{ i ( \bd{k}_1 + \bd{k}_2 + \bd{k}_3 )  \cdot \bd{a}_1 } V_{ \bd{2}},
 \\
 \Gamma^{ \bar{c} c \bar{b}   } ( \bd{1} , \bd{2} , \bd{3} ) & = & 
 \Gamma^{ c   \bar{c} \bar{b}  } ( \bd{1} , \bd{2} , \bd{3} )  =
 \Gamma^{ c \bar{c}   {b}   } ( \bd{1} , \bd{2} , \bd{3} )  = 
 \Gamma^{  \bar{c} c  {b}  } ( \bd{1} , \bd{2} , \bd{3} ) 
=   e^{ i ( \bd{k}_1 + \bd{k}_2 + \bd{k}_3 )  \cdot \bd{a}_1 } V_{ \bd{3}},
 \end{eqnarray}
 \end{subequations}
 \begin{subequations}
 \label{eq:threevertices4}
 \begin{eqnarray}
 \Gamma^{ \bar{a} \bar{d} d } ( \bd{1} , \bd{2} , \bd{3} ) & = & 
 \Gamma^{ \bar{a} d \bar{d}  } ( \bd{1} , \bd{2} , \bd{3} ) = 
  \Gamma^{ {a} d \bar{d}  } ( \bd{1} , \bd{2} , \bd{3} )  =  
 \Gamma^{  {c} \bar{b} b  } ( \bd{1} , \bd{2} , \bd{3} )
 =  - e^{ i ( \bd{k}_1 + \bd{k}_2 + \bd{k}_3 )  \cdot \bd{d}_x } V^{\ast}_{ \bd{1}},
 \\
 \Gamma^{ \bar{d} \bar{a}  d } ( \bd{1} , \bd{2} , \bd{3} ) & = & 
 \Gamma^{ d \bar{a}  \bar{d}  } ( \bd{1} , \bd{2} , \bd{3} )  = 
\Gamma^{  {d}  {a}  \bar{d} } ( \bd{1} , \bd{2} , \bd{3} )  =  
 \Gamma^{ \bar{d}  {a}   {d}  } ( \bd{1} , \bd{2} , \bd{3} ) =
 - e^{ i ( \bd{k}_1 + \bd{k}_2 + \bd{k}_3 )  \cdot \bd{d}_x } V^{\ast}_{ \bd{2}},
 \\
 \Gamma^{ \bar{d} d \bar{a}   } ( \bd{1} , \bd{2} , \bd{3} ) & = & 
 \Gamma^{ d   \bar{d} \bar{a}  } ( \bd{1} , \bd{2} , \bd{3} )  =
 \Gamma^{ d \bar{d}   {a}   } ( \bd{1} , \bd{2} , \bd{3} )  = 
 \Gamma^{  \bar{d} d {a}  } ( \bd{1} , \bd{2} , \bd{3} ) 
 =
 - e^{ i ( \bd{k}_1 + \bd{k}_2 + \bd{k}_3 )  \cdot \bd{d}_x } V^{\ast}_{ \bd{3}},
 \end{eqnarray}
 \end{subequations}
\end{widetext}
where  we have abbreviated the momentum labels $\bm{k}_i $ by ${\bm i}={\bf 1}, {\bf 2},{\bf 3}$, and
the interaction vertex $V_{\bd{k}}$ is defined in Eq.~(\ref{eq:V_01}).
For clarity, we have replaced the superscipts $\mu, \nu ,  \lambda \in \{ 1,2,3,4,5,6,7,8 \}$
by the associated field types $\{ a,b,c,d, \bar{a} , \bar{b} , \bar{c}, \bar{d} \}$.
Note that the cubic vertices are not periodic in the first magnetic Brillouin zone because 
of  our choice of the Fourier transformation in~(\ref{eq:FT_01}).
The corresponding vertices $\tilde{\Gamma}^{\mu \nu \lambda} ( \bd{k}_1 , \bd{k}_2 ,
 \bd{k}_3 )$ in the Bogoliubov basis can be obtained from Eq.~(\ref{eq:tildeGammadef}).

\section*{APPENDIX D: Magnon damping for $\Gamma = K $ and small $ J_3 >0$}
\setcounter{equation}{0}
\renewcommand{\theequation}{D\arabic{equation}}
 \label{sec:smallJ3}
\begin{figure}
	\includegraphics[width=0.80\columnwidth]{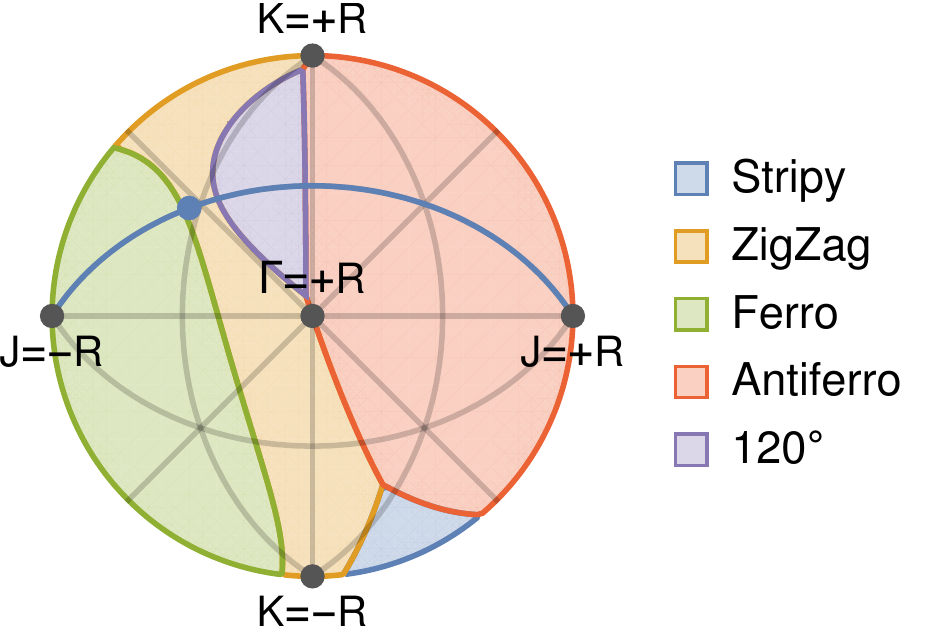}
	\caption{Phase diagram of the Kitaev-Heisenberg-$\Gamma$ model for 
		$J^2+K^2+\Gamma^2 = 177\text{\text{meV}}^2$ with additional third nearest neighbor Heisenberg exchange  
		$J_3 = 0.5\text{\text{meV}}$. We use the same parametrization and projection as in Fig.~\ref{fig:phases}(c). 
		We highlight the line $\Gamma = K > 0$ and the point~(\ref{eq:2nd_point}) in the parameter space, for which the 
		magnon damping is calculated.}
	\label{fig:2nd_point}
\end{figure}
\begin{figure}
	\begin{minipage}{.49\textwidth}
		\includegraphics[width=\columnwidth]{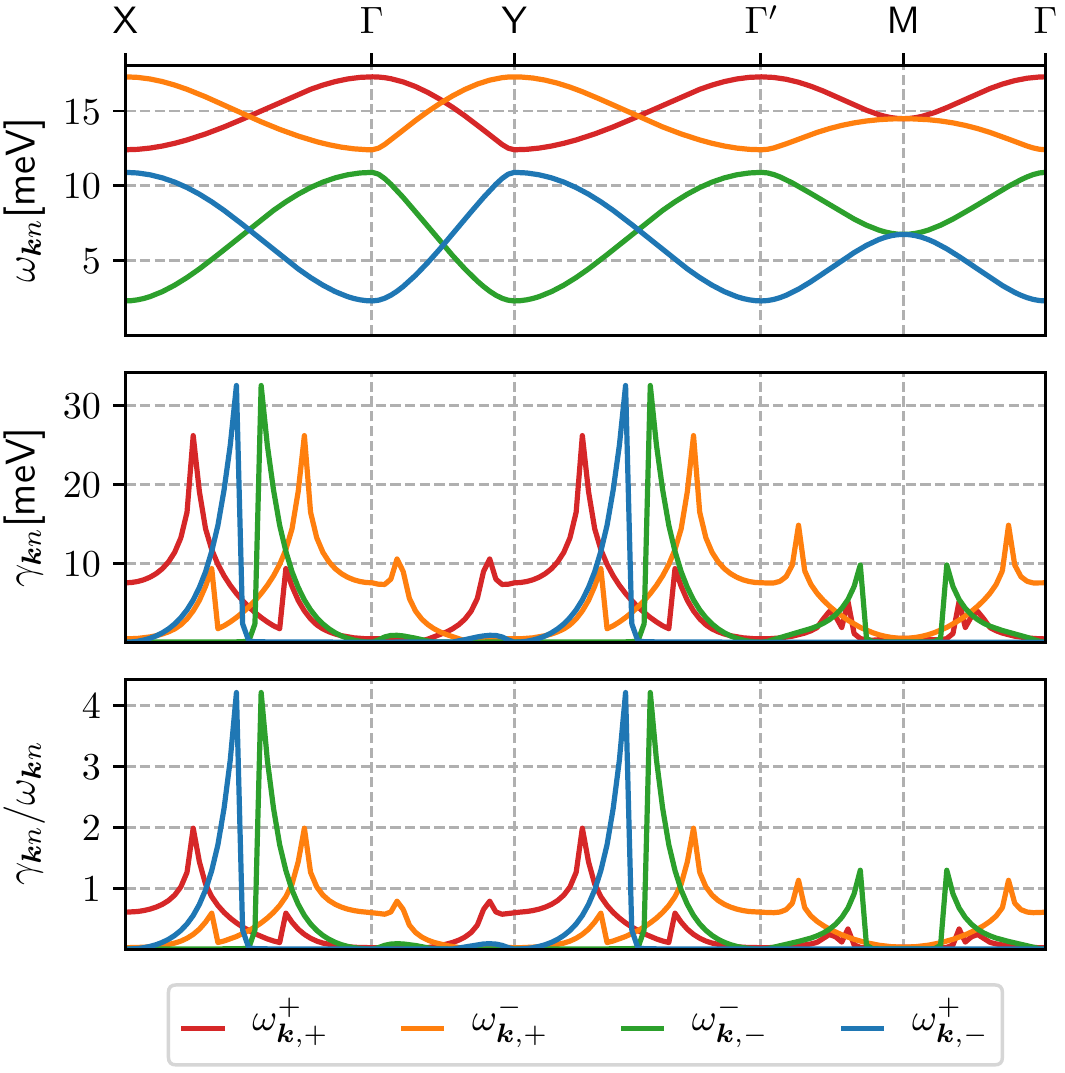}
		\caption{
			Magnon damping in the Born approximation (\ref{eq:damping_integral}). 
			(a) Magnon energies $\omega_{\bm{k}n}$, (b) 
			magnon damping $\gamma_{\bm{k}n}$, and  (c)  magnon damping rates, 
			$\gamma_{\bm{k}n}/\omega_{\bm{k}n}$, 
			for the model parameters given in Eqs.~(\ref{eq:2nd_point}) and along the momentum path 
			shown in Fig.~\ref{fig:contour}. The color-coding is the same as in Fig.~\ref{fig:damping_born}. 
		}
		\label{fig:damping_born_appendix}
	\end{minipage}
	\begin{minipage}{.49\textwidth}
		\vspace{5mm}
		\includegraphics[width=\columnwidth]{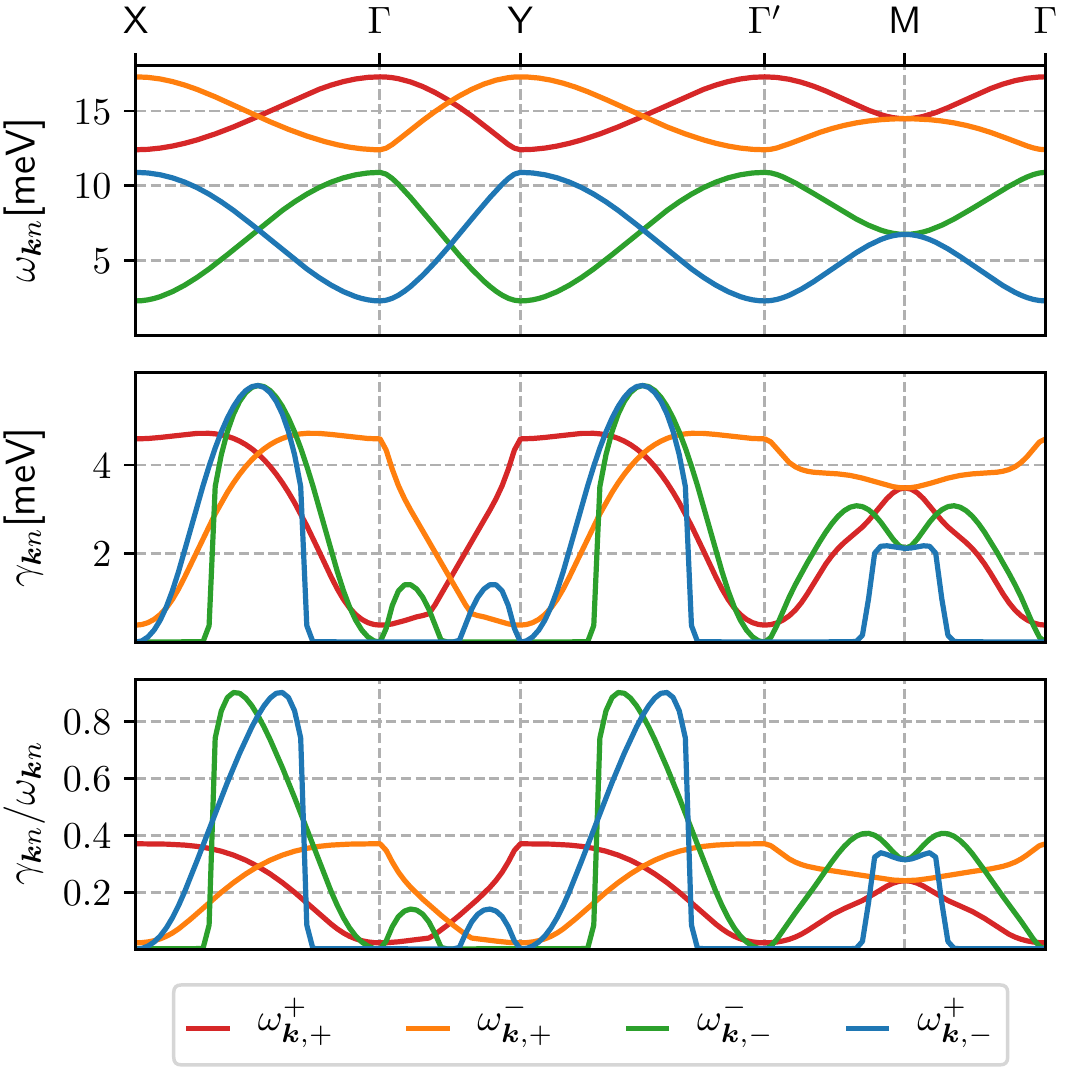}
		\caption{
			Same as in Fig.~\ref{fig:damping_born_appendix} in the self-consistent iDE approximation (\ref{eq:dampingiDE0}). 
		}
		\label{fig:damping_appendix}
	\end{minipage}
\end{figure}

	\begin{figure*}
		\includegraphics[width=.99\textwidth]{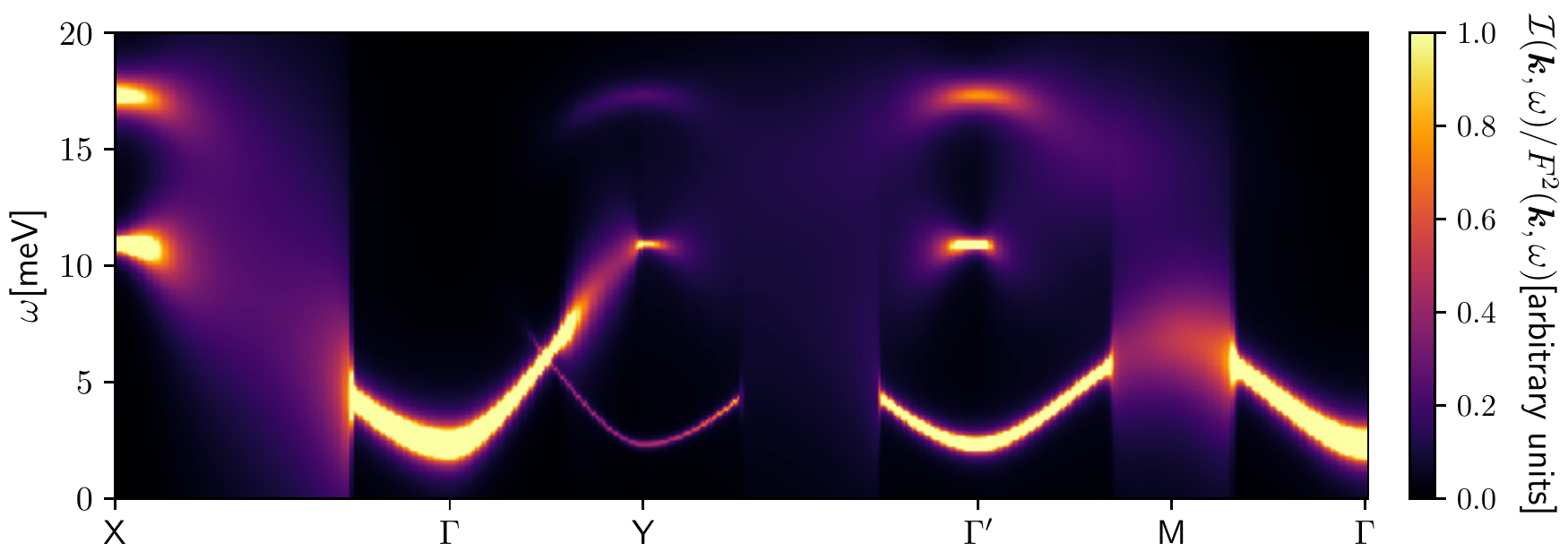}
		\caption{
			The neutron-scattering intensity $\mathcal{I}( \bm{k},\omega )$ 
			normalized by  the square of the atomic form factor as given in Eq.~(\ref{eq:I_div_f_01}) with magnon lifetime effects in the iDE approximation given by Eq.~(\ref{eq:dampingiDE}), evaluated for the model parameters given in Eqs.~(\ref{eq:2nd_point}).}
		\label{fig:intensity_appendix}
	\end{figure*}

The full analytical calculation of the matrix elements for the magnon-magnon interaction gives deviations in the magnon damping as compared with the constant matrix element approximation near certain symmetry points on the momentum plane, where the single-magnon modes become damped at all energies. This damping becomes even stronger if the values of the third-nearest-neighbor coupling $J_3$ are smaller than the other microscopic constants. In this appendix we present the numerical evaluation of the damping and of the neutron scattering intensity in this regime, for the following values of the microscopic parameters,
\begin{subequations}\label{eq:2nd_point}
	\begin{eqnarray}
		J &=& -8\text{\text{meV}}, \\ 
		K=\Gamma &=& \hphantom{-}7\text{\text{meV}}, \\ 
		J_3 &=& \hphantom{-}0.5\text{\text{meV}}.
	\end{eqnarray}
\end{subequations}

This set of parameters is shown in Fig.~\ref{fig:2nd_point} as a blue dot along the $K=\Gamma$ line.

The magnon damping evaluated numerically in the Born approximation (as described in Sec.~\ref{sec:numerical_Born}) for these parameters  is presented in Fig.~\ref{fig:damping_born_appendix}. The magnon damping evaluated numerically using the iDE approach (as described in Sec.~\ref{sec:iDE}) is presented in Fig.~\ref{fig:damping_appendix}. The neutron scattering intensity evaluated numerically (as described in Section \ref{sec:neutron}) is presented in Fig.~\ref{fig:intensity_appendix}. The amplitude of the damping increases in the middle of the  X-$\Gamma$ and Y-$\Gamma'$ lines as the ratio of $J_3$ to the other scales $K$, $\Gamma$, $J$ decreases. It results in very large broadening of the single magnon peaks in these areas of the momentum plane, while the single-magnon modes remain well-defined in all other areas of the momentum plane.

\end{appendix}

\FloatBarrier


\begin{thebibliography}{99}
	
	\bibitem{Witczak} W. Witczak-Krempa, G. Chen, Y. B. Kim, and L. Balents, {\it Correlated Quantum 
		Phenomena in the Strong Spin-Orbit Regime}, 
	Annu. Rev. Condens. Matter Phys. {\bf 5}, 57 (2014).
	
	\bibitem{Khal_ProgSupp} G. Khaliullin, {\it Orbital Order and Fluctuations in Mott
		Insulators}, Prog. Theor. Phys. Suppl. {\bf 160}, 155 (2005).
	
	\bibitem{JK09} G. Jackeli and G. Khaliullin, {\it Mott Insulators in the Strong Spin-Orbit Coupling Limit: 
		From Heisenberg to a Quantum Compass and Kitaev Models}, Phys. Rev. Lett. {\bf 102}, 017205 (2009).
	
	\bibitem{JK10} J. Chaloupka, G. Jackeli, and G. Khaliullin, {\it Kitaev-Heisenberg Model on a Honeycomb 
		Lattice: Possible Exotic Phases in Iridium Oxides A$_2$IrO$_3$}, Phys. Rev. Lett. {\bf 105}, 027204 (2010).
	
	\bibitem{RauKee_review} J. G. Rau, E. K.-H. Lee, and H.-Y. Kee, {\it Spin-Orbit Physics Giving Rise to 
		Novel Phases in Correlated Systems: Iridates and Related Materials}, Annu. Rev. Condens. Matter Phys. {\bf 7}, 195 (2016).
	
	\bibitem{Winter_review} S. M. Winter, A. A. Tsirlin, M. Daghofer, J. van den Brink, Y. Singh, P. Gegenwart, 
	and R. Valent\`{i}, {\it Models and materials for generalized Kitaev magnetism}, 
	J. Phys.: Condens. Matter {\bf 29}, 493002 (2017).
	
	\bibitem{YBKim_review} R. Schaffer, E. K.-H. Lee, B.-J. Yang, and Y. B. Kim, {\it Recent progress on correlated 
		electron systems with strong spin-orbit coupling}, Rep. Prog. Phys. {\bf 79}, 094504 (2016).
	
	\bibitem{Takagi_review} H. Takagi, T. Takayama, G. Jackeli, G. Khaliullin, and S. E. Nagler,
	{\it Concept and realization of Kitaev quantum spin liquids},
	Nat. Rev. Phys. {\bf 1}, 264 (2019).
	
	\bibitem{Chaloupka} J. Chaloupka and G. Khaliullin, {\it Hidden Symmetries of the Extended 
		Kitaev-Heisenberg Model: Implications for the Honeycomb-Lattice Iridates A$_2$IrO$_3$}, 
	Phys. Rev. B {\bf 92}, 024413 (2015).
	
	\bibitem{Plumb} K. W. Plumb, J. P. Clancy, L. J. Sandilands, V. V. Shankar, Y. F. Hu, K. S. Burch, H.-Y. Kee, and Y.- J. Kim, 
	{\it $\alpha$-RuCl$_3$: A spin-orbit assisted Mott insulator on a honeycomb lattice},
	Phys. Rev. B {\bf 90}, 041112 (2014).
	
	\bibitem{Coldea1} 
	A. Biffin, R. D. Johnson, I. Kimchi, R. Morris, A. Bombardi, J. G. Analytis, A. Vishwanath, and R. Coldea, 
	{\it Noncoplanar and counterrotating incommensurate magnetic order stabilized by Kitaev interactions 
		in $\gamma$-Li$_2$IrO$_3$}, 
	Phys. Rev. Lett. {\bf 113}, 197201 (2014).
	
	\bibitem{Rau14} J. G. Rau, E. K.-H. Lee, H.-Y. Kee, 
	{\it{Generic Spin Model for the Honeycomb Iridates beyond the Kitaev Limit}},
	Phys. Rev. Lett. {\bf{112}}, 077204 (2014).
	
	\bibitem{Kimchi14} I. Kimchi and A. Vishwanath, {\it Kitaev-Heisenberg Models for Iridates on the 
		Triangular, Hyperkagome, Kagome, fcc, and Pyrochlore lattices}, Phys. Rev. B {\bf 89}, 014414 (2014).
	
	\bibitem{Kee} H.-S. Kim, V. V. Shankar, A. Catuneanu, and H.-Y. Kee, 
	{\it Kitaev magnetism in honeycomb RuCl$_3$ with intermediate spin-orbit coupling}, 
	Phys. Rev. B {\bf 91}, 241110 (2015). 
	
	\bibitem{Winter18} 
	S. M. Winter, K.  Riedl, D. Kaib, R. Coldea, and R. Valent\`{i},
	{\it Probing $\alpha$-RuCl$_3$ Beyond Magnetic Order: Effects of Temperature and Magnetic Field},
	Phys. Rev. Lett. {\bf 120}, 077203 (2018).
	
	\bibitem{kaib19}
	A. Sahasrabudhe, D. A. S. Kaib, S. Reschke, R. German, T. C. Koethe, J. Buhot, D. Kamenskyi, C. Hickey, P. Becker, 
	V. Tsurkan, A. Loidl, S. H. Do, K. Y. Choi, M. Gr\"{u}ninger, S. M. Winter, Zhe Wang, R. Valent\`{i}, and P. H. M. van Loosdrecht,
	{\it High-Field Quantum Disordered State in $\alpha$-RuCl$_3$: Spin Flips, Bound States, and a Multi-Particle Continuum},
	 arXiv:1908.11617 (2019).
	
	\bibitem{Perkins14} J. Knolle, Gia-Wei Chern, D. L. Kovrizhin, R. Moessner, and N. B. Perkins,
	{\it Raman Scattering Signatures of Kitaev Spin Liquids in A$_2$IrO$_3$ Iridates},
	Phys. Rev. Lett. {\bf 113}, 187201 (2014).
	
	\bibitem{Chaloupka16}
	J. Chaloupka and G. Khaliullin,
	{\it Magnetic anisotropy in the Kitaev model systems Na$_2$IrO$_3$ and RuCl$_3$},
	Phys. Rev. B {\bf{94}}, 064435 (2016).
	
	
	\bibitem{NP7} I. Rousochatzakis, S. Kourtis, J. Knolle, R. Moessner, and N. B. Perkins,
	{\it Quantum spin liquid at finite temperature: proximate dynamics and persistent typicality},
	Phys. Rev. B {\bf 100}, 045117 (2019).
	
	
	\bibitem{NP19} G. B. Hal\'{a}sz, S. Kourtis, J. Knolle, and N. B. Perkins, 
	{\it Observing spin fractionalization in the Kitaev spin liquid via temperature evolution of indirect resonant inelastic 
		x-ray scattering}, Phys. Rev. B {\bf 99}, 184417 (2019).
	
	\bibitem{Vojta2} 
	P. Lampen-Kelley, L. Janssen, E. C. Andrade, S. Rachel, J.-Q. Yan, C. Balz, D. G. Mandrus, S. E. Nagler, and M. Vojta,
	{\it Field-induced intermediate phase in $\alpha$-RuCl$_3$: Non-coplanar order, phase diagram, and proximate spin liquid}, arXiv:1807.06192 (2018).
	
	\bibitem{Kitaev06} A. Kitaev,  {\it Anyons in an Exactly Solved Model and Beyond}, Ann. Phys. (Amsterdam) {\bf 321}, 2 (2006).
	
	
	\bibitem{Rau14a}  J. G. Rau and H.-Y. Kee, 
	{\it Trigonal distortion in the honeycomb iridates: Proximity of zigzag and spiral phases in Na$_2$IrO$_3$},
	arXiv:1408.4811 (2014).
	
	\bibitem{Rau_tr} A. Catuneanu, J. G. Rau, H.-S. Kim, and H.-Y. Kee,  {\it 
		Magnetic Orders Proximal to the Kitaev Limit in Frustrated Triangular Systems: Application to Ba$_3$IrTi$_2$O$_9$}, 
	Phys. Rev. B {\bf 92}, 165108 (2015).
	
	\bibitem{Ioannis} I. Rousochatzakis, U. K. R\"{o}ssler, J. van den Brink, and M. Daghofer,  {\it 
		Kitaev anisotropy induces mesoscopic $Z_2$ vortex crystals in frustrated hexagonal antiferromagnets}, 
	Phys. Rev. B {\bf 93}, 104417 (2016).
	
	\bibitem{Trebst_tr} M. Becker, M. Hermanns, B. Bauer, M. Garst, and S. Trebst, {\it Spin-Orbit Physics of $j=\frac12$ 
		Mott Insulators on the Triangular Lattice}, Phys. Rev. B {\bf 91}, 155135 (2015).
	
	\bibitem{multiQ} C. Liu, X. Wang, and R. Yu,  {\it Semiclassical Ground-State Phase Diagram and Multi-${\bf Q}$ 
		Phase of a Spin-Orbit- Coupled Model on Triangular Lattice}, Phys. Rev. B {\bf 94}, 174424 (2016).
	
	
	\bibitem{NP5} S. Ducatman, I. Rousochatzakis, and N. B. Perkins, 
	{\it Magnetic structure and excitation spectrum of the hyperhoneycomb Kitaev magnet $\beta$-Li$_2$IrO$_3$},
	Phys. Rev. B {\bf 97}, 125125 (2018).
	
	\bibitem{NP6} M. Li, N. B. Perkins, and I. Rousochatzakis,
	{\it Collective spin dynamics of $Z_2$ vortex crystals in triangular Kitaev-Heisenberg antiferromagnets},
	Phys. Rev. Research {\bf 1}, 013002 (2019).
	
	
	\bibitem{Baskaran}  G. Baskaran, D.  Sen, and R. Shankar,
	{\it Spin-$S$ Kitaev model: Classical ground states, order from disorder,
		and exact correlation functions},
	Phys. Rev. B {\bf 78}, 115116 (2008).
	
	\bibitem{Avella} G. Jackeli and A. Avella, {\it Quantum Order-by-Disorder in Kitaev Model on a Triangular Lattice}, 
	Phys. Rev. B {\bf 92}, 184416 (2015).
	
	\bibitem{NP2}  Y. Sizyuk, P. W\"{o}lfle, and N. B. Perkins,
	{\it Selection of direction of the ordered moments in 
		Na$_2$IrO$_3$ and $\alpha$-RuCl$_3$},
	Phys. Rev. B {\bf 94}, 085109  (2016).
	
	
	\bibitem{Balents_review10} L. Balents, {\it Spin liquids in frustrated magnets}, Nature (London) {\bf 464}, 199 (2010).
	
	\bibitem{Starykh10} O. A. Starykh, H. Katsura, and L. Balents, {\it Extreme sensitivity of a frustrated quantum magnet: Cs$_2$CuCl$_4$}, Phys. Rev. B {\bf 82}, 014421 (2010).
	
	\bibitem{Herfurth13} T. Herfurth, S. Streib, and P. Kopietz, {\it Majorana spin liquid and dimensional reduction in Cs$_2$CuCl$_4$}, Phys. Rev. B {\bf 88}, 174404 (2013).
	
	\bibitem{topography} Z. Zhu, P. A. Maksimov, S. R. White, and A. L. Chernyshev,  {\it 
		Topography of Spin Liquids on a Triangular Lattice}, Phys. Rev. Lett. {\bf 120}, 207203 (2018).
	
	\bibitem{Tutsch19} U. Tutsch, O. Tsyplyatyev, M. Kuhnt, L. Postulka, B. Wolf, P. T. Cong, F. Ritter, C. Krellner, W. A\ss mus, B. Schmidt, P. Thalmeier, P. Kopietz, and M. Lang, {\it Specific Heat Study of 1D and 2D Excitations in the Layered Frustrated Quantum Antiferromagnets Cs$_2$CuCl$_{4-x}$Br$_x$}, Phys. Rev. Lett. {\bf 123}, 147202 (2019).
	
	\bibitem{PRX} P. A. Maksimov, Z. Zhu,  S. R. White, and A. L. Chernyshev,  {\it 
		Anisotropic-exchange magnets on a triangular lattice: spin waves, accidental degeneracies, and dual spin liquids}, 
	Phys. Rev. X {\bf 9}, 021017 (2019).
	
	\bibitem{Chen_nonKitaev} Y.-D. Li, X. Yang, Y. Zhou, and G. Chen,
	{\it  Non-Kitaev spin liquids in Kitaev materials},
	Phys. Rev. B {\bf 99}, 205119 (2019).
	
	
	\bibitem{aRu_DFT}
	S. M. Winter, Y. Li, H. O. Jeschke, and R. Valent\`{i}, 
	{\it Challenges in design of Kitaev materials: Magnetic interactions from competing energy scales}, 
	Phys. Rev. B {\bf 93}, 214431 (2016).
	
	\bibitem{Coldea2} A. Biffin, R. D. Johnson, S. Choi, F. Freund, S. Manni, A. Bombardi, P. Manuel, P. Gegenwart, and R. Coldea, 
	{\it Unconventional magnetic order on the hyperhoneycomb Kitaev lattice in $\beta$-Li$_2$IrO$_3$: 
		Full solution via magnetic resonant x-ray diffraction}, Phys. Rev. B {\bf 90}, 205116 (2014).
	
	\bibitem{Coldea3}
	S. K. Choi, R. Coldea, A. N. Kolmogorov, T. Lancaster, I. I. Mazin, S. J. Blundell, P. G. Radaelli, Yogesh Singh, 
	P. Gegenwart, K. R. Choi, S.-W. Cheong, P. J. Baker, C. Stock, and J. Taylor,
	{\it Spin Waves and Revised Crystal Structure of Honeycomb Iridate 
		Na$_2$IrO$_3$},
	Phys. Rev. Lett. {\bf 108}, 127204 (2012).
	
	\bibitem{Coldea4} I. Kimchi, R. Coldea, and A. Vishwanath, 
	{\it Unified theory of spiral magnetism in the harmonic-honeycomb iridates $\alpha$, $\beta$, 
		and $\gamma$-Li$_2$IrO$_3$}, Phys. Rev. B {\bf 91}, 245134 (2015).
	
	\bibitem{Coldea5} R. D. Johnson, S. C. Williams, A. A. Haghighirad, J. Singleton, V. Zapf, P. Manuel, I. I. Mazin, 
	Y. Li, H. O. Jeschke, R. Valent\`{i}, and R. Coldea, 
	{\it Monoclinic crystal structure of $\alpha$-RuCl$_3$ and the zigzag antiferromagnetic
		ground state}, 
	Phys. Rev. B {\bf 92}, 235119 (2015).
	
	\bibitem{Modic} K. A. Modic, T. E. Smidt, I. Kimchi, N. P. Breznay, A. Biffin, S. Choi, R. D. Johnson, R. Coldea, 
	P. Watkins-Curry, G. T. McCandless, J. Y. Chan, F. Gandara, Z. Islam, A. Vishwanath, A. Shekhter, R. D. McDonald, 
	and J. G. Analytis, {\it Realization of a three-dimensional spin-anisotropic harmonic honeycomb iridate}, 
	Nat. Commun. {\bf 5}, 4203 (2014).
	
	\bibitem{Vojta1} L. Janssen, E. C. Andrade, and M. Vojta, 
	{\it Magnetization processes of zigzag states on the honeycomb lattice: Identifying spin models for $\alpha$-RuCl$_3$ 
		and Na$_2$IrO$_3$},
	Phys. Rev. B {\bf 96}, 064430 (2017)
	
	\bibitem{NP8} J. Knolle, R. Moessner, and N. B. Perkins,
	{\it Bond disordered spin liquid and the honeycomb iridate H$_3$LiIr$_2$O$_6$--abundant 
		low energy density of states from random Majorana hopping},
	Phys. Rev. Lett. {\bf 122}, 047202 (2019).
		%
	
	%
	\bibitem{Mahan90}
	G. D. Mahan, {\it{Many-Particle Physics}}, (2nd Edition, Plenum Press,
	New York, 1990).
	
	\bibitem{Nagler1} A. Banerjee, C. A. Bridges, J.-Q. Yan, A. A. Aczel, L. Li, M. B. Stone, G. E. Granroth, 
	M. D. Lumsden, Y. Yiu, J. Knolle, S. Bhattacharjee, D. L. Kovrizhin, R. Moessner, D. A. Tennant, G. Mandrus, 
	and S. E. Nagler, 
	{\it Proximate Kitaev quantum spin liquid behaviour in a honeycomb magnet},
	Nat. Mater. {\bf 15}, 733 (2016).
	
	\bibitem{Nagler2} A. Banerjee, J. Yan, J. Knolle, C. A. Bridges, M. B. Stone, M. D. Lumsden, D. G. Mandrus, 
	D. A. Tennant, R. Moessner, and S. E. Nagler, 
	{\it Neutron scattering in the proximate quantum spin liquid  $\alpha$-RuCl$_3$},
	Science {\bf 356}, 1055 (2017).
	
	
	\bibitem{K1K2} I. Rousochatzakis, J. Reuther, R. Thomale, S. Rachel, and N. B. Perkins,
	{\it Phase Diagram and Quantum Order by Disorder in the Kitaev $K_1$-$K_2$ Honeycomb Magnet},
	Phys. Rev. X {\bf 5}, 041035 (2015).
	
	\bibitem{numerics1} M. Gohlke, G. Wachtel, Y. Yamaji, F. Pollmann, and Y. B. Kim,
	{\it Quantum spin liquid signatures in Kitaev-like frustrated magnets},
	Phys. Rev. B {\bf 97}, 075126 (2018).
	
	\bibitem{numerics2} J. Knolle, S. Bhattacharjee, and R. Moessner,
	{\it Dynamics of a quantum spin liquid beyond integrability -- the Kitaev-Heisenberg-$\Gamma$ model in an augmented parton mean-field theory}, 
	Phys. Rev. B {\bf 97}, 134432 (2018).
	
	\bibitem{Moessner16} J. Nasu, J. Knolle, D. L. Kovrizhin, Y. Motome, and R. Moessner
	{\it Fermionic response from fractionalization in an insulating two-dimensional magnet},
	Nat. Phys. {\bf 12}, 912 (2016).
	
	
	\bibitem{Winter17}
	S. M. Winter, K. Riedl, P. A. Maksimov,
	A. L. Chernyshev, A. Honecker, and R. Valenti,
	Nat. Commun. {\bf{8}}, 1152 (2017).
	%
	
	
	\bibitem{tri06}
	A. L. Chernyshev and M. E. Zhitomirsky, 
	{\it Magnon Decay in Noncollinear Quantum Antiferromagnets}, 
	Phys. Rev. Lett. {\bf 97}, 207202 (2006). 
	
	\bibitem{tri09}
	A. L. Chernyshev and M. E. Zhitomirsky, 
	{\it Spin-waves in triangular lattice antiferromagnet: decays, spectrum renormalization, and singularities}, 
	Phys. Rev. B {\bf 79}, 144416 (2009). 
	
	\bibitem{tri_Sqw}
	M. Mourigal, W. T. Fuhrman, A. L. Chernyshev, and M. E. Zhitomirsky,
	{\it Dynamical structure factor of triangular-lattice antiferromagnet}, 
	Phys. Rev. B {\bf 87}, 094407 (2013).
	
	\bibitem{JeGeun1} J. Oh, M. D. Le, J. Jeong, J. H. Lee, H. Woo, W.-Y. Song, T. G. Perring, W. J. L. Buyers, S.-W. 
	Cheong, and J.-G. Park, 
	{\it Magnon Breakdown in a Two Dimensional Triangular Lattice Heisenberg Antiferromagnet of Multiferroic 
		LuMnO$_3$},
	Phys. Rev. Lett. {\bf 111}, 257202 (2013).
	
	\bibitem{Moessner_decay} R. Verresen, F. Pollmann, and R. Moessner,
	{\it Strong quantum interactions prevent quasiparticle decay},
	Nat. Phys. {\bf 15}, 750 (2019). 
	
	\bibitem{kagome1}
	A. L. Chernyshev,
	{\it Strong quantum effects in an almost classical antiferromagnet on a kagome lattice}, 
	Phys. Rev. B {\bf 92}, 094409 (2015).
	
	\bibitem{kagome2}
	A. L. Chernyshev and M. E. Zhitomirsky,
	{\it Order and excitations in large-$S$ kagome-lattice antiferromagnets}, 
	Phys. Rev. B {\bf 92}, 144415 (2015). 
	
	
	\bibitem{square99}
	M. E. Zhitomirsky and A. L. Chernyshev, 
	{\it Instability of antiferromagnetic magnons in strong fields}, 
	Phys. Rev. Lett. {\bf 82}, 4536 (1999).
	
	\bibitem{Japanese_5_2} T. Masuda, S. Kitaoka, S. Takamizawa, N. Metoki, K. Kaneko, K. C. Rule, K. Kiefer, 
	H. Manaka, and H. Nojiri, 
	{\it Instability of magnons in two-dimensional antiferromagnets at high magnetic fields},
	Phys. Rev. B {\bf 81}, 100402(R) (2010).
	
	\bibitem{Hong} T. Hong, Y. Qiu, M. Matsumoto, D. A. Tennant, K. Coester, K. P. Schmidt, F. F. Awwadi, 
	M. M. Turnbull, H. Agrawal, A. L. Chernyshev,
	{\it Field-induced spontaneous quasiparticle decay and renormalization of quasiparticle 
		dispersion in a quantum antiferromagnet},
	Nat. Commun. {\bf 8}, 15148 (2017).
	
	\bibitem{square1}
	M. Mourigal, M. E. Zhitomirsky, and A. L. Chernyshev, 
	{\it Field-induced decay dynamics in square-lattice antiferromagnet}, 
	Phys. Rev. B {\bf 82}, 144402 (2010).
	
	\bibitem{square2}
	W. T. Fuhrman, M. Mourigal, M. E. Zhitomirsky, and A. L. Chernyshev,
	{\it Dynamical structure factor of quasi-2D antiferromagnet in high fields}, 
	Phys. Rev. B {\bf 85}, 184405 (2012).
	
	\bibitem{tri_H}
	P. A. Maksimov, M. E. Zhitomirsky, and A. L. Chernyshev,
	{\it Field-induced decays in XXZ triangular-lattice antiferromagnets}, 
	Phys. Rev. B  {\bf 94}, 140407(R) (2016).
	
	\bibitem{hex_H}
	P. A. Maksimov and A. L. Chernyshev,
	{\it Field-induced dynamical properties of the XXZ model on a honeycomb lattice}, 
	Phys. Rev. B {\bf 93}, 014418 (2016).
	
	
	\bibitem{JeGeun2}
	J. Oh, M. D. Le, H.-H. Nahm, H. Sim, J. Jeong, T. G. Perring, H. Woo, K. Nakajima, S. Ohira-Kawamura, 
	Z. Yamani, Y. Yoshida, H. Eisaki, S.-W. Cheong, A. L. Chernyshev, and J.-G. Park,
	{\it Spontaneous decays of magneto-elastic excitations in noncollinear antiferromagnet (Y,Lu)MnO$_3$}, 
	Nat. Commun. {\bf 7}, 13146 (2016).
	
	
	\bibitem{FM_DM}
	A. L. Chernyshev and P. A. Maksimov,
	{\it Damped Topological Magnons in the Kagome-Lattice Ferromagnets}, 
	Phys. Rev. Lett. {\bf 117}, 187203 (2016).
	
	\bibitem{FM_YIG}
	A. L. Chernyshev,
	{\it Field-dependence of magnon decay in yttrium iron garnet thin films}, 
	Phys. Rev. B {\bf 86}, 060401(R) (2012).
	
	
	\bibitem{RMP}
	M. E. Zhitomirsky and A. L. Chernyshev,
	{\it Spontaneous Magnon Decays}, 
	Rev. Mod. Phys. {\bf 85}, 219 (2013).
	
	\bibitem{Zh_SL} M. E. Zhitomirsky,  
	{\it Decay of quasiparticles in quantum spin liquids},
	Phys. Rev. B {\bf 73}, 100404(R) (2006).
	
	\bibitem{Kolezhuk}  A. Kolezhuk and S. Sachdev, 
	{\it Magnon Decay in Gapped Quantum Spin Systems},
	Phys. Rev. Lett. {\bf 96}, 087203 (2006).
	
	\bibitem{Zheludev}  T. Masuda, A. Zheludev, H. Manaka, L.-P. Regnault, J.-H. Chung, and Y. Qiu, 
	{\it Dynamics of Composite Haldane Spin Chains in IPA-CuCl$_3$},
	Phys. Rev. Lett. {\bf 96}, 047210 (2006).
	
	\bibitem{Zaliznyak}  M. B. Stone, I. A. Zaliznyak, T. Hong, C. L. Broholm, and D. H. Reich, 
	{\it Quasiparticle breakdown in a quantum spin liquid},
	Nature (London) {\bf 440}, 187 (2006).
	
	\bibitem{Plumb_ladder}  K. W. Plumb, K.  Hwang, Y. Qiu, L.  W. Harriger, G. E. Granroth, G. J. Shu, 
	F. C. Chou, Ch. Ruegg, Y. B. Kim, and Y.-J. Kim,
	{\it Quasiparticle-continuum level repulsion in a quantum magnet},
	Nat. Phys. {\bf 12}, 224 (2016).
	
	\bibitem{YBKim_decay} 
	K. Hwang and Y. B. Kim, 
	{\it Theory of triplon dynamics in the quantum magnet BiCu$_2$PO$_6$},
	Phys. Rev. B {\bf 93}, 235130 (2016).

\bibitem{McClarty18}  P. A. McClarty, X.-Y. Dong, M. Gohlke, J. G. Rau, F. Pollmann, R. Moessner, and K. Penc,
	{\it Topological Magnons in Kitaev Magnets at High Field},
Phys. Rev. B {\bf 98}, 060404 (2018).


	\bibitem{Hasselmann06} 	N. Hasselmann and P. Kopietz,  
	{\it Spin-wave interactions in quantum antiferromagnets},
Europhys. Lett. {\bf{74}}, 1067 (2006).
	%
	\bibitem{Kreisel07}
	A. Kreisel, N. Hasselmann, and P. Kopietz,  
	{\it Probing Anomalous Longitudinal Fluctuations of the Interacting Bose Gas via Bose-Einstein Condensation of Magnons},
	Phys. Rev. Lett. {\bf{98}}, 067203 
	(2007).
	%
	\bibitem{Kreisel08}
	A. Kreisel, F. Sauli, N. Hasselmann, and P. Kopietz,
	{\it Quantum Heisenberg antiferromagnets in a uniform magnetic field: 
		nonanalytic magnetic field dependence of the magnon spectrum},
	Phys. Rev. B {\bf{78}}, 035127 (2008).
	%
	\bibitem{Kreisel11}
	A. Kreisel, P. Kopietz, P. T. Cong, B. Wolf, and M. Lang,
	{\it Elastic constants and ultrasonic attenuation in the cone state of the frustrated antiferromagnet Cs$_2$CuCl$_4$},
	Phys. Rev. B {\bf{84}}, 024414 (2011).
	%
	\bibitem{Kreisel14}
	A. Kreisel, M. Peter, and P. Kopietz,
	{\it Singular spin-wave theory and scattering continua in the cone state of Cs$_2$CuCl$_4$},
	Phys. Rev. B {\bf{90}}, 075130 (2014).

	
	\bibitem{Chaloupka13}
J. Chaloupka, G. Jackeli, and G. Khaliullin, 	{\it{Zigzag magnetic order in the Iridium Oxide
			Na$_2$IrO$_3$}}, Phys. Rev. Lett. {\bf{110}}, 097204 (2013).
		
	
	\bibitem{Imambekov09a} A. Imambekov and L. I. Glazman, {\it Phenomenology of One-Dimensional Quantum Liquids Beyond the Low-Energy Limit}, Phys. Rev. Lett. {\bf 102}, 126405 (2009).
	
	\bibitem{Imambekov09b} A. Imambekov and L. I. Glazman, {\it Universal Theory of Nonlinear Luttinger Liquids}, Science {\bf 323}, 228 (2009).
	
	\bibitem{Imambekov12} A. Imambekov, T. L. Schmidt, and L. I. Glazman, {\it One-dimensional quantum liquids: Beyond the Luttinger liquid paradigm}, Rev. Mod. Phys. {\bf 84}, 1253 (2012). 
	
	\bibitem{Jin19} Y. Jin, O. Tsyplyatyev, M. Moreno, A. Anthore, W. K. Tan, J. P. Griffiths, I. Farrer, D. A. Ritchie, L. I. Glazman, A. J. Schofield, and C.J.B. Ford, {\it Momentum-dependent power law measured in an interacting quantum wire beyond the Luttinger limit}, Nat. Commun. {\bf 10}, 2821 (2019).
	
	\bibitem{Tsyplyatyev15} O. Tsyplyatyev, A. J. Schofield, Y. Jin, M. Moreno, W. K. Tan, C. J. B. Ford, J. P. Griffiths, I. Farrer, G. A. C. Jones, and D. A. Ritchie, {\it Hierarchy of Modes in an Interacting One-Dimensional System}, Phys. Rev. Lett. {\bf 114}, 196401 (2015); {\it ibid., Nature of the many-body excitations in a quantum wire: theory and experiment}, Phys. Rev. B {\bf 93}, 075147 (2016).
	
	\bibitem{Tsyplyatyev14} O. Tsyplyatyev and A. J. Schofield, {\it Spectral-edge mode in interacting one-dimensional systems}, Phys. Rev. B {\bf 90}, 014309 (2014).
	
	\bibitem{Moreno16} M. Moreno, C. J. B. Ford, Y. Jin, J. P. Griffiths, I. Farrer, G. A. C. Jones, D. A. Ritchie, O. Tsyplyatyev, and A. J. Schofield, {\it Nonlinear spectra of spinons and holons in short GaAs quantum wires}, Nat. Commun. {\bf 7}, 12784 (2016).
	

	
	\bibitem{future}  P. A. Maksimov, P. Kopietz, and A. L. Chernyshev, to be published.  
	
	\bibitem{Holstein40}
	T. Holstein and H. Primakoff, 
	{\it Field Dependence of the Intrinsic Domain Magnetization of a Ferromagnet},
	Phys. Rev. {\bf{58}}, 1098 (1940).
	%
	\bibitem{Colpa78}
	J. H. P. Colpa, 
	{\it Diagonalization of the quadratic boson Hamiltonian},
	Physica A {\bf{93}}, 327 (1978).
	%
	
	\bibitem{Blaizot86}
	J. P. Blaizot and G. Ripka, {\it{Quantum theory of finite systems}}, (MIT Press, Cambridge, Massachusetts, 1986).
	%
	\bibitem{Maldonado93}
	O. Maldonado, 
	{\it On the Bogoliubov transformation for quadratic boson observables},
	J. Math. Phys. {\bf{34}}, 5016 (1993).
	%
	%
	\bibitem{Goldstein80}
	See, for example, H. Goldstein, J. Safko, and C. Poole, {\it{Classical Mechanics}}, 
	(Pearson Education Limited, Harlow (UK),  Pearson New International Edition, 2014).
	%
	\bibitem{Serga12}
	A. A. Serga, C. W. Sandweg, V. I. Vasyuchka, M. B. Jungfleisch, B. Hillebrands, A. Kreisel, P. Kopietz, and M. P. Kostylev,
	{\it Brillouin light scattering spectroscopy of parametrically excited dipole-exchange magnons},
	Phys. Rev. B {\bf{86}}, 134403 (2012).
	%
	
	
	
	
	
	\bibitem{Schuetz03}
	F. Sch\"{u}tz, M. Kollar, and P. Kopietz,
	{\it{Persistent spin currents in mesoscopic Heisenberg rings}},
	Phys. Rev. Lett. {\bf{91}}, 017205 (2003).
	%
	\bibitem{Spremo05}
	I. Spremo, F. Sch\"{u}tz, P. Kopietz, V. Pashchenko, B. Wolf, M. Lang,
	J. W. Bats, C. Hu, M. U. Schmidt,
	{\it{Magnetic properties of a metal-organic antiferromagnet on a distorted honeycomb lattice}},
	Phys. Rev. B {\bf{72}} , 174429 (2005).
	%
	\bibitem{Singh10}
	Y. Singh and P. Gegenwart, 
	{\it{Antiferromagnetic Mott insulating state in single crystals of the honeycomb lattice material
			Na$_2$IrO$_3$}}
	Phys. Rev. B {\bf{82}}, 064412 (2010).
	%

	%
	%

	%
	%
	%
\end{thebibliography}
\end{document}